\documentclass[fleqn,usenatbib]{mnras}

\usepackage{newtxtext,newtxmath}
\usepackage[T1]{fontenc}
\usepackage{booktabs}
\usepackage{array}

\DeclareRobustCommand{\VAN}[3]{#2}
\let\VANthebibliography\thebibliography
\def\thebibliography{\DeclareRobustCommand{\VAN}[3]{##3}\VANthebibliography}

\usepackage{todonotes}
\usepackage{graphicx}
\usepackage{subcaption}
\usepackage{amsmath}
\usepackage{xspace}	
\usepackage{afterpage}
\usepackage{longtable}
\usepackage{caption}
\usepackage{textcomp}
\usepackage[normalem]{ulem}
\usepackage{gensymb}
\usepackage{footnote}
\usepackage{float}
\usepackage{fix-cm}
\usepackage{threeparttable}
\usepackage{xcolor}

\newcommand{\grbssample}{GRBs~240122A, 240225B, 240619A, 240910A, 240916A, 241002B, and 241228B}

\newcommand{\fermi}{{\em Fermi}\xspace}
\newcommand{\maxi}{{\em MAXI}\xspace}

\newcommand{\fermiT}{T$_{0}$}
\newcommand{\maxiT}{T$_{0}$}
\newcommand{\keV}{{\rm keV}\xspace}
\newcommand{\meV}{{\rm MeV}\xspace}

\newcommand{\swift}{{\em Swift}\xspace}
\newcommand{\tninty}{{$T_{\rm 90}$}\xspace}

\title[GOTO-Discovered Afterglows of Seven LGRBs]{Discovery and Analysis of Afterglows from Poorly Localised GRBs with the Gravitational-wave Optical Transient Observer (GOTO) All-sky Survey}

\author[Kumar A., et al., 2025]{\href{https://orcid.org/0000-0002-4870-9436}{Amit~Kumar}$^{1,2}$\thanks{Contact: \href{mailto:amit.kumar@rhul.ac.uk}{amit.kumar@rhul.ac.uk}; \href{mailto:amitkundu515@gmail.com}{amitkundu515@gmail.com}},
\href{https://orcid.org/0000-0002-5826-0548}{B. P. Gompertz}$^{3,4}$,
\href{https://orcid.org/0000-0003-4876-7756}{B. Schneider}$^{5}$,
\href{https://orcid.org/0000-0002-9798-029X}{S. Belkin}$^{6}$,
\href{https://orcid.org/0009-0009-8473-3407}{M. E. Wortley}$^{3,4}$,
\href{https://orcid.org/0000-0002-6950-4587}{A. Saccardi}$^{7}$,
\href{https://orcid.org/0009-0001-1554-1868}{D. O'Neill}$^{3,4}$,
\newauthor
\href{https://orcid.org/0000-0002-8648-0767}{K. Ackley}$^{2}$,
\href{https://orcid.org/0009-0000-2285-8188}{B. Rayson}$^{8}$,
\href{https://orcid.org/0000-0001-7717-5085}{A. de Ugarte Postigo}$^{5}$,
\href{https://orcid.org/0000-0002-0786-7307}{A. Gulati}$^{9,10,11}$,
\href{https://orcid.org/0000-0003-0771-4746}{D. Steeghs}$^{2}$,
\href{https://orcid.org/0000-0002-7517-326X}{D. B. Malesani}$^{12,13,14}$,
\href{https://orcid.org/0000-0003-0733-7215}{J. R. Maund}$^{1}$, 
\newauthor
\href{https://orcid.org/0000-0003-3665-5482}{M. J. Dyer}$^{15,16}$,
\href{https://orcid.org/0000-0002-2815-7291}{S. Giarratana}$^{17}$,
M. Serino$^{18}$,  
\href{https://orcid.org/0000-0002-0774-2328}{Y.~Julakanti}$^{8}$,
\href{https://orcid.org/0000-0001-7225-2475}{B. Kumar}$^{19,20}$,
\href{https://orcid.org/0000-0003-3257-9435}{D. Xu}$^{21}$,
\href{https://orcid.org/0000-0002-8775-2365}{R. A. J. Eyles-Ferris}$^{8}$, 
\newauthor
\href{https://orcid.org/0000-0002-9022-1928}{Z.-P. Zhu}$^{21}$,
\href{https://orcid.org/0009-0005-8379-3871}{B. Warwick}$^{2}$,
\href{https://orcid.org/0000-0002-7400-4608}{Y.-D. Hu}$^{22}$,
I. Allen$^{3}$,
\href{https://orcid.org/0000-0001-8722-9710}{G. Ramsay}$^{23}$,
\href{https://orcid.org/0000-0001-5803-2038}{R. L. C. Starling}$^{8}$,
\href{http://orcid.org/0000-0002-3464-0642}{J. Lyman}$^{2}$, 
K. Ulaczyk$^2$, 
\newauthor
\href{https://orcid.org/0000-0003-3766-7266}{B. Godson}$^{2}$,
\href{https://orcid.org/0000-0002-6558-5121}{D. K. Galloway}$^{6, 24}$,
\href{https://orcid.org/0000-0003-4236-9642}{V. S. Dhillon}$^{15,25}$,  
\href{https://orcid.org/0000-0002-5128-1899}{P. O'Brien}$^{8}$, 
\href{https://orcid.org/0000-0001-9109-8311}{K. Noysena}$^{26}$, 
R. Kotak$^{27}$, 
R. P. Breton$^{28}$, 
\newauthor
\href{https://orcid.org/0000-0002-8599-8791}{L. K. Nuttall}$^{29}$, 
D. Pollacco$^{2}$, 
J. Casares$^{25,30}$, 
\href{https://orcid.org/0000-0001-5803-2038}{T.~L. Killestein}$^{2}$,
\href{https://orcid.org/0000-0001-6894-6044}{M. R. Kennedy}$^{31}$,
\href{https://orcid.org/0000-0003-0944-7194}{N. Habeeb}$^{8}$,
\newauthor
\href{https://orcid.org/0000-0001-5221-0243}{S. Moran}$^{8}$,
\href{https://orcid.org/0000-0002-9133-7957}{K. Wiersema}$^{32}$,
\href{https://orcid.org/0009-0007-3476-2272}{I. Worssam}$^{3,4}$,
\href{https://orcid.org/0000-0001-5126-6237}{D. L. Coppejans}$^{2}$,
\href{https://orcid.org/0000-0001-8905-4299}{C. A. Phillips}$^{2}$,
\href{https://orcid.org/0000-0001-5108-0627}{A. Martin-Carrillo}$^{33}$,
\newauthor
\href{https://orcid.org/0000-0003-0759-1141}{N. S. Pankov}$^{34,35}$,
\href{https://orcid.org/0000-0001-6991-7616}{J. F. Ag\"u\'i Fern\'andez}$^{36}$,
\href{https://orcid.org/0000-0002-5552-7681}{M. A. Aloy}$^{37,38}$,
\href{https://orcid.org/0009-0000-5068-3434}{J. An}$^{21}$,
\href{https://orcid.org/0000-0001-6544-8007}{G. E. Anderson}$^{39,9}$,
\href{https://orcid.org/0009-0008-2714-2507}{A. Bochenek}$^{40}$,
\newauthor
\href{https://orcid.org/0000-0003-2999-3563}{A. J. Castro-Tirado}$^{41,42}$,
\href{https://orcid.org/0009-0000-4068-1320}{X. Chen}$^{19,20}$,
\href{https://orcid.org/0000-0002-7910-6646}{L. Cotter}$^{33}$,
\href{https://orcid.org/0000-0001-6191-7160}{R. Dastidar}$^{43}$,
\href{https://orcid.org/0000-0002-4036-7419}{M. De Pasquale}$^{44}$,
\href{https://orcid.org/0000-0003-3703-4418}{V. D'Elia}$^{45}$,
\href{https://orcid.org/0009-0006-1010-1325}{Y. Fang}$^{19,20}$,
\newauthor
\href{https://orcid.org/0009-0002-7730-3985}{S. Y. Fu}$^{46}$,
\href{https://orcid.org/0000-0002-8149-8298}{J. P. U. Fynbo}$^{12,13}$, 
\href{https://orcid.org/0000-0002-8028-0991}{D. H. Hartmann}$^{47}$,
L. B. He$^{21}$,
\href{https://orcid.org/0000-0001-9695-8472}{L. Izzo}$^{48,13}$,
\href{https://orcid.org/0009-0001-8155-7905}{S. Q. Jiang}$^{21}$,
\href{https://orcid.org/0000-0002-2064-3164}{Y. Kawakubo}$^{18}$,
\newauthor
E. V. Klunko$^{49}$,
\href{https://orcid.org/0000-0001-7821-9369}{A. J. Levan}$^{14,2}$,
\href{https://orcid.org/0000-0003-1295-2909}{X.-W. Liu}$^{19,20}$,
\href{https://orcid.org/0000-0002-4072-6899}{X. Liu}$^{21}$,
\href{https://orcid.org/0000-0003-3412-0556}{G. Lombardi}$^{50,25}$,
\href{https://orcid.org/0000-0003-2593-4355}{E. Maiorano}$^{51}$,
\href{https://orcid.org/0000-0002-9408-1563}{J. T. Palmerio}$^{52}$,
\newauthor
\href{https://orcid.org/0000-0001-8472-1996}{D. A. Perley}$^{40}$,
\href{https://orcid.org/0000-0003-3114-2733}{D. L. A. Pieterse}$^{14}$,
\href{https://orcid.org/0000-0001-9435-1327}{A. S. Pozanenko}$^{34,35,53}$,
\href{https://orcid.org/0000-0003-3457-9375}{G. Pugliese}$^{54}$,
\href{https://orcid.org/0000-0002-8860-6538}{A. Rossi}$^{51}$,
\href{https://orcid.org/0000-0001-6620-8347}{B. Sbarufatti}$^{17}$,
\newauthor
S. Bijavara Seshashayana$^{55,56}$,
\href{https://orcid.org/0000-0003-3274-6336}{N.~R.~Tanvir}$^{8}$,
\href{https://orcid.org/0000-0002-7978-7648}{C. C. Th\"one}$^{57}$,
\href{https://orcid.org/0000-0001-9149-6707}{A. J. van der Horst}$^{58}$,
\href{https://orcid.org/0000-0001-9398-4907}{S. D. Vergani}$^{59}$,
\newauthor
A. A. Volnova$^{35}$,
\href{https://orcid.org/0000-0002-3101-1808}{R. A. M. J. Wijers}$^{54}$, and
\href{https://orcid.org/0000-0003-0733-2916}{J. L. Wise}$^{40}$
\\
\textit{Affiliations are listed at the end of the paper}
}

\date{Accepted 2025 September 26. in original form 2025 September 11}

\pubyear{\the\year{}}

\begin{document}
\label{firstpage}
\pagerange{\pageref{firstpage}--\pageref{lastpage}}
\maketitle

\begin{abstract}
Gamma-ray bursts (GRBs), particularly those detected by wide-field instruments such as the \fermi/GBM, pose challenges for optical follow-up because of their large initial localisation regions, leaving many GRBs without identified afterglows. The Gravitational-wave Optical Transient Observer (GOTO), with its wide field of view, dual-site coverage, and robotic rapid-response capability, bridges this gap by rapidly identifying and localising afterglows from alerts issued by space-based facilities including \fermi, \textit{SVOM}, \swift, and the \textit{EP}, providing early optical positions for coordinated multi-wavelength follow-up. In this paper, we present optical afterglow localisation and multi-band follow-up of five \fermi/GBM (240619A, 240910A, 240916A, 241002B, and 241228B) and two \maxi/GSC (240122A and 240225B) triggered long GRBs discovered by GOTO in 2024. Spectroscopy for six GRBs (no spectroscopy for GRB~241002B) with VLT/X-shooter and GTC/OSIRIS yields precise redshifts spanning $z\approx0.40$--3.16 and absorption-line diagnostics of hosts and intervening systems. Radio detections for four events (240122A, 240619A, 240910A, and 240916A) confirm the presence of long-lived synchrotron emission. Prompt-emission analysis with \fermi~and \maxi~data reveals a spectrally hard population, with two bursts lying $>3\sigma$ above the Amati relation. Although their optical afterglows resemble those of typical long GRBs, the prompt spectra are consistently harder than the long-GRB average. Broadband afterglow modelling of six GOTO-discovered GRBs yields jet half-opening angles of a few degrees and beaming-corrected kinetic energies \(E_{\rm jet}\sim10^{51}\)--\(10^{52}\)\,erg, consistent with the canonical long-GRB population. These findings suggest that optical discovery of poorly localised GRBs is likely subject to observational biases favouring luminous events with high spectral peak energy ($E_{\rm p}$), while also providing insight into jet microphysics and central engine diversity.
\end{abstract}


\section{Introduction}

The study of gamma-ray bursts (GRBs) has advanced significantly since their discovery in the 1960s \citep{Klebesadel1973, Strong1974}, driven by a combination of dedicated space-based surveys and ground-based follow-up observations. From confirming their cosmological origins \citep{Meegan1992, Costa1997, Metzger1997, vanParadijs1997} to uncovering possible progenitors for long and short GRBs \citep{Mazets1981, Kouveliotou1993, woosley_supernova_2006, Zhang2012a, Abbott2017b}, these high-energy events have now been recognised as among the most luminous explosive phenomena in the universe \citep{Meszaros2013, Kumar2015, Levan2016, LHAASO2023}. Observations across the electromagnetic spectrum have not only enhanced our understanding of GRB mechanisms but also established their ability to probe the distant universe \citep{Fiore2001, Tanvir2009, Petitjean2011, Saccardi2023, Saccardi2025} and constrain cosmological parameters \citep{Amati2013, Demianski2017, Luongo2021, Moresco2022}, marking them as invaluable tools in modern astrophysics.

A fundamental classification distinguishes GRBs into long-duration (\tninty\footnote{\tninty{} marks the period during which the central 90\% of a GRB's total detected emission is observed, from 5\% to 95\% cumulative count levels.} $\gtrsim 2$\,s) and short-duration (\tninty{} $\lesssim 2$\,s) bursts \citep{Kouveliotou1993}. Long GRBs (LGRBs) are typically associated with the collapse of massive, rapidly rotating stars \citep{Woosley1993, Zhang2004, Maeder2012, Zhang2019book, Aloy2021, Obergaulinger2022}, occasionally accompanied by broad-lined Type Ic supernovae \citep{Galama1998, woosley_supernova_2006, Cano2017, Kumar2024mnras}. In contrast, short GRBs (SGRBs) generally are thought to originate from compact binary mergers involving neutron stars and/or black holes \citep{Eichler1989, Narayan1992, Tanaka2016}, and are associated with kilonovae, a connection confirmed through GW170817/GRB~170817A/AT~2017gfo \citep{Abbott2017, Abbott2017b, Goldstein2017, Pian2017, Troja2017, Valenti2017, Wang2017d,Tanvir2017}, see also \cite{Metzger2019}. Although the long-short dichotomy holds in general, recent observations reveal notable exceptions, such as LGRBs~211211A and 230307A that exhibited signatures consistent with kilonova emission and compact object merger progenitors \citep{Rastinejad2022, Troja2022, Yang2022, Gompertz2023, Dai2024evidance, Levan2024, Sun2025}. Conversely, SGRB~200826A showed a possible association with a supernova, suggesting a massive star origin \citep{Ahumada2021, Zhang2021, Rossi2022}. These atypical cases challenge the traditional progenitor classification and motivate further systematic investigations into the diversity of GRB origins. However, in this work we adopt the conventional $T_{90} \lesssim 2$\,s and $T_{90} \gtrsim 2$\,s division as a working classification. While exceptions to this dichotomy are known \citep{Zhang2009, Bromberg2013, Kulkarni2017}, the 2\,s threshold remains the standard convention for comparability, and our sample lies comfortably above this boundary, where the risk of misclassification is lower.

The origin of GRBs and the understanding of the underlying physics can be probed using multi-wavelength afterglow observations (see \citealt{Miceli2022}). Synchrotron emission from relativistic jets that interact with the circumburst medium encodes information about the jet geometry, ambient density, and microphysical parameters \citep{Sari1998, Granot2002, Panaitescu2002}, see also \citet{ZhangLu2024}. Afterglow light curve features such as jet breaks, cooling breaks, and chromatic evolution offer insights into jet collimation and energy structure \citep{Rhoads1999, Sari1999}. Theoretical models provide further context: relativistic jet propagation in collapsars was first explored via simulations \citep{Aloy2000}, and more recent 3D magnetorotational core-collapse models further demonstrate jet collimation and dynamics in magnetised environments \citep{Obergaulinger2021}. Complementary hydrodynamical studies examine jet--cocoon mixing and structured outflow morphologies \citep{Gottlieb2020}. 

Early optical and multi-wavelength follow-up has revealed a broad diversity in afterglow behaviours, including evidence for reverse shocks \citep{Zhang2005, Mundell2007, Mimica2009, Mimica2010, Laskar2013, Yi2020}, energy injection episodes \citep{Bjornsson2004, Zhang2006, Laskar2015}, and structured jets \citep{Lamb2017, Beniamini2020, Oganesyan2020}. Combined with X-ray and radio data, optical observations enable comprehensive modelling of afterglows, shedding light on the energetics and structure of GRB jets \citep{Margutti2013}. However, a persistent challenge in GRB afterglow detection and follow-up arises from the poor initial localisation provided by wide-field gamma-ray monitors such as \fermi~Gamma-ray Burst Monitor (GBM; \citealt{Meegan2009}). With the highest GRB detection rate and strong sensitivity to prompt gamma-ray emission, \fermi/GBM enables detailed temporal and spectral studies \citep{Meegan2009, vonKienlin2020}, but typically provides localisation uncertainties spanning several square degrees. In contrast, missions like Neil Gehrels Swift Observatory (\swift~hereafter; \citealt{Gehrels2004}), Einstein Probe (\textit{EP}; \citealt{Yuan2015,Yuan2022}), and Space-based multi-band astronomical Variable Objects Monitor (\textit{SVOM}; \citealt{Wei2016}) offer arcsecond- to arcminute-level localisations but detect comparatively fewer bursts. Recovering counterparts to these poorly localised GRBs provides a valuable opportunity to expand our understanding of GRB diversity and reduce selection biases. However, the large error regions often exceed the field of view (FoV) of conventional optical telescopes, complicating timely afterglow identification. Furthermore, the intrinsic faintness of some afterglows \citep{Liang2007, Dereli2017} and circumburst extinction \citep{Savaglio2004, Schulze2011} can further hinder follow-up. Without alternative localisation strategies, a significant fraction of GRBs, particularly those detected by \fermi/GBM, remain uncharacterised, limiting our ability to probe jet physics, energetics, and progenitor properties.

Wide-field optical instruments, such as the Gravitational-wave Optical Transient Observer (GOTO\footnote{\url{https://goto-observatory.org/}}; \citealt{Dyer2020SPIE, Steeghs2022}), have emerged as powerful tools to address this gap. GOTO, comprising 32 robotic telescopes across two sites, Roque de los Muchachos Observatory (La Palma, Canary Islands) and Siding Spring Observatory (New South Wales, Australia), enables near-continuous coverage of both hemispheres \citep{Dyer2024SPIE}. Unlike traditional follow-up facilities, the GOTO instruments can ``tile'' the large error regions of GBM-like triggers in near real-time, providing a complementary discovery channel to narrow-field missions and helping to overcome localisation-driven selection effects. Its fast-response capabilities and wide FoV make it well-suited to bridging the gap between gamma-ray detection and multi-wavelength characterisation, particularly for poorly localised but scientifically valuable GRBs.

In this work, we present a systematic study of seven poorly localised L\grbssample{}, whose optical afterglows were discovered by GOTO in response to alerts from \fermi/GBM and the Monitor of All-sky X-ray Image (\maxi)~Gas Slit Camera (GSC; \citealt{Matsuoka2009, Mihara2011PASJ}), with localisation uncertainties ranging from a few arcminutes to several square degrees in radius. The structure of this paper is as follows. Section~\ref{sec:goto_stra} provides an overview of GOTO, outlining its observational strategy and summarising its past observation records. Section~\ref{sec:sample} introduces the sample and describes the discovery of their afterglows in terms of localisation coverage and optical afterglow detections. Section~\ref{sec:afterglow_follow_ups} presents the multi-wavelength (X-ray to radio) follow-up observations of the afterglows. Section~\ref{sec:prompt_prop} details the prompt gamma-ray analyses and properties of the GRBs in our sample, along with comparisons to other GRBs. Section~\ref{sec:analysis_afterglow} discusses the multi-wavelength afterglow properties and compares them with literature data for other GRBs, including optical spectroscopic analyses of the afterglows and precise redshift estimates. Section~\ref{sec:afterglow_modelling} presents the afterglow modelling of six GRBs from our sample using \texttt{afterglowpy} and Bayesian inference with \texttt{dynesty} nested sampling. Finally, Section~\ref{sec:summary_conclusion} summarises our findings and presents the main conclusions of this study. 

All magnitudes reported in this work are given in the AB photometric system. We define $T_{0}$ as the trigger time reported by the detecting satellite, which serves as the reference epoch for our temporal analysis. For GRBs~240122A and 240225B this corresponds to the \maxi/GSC trigger, while for the remaining five events it corresponds to the \fermi/GBM trigger.

\section{GOTO and its Approach to GRB Counterpart Searches}\label{sec:goto_stra}

\subsection{GOTO Overview}

\begin{figure}
    \centering
    \includegraphics[width=\linewidth]{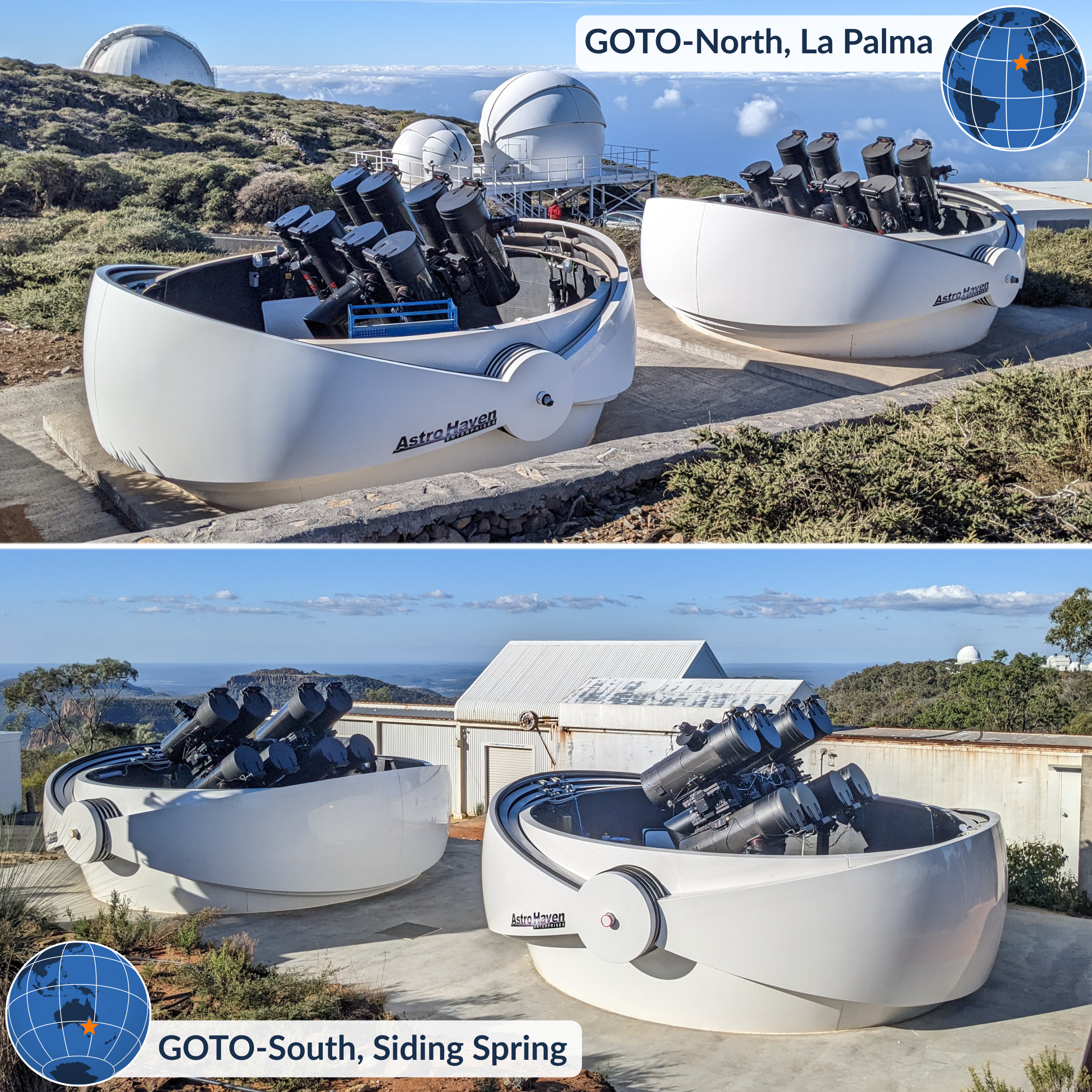}
    \caption{The full configuration of the GOTO telescope network in April 2023, comprising 32 robotic unit telescopes distributed across four domes, two domes at each of the two sites. Top: GOTO-N, located at the Observatorio del Roque de los Muchachos on La Palma, comprising GOTO-1 (left) and GOTO-2 (right). Bottom: GOTO-S, hosted at Siding Spring Observatory in Australia, consisting of GOTO-3 (left) and GOTO-4 (right). Figure credit: \citealt{Dyer2024SPIE}.}
    \label{fig:goto_north_south}
\end{figure}

The Gravitational-wave Optical Transient Observer \citep[GOTO;][]{Dyer2020SPIE, Steeghs2022, Dyer2024SPIE} is a global network of 32 robotic unit telescopes (UTs) distributed over two sites, with two domes at each site: the Roque de los Muchachos Observatory on La Palma, Canary Islands, and the Siding Spring Observatory in New South Wales, Australia (see Figure~\ref{fig:goto_north_south}). This configuration enables near-continuous monitoring of both the northern and southern skies.

Each site hosts two independent mounts, with the eight UTs on each mount aligned to form a tiled array with small overlaps, yielding a combined FoV of $\approx44\,\mathrm{deg}^2$ per mount. With two mounts per site, this provides $\approx88\,\mathrm{deg}^2$ of instantaneous coverage, and across both sites the network spans $\approx176\,\mathrm{deg}^2$. This wide coverage makes GOTO particularly well-suited to search for optical counterparts of poorly localised transients such as gravitational-wave events \citep[e.g.,][]{Gompertz2020}, GRBs detected by facilities such as \fermi, \textit{SVOM} and \textit{EP} \citep[e.g.,][]{Mong2021, Belkin2024RNAAS}, as well as other fast and exotic transients including rapidly-evolving supernovae and tidal disruption events. The telescopes are equipped with ON Semiconductor KAF-50100 CCDs, which provide broad sensitivity across the optical range, with the deployed Baader filters setting the effective bandpass. In survey mode, a wide $L-$band ($400-700$\,nm) encompassing the Sloan \textit{gri} filters is used, providing sensitivity to a wide range of transients and maximising discovery potential.

In ``responsive'' mode, the GOTO instruments autonomously observe large sky regions associated with poorly localised transient events, such as GW events, GRBs, and high-energy neutrino alerts, to search for their optical counterparts. The exposure time and cadence in this mode are adapted to the nature of the event and vary accordingly across different source types. This study focuses specifically on the discovery of optical afterglows from poorly localised GRBs using GOTO. The following section outlines the observational strategies employed by GOTO to identify and confirm these afterglows.

\subsection{Follow-up Strategies to Discover GRBs' Optical Afterglows}\label{sec:GOTO_follow-up_strategies}

\begin{figure}
    \centering
    \includegraphics[width=1.05\linewidth]{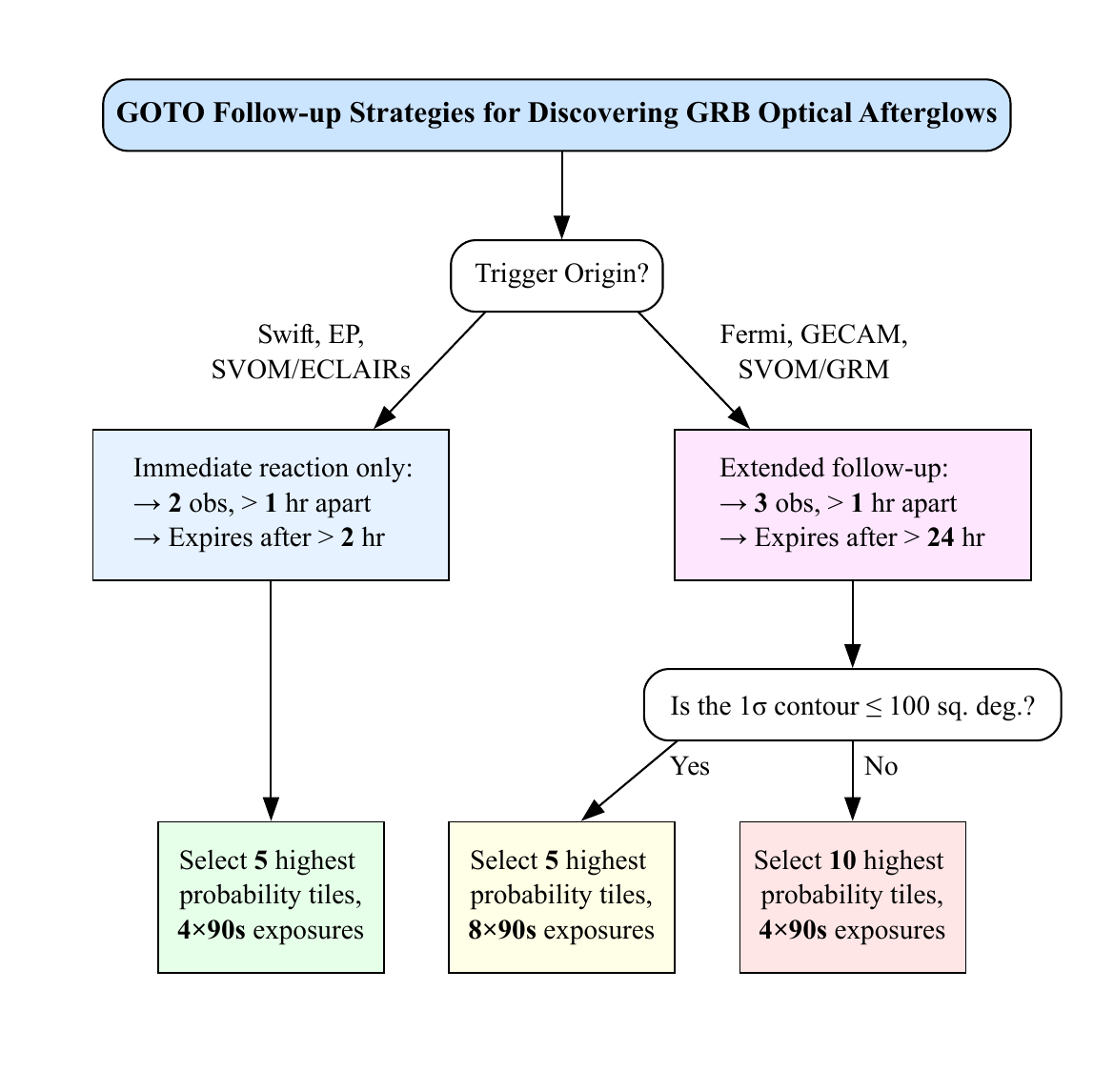}
    \caption{A summary of the GOTO GRB follow-up strategy. 
    }
    \label{fig:grb_flowchart}
\end{figure}

In responsive mode, if triggered by a GRB alert, GOTO pauses its survey operations to target the localisation region. Follow-up strategies are tailored based on the source of the trigger and the localisation uncertainty, as described below and illustrated in the flowchart in Figure~\ref{fig:grb_flowchart}. 

{\it Swift/BAT, SVOM/ECLAIRs, and EP triggers:} \swift, \textit{SVOM}/ECLAIRs, and \textit{EP} GRB detections generally come with precise localisation (arcsecs to arcmins), far smaller than a single GOTO tile, and which can be easily covered by other observatories with a lower FoV. Therefore, the primary reason for GOTO to follow up these events is to take advantage of its fast, robotic nature to get rapid coverage of the localisation region immediately following the trigger, in order to capture any optical afterglow while it is still bright and young. For these triggers, up to five tiles on the GOTO survey grid are permitted to be selected; however, given the well-localised nature of these sources, the search region is almost always within a single tile (more than one tile is allowed to be selected for rare cases where a source falls within the overlapping region on the edge of multiple tiles). Two observations are scheduled for each tile, spaced one hour apart, each using the standard set of $4\times90$\,s exposures, which typically reaches a depth of $\sim19.8$~AB~mag in GOTO $L-$band. However, these targets are only valid in the GOTO scheduling queue for the first two hours after the trigger time. This ensures that rapid observations will be taken if any of the GOTO telescopes are available immediately after the trigger; however, after the two-hour window, GOTO's usefulness is lessened, and any observations are left to other observatories.

{\it Fermi/GBM, SVOM/GRM, and GECAM triggers:} Detections from \fermi/GBM, \textit{SVOM}~Gamma Ray burst Monitor (GRM; \citealt{He2025}) and Gravitational Wave Electromagnetic Counterpart All-sky Monitor (\textit{GECAM}; \citealt{Chen2020}) often have poorer localisation areas (radial uncertainties of a few to tens of degrees), and therefore GOTO's wide FoV is well suited to locating any afterglow. As such, observations for these follow-up campaigns are valid for a 24-hour period from the GRB trigger time, with three epochs scheduled spaced at least one hour apart, to ensure the best chance of discovering and observing the evolution of the optical afterglow during its peak brightness phase. As these localisation regions can stretch to cover large areas of sky, a limit is imposed to target only the 10 highest tiles sorted by the contained localisation probability. This limit was picked based on a recovery rate of $75.5\%$ when applied to 102 historical \fermi/GBM triggers with corresponding \swift/X-Ray Telescope (XRT; \citealt{Burrows2005}) detections. The selected tiles are then each scheduled for three observations using the standard 4$\times$90\,s exposure set. However, in 2024 an improved strategy was developed: for well-localised events --- where the $1\sigma$ localisation region covers less than 100 square degrees --- a more focused strategy was created which selects a maximum of five grid tiles for observations but with a double set of $8\times90$\,s exposures, reaching a $5\sigma$ depth of $L \sim 20.5$~mag. From simulating 50,000 artificial GRB afterglows of varied localisation regions, the likelihood of detection was maintained when trading spatial coverage for increased depth and, in certain cases, can result in an approximate $10\%$ increase in the afterglow detection chance. This new focused strategy, along with the 10 tile limit for larger areas, was introduced in July 2024, prior to which all triggers selected only the five highest tiles for $4\times90$\,s exposures. Of the five \fermi-triggered campaigns described in Section~\ref{sec:sample}, only GRB~240619A used the old selection criteria, meaning it was limited to only five triggered pointings. However, all five events have localisation regions of larger than 100 square degrees, so none would have used the deeper $8\times90$\,s sets.

\subsection{Identifying the GRB Optical Counterpart}\label{sec:id_opt_ctpt}

For each GOTO observing sequence, whether taken in responsive or survey mode, images are processed in near real-time by the GOTO pipeline (Lyman et al., in prep.), which includes calibration, astrometric solution, and difference imaging against archival deep GOTO templates. Transient candidates are then ranked by a machine-learning classifier \citep{Killestein2021}, cross-matched to contextual catalogues (e.g., SDSS~\citealt{York2000}; Pan-STARRS~\citealt{Chambers2016}), and Solar System ephemerides, and subsequently passed through automated real/bogus and contextual filters. 

The Burst Advocate (BA) monitors GRB alerts, confirms that GOTO follow-up has been executed, and initiates candidate vetting in the GOTO marshall. Promising sources are inspected by the working group, including checks against archival imaging and forced photometry, before being promoted for group review. A candidate is classified as the optical afterglow counterpart if it satisfies the following:

\begin{enumerate}
    \item Spatial consistency: positionally coincident with the high-probability GRB localisation region, with a point-like PSF and no association with known artefacts or moving objects (minor-planet checks performed). If present, the location relative to a plausible host galaxy is also considered.
    \item Temporal behaviour: evidence of fading between successive GOTO epochs, or a later non-detection deeper than the initial detection; where possible, the decline should be consistent with a power-law afterglow behaviour.
    \item Contextual screening: absence of a persistent source in archival templates; no counterpart in variable-star catalogues, and not coincident with a known AGN nucleus.
    \item Multi-wavelength corroboration: spatial consistency with a \swift/XRT source strengthens the association, but is not required.
\end{enumerate}

Candidates satisfying criteria (i)–(iii) are promoted to the transient stream and considered GRB afterglow counterparts, while those also fulfilling (iv) are prioritised for rapid spectroscopy and additional ToO follow-up. Confirmed counterparts are reported in General Coordinates Network (GCN\footnote{\url{https://gcn.nasa.gov/circulars}}) Circulars and logged in the Transient Name Server (TNS\footnote{\url{https://www.wis-tns.org/}}).

\subsection{Observational Performance and Discovery Statistics}\label{sec:Discovery_Statistics}

The GOTO GRB follow-up programme has developed from its prototype stage to a fully operational dual-site facility \citep{Dyer2024SPIE}, delivering a series of notable discoveries. The prototype system (GOTO-4), comprising four UTs at the La Palma site, achieved first light in June 2017 and was officially inaugurated in July 2017, initiating routine operations \citep{Steeghs2022}. During this early phase, GOTO secured its first GRB optical afterglow detection with GRB~171205A, associated with SN~2017iuk \citep{Steeghs2017GCN22190, Izzo2019}. In the initial 3 years between June 2017 and June 2020, the GOTO-4 system responded to 77 \fermi/GBM and 29 {\em Swift}/BAT triggers \citep[see][]{Mong2021, Steeghs2022}, demonstrating the scientific potential of the facility even in its prototype configuration. 

With full deployment at both sites, as of 31 December 2024 (with GRBs observed thereafter will be included in a future study), GOTO has conducted follow-up observations for over 257 \fermi\nobreakdash-, 43 \swift\nobreakdash-, 28 \textit{EP}\nobreakdash-, and 7 \textit{GECAM}\nobreakdash-triggered events. No targeted observations were conducted for \textit{SVOM} events during this period, as GOTO began following up \textit{SVOM} triggers in 2025, coinciding with the scheduled start of \textit{SVOM}'s science operations in February 2025. On average, GOTO's first targeted observation latency was approximately 11.2~hr, with response times ranging from 270~s--69.77~hr post-trigger. The average latency for the last observation in each follow-up series was 26.36~hr, while the alert latency averaged 4.8~hr, underscoring the challenges in achieving timely and efficient follow-ups.

\begin{table*}
\centering
\begin{threeparttable}
    \caption{Summary of the LGRBs analysed in this work. For each burst, we list the high-energy and optical afterglow discoverers, the time from the high-energy trigger to the optical discovery ($T-T_0$), the GOTO internal afterglow name, J2000 coordinates, discovery $L-$band magnitude, Galactic extinction $E(B-V)$, and spectroscopic redshift.}
    \label{tab:sample}
    \addtolength{\tabcolsep}{-1.5pt}
    \renewcommand{\arraystretch}{1.1}
    \begin{tabular}{|c|c|c|c|c|c|c|c|c|c|}
        \hline
        GRB & High-energy & Optical & Discovery & GOTO & RA & Dec & Discovery & $E(B-V)^b$ & Redshift$^c$ \\
            & discoverer & discoverer & $T-T_0$ (hrs)$^a$ & internal name & ($^{\mathrm{h}}$:$^{\mathrm{m}}$:$^{\mathrm{s}}$) & ($^{\circ}$:$'$:$''$) & $L$-band mag & (mag) & \\
        \hline
        240122A & \maxi/GSC   & GOTO-S & 0.73 & GOTO24eu  & 06:12:12.91 & $-$19:08:38.81 & $17.58 \pm 0.04$ & 0.0651 & $3.1634 \pm 0.0003$ \\
        240225B & \maxi/GSC   & GOTO-N & 1.50 & GOTO24tz  & 08:33:26.67 &   +27:04:32.71 & $17.12 \pm 0.04$ & 0.0354 & $0.9462 \pm 0.0002$ \\
        240619A & \fermi/GBM  & GOTO-S & 4.69 & GOTO24cvn & 10:49:34.70 &   +17:16:58.07 & $17.17 \pm 0.17$ & 0.0253 & $0.3960 \pm 0.0001$ \\
        240910A & \fermi/GBM  & GOTO-S & 9.43 & GOTO24fvl & 01:36:23.45 & $-$00:12:17.86 & $19.33 \pm 0.13$ & 0.0247 & $1.4605 \pm 0.0007$ \\
        240916A & \fermi/GBM  & GOTO-S & 7.73 & GOTO24fzn & 15:43:39.23 & $-$07:45:53.21 & $17.80 \pm 0.06$ & 0.1359 & $2.6100 \pm 0.0002$ \\
        241002B & \fermi/GBM  & GOTO-S & 3.05 & GOTO24gpc & 21:53:16.56 & $-$58:56:51.98 & $19.53 \pm 0.09$ & 0.0268 & --- \\
        241228B & \fermi/GBM  & GOTO-N & 0.32 & GOTO24jmz & 08:31:05.46 &   +06:50:54.07 & $14.54 \pm 0.01$ & 0.0290 & $2.6745 \pm 0.0004$ \\
        \hline
    \end{tabular}
    \begin{tablenotes}
        \footnotesize
        \item[a] For \maxi~GRBs, $T_0$ denotes the \maxi/GSC trigger time; for \fermi~GRBs, $T_0$ denotes the \fermi/GBM trigger time.
        \item[b] Galactic extinction values are estimated following recalibrated dust maps of \cite{Schlafly2011}.
        \item[c] see Section~\ref{sec:spec_afterglow_analysis}.
    \end{tablenotes}
\end{threeparttable}
\end{table*}

The GOTO collaboration has reported nearly 80 GCN circulars to date based on GOTO observations, including the detection and upper-limit constraints of GRB afterglows. To date, nearly 28 successful detections have been reported, including GRB 230818A, detected 4.43~min after its trigger \citep{Gompertz2023GCN34480}. In addition, GOTO has provided numerous upper-limit constraints and contributed to serendipitous discoveries like orphan afterglow AT2023lcr \citep{Gompertz2023GCN34023}, with early-phase observations aiding in the refinement of transient properties (Martin-Carrillo et al., in prep.). GOTO’s follow-up capabilities continue to evolve in response to operational experience and scientific objectives.

GRB~230911A was the first LGRB for which GOTO discovered an optical afterglow \citep{Belkin2023GCN34681}; details are published in \cite{Belkin2024RNAAS}. After this first case, in 2024, GOTO discovered optical afterglows of 2 \maxi-triggered (GRB 240122A and GRB 240225B) serendipitously and 5 \fermi-triggered (GRBs 240619A, 240910A, 240916A, 241002B and 241228B) LGRBs in responsive mode, which are studied in detail in this work. The details of these 7 GRBs in our sample are discussed in the following section. In addition to these LGRBs, within 2024 itself, GOTO also identified the optical afterglow of the SGRB~241105A \citep{Julakanti2024GCN38088}, which was localised by \fermi/GBM with an uncertainty of $\sim4\,\mathrm{deg}$ \citep{Fermi2024GCN38085}. GOTO rapidly responded to the trigger, tiling 277.9 square degrees within the 90\% GBM localisation region and covering $\sim 84\%$ of the total probability within 1.6~hr. The afterglow was discovered at $L \sim 17.2$~mag, later confirmed through multi-wavelength follow-up and spectroscopy to lie at a redshift of $z = 2.681$ \citep{Izzo2024GCN38167}. Although this event is not part of the core LGRB sample analysed here, it highlights GOTO’s capabilities to detect optical afterglows from both long and short GRBs, even under challenging localisation conditions. A detailed analysis of GRB~241105A is presented in a separate paper by \cite{Dimple2025}.

\section{The GRB sample}\label{sec:sample}
Our sample comprises seven LGRBs whose optical afterglows were discovered by GOTO in 2024: two \maxi--triggered events (GRBs~240122A and 240225B) identified serendipitously, and five \fermi--triggered events (GRBs~240619A, 240910A, 240916A, 241002B, and 241228B) detected in responsive mode. This section begins by discussing the prompt high-energy triggers and observations of these seven GRBs. The basic properties of the GRBs in our sample are listed in Table~\ref{tab:sample}. The detection circumstances provide essential context --- in particular, the trigger times, localisation accuracy, and alert distribution --- that directly influenced GOTO’s follow-up strategy, as discussed above in Section~\ref{sec:GOTO_follow-up_strategies}. The subsequent subsections describe the localisation, follow-up coverage, and an observational summary of the seven LGRBs in our sample.  

\subsection{High-Energy Triggers}\label{sec:High-Energy_Triggers}
The GOTO follow-up campaigns for GRBs in our sample were initiated by triggers from the \maxi/GSC \citep{Matsuoka2009, Mihara2011PASJ} and the \fermi/GBM \citep{Meegan2009}. We briefly summarise below the specific GRBs that prompted these observations. Figure~\ref{fig:locs} shows the 90\% containment localisation regions provided by the triggering satellites.

\subsubsection{MAXI/GSC GRBs~240122A and 240225B}

GRB~240122A was detected by the \maxi/GSC on 2024-01-22 at 10:28:03 UT in the 2--10~\keV{} range \citep{Negoro2024GCN35593}. The burst was localized to RA = 06$^{\mathrm{h}}$11$^{\mathrm{m}}$18$^{\mathrm{s}}$ and Dec = $-$19$^{\circ}$01$^{\prime}$51$^{\prime\prime}$ (J2000), with an uncertainty of 30 arcmin. This event was detected solely by \maxi/GSC, with no additional high-energy instruments reporting a detection. Due to the large localisation uncertainty, it was not followed up by narrow-field optical instruments but represented a good candidate for wide-field facilities such as GOTO.  

\maxi/GSC triggered GRB~240225B on 2024-02-25 at 20:15:46 UT in the 4--10~\keV{} range \citep{Nakajima2024GCN35796}. The burst was localized to RA = 08$^{\mathrm{h}}$33$^{\mathrm{m}}$49$^{\mathrm{s}}$ and Dec = +27$^{\circ}$29$^{\prime}$13$^{\prime\prime}$ (J2000), with a statistical 90\% confidence level elliptical error region, where the semi-major and semi-minor axes have radii of 0.13\degree\ and 0.11\degree, respectively (see Figure~\ref{fig:locs}). The burst was also observed by several other high-energy instruments, including \textit{AstroSat}'s Cadmium Zinc Telluride Imager (CZTI) in the 20--200~\keV{} band, and the CsI anticoincidence detectors in the 100--500~\keV{} band \citep{Joshi2024GCN35798J}. Additional detections include \textit{INTEGRAL} SPI--Anti-Coincidence Shield (SPI--ACS) in energies $\gtrsim$80~\keV, the CALET Gamma-ray Burst Monitor (CGBM) in the 100–500~\keV{} range and coverage up to 40--1000~\keV{} \citep{Kawakubo2024GCN35811}, \textit{Konus}-Wind up to $\sim$3~\meV{} \citep{Frederiks2024GCN35835}, and the \textit{Glowbug} gamma-ray telescope in the 10--10000~\keV{} range \citep{Cheung2024GCN35848}.

\subsubsection{Fermi/GBM GRBs~240619A,~240910A,~240916A,~241002B, and~241228B}

GRB~240619A triggered the \fermi/GBM on 2024-06-19 at 03:43:31~UT in the 50--300~\keV{} band \citep{FermiGCN36694, Dalessi2024GCN36717}. The burst was localised to RA = 10$^{\mathrm{h}}$48$^{\mathrm{m}}$00$^{\mathrm{s}}$ and Dec = +17$^{\circ}$18$^{\prime}$00$^{\prime\prime}$ (J2000), with a statistical uncertainty of 1.6\degree, which is circularised 90\% containment radius $R_{\mathrm{err,90}}$ in degrees (see Figure~\ref{fig:locs} and column two of Table~\ref{tab:coverage_fermi}). In addition to \fermi/GBM, the burst was also detected by the CALET CGBM in the 40--1000~\keV{} band \citep{Torii2024GCN36719}, \textit{Konus}--Wind up to $\sim$10~MeV{} \citep{Svinkin2024GCN36768}, and the 1U-sized CubeSat \textit{GRBAlpha} in the 70--890~\keV{} band \citep{Dafcikova2024GCN36724}.

GRB~240910A triggered the \fermi/GBM on 2024-09-10 at 04:00:44~UT \citep{Fermi2024GCN37441}. The event was localised to RA = 01$^{\mathrm{h}}$00$^{\mathrm{m}}$00$^{\mathrm{s}}$ and Dec = +04$^{\circ}$30$^{\prime}$00$^{\prime\prime}$ (J2000), with a statistical uncertainty of 4.5\degree. This burst was also picked up by the \textit{SVOM}/GRM operating in the 15~\keV{}--5~MeV{} range \citep{SVOM2024GCN37445}, as well as by the 1U CubeSat \textit{GRBAlpha} \citep{Ripa2024GCN37450}.

GRB~240916A, detected by the \fermi/GBM at 01:22:56~UT on 2024-09-16 \citep{Fermi2024GCN37518, Roberts2024GCN37535}, was localised to RA = 15$^{\mathrm{h}}$32$^{\mathrm{m}}$00$^{\mathrm{s}}$ and Dec = $-$07$^{\circ}$05$^{\prime}$00$^{\prime\prime}$ (J2000) with a statistical uncertainty of 1.2\degree. Additional high-energy observations were made by the \textit{INTEGRAL}/SPI--ACS \citep{Pawar2024GCN37519} and by the 1U CubeSat \textit{GRBAlpha} \citep{Dafcikova2024GCN37543}.

GRB~241002B triggered the \fermi/GBM at 06:14:18.76~UT on 2024-10-02 \citep{Fermi2024GCN37668, Roberts2024GCN37711}. The burst was localised to RA = 22$^{\mathrm{h}}$15$^{\mathrm{m}}$00$^{\mathrm{s}}$ and Dec = $-$64$^{\circ}$17$^{\prime}$00$^{\prime\prime}$ (J2000), with a statistical uncertainty of 3.7\degree. It was also observed by the \swift~Burst Alert Telescope -- Gamma-ray Urgent Archiver for Novel Opportunities (\swift/BAT--GUANO; \citealt{DeLaunay2024GCN37704}).

GRB~241228B triggered the \fermi/GBM at 04:13:05.39~UT on 2024-12-28 \citep{Fermi2024GCN38682, Scotton2024GCN38714}. The burst was localised to RA = 08$^{\mathrm{h}}$08$^{\mathrm{m}}$00$^{\mathrm{s}}$ and Dec = +14$^{\circ}$00$^{\prime}$00$^{\prime\prime}$ (J2000), with a statistical uncertainty of 1.7\degree. It was also detected by the \swift/BAT--NITRATES system \citep{DeLaunay2024GCN38700}, with the position consistent with the GBM localisation. In addition, the \fermi~Large Area Telescope (LAT) observed high-energy emission ($>100$~MeV) from this burst, including a 16~GeV photon detected 31~s after the trigger \citep{DiLalla2025GCN38843}.

These prompt high-energy detections by \fermi/GBM provided the initial localisation constraints and trigger alerts that enabled rapid optical follow-up by GOTO. The diversity in localisation uncertainties influenced the choice of tiling patterns and observing cadences. In the following section, we present the optical afterglow localisations and follow-up coverage, highlighting how the prompt trigger information shaped the subsequent GOTO observations.

\subsection{Localisation Coverage and Optical Afterglow Discoveries}\label{sec:goto_discoveries_optical}
Figure~\ref{fig:locs} provides an overview of the sky localisation and follow-up coverage for the GRBs analysed in this study (\grbssample), based on GOTO observations obtained within the first 24\,h post-trigger. The dark and light grey contours represent the $1\sigma$ and $2\sigma$ localisation regions extracted from the \textsc{HEALPix} \citep{healpix} probability skymaps. Light blue shading denotes the GOTO fields observed within 10~hr of each burst, demonstrating the system’s wide-area and rapid-response capabilities. Cyan stars mark the locations of confirmed optical afterglows, showing that GOTO’s coverage either enclosed or closely bordered the true source positions in all cases. In addition, the finding charts highlighting the confirmed afterglows, derived from GOTO observations, are presented in panels (a)–(g) of Figure~\ref{fig:finding_chart}. For context, each GOTO cutout is paired with a corresponding image from Legacy Survey (LS) DR10 covering the same $\sim6.3^{\prime} \times 6.3^{\prime}$ FoV; for GRB~240122A, a Pan-STARRS DR1 image is shown instead.

\subsubsection{MAXI/GSC GRBs~240122A and 240225B: arcminute-scale localisation}
\begin{figure}
    \centering
    \includegraphics[width=1.0\linewidth]{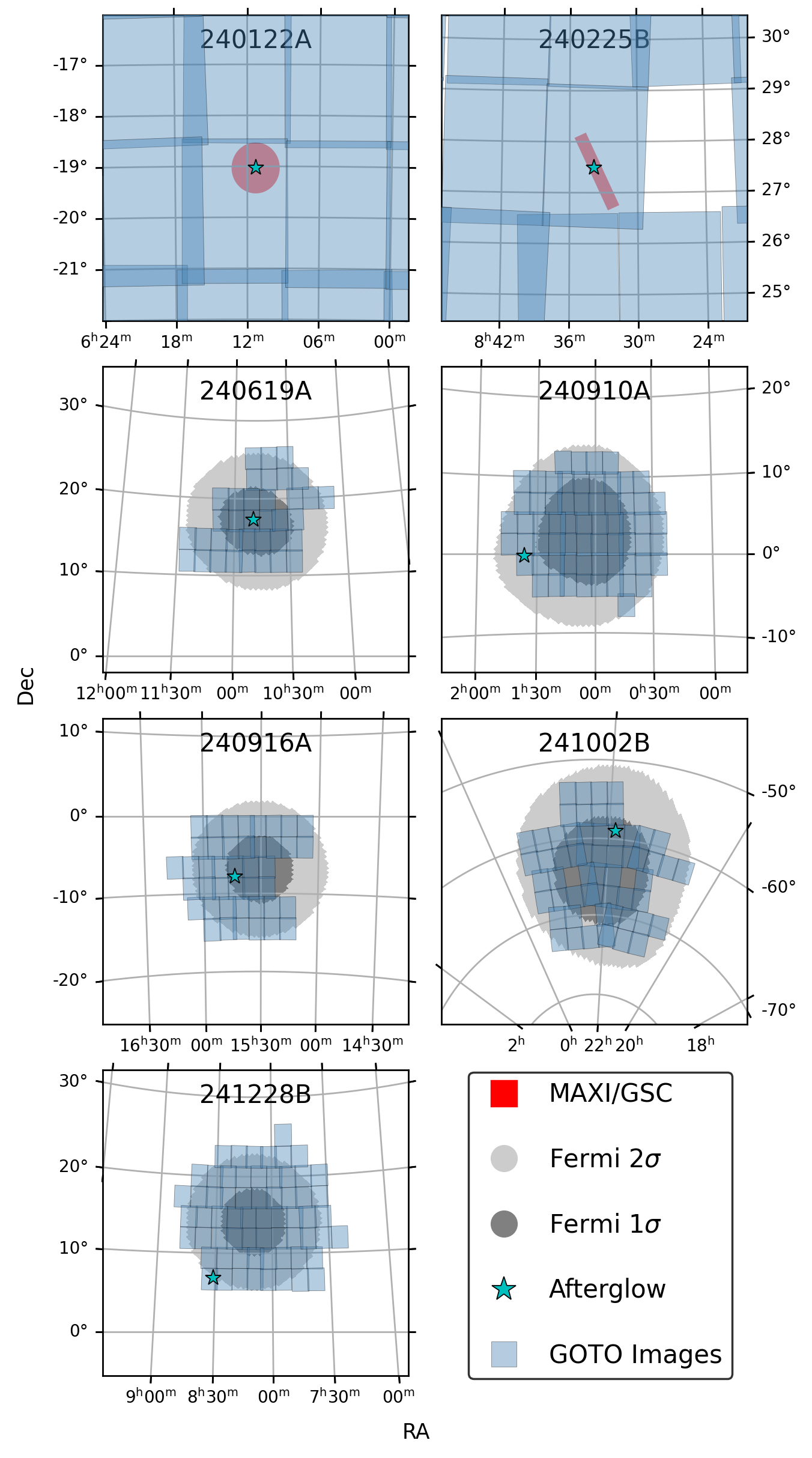}
    \caption{GOTO coverage of each of the GRBs in the sample. The first two plots denote the 90\% containment \maxi/GSC localisations (red) in a $3\degree\times3\degree$ field. The localisation areas are generated based on information from their discovery GCNs. The following five plots show \fermi/GBM localisations (grey) and the 1 and 2 $\sigma$ contours from their respective \textsc{HEALPix} skymaps in a $20\degree\times20\degree$ field. In all plots, the 2D footprint of GOTO images taken in the first 10~hr post-trigger that overlap the localisations are shown in light blue. The corresponding afterglow positions are marked with a cyan star.}
    \label{fig:locs}
\end{figure}

GOTO does not follow up \maxi~triggers in responsive mode, as \maxi~alerts are not distributed in a machine-readable format that can be ingested automatically by the sentinel. Instead, these fields are only covered serendipitously in survey mode. In Figure~\ref{fig:locs}, the top row shows the two events detected by \maxi/GSC, GRB~240122A and GRB~240225B, each displayed within a $3\degree \times 3\degree$ FoV. The red shaded regions indicate the approximate localisation areas from \maxi/GSC reports. For GRB~240122A, the afterglow is well centred within the localisation and fully encompassed by the GOTO field. In contrast, for GRB~240225B, the GOTO tiling intersected the elongated error region, providing timely coverage that included the eventual afterglow position. 

\paragraph{GRB~240122A}
GOTO-S serendipitously observed the localisation region of GRB~240122A during its routine all-sky survey on 2024-01-22, discovering the optical afterglow (GOTO24eu/\href{https://www.wis-tns.org/object/2024apy}{AT2024apy}) at J2000 coordinates RA = 06$^{\mathrm{h}}$12$^{\mathrm{m}}$12$^{\mathrm{s}}$.91 and Dec = $-$19$^{\circ}$08$^{\prime}$38$^{\prime\prime}$.81. The afterglow was detected at 11:11:43\,UT (\maxiT+43.68\,min) with an $L-$band magnitude of $17.58 \pm 0.04$\,mag \citep{Kumar2024GCN35596}, see also Table~\ref{tab:sample} for details. As shown in Figure~\ref{fig:locs} (top-left), the GOTO FoV comfortably covers the compact \maxi/GSC localisation, placing the afterglow well within the imaged area. This case demonstrates the ability of GOTO to capture transient counterparts during its high-cadence survey mode, even without a targeted trigger.

\begin{table*}
\centering
\begin{threeparttable}
    \caption{Summary of GOTO’s coverage of the \fermi/GBM GRBs in our sample. For each event, we list the \fermi/GBM-reported 90\% localisation uncertainty ($R_{\mathrm{err,90}}$), number of images, sky coverage area, enclosed probability, and mean $5\sigma$ $L-$band limiting magnitude.}
    \label{tab:coverage_fermi}
    \addtolength{\tabcolsep}{20pt}
    \begin{tabular}{|c|c|c|c|c|c|}
        \hline
        GRB & $R_{\mathrm{err,90}}$ & No. of  & Coverage   & Prob. & Mean $5\sigma$ \\
         $ $   & (deg) & images & (deg$^2$) & enclosed (\%) &  depth (mag) \\
        \hline
        240619A & 1.6 & 56  & 151.5 & 85.7 & 18.6 \\
        240910A & 4.5 & 191 & 295.3 & 90.3 & 20.0 \\
        240916A & 1.2 & 41  & 178.6 & 78.3 & 19.1 \\
        241002B & 3.7 & 58  & 273.1 & 84.9 & 20.3 \\
        241228B & 1.7 & 165 & 214.0 & 89.9 & 19.8 \\
        \hline
    \end{tabular}
\end{threeparttable}
\end{table*}

\paragraph{GRB~240225B}
Similarly, GOTO-N serendipitously covered the field of GRB~240225B and discovered its optical afterglow (GOTO24tz/\href{https://www.wis-tns.org/object/2024dgu}{AT2024dgu}) at J2000 coordinates RA = 08$^{\mathrm{h}}$33$^{\mathrm{m}}$26$^{\mathrm{s}}$.67 and Dec = +27$^{\circ}$04$^{\prime}$32$^{\prime\prime}$.71. The counterpart was first detected on 2024-02-25 at 21:45:51\,UT (\maxiT+1.50\,h) at $17.12 \pm 0.04$\,mag ($L-$band) and was last detected the following night at 22:10:38\,UT (\maxiT+25.91\,h) at $19.69 \pm 0.18$\,mag \citep{Gompertz2024GCN35805}. As shown in Figure~\ref{fig:locs} (top-right), the GOTO tiling intersected the elongated \maxi/GSC localisation, with the afterglow located near the centre of the observed field.

\begin{figure*}
\centering
\setlength{\tabcolsep}{0pt}
\renewcommand{\arraystretch}{0}
\begin{tabular}{@{} c c c c @{}}
    \multicolumn{2}{c}{\small GRB 240122A  \hspace{0.2cm}   (a)} &
    \multicolumn{2}{c}{\small GRB 240225B  \hspace{0.2cm}   (b)} \\[4pt]
    \includegraphics[width=0.26\textwidth]{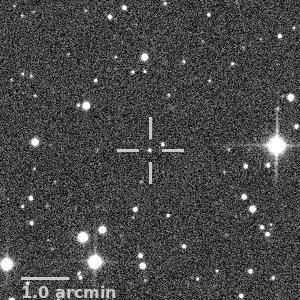} &
    \includegraphics[width=0.26\textwidth]{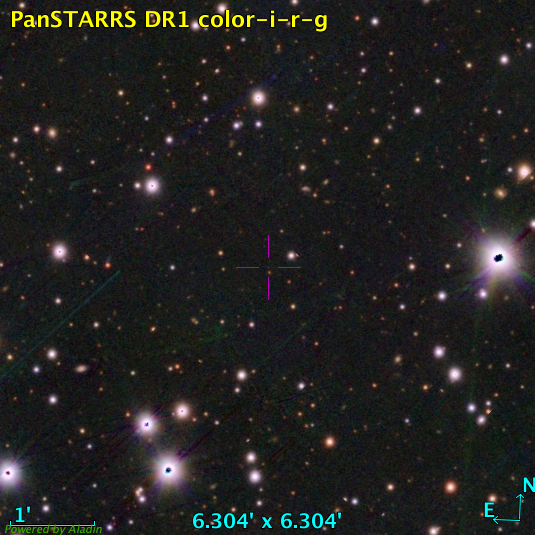} & \hspace{4pt}
    \includegraphics[width=0.26\textwidth]{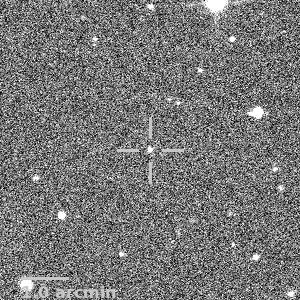} &
    \includegraphics[width=0.26\textwidth]{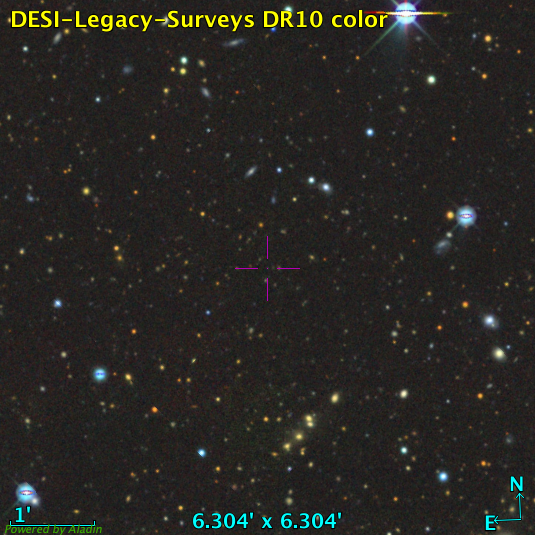} \\[8pt]
    \multicolumn{2}{c}{\small GRB 240619A  \hspace{0.2cm}   (c)} &
    \multicolumn{2}{c}{\small GRB 240910A  \hspace{0.2cm}   (d)} \\[4pt]
    \includegraphics[width=0.26\textwidth]{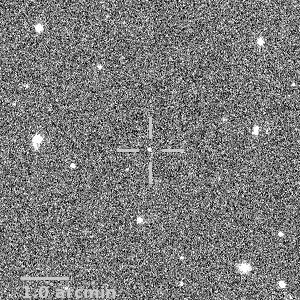} &
    \includegraphics[width=0.26\textwidth]{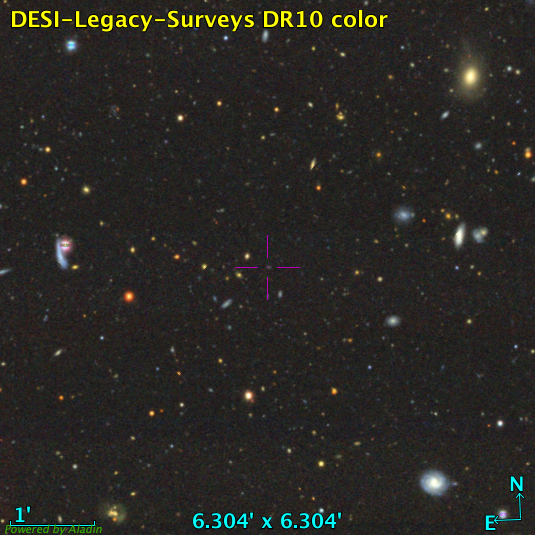} & \hspace{4pt}
    \includegraphics[width=0.26\textwidth]{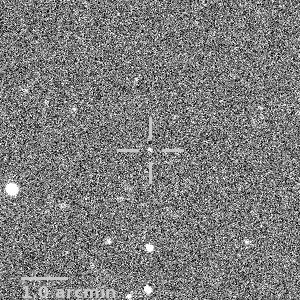} &
    \includegraphics[width=0.26\textwidth]{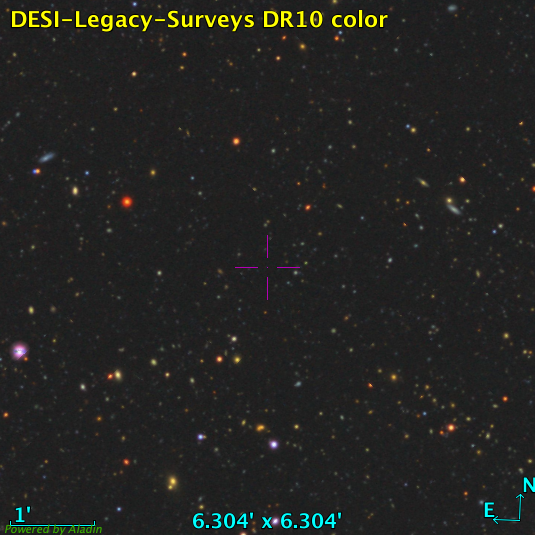} \\[8pt]
    \multicolumn{2}{c}{\small GRB 240916A  \hspace{0.2cm}   (e)} &
    \multicolumn{2}{c}{\small GRB 241002B  \hspace{0.2cm}   (f)} \\[4pt]
    \includegraphics[width=0.26\textwidth]{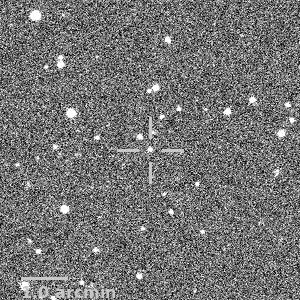} &
    \includegraphics[width=0.26\textwidth]{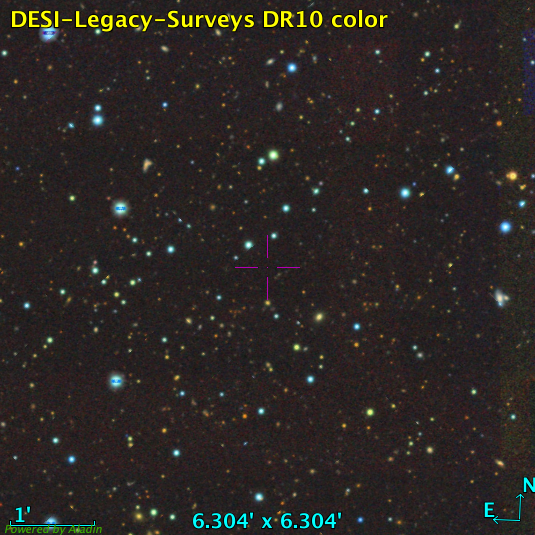} & \hspace{4pt}
    \includegraphics[width=0.26\textwidth]{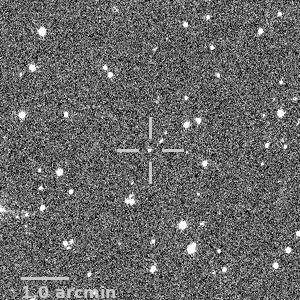} &
    \includegraphics[width=0.26\textwidth]{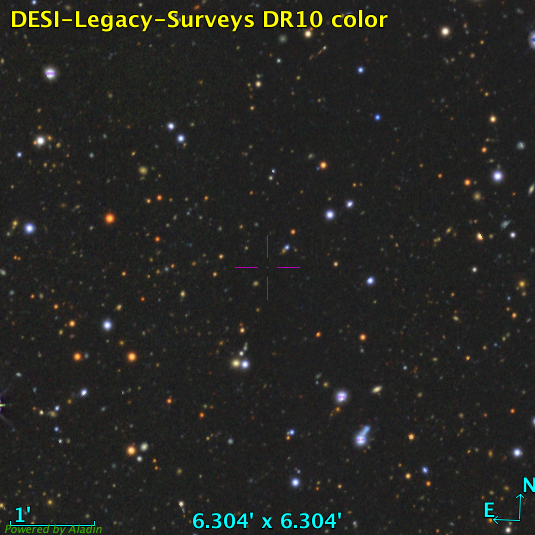} \\[8pt]
    \multicolumn{4}{c}{\small GRB 241228B  \hspace{0.2cm}   (g)} \\[4pt]
    \multicolumn{4}{c}{
            \includegraphics[width=0.26\textwidth]{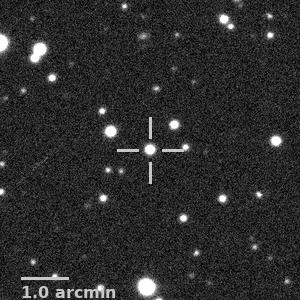} \hspace{-4pt}
            \includegraphics[width=0.26\textwidth]{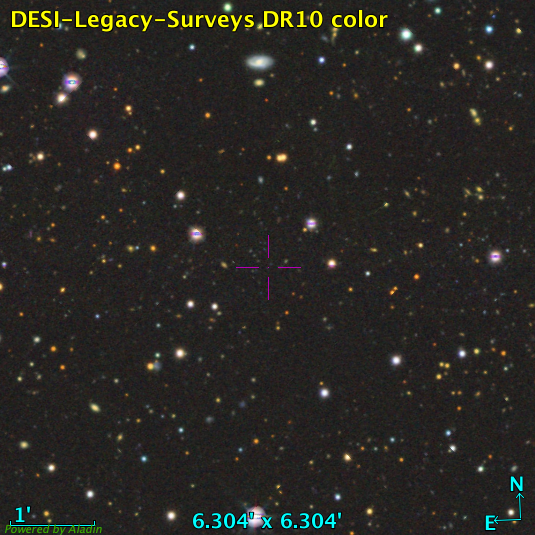}
    } \\
\end{tabular}
\caption{Finding charts of \grbssample{} in the GOTO $L-$band (400--700\,nm) observed by GOTO. Each cutout is a $300 \times 300$ pixel region centred on the transient, corresponding to a FoV of $\sim6.3\arcmin \times 6.3\arcmin$ at the GOTO pixel scale of $1.26\arcsec$/pix. For comparison, survey images from the Legacy Survey DR10 are shown (except for GRB~240122A, where a Pan-STARRS DR1 image is used), matched to the same FoV. Details of each object are listed in Table~\ref{tab:sample}.}
\label{fig:finding_chart}
\end{figure*}

\subsubsection{Fermi/GBM GRBs~240619A,~240910A,~240916A,~241002B, and 241228B: degree-scale localisation}
\fermi/GBM triggers are distributed in real time via machine-readable GCN notices, which the GOTO sentinel ingests automatically. This enables fully responsive follow-up, with observations scheduled immediately after the alert is received. Follow-up observations are carried out as soon as observing conditions and visibility constraints permit. The five different panels in Figure~\ref{fig:locs} correspond to events detected by \fermi/GBM GRBs~240619A,~240910A,~240916A,~241002B, and~241228B, each displayed over a wider $20\degree \times 20\degree$ field. A quantitative summary of GOTO follow-up coverage for the \fermi/GBM GRBs in our sample is provided in Table~\ref{tab:coverage_fermi}. For each event, we list the GBM localisation uncertainty (expressed as the circularised 90\% containment radius in degrees), number of images obtained to cover the 90\% GBM localisation region, the total sky area imaged, the fraction of localisation probability enclosed within the observed fields and the mean $5\sigma$ limiting magnitude. The two \maxi/GSC GRBs are not included here, as their compact localisations were fully covered serendipitously by single $\sim9$ deg$^2$ survey pointings.

\paragraph{GRB~240619A}
GOTO-S initiated targeted follow-up observations of GRB~240619A on 2024-06-19 at 08:24:01 UT (\fermiT+4.68\,h), continuing until 21:48:35 UT (\fermiT+18.08\,h). As illustrated in Figure~\ref{fig:locs} (second row; left panel), the GOTO tiling successfully overlapped the $1\sigma$ and $2\sigma$ \textsc{HEALPix} contours from the \fermi/GBM localisation, with the afterglow position (blue star) falling within the observed fields obtained in the first 10~hr. The afterglow (GOTO24cvn/\href{https://www.wis-tns.org/object/2024lwv}{AT2024lwv}) was identified in these data at J2000 coordinates RA = 10$^{\mathrm{h}}$49$^{\mathrm{m}}$34$^{\mathrm{s}}$.70 and Dec = +17$^{\circ}$16$^{\prime}$58$^{\prime\prime}$.07, with detections by GOTO-S at 08:24:50 UT (\fermiT+4.69\,h) and by GOTO-N at 21:40:50 UT (\fermiT+18.00\,h), exhibiting $L-$band magnitudes of $17.17 \pm 0.17$ and $18.38 \pm 0.09$\,mag, respectively \citep{Gompertz2024GCN36715}, details are tabulated in Table~\ref{tab:sample}. Although the afterglow was first discovered by GOTO, the position was also serendipitously covered by the ATLAS all-sky survey \citep{Tonry2018}, which provides a forced photometric detection at an earlier epoch. The source is detected in the ATLAS forced photometry data with an $o$-band (560--820\,nm) magnitude of $16.24 \pm 0.01$ at \fermiT+2.53\,hr and $18.72 \pm 0.12$ at \fermiT+26.71\,hr, retrieved from the ATLAS Forced Photometry Server \citep{Shingles2021}. While ATLAS did not identify the transient in real time, its archival data proved valuable in constraining the early-time brightness and confirming the fading behaviour consistent with an optical afterglow \citep{Gompertz2024GCN36715}.

\paragraph{GRB~240910A}
GOTO-S began targeted follow-up observations of GRB~240910A on 2024-09-10 at 12:25:28 UT (\fermiT+8.41\,h), continuing through to 16:33:57 UT (\fermiT+12.55\,h). As shown in Figure~\ref{fig:locs} (second row; right panel), the GOTO coverage successfully intersected the high-probability localisation contours, with the afterglow position clearly lying within the imaged area. The afterglow (GOTO24fvl/\href{https://www.wis-tns.org/object/2024vfp}{AT2024vfp}) was discovered at J2000 coordinates RA = 01$^{\mathrm{h}}$36$^{\mathrm{m}}$23$^{\mathrm{s}}$.45 and Dec = $-$00$^{\circ}$12$^{\prime}$17$^{\prime\prime}$.86, with detections spanning from 13:26:31.776 UT (\fermiT+9.43\,h) to 16:01:36 UT (\fermiT+12.01\,h), yielding $L-$band magnitudes of $19.32 \pm 0.13$ and $19.74 \pm 0.12$\,mag, respectively \citep{Julakanti2024GCN37459}.

\paragraph{GRB~240916A}
GOTO-S began targeted follow-up observations of GRB~240916A on 2024-09-16 at 09:06:47 UT (\fermiT+7.73\,h), continuing until 09:23:05 UT (\fermiT+8.00\,h). As seen in Figure~\ref{fig:locs} (third row; left panel), the GOTO tiling intersected the high-probability regions of the localisation, successfully encompassing the afterglow site. The afterglow (GOTO24fzn/\href{https://www.wis-tns.org/object/2024vlp}{AT2024vlp}) was identified at J2000 coordinates RA = 15$^{\mathrm{h}}$43$^{\mathrm{m}}$39$^{\mathrm{s}}$.229 and Dec = $-$07$^{\circ}$45$^{\prime}$53$^{\prime\prime}$.22, with a detection at 09:06:47.81 UT (\fermiT+7.73\,h) at an $L-$band magnitude of $17.80 \pm 0.05$\,mag \citep{Gompertz2024GCN37522}.

\paragraph{GRB~241002B}
GOTO-S conducted targeted follow-up observations of GRB~241002B starting on 2024-10-02 at 09:17:03 UT (\fermiT+3.05\,h), concluding at 09:40:00 UT (\fermiT+3.43\,h). As shown in Figure~\ref{fig:locs} (third row; right panel), the observed GOTO fields overlapped the high-probability regions of the \fermi/GBM localisation, with the afterglow position included within the footprint. The afterglow (GOTO24gpc/\href{https://www.wis-tns.org/object/2024xbg}{AT2024xbg}) was discovered at J2000 coordinates RA = 21$^{\mathrm{h}}$53$^{\mathrm{m}}$16$^{\mathrm{s}}$.56 and Dec = $-$58$^{\circ}$56$^{\prime}$51$^{\prime\prime}$.98, with a detection at 09:17:20 UT (\fermiT+3.05\,h) at an $L-$band magnitude of $19.53 \pm 0.09$\,mag \citep{Kumar2024GCN37676}.

\paragraph{GRB~241228B}
GOTO-N initiated follow-up observations of GRB~241228B on 2024-12-28 at 04:26:19 UT (\fermiT+0.22\,h), continuing through to 23:24:22 UT (\fermiT+19.19\,h). As depicted in Figure~\ref{fig:locs} (bottom-left panel), the GOTO tiling efficiently covered the high-probability localisation region, including the afterglow position. The afterglow (GOTO24jmz/\href{https://www.wis-tns.org/object/2024afgu}{AT2024afgu}) was identified by GOTO-N at J2000 coordinates RA = 08$^{\mathrm{h}}$31$^{\mathrm{m}}$05$^{\mathrm{s}}$.46 and Dec = +06$^{\circ}$50$^{\prime}$54$^{\prime\prime}$.07, with an initial detection at 04:32:24 UT (\fermiT+0.32\,h) at an $L-$band magnitude of $14.54 \pm 0.01$\,mag. Multiple detections followed throughout the observing sequence, with the final GOTO-N detection recorded at 13:00:42 UT (\fermiT+8.79\,h) at $19.70 \pm 0.10$\,mag \citep{Kumar2024GCN38684}. The afterglow candidate for GRB~241228B falls on the 94.5\% probability contour, formally outside the GBM 90\% localisation region. While most GRB afterglows are found within the 90\% contour, a small fraction are expected to lie just beyond it, making GRB~241228B a noteworthy case.

\begin{figure*}
     \centering
     \includegraphics[width=1.0\linewidth]{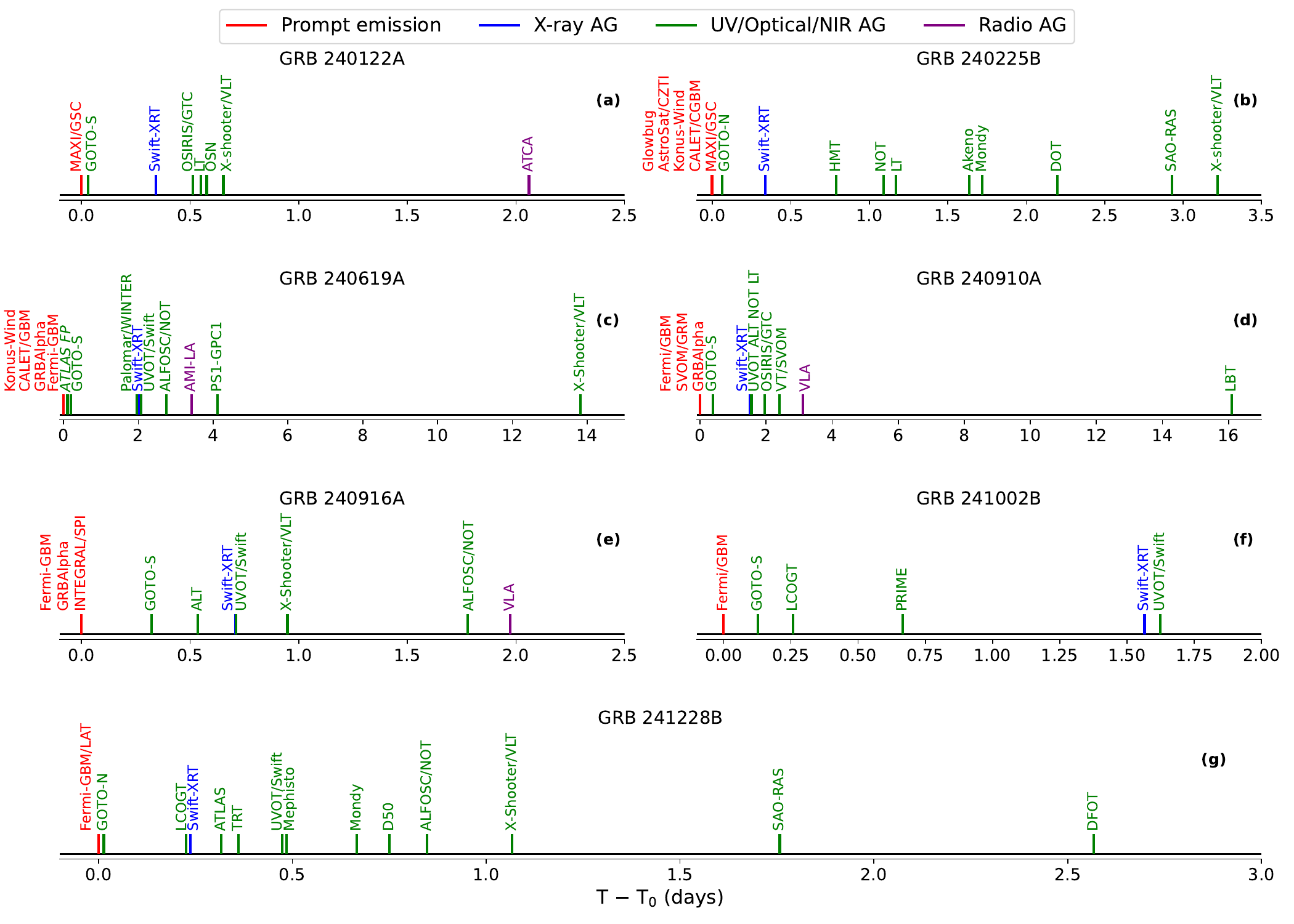}
     \caption{First detection times across various observatories for GRBs in our sample. Shown are prompt (red), X-ray afterglow (blue), UV/optical/NIR afterglow (green), and radio afterglow (purple) observations from both space- and ground-based facilities. In all cases, GOTO discovered the optical afterglow following the prompt emission. For GRB~240619A, although ATLAS has the earliest epoch, the afterglow was first discovered by GOTO, and the ATLAS data were serendipitously pre-covered and used in the GOTO discovery report \citep{Gompertz2024GCN36715}.}
     \label{fig:timeline_plot}
\end{figure*}

\section{Afterglow Follow-up Observations}\label{sec:afterglow_follow_ups}
In addition to the discovery imaging provided by GOTO, we carried out a coordinated programme of multi-wavelength follow-up to characterise the afterglows of our GRB sample. These observations span the X-ray, UV, optical, and radio regimes, enabling us to track the temporal evolution and spectral energy distributions of the counterparts. The combined dataset allows us to constrain the physical properties of the bursts, verify their association with the optical transients identified by GOTO, and provide essential input for modelling their afterglow emission.

Figures~\ref{fig:timeline_plot}(a--g) illustrate the timeline of the multi-wavelength follow-up campaigns for all GRBs in the sample, marking the epochs at which each facility recorded its first observation relative to the trigger time. The various phases of the events are colour-coded as follows: X-ray afterglow (blue), UV/optical/NIR afterglow (green), radio afterglow (purple), and prompt emission (red) for completeness. This timeline highlights the wide temporal coverage and the rapid, coordinated response from both ground- and space-based observatories across the electromagnetic spectrum for the GRBs in our sample. More observation details are provided below. A full log of photometric measurements, combining GCN Circular reports and data from this work, is presented in Tables~\ref{tab:phot_data} and~\ref{tab:radio}, while Table~\ref{tab:spec_data} summarises the spectroscopic campaigns.

\subsection{X-ray}
The XRT onboard \swift~\citep{Burrows2005} performed follow-up observations in response to our target-of-opportunity (ToO) requests, with exposures ranging from 1.3 to 3.6~ks depending on the GRB, and all data were collected in Photon Counting (PC) mode. X-ray data, including light curves, calibrated event files, and spectra, were retrieved from the public \swift/XRT GRB Catalogue hosted by the UK Swift Science Data Centre\footnote{\url{https://www.swift.ac.uk/}} and processed using the standard XRT pipeline as described in \citet{Evans2007,Evans2009}. All light curves and spectra were generated using the automated tools provided by the XRT team.

XRT detected X-ray afterglows for all seven events, with uncatalogued sources coincident with or close to the GOTO optical transient locations in each case. The corresponding 0.3--10~keV light curves (in counts\,s$^{-1}$) for the seven \grbssample, together with comparison GRBs, and spectra (in counts\,s$^{-1}$\,keV$^{-1}$) for our sample are discussed later in Section~\ref{sec:X-ray}.

\subsection{UV/Optical/NIR Photometric Observations}
Follow-up afterglow observations in the UV/optical/NIR, including data from both the GCN Circulars and this work, are compiled in Table~\ref{tab:phot_data}. The table provides details on the observing facilities, instruments used, measured magnitudes, and other relevant parameters. Figure~\ref{fig:optical_lcs} shows the extinction-corrected UV/optical/NIR afterglow light curves and the wavelength coverage of all filters used (passbands in nm).

While GOTO primarily contributed discovery optical observations for the GRBs in our sample, its relatively small aperture and lack of multiple band observations limit its utility for extended follow-up. Details of the GOTO observations and initial detection magnitudes have been presented in Section~\ref{sec:goto_discoveries_optical}. Therefore, GOTO observations are not included in this section. Instead, this section focuses on subsequent optical afterglow follow-up observations obtained with a range of facilities situated around the globe.

\subsubsection{Swift/UVOT}
In addition to XRT, \swift~simultaneously observes with its UV-Optical Telescope (UVOT; \citealt{Roming2005}). We obtained the resulting data from the UK \swift~Science Data Centre\footnote{\url{https://www.swift.ac.uk/index.php}} (UKSSDC) and used \textsc{uvotproduct v2.9}\footnote{As part of \textsc{HEASOFT v6.32} \citep{heasoft}.} to measure the photometry of the afterglow. We used a 5\arcsec~radius circular aperture centred at the positions noted in Section \ref{sec:goto_discoveries_optical} and a detection threshold of $3\sigma$. The measured magnitudes were converted from the UVOT photometric system to AB using the standard UVOT zeropoints \citep{Breeveld11}. The afterglow was detected in at least one epoch for five of the seven sources in our sample. The exceptions are GRB~240122A, where the UVOT FoV did not cover the source position, and GRB~240225B, for which no UVOT data were obtained. The UVOT follow-up observations are listed in Table~\ref{tab:phot_data}, and the corresponding light curves are presented in Figure~\ref{fig:optical_lcs}.

\subsubsection{LT}
\begin{figure*}
\centering
\includegraphics[angle=0,scale=0.45]{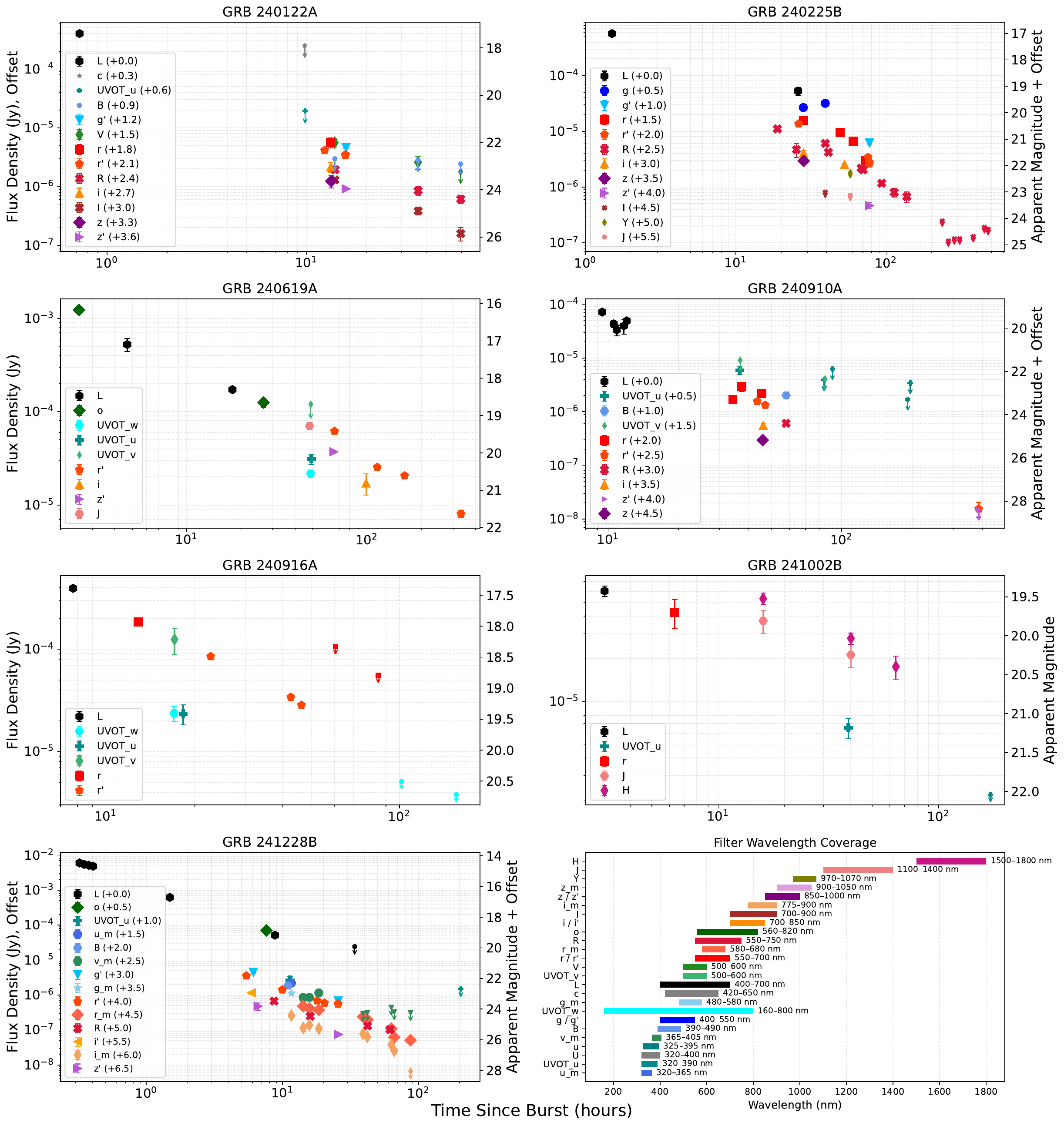}
\caption{UV/optical/NIR afterglow photometry for \grbssample{}, combining our measurements with values compiled from GCN Circulars (see Table~\ref{tab:phot_data}). Left ordinates show flux density (Jy) and right ordinates show AB magnitude; corrected for foreground extinction. Downward arrows indicate non-detections ($3\sigma$ upper limits). To reduce crowding, per-band vertical magnitude offsets are applied as noted in each legend; an offset of $\Delta m$ corresponds to a multiplicative factor of $10^{-0.4\Delta m}$ on the flux axis. Time is measured relative to the trigger ($T{-}T_0$): we adopt the \maxi~$T_0$ for GRBs~240122A and 240225B and the \fermi~$T_0$ for the remaining five bursts. The bottom-right panel summarises the approximate passband coverage (nm) of all UV/optical/NIR filters used.}
\label{fig:optical_lcs}
\end{figure*}

The IO:O Imager at the robotic 2m Liverpool Telescope \citep[LT;][]{Steele2004} located at the international Observatorio del Roque de los Muchachos, La Palma, was triggered for follow-up of GRBs~240122A, 240225B, and 240910A. For GRBs~240122A and 240910A, one epoch was obtained in each of $riz$ bands. For GRB~240225B, one epoch is in $griz$, and three later epochs were secured in $r-$band. All LT data were pre-reduced for bias, dark, and flat-field corrections using the facility pipeline. The photometry was then extracted with the photometry-sans-frustration pipeline \citep[\texttt{psf};][]{Nicholl2023}, making use of its built-in template subtraction. The observations are summarised in Table~\ref{tab:phot_data}, and the corresponding light curves are shown in Figure~\ref{fig:optical_lcs}.

\subsubsection{NOT}
The Nordic Optical Telescope (NOT; \citealt{Djupvik2010}) is a 2.56 m telescope located at the Observatorio del Roque de los Muchachos in La Palma (Canary Islands, Spain). The NOT routinely performs ToO observations of GRB and FXT afterglows. The NOT observed and detected the counterparts of GRBs~240225B,~240619A,~240910A,~240916A, and~241228B using the Alhambra Faint Object Spectrograph and Camera (ALFOSC) optical imager. The reduction of the NOT data follows standard procedures, including bias and flat-field correction. The photometric calibration was computed against the Pan-STARRS catalogue. Details of these measurements are provided in Table~\ref{tab:phot_data}, with their temporal evolution illustrated in Figure~\ref{fig:optical_lcs}.

\subsubsection{VLT/X-shooter acquisition camera}
The X-shooter spectrograph \citep{vernet2011a} mounted on Unit Telescope 3 (Melipal) of the Very Large Telescope (VLT) at Cerro Paranal Observatory was triggered for follow-up of a subset of GRBs in our sample. 

Prior to the spectroscopic observations (see Section~\ref{subsec:spec-vlt-xshoot}), acquisition images were obtained with the acquisition and guiding (A\&G) camera in the Sloan $g^\prime$, $r^\prime$, and $z^\prime$ bands, depending on the target. These images were used to verify the target acquisition and also provide valuable photometric information on the afterglow. The raw frames were reduced using a custom pipeline based on \texttt{ccdproc}~\citep{matt_craig_2017_1069648}, including bias subtraction and flat-field correction. 

Astrometric calibration was applied using \textit{Astrometry.net}~\citep{Lang+2010}, and the images were aligned and stacked where appropriate. Aperture photometry was performed using \texttt{photutils}~\citep{Larry2024}, and the zero-point was calibrated against Pan-STARRS DR2 field stars. A log of the observations is given in Table~\ref{tab:phot_data}, and the resulting light curves are presented in Figure~\ref{fig:optical_lcs}.

\subsubsection{BOOTES}
The Burst Observer and Optical Transient Exploring System (BOOTES)\footnote{\url{https://bootesnetwork.com/}} observed GRB~240122A using the 60\,cm robotic telescope at the BOOTES-2/TELMA station in La Mayora, Málaga, Spain. Observations began on 2024 January 22 at 19:44:56~UT, approximately 9.3~hours after the trigger. A series of 60\,s exposures were obtained with a clear filter. The afterglow was faint and remained undetected in both individual frames and the stacked image. A $3\sigma$ upper limit was derived, as listed in Table~\ref{tab:phot_data}.

\subsubsection{OSN}
The follow-up observations of GRB~240122A are also performed with the 1.5\,m telescope at the Sierra Nevada Observatory (OSN, Granada, southern Spain)\footnote{\url{http://www.osn.iaa.es/}}, targeting the burst position starting on 2024 January 22 at 23:45:57~UT (13.3hr post-trigger). The afterglow was clearly detected during the first night, prompting continued monitoring over the following two nights (January 23 and 24). Observations across all three epochs were performed in the Johnson--Cousins $B$, $V$, $R$, and $I$ bands, with exposure times of 90\,s and 150\,s. The afterglow remained clearly visible in the stacked images. Photometric measurements were obtained via aperture photometry using standard procedures in the \texttt{IRAF} software package \citep{Tody+1986}, following bias subtraction and flat-field correction. Magnitudes were calibrated against nearby reference stars in the field, listed in the SDSS catalogue, using the transformation equations from \citet{Lupton2005}\footnote{\url{http://www.sdss.org/dr4/algorithms/sdssUBVRITransform.htm}}. The resulting magnitudes are reported in Table~\ref{tab:phot_data}, while Figure~\ref{fig:optical_lcs} displays the corresponding light curves.

\subsubsection{1.5m AZT-33IK Mondy}
The AZT-33IK 1.5-meter telescope at the Sayan Solar Observatory (ISTP SB RAS), located near the village Mondy in Buryatia, was triggered for the follow-up observations of GRB~240225B on 2024-02-27 at 13:01:23 UT. A series of 30 images with an individual exposure of 120 seconds was obtained using the Andor NEO CMOS photometer attached to the telescope. The observations were carried out using the Johnson R-filter. Aperture photometry of the stacked image from the entire series yielded a magnitude for the optical afterglow of $R = 19.72 \pm 0.07$ (AB). We continued observations with AZT-33IK until the 2024-03-17 epoch. 

We processed all our observations using the APEX pipelines~\citep{Devyatkin2010, Kouprianov2012, PankovCCIS2022}. The process involved image calibration (dark frame subtraction, flat-fielding, cosmic-ray removal), image quality control, image stacking, and source extraction. The \texttt{apex\_forced\_phot} pipeline was utilised for the forced photometry of the GRB~240225B afterglow on difference images. Image subtraction with the method described in \citealt{TomaneyAJ1996}, was performed by \texttt{apex\_subtract} pipeline, using Pan-STARRS-DR1 survey images as a reference obtained from \texttt{HIPS2FITS} service\footnote{\url{https://alasky.cds.unistra.fr/hips-image-services/hips2fits}} \citep{Boch2020HIPS2FITS}. This step ensured that the underlying host galaxy, presented in the Legacy Survey DR9 with a magnitude of $r \sim 24.2$ and a photometric redshift of $z \sim 0.9$, did not affect the afterglow measurement. 

We note an LS DR10 source at the coordinates RA = 08$^{\mathrm{h}}$33$^{\mathrm{m}}$26$^{\mathrm{s}}$.06 and Dec = +27$^{\circ}$04$^{\prime}$32$^{\prime\prime}$.9 (8$''$ West of the afterglow position) with a magnitude of $r \sim 22.4$ that may affect the photometry in the images with poor seeing. The apparent magnitudes were initially calibrated against three nearby USNO-B1.0 stars (identifiers are 1171-0194062, 1171-0194079, and 1171-0194031) in the Vega system and then converted to the AB system using standard Vega-to-AB magnitude conversion\footnote{\url{https://www.astronomy.ohio-state.edu/martini.10/usefuldata.html}}. The complete record of these observations is compiled in Table~\ref{tab:phot_data}, and their light curves are plotted in Figure~\ref{fig:optical_lcs}.

\subsubsection{LBT}
We obtained late-time $r^\prime z^\prime$ imaging of GRB 240910A with the Large Binocular Cameras \citep[LBCs;][]{Giallongo2008a} mounted on the Large Binocular Telescope (LBT) on Mt. Graham, Arizona, USA (Program ID: IT-2024B-023). LBC imaging data were reduced using the dedicated data reduction pipeline \citep{Fontana2014a}. Aperture photometry was performed via \texttt{IRAF} tools and calibrated against SDSS field stars. Observation logs are reported in Table~\ref{tab:phot_data}, with the associated light-curve behaviour shown in Figure~\ref{fig:optical_lcs}.

\subsubsection{HMT}
The Half-Meter Telescope (HMT) is a 50~cm wide-field telescope, located at Nanshan Station of Xinjiang Astronomical Observatory, Chinese Academy of Sciences. HMT conducted two observations during the night of 2024-02-26 (UT), 18.40 and 24.87\,hr since the GRB~240225B trigger, respectively. Based on the standard data processing of \texttt{IRAF} and aperture photometry, the measured brightness in R-band is listed in Table~\ref{tab:phot_data}, and plotted in Figure~\ref{fig:optical_lcs}.

\subsubsection{TRT}
The Thai Robotic Telescope (TRT) is an automated telescope network comprising four 70~cm CDK700 Telescopes equipped with Andor CCD cameras, distributed in the United States (SRO), Chile (CTO), Australia (SBO), and China (GAO). The telescope located at SBO started observations at 8.7\,hr after the GRB~241228B trigger and obtained $4\times300$\,s frames in the $R-$band. The Johnson-Cousin filters were calibrated with the converted magnitude from the Sloan system\footnote{\url{https://www.sdss4.org/dr12/algorithms/sdssUBVRITransform/\#Lupton}}. The aperture photometry is calibrated with the Pan-STARRS Data Release 2~\citep{Chambers2016, Flewelling+2018} and listed in Table~\ref{tab:phot_data}.

\subsubsection{Altay}
The Altay Telescopes are located at the Altay Observatory, Xinjiang, China, as part of the Altay Astronomical Time-domain Project (also known as JinShan Project). This project consists of four 50~cm telescopes with a FoV of $1.7\degree\times1.7\degree$, which are named from 50A to 50B, two 100~cm telescopes with a FoV of $1.4\degree\times1.4\degree$, which are named 100A and 100B, and one 100~cm telescope with a FoV of $14^\prime\times14^\prime$, which is named 100C. In the early commissioning stage of the project, we triggered the GRB~240910A and GRB~240916A with the 100~cm telescopes using the Sloan $r-$filter. 

The obtained images were processed with the standard \texttt{IRAF} procedures, including bias and dark subtraction, flat correction, and image combination. After the astrometric calibration by \textit{Astrometry.net}~\citep{Lang+2010}, the apparent photometric data were calibrated with the Pan-STARRS Data Release 2~\citep{Chambers2016, Flewelling+2018}. A comprehensive summary of the results is given in Table~\ref{tab:phot_data}, and light curves are shown in Figure~\ref{fig:optical_lcs}.

\subsubsection{1.6m Mephisto}
The 1.6m Multi-channel Photometric Survey Telescope (Mephisto) is a wide-field multichannel telescope \citep{Yuan2020}. It is located at Lijiang Observatory of Yunnan Astronomical Observatories, Chinese Academy of Sciences, and is operated by the South-Western Institute for Astronomy Research, Yunnan University. Equipped with three-channel CCD cameras (blue $uv$, yellow $gr$, and red $iz$ channels), Mephisto can perform simultaneous observations in $ugi_m$ or $vrz_m$ optical bands at a particular moment. The wavelength coverage of the $u_m, v_m, g_m, r_m, i_m$, and $z_m$ filters is 320--365, 365--405, 480--580, 580--680, 775--900, and 900--1050\,nm with central wavelengths at 345, 385, 529, 628, 835, and 944\,nm, respectively \citep[see e.g.,][]{Yang2024ApJ...969..126Y, Cheng2025ApJ}. Presently, the facility is in an advanced stage of commissioning.

Mephisto was triggered to observe GRB~241228B on 2024-12-28 (15:38:16) UT and continued until 2024-12-31. Multiple frames with an exposure time of 300\,s were obtained at different epochs during the follow-up. The pre-processing of raw frames was performed using a specialised pipeline developed for the Mephisto observational data (Fang et al., in prep.). To obtain the instrumental magnitudes of the GRB, Point Spread Function photometry was performed on the stacked images. The corrected Gaia XP low-resolution spectra \citep{2024ApJS..271...13H} were utilised for the photometric calibration. Considering that the Mephisto bands are not fully covered by the corrected Gaia XP spectra (336--1020\,nm), it was extrapolated partially in the $u$ and $z$ Mephisto bands. Each band’s synthetic magnitude in the AB system was calculated by convolving the spectra with the transmission efficiency. The median of the magnitude offset between the instrumental and synthetic magnitudes of the non-variable stars in the field was used to finally calibrate the Mephisto photometric measurements \citep[for details see][]{Chen2024ApJ...971L...2C, ZouKumar2025}. The overall uncertainties in the photometric calibration were constrained to be within 0.03, 0.01, and 0.005 mag in the $u_m$, $v_m$, and $griz_m$ bands, respectively. The detailed dataset is compiled in Table~\ref{tab:phot_data}, and the corresponding light curves are shown in Figure~\ref{fig:optical_lcs}.

\subsection{Spectroscopic Observations}\label{sec:afterglow_spec}
For the GRBs in our sample, we acquired spectra using the X-shooter instrument mounted on the VLT \citep{vernet2011a}, and OSIRIS (Optical System for Imaging and low-Intermediate-Resolution Integrated Spectroscopy) on the Gran Telescopio Canarias (GTC; \citealt{Cepa1998}), except for GRB~241002B, for which we did not get any spectroscopic observation due to scheduling constraints. A complete summary of the VLT/X-shooter and GTC/OSIRIS spectroscopic configurations, exposure details, and observing conditions is provided in Table~\ref{tab:spec_data}, whereas spectra are shown in Figures~\ref{fig:VLT_spec_22A}--\ref{fig:VLT_spec_28B}.

\begin{table*}
    \caption{Log of spectroscopic afterglow observations for the GRBs analysed in this work. Redshift values are included here for completeness; detailed measurement methods behind estimating these values are described in Section~\ref{sec:spec_afterglow_analysis}.}\label{tab:spec_data}
    \addtolength{\tabcolsep}{-2pt}
    \renewcommand{\arraystretch}{1.1}
    \begin{tabular}{ |c|c|c|c|c|c|c|c|c|c|}
    \hline
    GRB & Date-Obs & T-T$_0$ & Telescope & Instrument & Exp. Time & Slit width & Airmass & Seeing & Redshift \\
      &   & (hours) &  &  &  & ($''$) &  & ($''$) &  \\

    \hline
    240122A & 2024-01-22 22:56:23 UT & 12.47 & GTC & OSIRIS & $3\times900$~s & 1.0 & 1.50 & 1.70 & $3.1634 \pm 0.0003$ \\
    240122A & 2024-01-23 02:25:01 UT & 15.95 & VLT & X-shooter & $4\times1200$~s & 1.0$^a$-0.9$^b$ & 1.00-1.01 & 0.45-0.49 & $3.1634 \pm 0.0003$ \\
    240225B & 2024-02-29 01:45:36 UT & 77.52 & VLT & X-shooter & $4\times1200$~s & 1.0$^a$-0.9$^b$ & 1.61-1.65 & 0.66-0.70 &  $0.9462 \pm 0.0002$ \\    
    240619A & 2024-07-02 23:20:52 UT & 331.68 & VLT & X-shooter & $4\times600$~s & 1.0$^a$-0.9$^b$ & 1.90-2.00 & 0.94-0.96 &  $0.3960 \pm 0.0001$ \\
    240910A & 2024-09-12 03:08:33 UT & 47.13 & GTC & OSIRIS & $3\times1200$~s & 1.0 & 1.14 & 0.80 & $1.4605 \pm 0.0007$ \\
    240916A & 2024-09-17 00:08:31 UT & 22.76 & VLT & X-shooter & $4\times600$~s & 1.0$^a$-0.9$^b$ & 1.74-1.99 & 1.07-1.09 &  $2.6100 \pm 0.0002$ \\
    241228B & 2024-12-29 06:00:45 UT & 25.68 & VLT & X-shooter & $4\times1200$~s & 1.0$^a$-0.9$^b$ & 1.17-1.18 & 0.53-0.54 &  $2.6745 \pm 0.0004$ \\
    \hline
    \multicolumn{10}{l}{NOTE: $^a$UVB arm, $^b$VIS and NIR arms.}
    \end{tabular}
\end{table*}

\subsubsection{VLT/X-shooter spectrograph}
\label{subsec:spec-vlt-xshoot}
Spectroscopic observations of GRBs~240122A, 240225B, 240619A, 240916A, and 241228B were performed as part of the ``Stargate'' GRB program at ESO, using the X-shooter spectrograph \citep{vernet2011a}, installed on the ESO VLT UT3 at Cerro Paranal, Chile. X-shooter simultaneously covers the ultraviolet-blue (UVB; 300--560\,nm), visible (VIS; 550--1020\,nm), and near-infrared (NIR; 1020--2100\,nm) wavelength ranges, with resolving powers of $\lambda/\Delta\lambda = 5400$, $8900$, and $5600$, respectively. Observations were carried out in the ABBA nodding mode along the slit to enable effective subtraction of the sky emission, especially in the NIR. In addition, a K-band blocking filter was employed to reduce thermal background contamination in the NIR. The data were reduced in STARE mode using calibration files from the night and the standard ESO X-shooter pipeline \citep{goldoni2006a, modigliani2010a}, which performs bias and dark correction, flat-fielding, wavelength calibration via arc lamps, and flux calibration based on standard star observations. Following the method described in \citet{selsing2019a}, individual reduced exposures were directly co-added for the UVB and VIS arms, while A–B nod pairs were pair-subtracted prior to combination for the NIR arm. All reported wavelengths are given as observed in vacuum and corrected for the barycentric motion of the Earth.

The X-shooter observations of GRB~240122A began on 2024 January 23 at 02:25:01 UT ($T_0 + 15.95$\,hr) under excellent seeing conditions of $0.49^{\prime\prime}$ \citep{Saccardi2024GCN35599}. GRB~240225B was observed on 2024 February 29, starting at 01:45:36 UT ($T_0 + 3.23$\,d), with a seeing of 0.66$^{\prime\prime}$ \citep{Schneider2024GCN35832}. We observed GRB~240619A on 2024 July 02 starting at 23:20:52 UT ($T_0 + 13.82$\,d); our observations targeted the catalogued galaxy PSO J162.3946+17.2828 in spatial coincidence with the optical afterglow and were obtained with a seeing of $0.94^{\prime\prime}$ \citep{Cotter2024GCN36813}. Spectroscopic observations of GRB~240916A were conducted on 2024 September 16, beginning at 00:08:31 UT ($T_0 + 22.76$\,hr), under a seeing of $1.07^{\prime\prime}$ \citep{Pieterse2024GCN37532}. Finally, X-shooter observations of GRB~241228B were carried out on 2024 December 29, starting at 06:00:45 UT ($T_0 + 1.07$ d), with a seeing of $0.54^{\prime\prime}$ \citep{An2024GCN38704}.

In all, for GRBs~240122A, 240225B, 240619A, 240916A, 241228B observed with VLT/X-shooter, the start times of the observations spanned across the sample from $T_0+15.95$~hr to $T_0+13.82$~d, seeing ranged from $0.49^{\prime\prime}$ to $1.07^{\prime\prime}$, and each target was obtained in four exposures per arm with per-exposure times of either 600~s or 1200~s; for GRB~240619A the slit also encompassed a nearby second galaxy. The full spectroscopic observing log, along with observing conditions, is listed in Table~\ref{tab:spec_data}. The final column lists the precise redshifts estimated for these GRBs, derived from the analysis of their afterglow spectra. These values were determined through the identification of absorption and/or emission features associated with the host galaxies. The full methodology, including line identification, fitting procedures, and associated uncertainties, is described in detail in Section~\ref{sec:spec_afterglow_analysis}.

\subsubsection{GTC/OSIRIS}
Spectroscopic observations of GRBs~240122A and 240910A were performed using the OSIRIS mounted on the 10.4\,m GTC \citep{Cepa1998} at the Observatorio del Roque de los Muchachos (ORM), La Palma (see full spectroscopic log in Table~\ref{tab:spec_data}). The observations were obtained as part of GTC programs GTCMULTIPLE2J-23B and GTCMULTIPLE4G-24B (PI: J. F. Ag\"u\'i Fern\'andez). For both targets, the R1000B grism was used in long-slit spectroscopy mode (LSS) with a slit width of $1.0^{\prime\prime}$ and binning of $2 \times 2$ pixels, providing a resolving power of $R \sim 600$ and a wavelength coverage of 3650–7800\,\AA.

The data were acquired in a sequence of three individual exposures, nodding along the slit to cancel the effect of possible artefacts or defects and provide a clean, final reduced product. The OSIRIS spectrum of GRB~240122A was began on 2024 January 22 at 22:56:23 UT ($T_0 + 12.47$\,hr), approximately 3.5\,hr prior to the VLT/X-shooter observations \citep{Thoene2024GCN35598}, , with $3\times900$\,s exposures. The GTC/OSIRIS spectrum of GRB~240910A was started on 2024 September 12 at 03:08:33 UT ($T_0 + 47.13$\,hr), using $3\times1200$\,s exposures, under good observing conditions (airmass $\sim1.14$ and seeing of $0.8^{\prime\prime}$, \citealt{Postigo2024GCN37467}).

Data reduction was carried out using a combination of \texttt{IRAF}-based tasks and custom Python scripts developed for OSIRIS, which included bias subtraction, flat-fielding, wavelength calibration using arc lamps, and flux calibration using spectrophotometric standards. Accurate one-dimensional spectra were extracted using optimal extraction techniques and corrected for instrumental response across the full wavelength range. Later on, Section~\ref{sec:spec_afterglow_analysis} details the methodology, including line identification, fitting, and uncertainty estimation.

\subsection{Radio}
We observed radio afterglows of GRBs~240122A and 240910A utilising the Australia Telescope Compact Array (ATCA) and the Karl G. Jansky Very Large Array (VLA), respectively. Furthermore, of the GRBs in our sample, GRBs~240225B and 241002B had no radio observations. GRB~240619A was detected in the radio at 15.5 GHz at $\sim$3.4 days post-burst using the Arcminute Microkelvin Imager Large Array (AMI-LA; \citealt{Rhodes2024GCN36744}). GRB~240916A was observed with the VLA at central frequencies of 6, 10, and 15~GHz, yielding surface peak brightnesses of 35, 44, and 135~$\mu$Jy/beam, respectively \citep{Giarratana202437788}. The observation details are tabulated in Table~\ref{tab:radio}.

\subsubsection{ATCA}
Following its optical localisation, GRB 240122A was observed with the ATCA under the PanRadio GRB programme C3542 (PI: Anderson) on 2024 January 24, 26, 28, and February 12. This program aims to perform high-cadence multi-frequency radio monitoring of a large sample of LGRBs in the southern hemisphere (Declinations $<-10$ deg) between minutes to years after the burst to explore the evolution and properties of their afterglows (Leung et al., in prep.; Anderson et al., in prep.). GRB 240122A was observed with a wide range of frequencies centred on  5.5, 9.0, 16.7, and 21.2\,GHz, each with a 2048 MHz-wide band. We reduced the visibility data using standard routines in \texttt{MIRIAD} \citep{1995ASPC...77..433S}. We used a combination of manual and automatic RFI flagging before calibration, conducted with \texttt{MIRIAD} tasks \mbox{} \texttt{uvflag} and \texttt{pgflag}, respectively. We used PKS~1934$-$63 to determine the bandpass response and to calibrate the flux density scale for all frequency bands. We used PKS~0607$-$157 to calibrate the time-variable complex gains for all epochs and frequency bands. After calibration, we inverted the visibilities using a robust weighting of 0.5 and then used the CLEAN algorithm \citep{1980A&A....89..377C} to the target source field using standard \texttt{MIRIAD} tasks \texttt{INVERT}, \texttt{CLEAN}, and \texttt{RESTOR} to obtain the final images. For each observation, we measure the flux density of a detected source by fitting a point-source model to the restored image using the {\sc Miriad} task {\sc imfit} and report a non-detection using the rms sensitivity obtained from the residual image. The $1\sigma$ errors reported are purely statistical, as the systematic errors are expected to be much smaller \citep[$\lesssim 5$\%; e.g.,][]{Reynolds1994,Tingay2003}. We detected the radio counterpart at 9~GHz on both 2024 January 24 and 28 at a position consistent with the GOTO optical counterpart \citep{Kumar2024GCN35596}. For all other frequencies, we estimated $3\sigma$ upper limits (see Table~\ref{tab:radio}). 

\subsubsection{VLA}
We observed GRB~240910A with the VLA 3.1 (2024 September 13, \citealt{Giarratana2024GCN37569}), 9.1 (2024 September 19), 21.3 (2024 October 1) and 46.2 (2024 October 26)~days post-burst (Project code: SF171028) at the central frequencies of 6 (C band), 10 (X band) and 15\,GHz (Ku band), with a bandwidth of 4, 4 and 6\,GHz, respectively. The VLA source J0125$-$0005 was used as a phase calibrator. The distance between the target and the phase calibrator was about 2.7$^{\circ}$. Each observation included scans on the flux and bandpass calibrator 3C48. The data were calibrated using the custom \textsc{casa} pipeline \citep[Version 6.5.4;][]{McMullin2007} and visually inspected for possible radio frequency interference. The final images were produced with the \texttt{tclean} task in \textsc{casa} (Version 6.5.4) using a Briggs weighting scheme (robust $=0.5$). Results from the campaign are reported in Table~\ref{tab:radio}. The GRB is detected at all frequencies during the first two epochs, and at 6\,GHz it is also detected in the third epoch. The maximum flux densities, measured in the first epoch, are $137 \pm 10$, $114 \pm 9$, and $86 \pm 10~\mu$Jy at 6, 10, and 15~GHz, respectively. For each detection, the flux density was measured by fitting a Gaussian to the cleaned image using the \texttt{imview} task in \textsc{casa}. The final flux density error was estimated as the squared sum of the root mean square (RMS) and a typical 5\% accuracy for the amplitude scale calibration. Upper limits are reported with a $5\sigma$ confidence level.

\begin{table*}
\caption{Results from fitting the prompt emission spectra of \maxi/GSC and \fermi/GBM detected GRBs. The models used are those preferred based on the Akaike Information Criterion (AIC, \citealt{Akaike1974}). In order to maximise the number of measured peak energies, the cutoff power-law (CPL) and Band \citep{Band1993} models were chosen when AIC was agnostic between them and the PL model. Hardness ratios (HR) are calculated using the 50--300~\keV{} and 10--50~\keV{} bands. HR values for \maxi/GSC bursts are from extrapolating the 2--20~keV fits to these energies.} \label{tab:fermi_prompt}
\addtolength{\tabcolsep}{-1pt} 
\renewcommand{\arraystretch}{1.3} 
\begin{tabular}{|c|c|c|c|c|c|c|c|c|c|c|}
\hline
GRB & $z$ & \tninty & Model & Fluence (2 -- 20\,keV) & $\alpha$ & $\beta$ & $E_{\rm p}$ & PG-Statistic & DoF & HR \\
(\maxi/GSC) & & (s) & & ($10^{-7}$\,erg\,cm$^{-2}$) & & & (keV) & & & \\
\hline
240122A & $3.163$ & $\approx 36$ & PL & $1.26^{+0.47}_{-0.94}$ & $1.9 \pm 0.4$ & -- & -- & -- & -- & $1.32^{+1.30}_{-0.65}$ \\
240225B & $0.946$ & $\approx 21$ & PL & $10.7^{+0.63}_{-1.26}$ & $1.9 \pm 0.1$ & -- & -- & -- & -- & $1.32^{+0.25}_{-0.21}$ \\

\hline  \hline

GRB & $z$ & \tninty & Model & Fluence (10 -- 1000\,keV) & $\alpha$ & $\beta$ & $E_{\rm p}$ & PG-Statistic & DoF & HR \\
(\fermi/GBM) & & (s) & & ($10^{-5}$\,erg\,cm$^{-2}$) & & & (keV) & & & \\
\hline

240619A & $0.396$ & $36.13 \pm 0.59$ & Band & $1.29_{-0.08}^{+0.05}$ & $1.36^{+0.14}_{-0.17}$ & $1.65^{+0.09}_{-0.04}$ & $149.5_{-143.5}^{+423.7}$ & 241.14 & 436 & $2.28_{-0.07}^{+0.06}$ \\
240910A & $1.460$ & $272.39 \pm 2.61$ & Band & $2.16_{-0.24}^{+0.18}$ & $1.25^{+0.08}_{-0.05}$ & $2.81^{+2.83}_{-0.44}$ & $113.3_{-22.7}^{+27.5}$ & 271.38 & 324 & $1.81_{-0.07}^{+0.06}$ \\
240916A & $2.610$ & $32.00 \pm 0.81$ & Band & $2.51_{-0.15}^{+0.11}$ & $1.11^{+0.04}_{-0.06}$ & $2.16^{+0.32}_{-0.14}$ & $665.6_{-186.1}^{+231.0}$ & 231.02 & 325 & $4.26_{-0.11}^{+0.08}$ \\
241002B & -- & $64.26 \pm 4.38$ & PL & $0.43_{-0.07}^{+0.06}$ & $1.97^{+0.06}_{-0.08}$ & -- & -- & 219.85 & 326 & $1.18_{-0.11}^{+0.09}$ \\
241228B & $2.674$ & $19.46 \pm 0.36$ & CPL & $4.19_{-0.13}^{+0.10}$ & $0.81^{+0.03}_{-0.02}$ & -- & $350.6_{-22.9}^{+17.9}$ & 310.74 & 325 & $5.72_{-0.07}^{+0.06}$ \\
\hline
\end{tabular}
\end{table*}

\section{Prompt Emission Analysis and Properties}\label{sec:prompt_prop}
The prompt gamma-ray emission encodes the immediate output of the central engine and provides key diagnostics of the physical conditions in the relativistic outflow. For the seven GRBs in our sample, we analysed \maxi/GSC and \fermi/GBM data to measure basic spectral and temporal properties, including the photon indices, peak energies ($E_{\rm p}$), isotropic-equivalent gamma-ray energies ($E_{\rm iso}$), and \tninty; analyses using other high-energy instruments discussed in Section~\ref{sec:High-Energy_Triggers} are beyond the scope of this work. These quantities are critical for placing the bursts in the broader GRB population, identifying any outliers, and examining potential links between the prompt emission and the optical afterglows recovered by GOTO. In particular, we aim to investigate whether the unusually hard spectra and high $E_{\rm p}$ values observed in several cases are connected to the optical detectability of these poorly localised events.

\subsection{Prompt Emission Analysis}\label{sec:prompt_analysis}
Here, we describe the methods used to extract and analyse the prompt emission properties for each GRB, using data from the relevant high-energy instruments. The analysis is divided into two parts: \maxi/GSC events and the \fermi/GBM events.

\subsubsection{GRBs~240122A and 240225B with \maxi/GSC}

We analysed the prompt emission of GRBs 240122A and 240225B with archival data of \maxi/GSC using High Energy Astrophysics Software (\textsc{HEASOFT}\footnote{\url{https://heasarc.gsfc.nasa.gov/docs/software/heasoft/}}). X-ray events of \textit{gsc\_med} type are processed with \textsc{mxproduct}. Because light curves and spectra produced by \textsc{mxproduct} are not suitable for short and variable transients like GRBs, we performed an additional step to extract light curves with a 1-second time resolution and applied effective area correction. Note that the process is identical to the process used for the \maxi~GRB catalogue.

Figure~\ref{fig:MAXI_LCs} shows the GSC light curves of GRBs 240122A and 240225B in the 2--20, 2--4, 4--10, and 10--20~keV energy bands. We estimated the \tninty duration in the 2--20~keV band to be $\approx36$~s and $\approx21$~s for GRBs 240122A and 240225B, respectively. For GRB~240225B, the duration was calculated using data from a single \maxi/GSC scan. This value differs from those obtained by other instruments because the scan began at 20:15:30~UT, about 200~s later than the trigger times reported by the others.

Then, we extracted a spectrum of the \tninty interval and corrected for variations in the effective area. The spectra of GRBs 240122A and 240225B are fit with a single power-law model, and the photon indices are found to be $1.9\pm0.4$ and $1.9\pm0.1$ ($1\sigma$ error), respectively. The energy flux in 2--20~\keV{} was $3.5_{-2.6}^{+1.3} \times $10$^{-9}$ erg cm$^{-2}$ s$^{-1}$ and $5.1_{-0.6}^{+0.3} \times $10$^{-8}$ erg cm$^{-2}$ s$^{-1}$. Figure~\ref{fig:maxi-GBM_Specs}  shows the spectrum with the best-fitting model. The results of each fit are shown in Table~\ref{tab:fermi_prompt}.

\subsubsection{GRBs~240619A, 240910A, 240916A, 241002B, and 241228B with \fermi/GBM}

We analysed the prompt emission of GRBs~240619A, 240910A, 240916A, 241002B, and 241228B using the \fermi/GBM data available from the High Energy Astrophysics Science Archive Research Center (\textsc{HEASARC}\footnote{\url{https://heasarc.gsfc.nasa.gov}}) archive \citep{Gruber2014, vonKienlin2014, Bhat2016, vonKienlin2020}. Using the \textsc{HEASOFT} and the \fermi Gamma-ray Data Tools \citep{GDT-Fermi}, we took the time-tagged event (TTE) data in the \tninty interval for each burst to use as the source. We used the data from the brightest NaI detectors and the corresponding BGO detectors. These were NaI 0, NaI 1, NaI 2 and BGO 0 for GRB 240619A; NaI 8, NaI 11, and BGO 1 for GRB 240910A; NaI 3, NaI 4, and BGO 0 for GRB 240916A; NaI 10, NaI 11, and BGO 1 for GRB 241002B; and NaI 6, NaI 7, and BGO 1 for GRB 241228B. The background was modelled in the standard way, using a polynomial function fit to the CSPEC data for each burst. Polynomial order increased until the reduced fit statistic was $ < 1.15 $, resulting in the models fitting each GRB suitably well. Once the background model was obtained, we interpolated it across the source interval and exported this as the background to be used for spectral analysis. Additionally, we extracted the necessary response files. 

We performed our analysis of the spectra using PyXspec \citep{PyXspec} using three models of varying complexity: a simple power-law (PL), which measures a photon index, $\alpha$; a power-law with a high-energy exponential cutoff (CPL), which measures $\alpha$ and the spectral peak energy, $E_{\rm p}$; and the Band model \citep{Band1993}, which measures two photon indices $\alpha$ and $\beta$, smoothly connected at a characteristic break energy, $E_c$. This break energy is converted to a peak energy using $E_p = E_c(2-\alpha)$. We used the PG-Statistic in our analysis, which is appropriate for Poisson data with a Gaussian background. The results of each fit are shown in Table~\ref{tab:fermi_prompt}. Figures~\ref{fig:fermi-GBM_LCs} and \ref{fig:fermi-GBM_Specs} present the \fermi/GBM observations of GRBs 240619A, 240910A, 240916A, 241002B, and 241228B, showing the light curves and the corresponding fitted spectra, respectively.

\subsection{Prompt Emission Properties}

\begin{figure}
    \centering
    \includegraphics[width=1.0\linewidth]{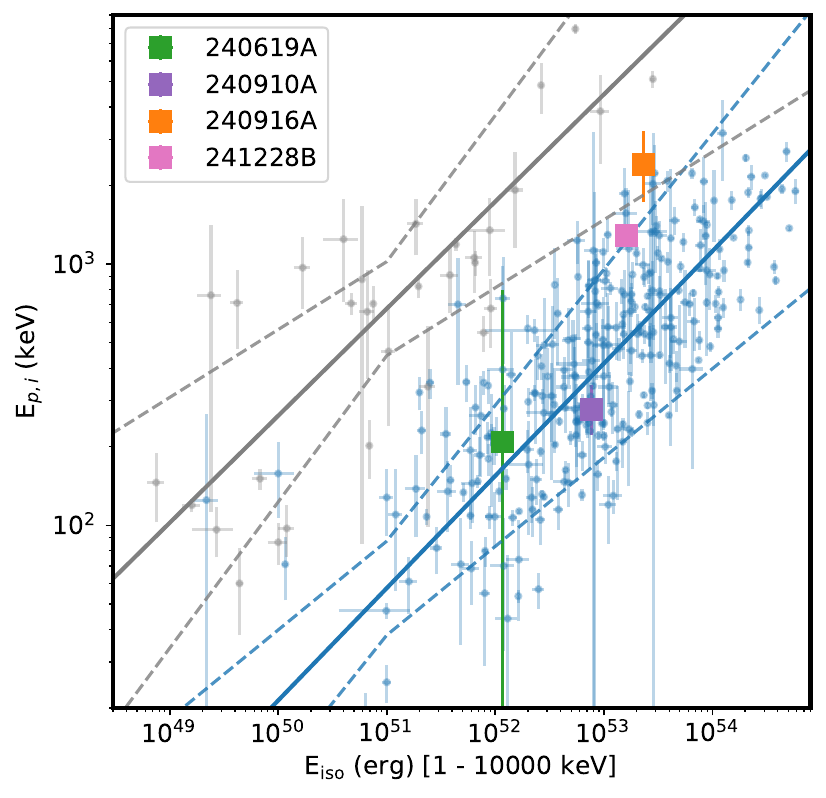}
    \caption{\fermi~GRBs in our sample plotted in the Amati plane \citep{Amati02,Amati06}, showing the relationship between the intrinsic (redshift corrected) peak energy ($E_{p,i}$) and the isotropic-equivalent gamma-ray energy release ($E_{iso}$). Lines show the best-fitting correlation (solid) and their $3\sigma$ bounds (dashed) for the long (blue) and short (grey) GRB populations. Correlation fits and comparison data are from \citet{Minaev20}.}
    \label{fig:Ep_Eiso}
\end{figure}

We plot the four of five \fermi~bursts where it was possible to measure both $E_{\rm p}$ and $E_{iso}$ (see Table~\ref{tab:fermi_prompt}) on the Amati plane \citep{Amati02,Amati06} in Figure~\ref{fig:Ep_Eiso}. GRBs~240916A and 241228B appear to show unusually high redshift-corrected peak energies ($E_{p,i}$) relative to their $E_{iso}$ measurements, and are inconsistent with the Amati relation at the $3\sigma$ level. Both bursts lie at the high end of the $E_{iso}$ distribution, indicating intrinsically powerful GRBs. In addition to this, the high energy photon index ($\beta$) measured for GRB~240619A is less than 2, indicating that the true peak of the spectrum is at an even higher energy. This may make 240619A an outlier of the Amati relation too. Only GRB~240910A appears to be typical in terms of its measured prompt properties.

\begin{figure}
    \centering
    \includegraphics[width=1.0\linewidth]{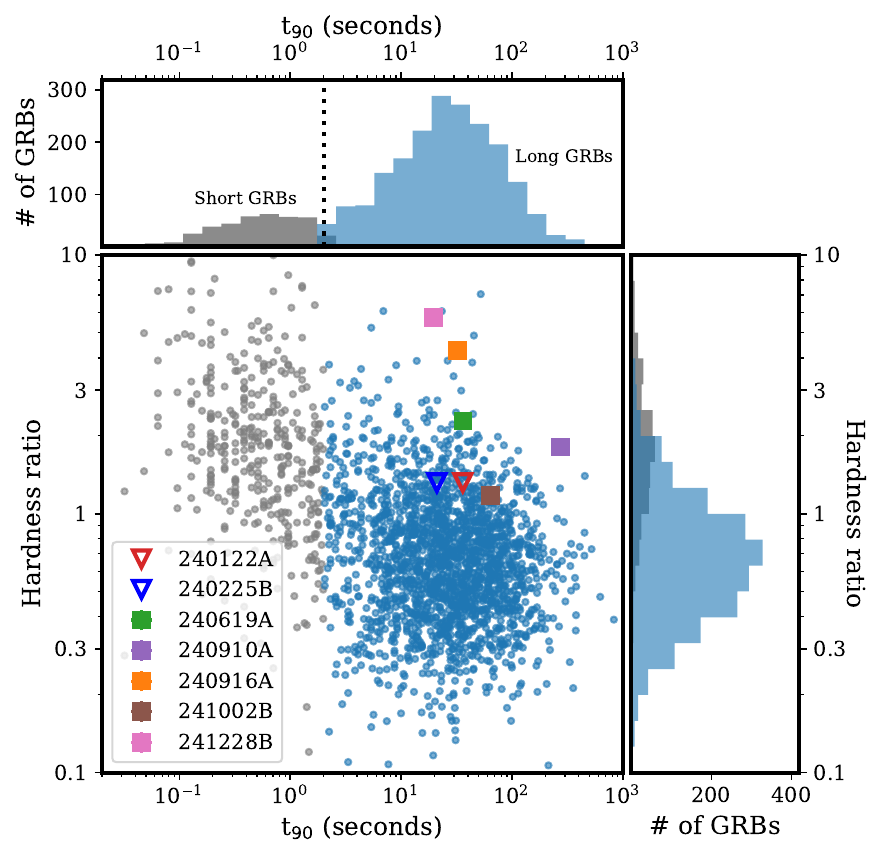}
    \caption{Hardness ratio vs \tninty for the five \fermi/GBM bursts in our sample. The hardness ratio is the 50 -- 300\,keV fluence over the 10 -- 50\,keV fluence. Comparison data is taken from \citet{vonKienlin2020}.}
    \label{fig:HR_t90}
\end{figure}

We also plot all five \fermi~and two \maxi~GRBs on an HR -- \tninty diagram \citep[cf.][]{Kouveliotou1993}, see Figure~\ref{fig:HR_t90}. It is immediately apparent that the \fermi~GRBs are unusually spectrally hard. Only GRB~241002B (best fit with a power-law model with a photon index of $\alpha = 1.97^{+0.08}_{-0.06}$) sits within the main `cloud' of LGRBs. The others are either longer in \tninty (GRB~240910A), or driven to high HR by their abnormally hard (for collapsar GRBs) photon indices (GRBs~240619A, 241228B) and/or high $E_{\rm p}$ (GRB~240916A). The \maxi~GRBs also sit at the high end of the GBM HR distribution, but these values are obtained by extrapolating the 2--20\,keV spectral fits to 300\,keV, and so should be considered upper limits because we have no constraints on any breaks in the spectrum.

\begin{figure}
\centering
\includegraphics[angle=0,scale=0.57]{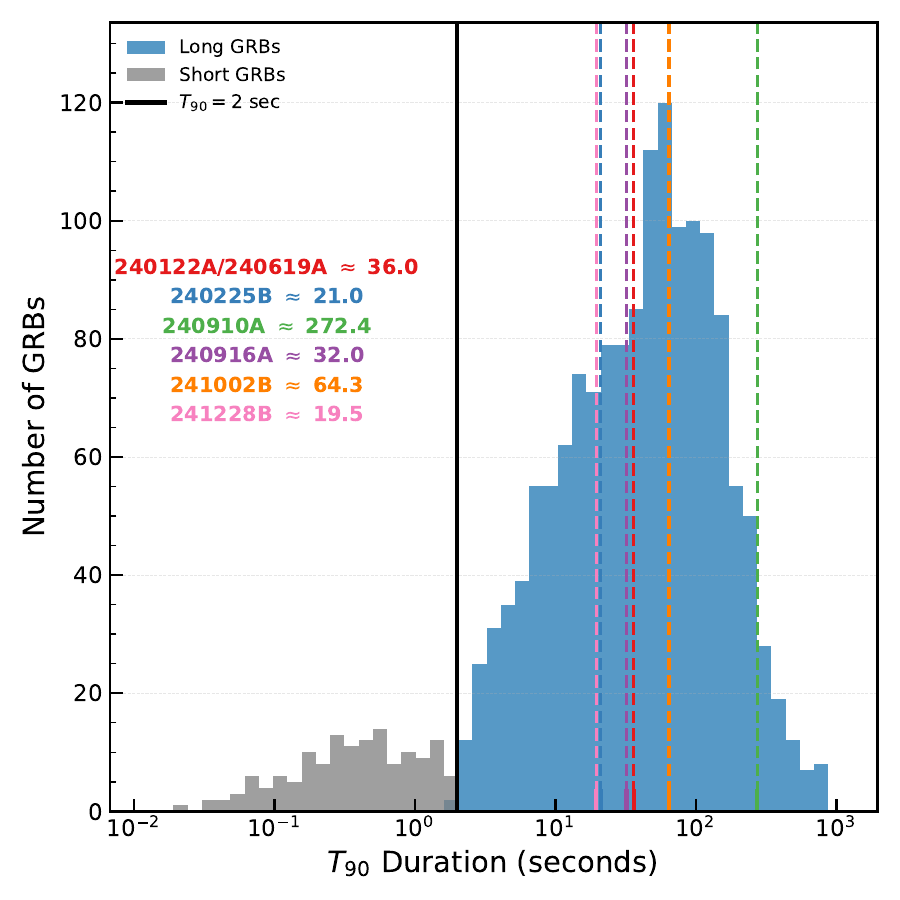}
\caption{Distribution of \tninty durations for a sample of GRBs from the BAT catalogue, divided into long (blue) and short (grey) populations. The dashed vertical lines mark the durations of GOTO GRBs, labelled with their respective names and \tninty values. All GOTO GRBs fall in the long-duration class and cover a wide range of durations.}
\label{fig:T90_comp}
\end{figure}

Given that the GRBs in our sample were selected based on the recovery of their optical afterglows by a relatively shallow telescope like GOTO, it is perhaps not surprising that they're outliers with respect to the wider sample of GRBs. GRBs~240916A and 241228B both appear to be straightforward cases of intrinsically powerful GRBs, as evidenced by their higher than average $E_{iso}$. However, their measured $E_{p,i}$ is high relative to the Amati relation even when accounting for the large $E_{iso}$.

In contrast, GRB~240619A appears to be a much more energetically typical GRB, but at a low redshift ($z = 0.3960$). The higher hardness ratio in this case may simply be the result of less redshifting than the majority of the detected population, resulting in a harder portion of the synchrotron spectrum falling in the 10--300\,keV bandpass than usual. GRB~240910A also appears to be more typical energetically. In this case, a longer central engine duration ($T_{90} = 272.39 \pm 2.61$\,s) may be responsible for producing the bright afterglow. 

All four of the above GRBs (240619A, 240910A, 240916A and 241228B) have measured low-energy photon indices of $\alpha < 1.5$. This indicates that the low-energy synchrotron break was likely in the bandpass, causing $\alpha$ to be an unusually hard blend of two portions of the synchrotron spectrum \citep[see, e.g.,][]{Ravasio18, Ravasio19}. The positions of these breaks are functions of the underlying physical parameters of the jet, and unusual parameter values may be responsible for the abnormally high hardness ratio. An alternative explanation is that the unusually hard prompt spectra in GRB 240916A and GRB~241228B may reflect jet--stellar-envelope interactions, supported by structured-jet and radiative-transfer simulations of LGRBs \citep{Lazzati2005,Lazzati2013, Lundman2013} and even low-luminosity jets \citep{Aloy2018}.

\subsection{\texorpdfstring{$T_{90}$}{T90} comparison}
The \tninty{} duration is a key parameter for classifying GRBs into long and short populations, with a conventional threshold at $T_{90} = 2$~s. Figure~\ref{fig:T90_comp} shows the distribution of \tninty{} values from the BAT GRB catalogue\footnote{\url{https://swift.gsfc.nasa.gov/results/batgrbcat/index_tables.html}}, plotted on a logarithmic scale and separated into long (blue) and short (grey) classes. The histogram illustrates the well-known bimodality in the GRB population, with the majority of events falling in the long-duration category.  

Overlaid on this distribution are the durations of the seven GOTO-detected GRBs in our sample, marked with vertical dashed lines and annotated with their names and \tninty values. All seven events lie securely within the long-duration class, with durations ranging from $\sim20$~s to $\sim272$~s. GRB~241228B ($19.5$~s) and GRB~240225B ($21$~s) sit at the lower end of the LGRB population, while GRBs~240122A, 240619A, 240916A, and 241002B ($32-64$~s) are closer to the peak of the LGRB distribution. GRB~240910A ($272$~s) lies toward the higher end, placing it among the longest events in the Swift sample.

\begin{table*}
\centering
\caption{Summary of \swift/XRT afterglow spectral and temporal properties for our GRB sample. Exp.~denotes the spectral extraction time interval. $N_{\rm H,Gal}$ is fixed to the Galactic line-of-sight value, while $N_{\rm H,int}$ is the intrinsic absorption component derived from spectral fitting. Fluxes are unabsorbed values in the 0.3--10.0~keV band.  $\alpha_X$ is the X-ray temporal index, where negative values indicate apparent rising trends likely caused by low-count statistics.}
\label{tab:XRT_summary}
\addtolength{\tabcolsep}{5pt}
\renewcommand{\arraystretch}{1.6}
\begin{tabular}{|c|c|c|c|c|c|c|c|c|}
\hline
GRB & $T - T_0$ & $\alpha_X$ & $\Gamma$ & $N_{\rm H,g}$ & $N_{\rm H,intr}$ & Flux$_{\rm 0.3-10\,keV}$ & Exp. & $z$ \\
    & ($10^{3}$ s) &       &         & ($10^{22}$ cm$^{-2}$) & ($10^{22}$ cm$^{-2}$) & ($10^{-12}$ erg cm$^{-2}$ s$^{-1}$) & (s) & \\
\hline
240122A & 29.5   & $2.50^{+0.70}_{-2.10}$ & $1.88^{+0.31}_{-0.24}$ & 0.105  & $\sim$0.0 & $6.54^{+1.41}_{-1.16}$ & 926.5 & 3.163 \\
240225B & 461.7  & $1.09^{+0.16}_{-0.13}$ & $2.19^{+0.72}_{-0.59}$ & 0.040  & $0.29^{+0.49}_{-0.29}$ & $12.6^{+6.5}_{-3.2}$ & 289.7 & 0.946 \\
240619A & 174.0    & $-0.94^{+3.69}_{-0.05}$ & $1.61^{+0.45}_{-0.71}$ & 0.028  & $0.038^{+0.034}_{-0.038}$ & $1.23^{+0.54}_{-0.36}$ & 1978.0 & 0.396 \\
240910A & 128.3  & $1.10^{+0.40}_{-0.30}$ & $1.75^{+2.22}_{-0.61}$ & 0.027  & $0.006^{+0.005}_{-0.006}$ & $0.69^{+0.49}_{-0.27}$ & 1983.0 & 1.460 \\
240916A & 61.0     & $1.29^{+0.34}_{-0.23}$ & $2.90^{+0.80}_{-0.60}$ & 0.140  & $2.17^{+2.63}_{-1.92}$ & $4.34^{+3.92}_{-1.52}$ & 1643.0 & 2.610 \\
241002B & 134.5  & $1.10^{+0.40}_{-0.30}$ & $1.82^{+1.06}_{-0.70}$ & 0.030  & $0.096^{+0.251}_{-0.096}$ & $1.34^{+0.68}_{-0.37}$ & 2702.0 & --\\
241228B & 40.0   & $-1.28^{+4.49}_{-0.03}$ & $1.44^{+0.46}_{-0.32}$ & 0.003 & $< 2.37$ & $2.81^{+1.65}_{-0.69}$ & 1809.0 & 2.674 \\
\hline
\end{tabular}
\end{table*}

\section{Afterglow Analysis and Properties}\label{sec:analysis_afterglow}
In this section, we analyse the afterglow properties of the GOTO-discovered GRBs across X-ray, UV and optical (photometric and spectroscopic), and radio wavelengths. We characterise their temporal and spectral behaviour, and place the results in the context of the broader GRB afterglow population.

\subsection{X-ray}\label{sec:X-ray}
\begin{figure}
\centering
\includegraphics[angle=0,scale=0.47]{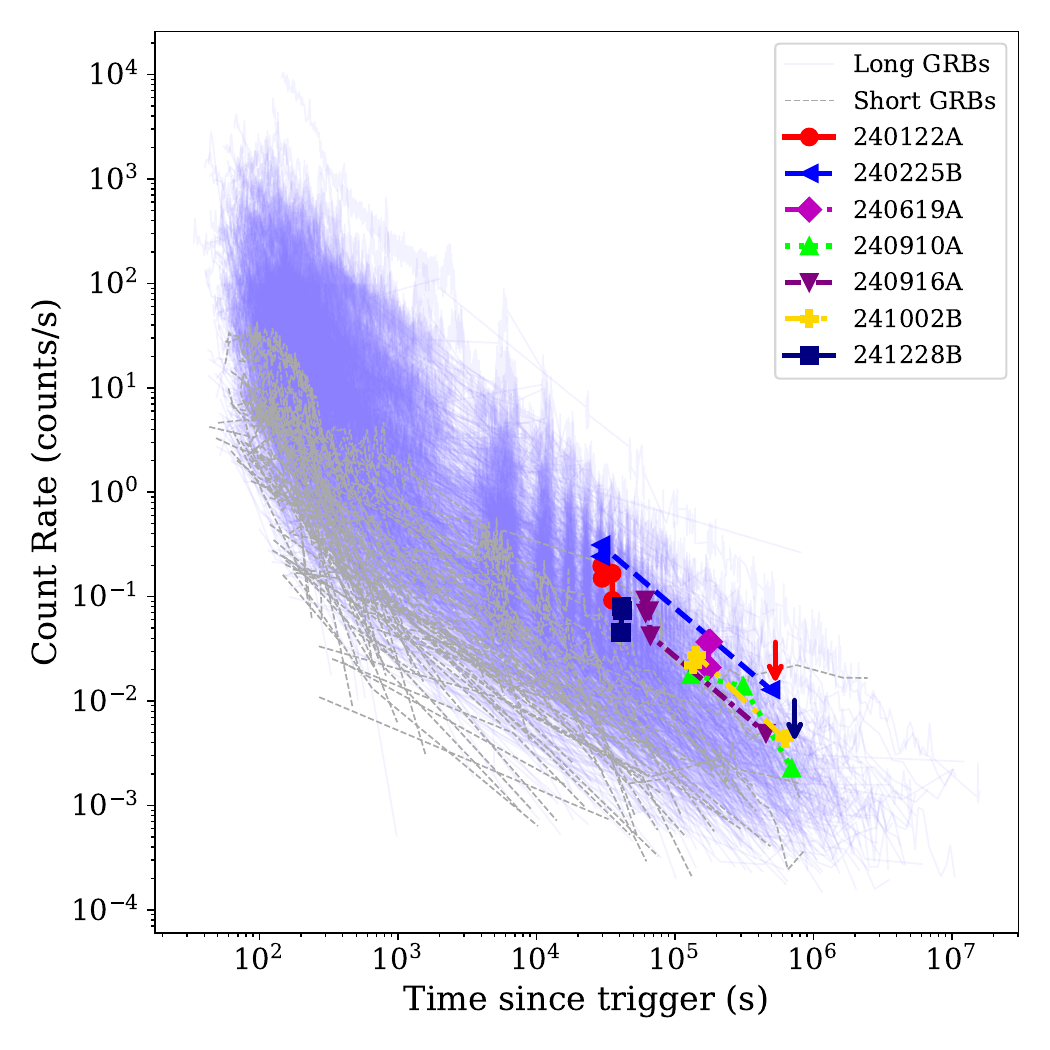}
\caption{\swift/XRT 0.3--10\,keV light curves. Archival LGRBs (grey) and SGRBs (blue dashed) are plotted as background. The seven GOTO GRBs are shown with distinct colours and styles.}
\label{fig:AG_comp_XRT}
\end{figure}

Each \swift/XRT-detected source confirmed its association with the GOTO optical GRB counterpart on the basis of spatial coincidence and, in most cases, temporal fading. For time-domain analysis, we utilised the automated \swift/XRT light curve fits from the UKSSDC pipeline (\citet{Evans2007,Evans2009}). The X-ray light curves were adequately described by single power-law decays (see Figure~\ref{fig:AG_comp_XRT} for the light curves in counts per second).

Spectra for both source and background regions, along with corresponding ancillary and response files extracted in PC mode from the \swift/XRT repository, were grouped to a minimum of one count per bin using \texttt{grppha} task. Spectral fitting was performed in \texttt{XSPEC} using Cash statistics \citep{Cash1979},  appropriate for low-count Poisson-distributed data. Each spectrum was modelled with a simple absorbed powerlaw, \texttt{tbabs*ztbabs(zpowerlw)} \citep{Wilms2000}, accounting for both Galactic foreground and intrinsic absorption in the host galaxy. Galactic column density ($N_{\rm H,g}$) for each GRB was fixed at a value obtained from \swift~Galactic $N_{\rm H}$ tool\footnote{\url{https://www.swift.ac.uk/analysis/nhtot/}} \citep{Willingale2013}, while the intrinsic host galactic absorption ($N_{\rm H,intr}$) and photon index ($\Gamma$) were left free to vary. All fits were performed in the 0.3--10.0\,keV energy range, and the errors reported here correspond to 90\% confidence intervals. Figure~\ref{fig:swift-XRT_Spec} presents the 0.3--10.0\,keV XRT count rate spectra for all seven GRBs in our sample, overlaid (red solid line) with their respective best-fitting absorbed powerlaw models.

Table~\ref{tab:XRT_summary} summarises the temporal and spectral properties derived from \swift/XRT observations of our GRB sample. Redshifts used for spectral fitting are reported in later sections. Photon indices lie in the range $\Gamma \simeq 1.4 - 2.9$, consistent with typical afterglow spectra. The intrinsic absorption shows substantial variation: GRBs~240122A and 240910A are consistent with negligible additional absorption beyond the Galactic foreground, whereas others, most notably GRB~240916A, require higher column densities of the order of $10^{22}$\,cm$^{-2}$. The inferred unabsorbed 0.3--10\,keV fluxes span nearly two orders of magnitude, from $\sim7\times10^{-13}$ to $\sim1.3\times10^{-11}$\,erg\,cm$^{-2}$\,s$^{-1}$, with the highest values observed for GRBs 240225B and 240122A, despite their relatively short spectral extraction intervals. Temporal decay slopes cluster around $\alpha_X \approx 1.0 - 1.3$ for most afterglows, though three cases (GRBs~240619A, 240910A, and 241228B) show formally flat or rising indices; in each case, the uncertainties are large and the behaviour is consistent with constant flux within errors. At such late phases, exposure times were modest (typically 1--3~ks), naturally limiting the statistical precision of the spectral fits.

\begin{figure*}
\centering
\includegraphics[angle=0,scale=0.5]{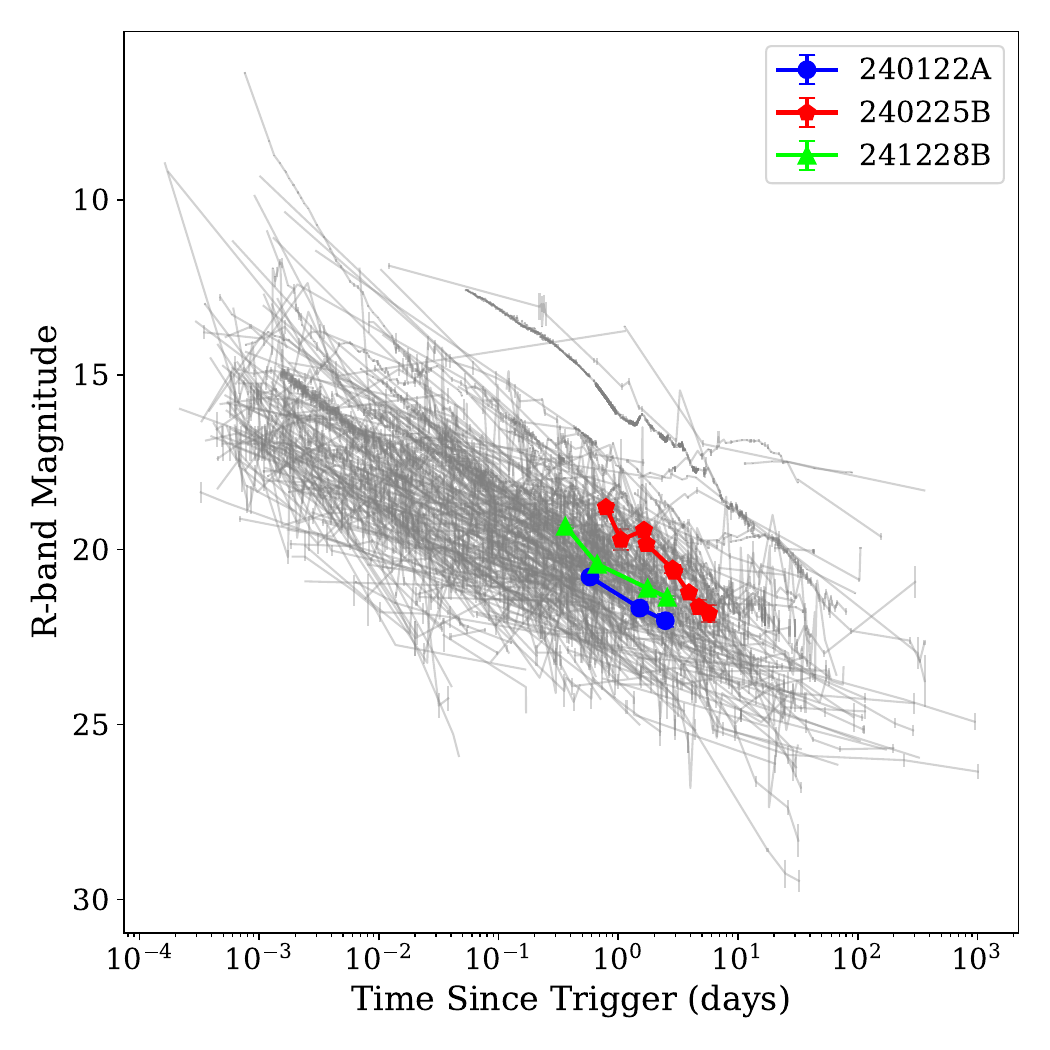}
\includegraphics[angle=0,scale=0.5]{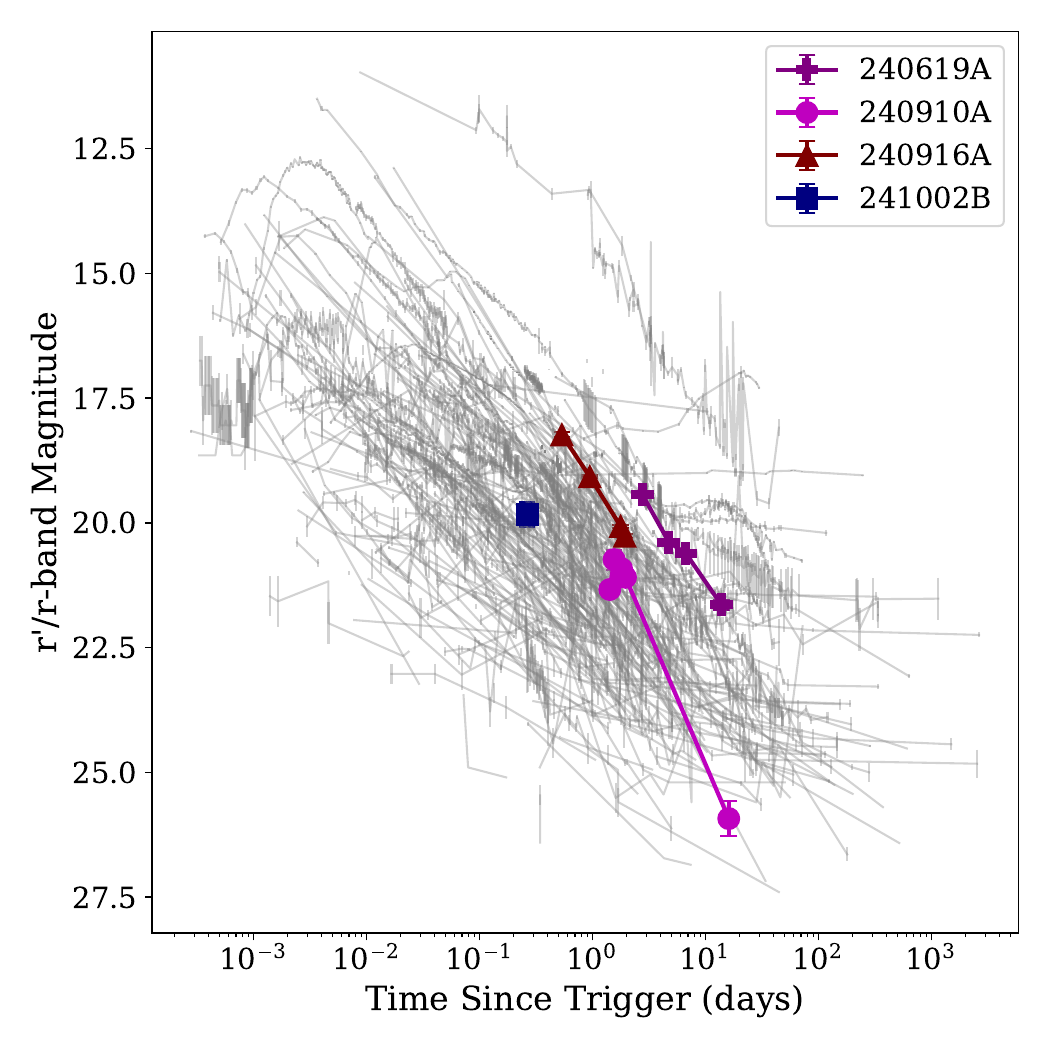}
\caption{$R$- and $r$/$r^\prime$-band afterglow light curves for the seven GRBs in our sample compared against archival GRB afterglows compiled by \citet{Dainotti2024}, known as the "Kann plot".}
\label{fig:AG_comp_R_r}
\end{figure*}

\subsubsection{Swift/XRT light curves comparison}\label{sec:XRT_LC_comparison}
To investigate the temporal behaviour of GRB afterglows and compare our GOTO-discovered GRBs against known populations, we compiled a comprehensive sample of \swift/XRT light curves, combining both archival GRBs and the GOTO sample (Figure~\ref{fig:AG_comp_XRT}). Light curves were obtained from the UK Swift Science Data Centre (UKSSDC)\footnote{\url{https://www.swift.ac.uk/xrt_products/bulk.php}} bulk access portal, using a custom notebook provided by the UKSSDC team to automate downloads. In our visualisation, archival LGRBs are shown in blue solid lines, SGRBs in grey dashed lines, and the seven GOTO-discovered GRBs (240122A, 240225B, 240619A, 240910A, 240916A, 241002B, 241228B) are overplotted with distinct markers and colours. 

As seen in Figure~\ref{fig:AG_comp_XRT}, the GOTO GRBs lie well within the canonical LGRB distribution, with count rates and temporal slopes consistent with typical long-burst afterglows. Their X-ray light curves track the faint to intermediate range of the LGRB population, showing no evidence for extreme behaviour. GRBs~240225B, 240910A, 240916A and 241002B in particular seems to follow smooth declines, while others (e.g., GRB~241228B) are represented only by a few points, underscoring the sparse nature of the coverage. This sparseness arises not from the afterglows themselves but from the fact that these bursts were not initially triggered by \swift/BAT or XRT, but instead by wide-field, poorly localised instruments such as \fermi/GBM and \maxi/GSC.

\subsection{UV/Optical/NIR}
UV/Optical/NIR afterglow light curves for all seven \grbssample~in our sample are shown in Figure~\ref{fig:optical_lcs} (see Table~\ref{tab:phot_data} for complete photometric observations), plotted in both flux density and apparent magnitude space. For visual clarity, magnitudes in different filters have been offset vertically where indicated in the legends. All magnitudes are reported in the AB system and have been corrected for Galactic extinction using the $E(B-V)$ values listed in Table~\ref{tab:sample}, based on the recalibrated dust maps of \citet{Schlafly2011}.

Among the GRBs in our sample, GRB~241228B shows the highest cadence multi-colour coverage, with early bright detections in GOTO $L$-band and a well-sampled decline. GRBs~240122A and 240225B each have moderate multi-band coverage, while events such as GRBs~241002B and 240916A have sparser datasets but still provide key temporal constraints. The sparse and uneven temporal coverage of the light curves prevents us from robustly constraining decay slopes, break times, or colour evolution. In most cases, only a few photometric points are available per GRB, which precludes detailed afterglow fitting. Nevertheless, these datasets enable comparison with the extended sample of GRB afterglow light curves discussed in the following subsection and provide the basis for the afterglow modelling presented in Section~\ref{sec:afterglow_modelling}.

\subsubsection{Optical light curves comparison - Kann plot}
To compare the optical behaviour of the GOTO-detected GRBs in our sample with previously well-observed GRB afterglows, we plotted their light curves alongside a reference sample from \citealt{Dainotti2024} (Figure~\ref{fig:AG_comp_R_r}). Three GRBs (240122A, 240225B, 241228B) have well-sampled $R$-band data, while the remaining four (240619A, 240910A, 240916A, 241002B) are well-observed in the $r$/$r^\prime$ band (see Figure~\ref{fig:optical_lcs}). In each case, the same filter band from the reference sample is used to minimise colour offsets, and all magnitudes are corrected for Galactic extinction. The reference light curves span a wide range of brightness and decay behaviours, and are plotted in grey for comparison.

The left panel of Figure~\ref{fig:AG_comp_R_r} shows the $R$-band events (240122A, 240225B, 241228B) in blue, red, and lime, while the right panel shows the $r/r'$-band events (240619A, 240910A, 240916A, 241002B) in purple, magenta, maroon, and navy. The afterglow light curves comparison in $R$- and $r$/$r^\prime$ bands demonstrates that the afterglows of GOTO-detected GRBs lie within the overall distribution of known GRB afterglows in both brightness and decay behaviour. The $R$-band events are consistent with the median properties of the sample, whereas the $r$/$r^\prime$-band events span a broader range in brightness and decline rates.

To place our events quantitatively within the broader population, we interpolated the extinction–corrected light curves at fixed epochs, using only those with data coverage near the corresponding epochs of our GRBs. In $R$ band at $t=0.79$~d (close to the first data point for GRB~240225B), the comparison sample spans $14.14$–$23.93$~mag (median $20.45$~mag). Our GRBs fall within this range: GRB~240225B is relatively bright ($18.79$~mag), while GRBs~241228B and 240122A are near the population median ($20.49$ and $20.98$~mag, respectively). In $r$/$r^\prime$ band at $t=1.64$~d (a phase where both GRBs~240910A and 240916A have measurements), the sample spans $14.20$–$25.50$~mag (median $21.13$~mag). GRB~240916A is somewhat brighter than average ($19.90$~mag), while GRB~240910A lies close to the median ($20.80$~mag) but shows an unusually steep decline, fading by $\Delta m \approx 4.7$~mag over $\Delta t \approx 14.7$~d ($\sim$0.32~mag~d$^{-1}$). By contrast, GRB~241002B appears relatively faint even at early times. Overall, these comparisons confirm that the GOTO afterglows occupy the central brightness distribution of known GRB afterglows while also sampling the diversity of decline rates and brightness within the population.

\subsection{Spectroscopic Analysis -- Redshift Estimation}\label{sec:spec_afterglow_analysis}
When a GRB explodes, the resulting afterglow light passes through both the interstellar medium of its host galaxy and any intervening material along the line of sight, imprinting a series of absorption features onto the spectrum. In our X-shooter spectra, only the highest redshift absorption system is identified and assigned as the redshift of the GRB since no higher redshift intervening material is physically possible. While additional foreground absorption systems may be present, a detailed analysis and characterisation of these intervening absorbers is beyond the scope of this paper and will be addressed in future work.
The redshifts are estimated by identifying common absorption lines in GRB afterglows using the line lists of \cite{2009ApJS..185..526F} and \cite{2011ApJ...727...73C}, and/or emission lines from their host galaxies. The redshift and its associated uncertainty are then derived by fitting Voigt profiles \cite{2018arXiv180301187K} to the absorption features, prioritizing low-ionization, unsaturated, and unblended transitions, and Gaussian profiles to the emission lines.

\begin{figure}
\centering
\includegraphics[angle=0,scale=0.57]{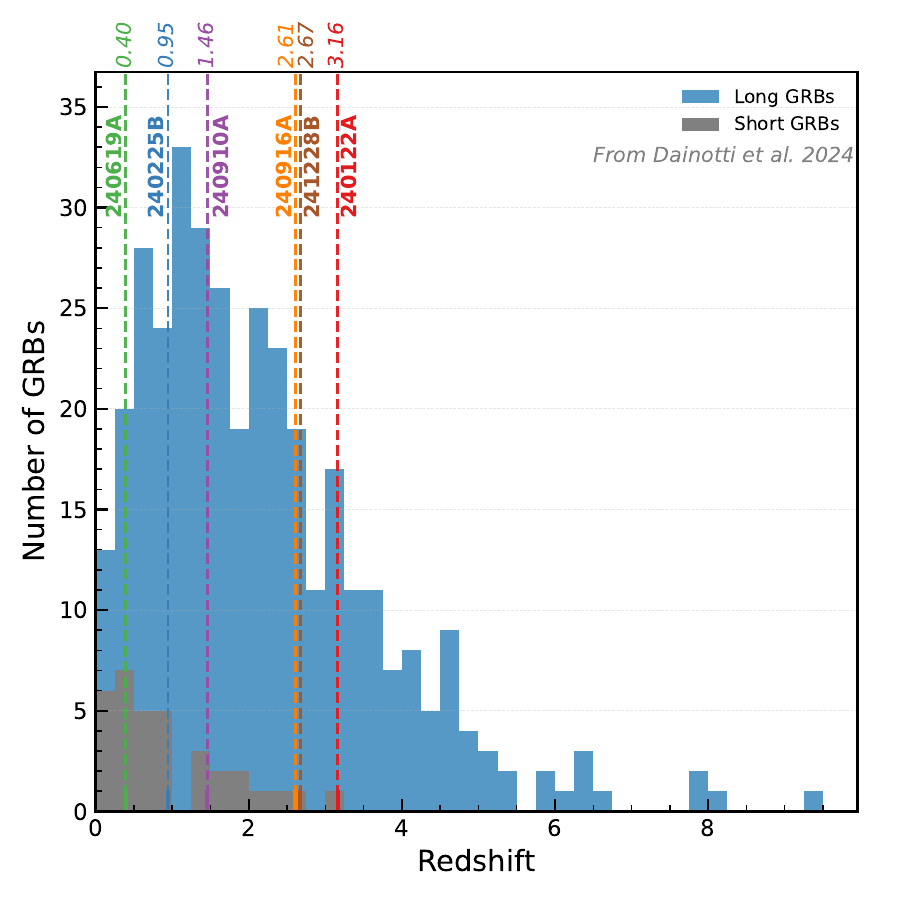}
\caption{Histogram of redshift distribution for long (blue) and short (grey) GRBs from the sample of \citet{Dainotti2024}. Overlaid are the redshifts of six GRBs with optical afterglows discovered by GOTO in 2024, marked with vertical dashed lines, labelled by GRB name within the plot, and redshift values above. This visual comparison highlights the diversity in redshift of GOTO-discovered GRBs and demonstrates their placement within the broader GRB population.}
\label{fig:redshift_comp}
\end{figure}

\begin{figure*}
    \centering
    \includegraphics[width=0.44\linewidth]{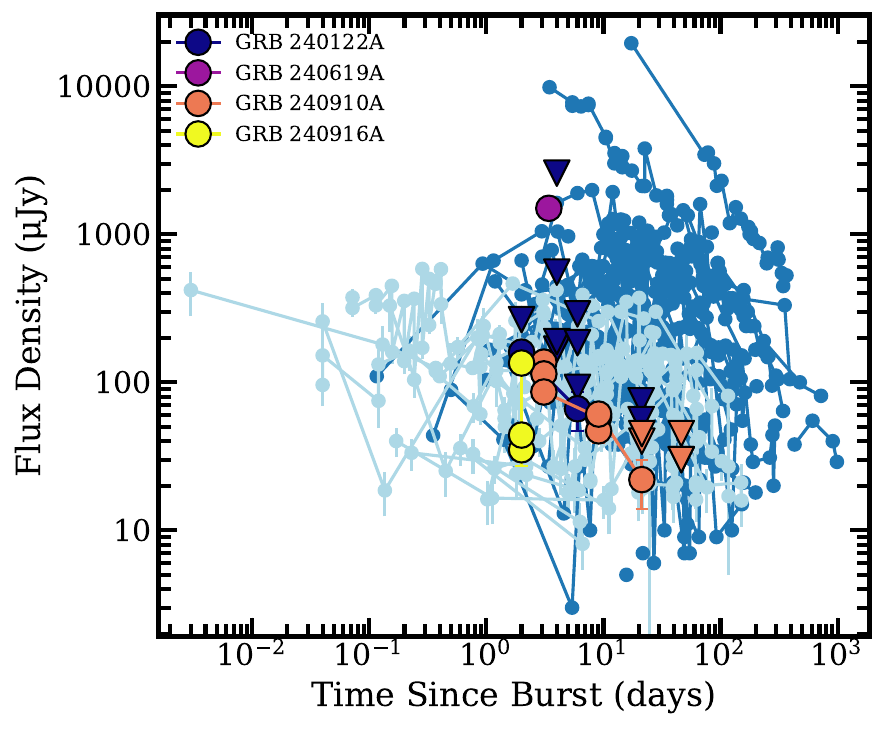}
    \includegraphics[width=0.44\linewidth]{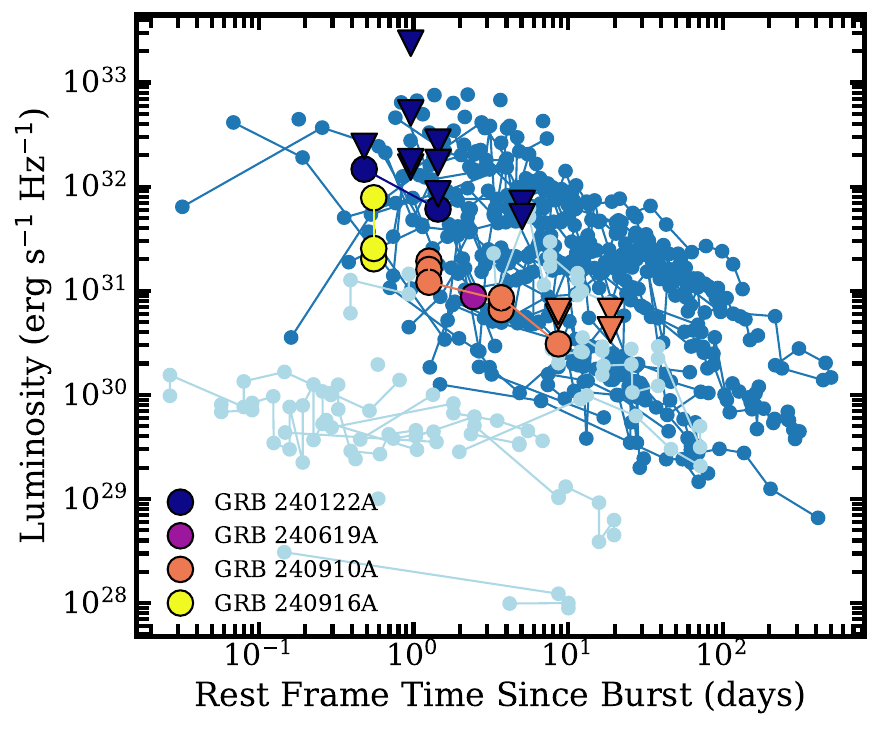}
    \caption{Left panel: 8-10 GHz radio afterglow light curves of GRBs, showing flux density (in $\mu$Jy) as a function of observer-frame time since burst. The coloured markers represent four LGRBs discovered by GOTO with available radio follow-up observations: GRB~240122A (dark blue), GRB~240619A (magenta), GRB~240910A (orange), and GRB~240916A (yellow), from a sample of seven GOTO GRBs. For comparison, literature LGRB afterglows are shown in dark blue, and SGRBs in grey. Right panel: rest-frame radio luminosity light curves of the same GRBs, showing monochromatic luminosity (in erg\,s$^{-1}$\,Hz$^{-1}$) as a function of rest-frame time since explosion. The GOTO events are consistent with the broader LGRB population in terms of both luminosity and temporal evolution, highlighting the capability of GOTO to detect GRBs with typical radio afterglow properties. 
    References: \citet{taylor_discovery_1998,frail_radio_1999,harrison_optical_1999,berger_jet_2000,frail_enigmatic_2000,galama_bright_2000,berger_host_2001,djorgovski_afterglow_2001,harrison_broadband_2001,berger_grb_2001,2002ApJ...572L..51P,galama_continued_2003,soderberg_constraints_2004,soderberg_redshift_2004,frail_accurate_2005,2005Natur.438..988B,cenko_multiwavelength_2006,frail_energetic_2006,2006ApJ...650..261S,rol_grb_2007,chandra_comprehensive_2008,perley_grb_2008,van_der_horst_detailed_2008,chandra_discovery_2010,cenko_afterglow_2011,cenko_swift_2012,2012GCN.12804....1H,greiner_unusual_2013,moin_radio_2013,perley_afterglow_2014,2014ApJ...780..118F,horesh_unusual_2015,laskar_reverse_2016,laskar_vla_2018,2019ApJ...883...48L,2021ApJ...906..127F,2022ApJ...935L..11L,oconnor_structured_2023,laskar_radio_2023,2023GCN.33475....1A,rhodes_rocking_2024,schroeder_radio_2024,2024ApJ...975L..13A,2024Natur.626..737L,anderson2025arXiv250814650A}}
    \label{fig:radio_comp_lcs}
\end{figure*}

\subsubsection{VLT/X-shooter}
Here, we summarise the results from our VLT/X-shooter spectra of GRBs~240122A, 240225B, 240619A, 240916A and 241228B:

\paragraph{GRB~240122A}
The VLT/X-shooter of spectrum GRB~240122A exhibits a strong Ly$\alpha$ absorption feature near 5060~\AA, accompanied by a set of metal lines, including Si~\textsc{ii}, Fe~\textsc{ii}, C~\textsc{ii}, Si~\textsc{iv}, C~\textsc{iv}, Al~\textsc{ii}, and Mg~\textsc{ii}. All features are consistent with a common redshift of $z = 3.1634 \pm 0.0003$. The spectrum is shown in Figure~\ref{fig:VLT_spec_22A}, and a detailed list of the identified lines is provided in Table~\ref{tab:240122A}. We note the presence of multiple intervening absorbers at the following redshifts: $z = 2.7583$, 2.5384, 2.4879, 2.4230, 1.5111 and 1.4618.

\paragraph{GRB~240225B}
In the case of GRB 240225B, a continuum is detected over the entire wavelength range (from 3300 to 20400~\AA) and the following strong absorption features are identified: Al\,\textsc{iii}, Cr\,\textsc{ii}, Fe\,\textsc{ii}, Mn\,\textsc{ii}, Mg\,\textsc{ii}, Mg\,\textsc{i}, and Ca\,\textsc{ii} at a common redshift of $z = 0.9462 \pm 0.0002$. At the same redshift, three emission lines ([O \textsc{ii}] $\lambda\lambda3727,\,3730$ and [O \textsc{iii}] $\lambda5008$) are identified from the host galaxy. The complete list of identified lines is provided in Table~\ref{tab:240225B}. One intervening system is identified at $z = 0.7056$.

\paragraph{GRB~240619A}
For GRB~240619A, we identified several strong emission lines as due to [O \textsc{ii}] $\lambda\lambda3727,\,3730$, [Ne \textsc{iii}] $\lambda3870$, H$\gamma$, H$\beta$, [O \textsc{iii}] $\lambda\lambda4960,\,5008$ and H$\alpha$ at a common redshift of $z = 0.3960 \pm 0.0001$. A second fainter object is visible in the Legacy Survey images, located about $1.7^{\prime\prime}$ west of the GRB afterglow position. This source was also covered by the X-shooter slit, and a redshift of $z = 1.34$ was derived from the detection of the emission lines of [O \textsc{ii}] doublet and H$\alpha$. Due to its larger angular offset, we consider this galaxy to be unrelated to the GRB. The spectrum of the host galaxy and the identified lines are shown in Figure~\ref{fig:VLT_spec_19A} and Table~\ref{tab:240619A}.

\paragraph{GRB~240916A}
The reduced spectrum of GRB~240916A reveals a prominent H\textsc{i} Ly$\alpha$ absorption feature at $\sim$4400~\AA, along with a rich set of metal absorption lines including Si\,\textsc{ii}, Al\,\textsc{iii}, and Fe\,\textsc{ii}. In addition, we detect several fine-structure transitions such as Fe\,\textsc{ii}$^{*}$ and Ni\,\textsc{ii}$^{*}$. From these features, we derive a redshift of $z = 2.6100 \pm 0.0002$. The spectrum and line identifications are shown in Figure~\ref{fig:VLT_spec_16A} and listed in Table~\ref{tab:240916A}. Two intervening absorbers are identified at $z = 2.2904$ and $z = 2.2140$.

\paragraph{GRB~241228B}
GRB~241228B spectrum displays a clear Ly$\alpha$ absorption line at $\sim$4470~\AA, along with numerous metal absorption features, including N~\textsc{v}, S~\textsc{ii}, Si~\textsc{ii}, Fe~\textsc{ii}, and O~\textsc{i}. Several fine-structure transitions such as Si~\textsc{ii}$^{*}$, O~\textsc{i}$^{*}$, C~\textsc{ii}$^{*}$, Fe~\textsc{ii}$^{*}$, and Ni~\textsc{ii}$^{*}$ are also detected. In addition, a strong Ly$\alpha$ emission line is observed from the host galaxy. These features indicate a redshift of $z = 2.6745 \pm 0.0004$. The spectrum and complete line identifications are presented in Figure~\ref{fig:VLT_spec_28B} and Tables~\ref{tab:241228B} and~\ref{tab:241228B-2}. Absorption features corresponding to the intervening systems at $z = 2.4576$, 2.0004, 1.8244, 0.9504 are also observed.

\subsubsection{GTC/OSIRIS}
Here, we summarise the results from our GTC spectra of GRBs~240122A and 240910A:

\paragraph{GRB~240122A}
Despite poorer observing conditions (seeing of $\sim$1.7$^{\prime\prime}$), the GTC/OSIRIS spectrum of GRB~240122A clearly reveals a strong Ly$\alpha$ absorption feature near 5060\,\AA, along with a consistent set of metal lines, including Si\,\textsc{ii}, Fe\,\textsc{ii}, C\,\textsc{ii}, Si\,\textsc{iv}, C\,\textsc{iv}, Al\,\textsc{ii}, and Mg\,\textsc{ii}. These features confirm a redshift similar to that derived from the higher-resolution VLT spectrum ($z = 3.163 \pm 0.003$). The reduced OSIRIS spectrum is shown in Figure~\ref{fig:GTC_spec_22A}, with identified features listed in Table~\ref{tab:240122A}.

\paragraph{GRB~240910A}
In the case of GRB~240910A, the afterglow continuum is clearly detected across the full spectral range, and the spectrum reveals a rich set of absorption features. Prominent lines include Si\,\textsc{ii}, C\,\textsc{iv}, Fe\,\textsc{ii}, Al\,\textsc{ii}, Al\,\textsc{iii}, Cr\,\textsc{ii}, Mn\,\textsc{i}, Mn\,\textsc{ii}, Ni\,\textsc{ii}$^{*}$, Mg\,\textsc{ii}, and Mg\,\textsc{i}, along with several fine-structure transitions such as Fe\,\textsc{ii}$^{*}$ and Ni\,\textsc{ii}$^{*}$. All lines are consistent with a common redshift of $z = 1.4605 \pm 0.0007$. The reduced spectrum is presented in Figure~\ref{fig:GTC_spec_10A}, and a complete list of identified lines is provided in Table~\ref{tab:240910A}.

In summary, our spectroscopic follow-up of seven GRBs using VLT/X-shooter and GTC/OSIRIS reveals a wide range of redshifts ($z \sim 0.40$ to $z \sim 3.16$), with absorption and emission features tracing both the GRB host environments and the intervening interstellar medium. High-quality afterglow spectra enable precise redshift measurements and identification of various ionic species, including fine-structure transitions. These results provide critical context for understanding the physical conditions in GRB host galaxies and lay the foundation for future studies of metallicity, dust content, and kinematics in GRB environments.

\subsubsection{Redshift Comparison}
To show the redshift distribution of GRBs in our sample, we compare them against the broader GRB population presented in the comprehensive compilation by \citet{Dainotti2024}. Figure~\ref{fig:redshift_comp} shows a histogram of GRBs with measured redshifts from that sample, classified into long and short categories. The majority of GRBs in the \citet{Dainotti2024} sample are LGRBs, with a redshift distribution peaking around $z \sim 0.5 - 2$, consistent with the star formation history of the universe. SGRBs appear more frequently at lower redshifts, consistent with their likely origin from compact object mergers with longer delay times.

Overlaid on this distribution are the measured redshifts of the GOTO GRBs, shown as vertical dashed lines with annotations above the axis. Our sample spans from $z = 0.40$ (GRB~240619A) to $z = 3.16$ (GRB~240122A). It is worth noting that the optical selection imposed by GOTO inherently limits detections to $z \lesssim 5$, since at higher redshifts the Lyman forest progressively enters and then highly absorbs flux across the GOTO $L$-band (400--700\,nm). GRBs~240225B ($z=0.95$) and 240910A ($z=1.46$) fall near the central peak of the LGRB distribution, while GRB~240916A ($z=2.61$) and GRB~241228B ($z=2.67$) occupy the higher-redshift tail together with GRB~240122A at $z=3.16$. This spread highlights that the GOTO sample encompasses both low- and high-redshift GRBs, demonstrating the survey’s capability to probe the wide observed redshift range of the LGRB population.

From a physical perspective, the low-redshift events, such as GRB~240619A, are particularly valuable for detailed host-galaxy and supernova connection studies, where high signal-to-noise follow-up is achievable. Conversely, the higher-redshift events (e.g., GRBs~240916A, 241228B, and 240122A) provide critical leverage for probing star-forming environments in the early universe and for constraining the role of GRBs as tracers of cosmic star formation beyond $z>2$.

\begin{table*}
\caption{Parameter estimation priors and marginalised posteriors for the GOTO-discovered GRBs using the \texttt{afterglowpy} TopHat model. Posteriors are medians with $16-84\%$ credible intervals.} \label{tab:afterglow_modelling}
\addtolength{\tabcolsep}{-1pt} 
\renewcommand{\arraystretch}{1.3}
\begin{tabular}{|c|c|c|c|c|c|c|c|c|c|}
\hline
\multicolumn{10}{|c|}{GRB~240122A; GOTO24eu} \\
\hline
 & $\theta_v$ (rad) & $\log_{10}(E_{0})$ (erg) & $\theta_c$ (rad) & $\log_{10}(n_0)$ (cm$^{-3}$) & $p$ & $\log_{10}\epsilon_e$ & $\log_{10}\epsilon_B$ & $\xi_N$ & $d_L$ (Mpc)\\
Priors & U(0.0; 0.5) & U(49; 57) & U(0.0; 0.5) & -0.379 & 2.119 & -1.246 & -4.290 & 1.0 & 27708.87 \\
Posteriors & $2.15_{-0.10}^{+0.09} \times 10^{-2}$ & $54.97_{-0.02}^{+0.02}$ & $2.39_{-0.12}^{+0.12} \times 10^{-2}$ & -0.379 & 2.119 & -1.246 & -4.290 & 1.0 & 27708.87 \\
\hline
\multicolumn{10}{|c|}{GRB~240225B; GOTO24tz} \\
\hline
 & $\theta_v$ (rad) & $\log_{10}(E_{0})$ (erg) & $\theta_c$ (rad) & $\log_{10}(n_0)$ (cm$^{-3}$) & $p$ & $\log_{10}\epsilon_e$ & $\log_{10}\epsilon_B$ & $\xi_N$ & $d_L$ (Mpc)\\
Priors & U(0.0; 0.5) & U(49; 57) & U(0.0; 0.5) & U(-5.0; 3.0) & 2.119 & -1.246 & -4.290 & 1.0 & 6342.55 \\
Posteriors & $9.43_{-0.41}^{+0.38} \times 10^{-2}$ & $54.27_{-0.04}^{+0.04}$ & $7.23_{-0.31}^{+0.31} \times 10^{-2}$ & $1.87_{-0.11}^{+0.09}$ & 2.119 & -1.246 & -4.290 & 1.0 & 6342.55 \\
\hline
\multicolumn{10}{|c|}{GRB~240619A; GOTO24cvn} \\
\hline
 & $\theta_v$ (rad) & $\log_{10}(E_{0})$ (erg) & $\theta_c$ (rad) & $\log_{10}(n_0)$ (cm$^{-3}$) & $p$ & $\log_{10}\epsilon_e$ & $\log_{10}\epsilon_B$ & $\xi_N$ & $d_L$ (Mpc)\\
Priors & U(0.0; 0.5) & U(49; 57) & U(0.0; 0.5) & -0.379 & 2.119 & U(-5.0; 0.0) & U(-5.0; 0.0) & 1.0 & 2217.13 \\
Posteriors & $0.41_{-0.07}^{+0.04}$ & $52.70_{-0.35}^{+0.42}$ & $0.45_{-0.07}^{+0.04}$ & -0.379 & 2.119 & $-0.85_{-0.39}^{+0.32}$ & $-2.18_{-0.15}^{+0.12}$ & 1.0 & 2217.13 \\
\hline
\multicolumn{10}{|c|}{GRB~240910A; GOTO24fvl} \\
\hline
 & $\theta_v$ (rad) & $\log_{10}(E_{0})$ (erg) & $\theta_c$ (rad) & $\log_{10}(n_0)$ (cm$^{-3}$) & $p$ & $\log_{10}\epsilon_e$ & $\log_{10}\epsilon_B$ & $\xi_N$ & $d_L$ (Mpc)\\
Priors & U(0.0; 0.5) & U(49; 57) & U(0.0; 0.5) & -0.379 & 2.119 & -1.246 & -4.290 & 0.180 & 10826.91 \\
Posteriors & $4.25_{-0.12}^{+0.07} \times 10^{-2}$ & $54.77_{-0.01}^{+0.01}$ & $4.07_{-0.12}^{+0.08} \times 10^{-2}$ & -0.379 & 2.119 & -1.246 & -4.290 & 1.0 & 10826.91 \\
\hline
\multicolumn{10}{|c|}{GRB~240916A; GOTO24fzn} \\
\hline
 & $\theta_v$ (rad) & $\log_{10}(E_{0})$ (erg) & $\theta_c$ (rad) & $\log_{10}(n_0)$ (cm$^{-3}$) & $p$ & $\log_{10}\epsilon_e$ & $\log_{10}\epsilon_B$ & $\xi_N$ & $d_L$ (Mpc)\\
Priors & U(0.0; 0.5) & U(49; 57) & U(0.0; 0.5) & -0.379 & 2.119 & U(-5; 0) & U(-5; 0) & 0.180 & 22000.43 \\
Posteriors & $2.29^{+0.71}_{-0.78} \times 10^{-2}$ & $54.40^{+1.03}_{-0.63}$ & $4.73^{+1.16}_{-1.26} \times 10^{-2}$ & -0.379 & 2.119 & $-1.34^{+0.59}_{-0.95}$ & $-1.47^{+0.91}_{-0.73}$ & 1.0 & 22000.43 \\
\hline
\multicolumn{10}{|c|}{GRB~241228B; GOTO24jmz} \\
\hline
 & $\theta_v$ (rad) & $\log_{10}(E_{0})$ (erg) & $\theta_c$ (rad) & $\log_{10}(n_0)$ (cm$^{-3}$) & $p$ & $\log_{10}\epsilon_e$ & $\log_{10}\epsilon_B$ & $\xi_N$ & $d_L$ (Mpc)\\
Priors & U(0.0; 0.5) & U(49; 57) & U(0.0; 0.5) & -0.379 & 2.119 & -1.246 & -4.290 & 0.180 & 22653.17 \\
Posteriors & $4.86^{+0.14}_{-0.11} \times 10^{-3}$ & $55.429^{+0.004}_{-0.004}$ & $10.56^{+0.05}_{-0.07} \times 10^{-3}$ & -0.379 & 2.119 & -1.246 & -4.290 & 0.180 & 22653.17 \\
\hline
\multicolumn{10}{l}{NOTE: $\theta_v$ -- viewing angle; $E_{0}$ -- isotropic-equivalent kinetic energy; $\theta_c$ -- half-opening angle of jet core; $n_0$ -- density of the surrounding ISM; $p$ -- electron energy}\\
\multicolumn{10}{l}{distribution power law index; $\epsilon_e$ -- fraction of energy that goes into electrons; $\epsilon_B$ -- fraction of energy that goes into the magnetic field; $\xi_N$ -- fraction of}\\
\multicolumn{10}{l}{shock-accelerated electrons; $d_L$ --  luminosity distance.}
\end{tabular}
\end{table*}

\begin{figure*}
\centering
\begin{subfigure}[t]{0.48\textwidth}
    \centering
    \captionsetup{justification=centering, font=small, skip=2pt}
    \caption*{GRB~240122A}
    \includegraphics[width=\textwidth]{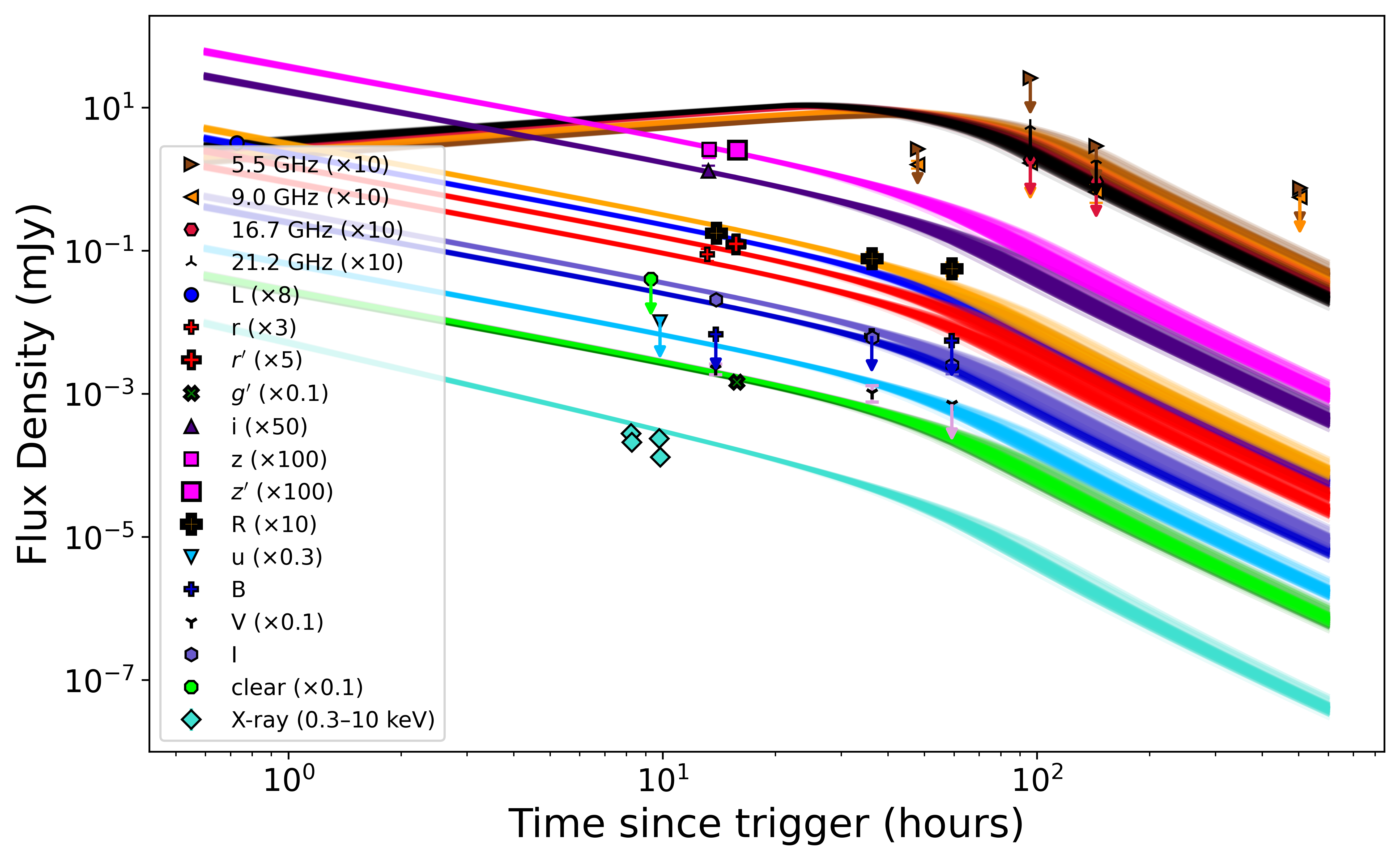}
\end{subfigure}
\hfill
\begin{subfigure}[t]{0.48\textwidth}
    \centering
    \captionsetup{justification=centering, font=small, skip=2pt}
    \caption*{GRB~240225B}
    \includegraphics[width=\textwidth]{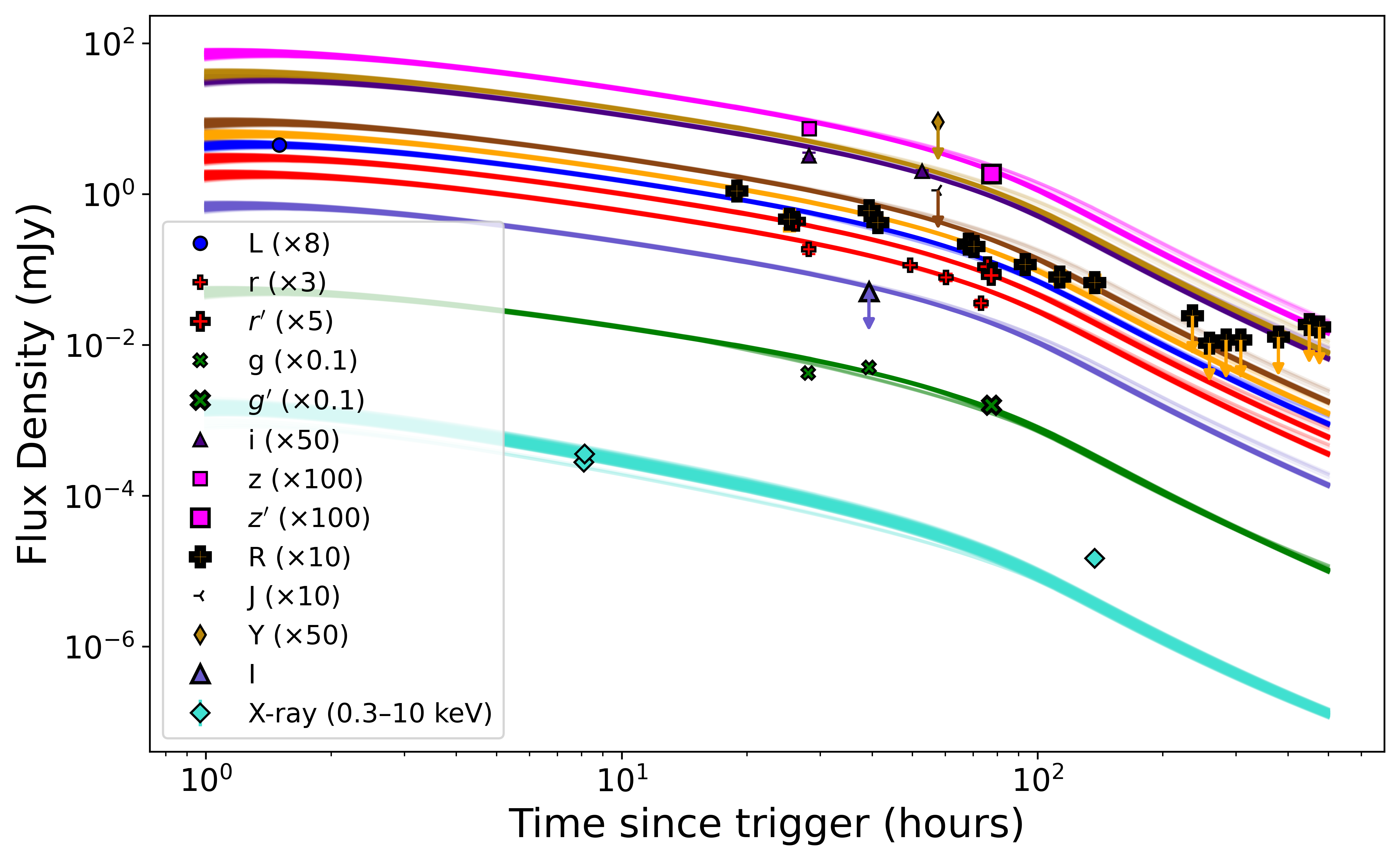}
\end{subfigure}
\vspace{1mm}
\begin{subfigure}[t]{0.48\textwidth}
    \centering
    \captionsetup{justification=centering, font=small, skip=2pt}
    \caption*{GRB~240619A}
    \includegraphics[width=\textwidth]{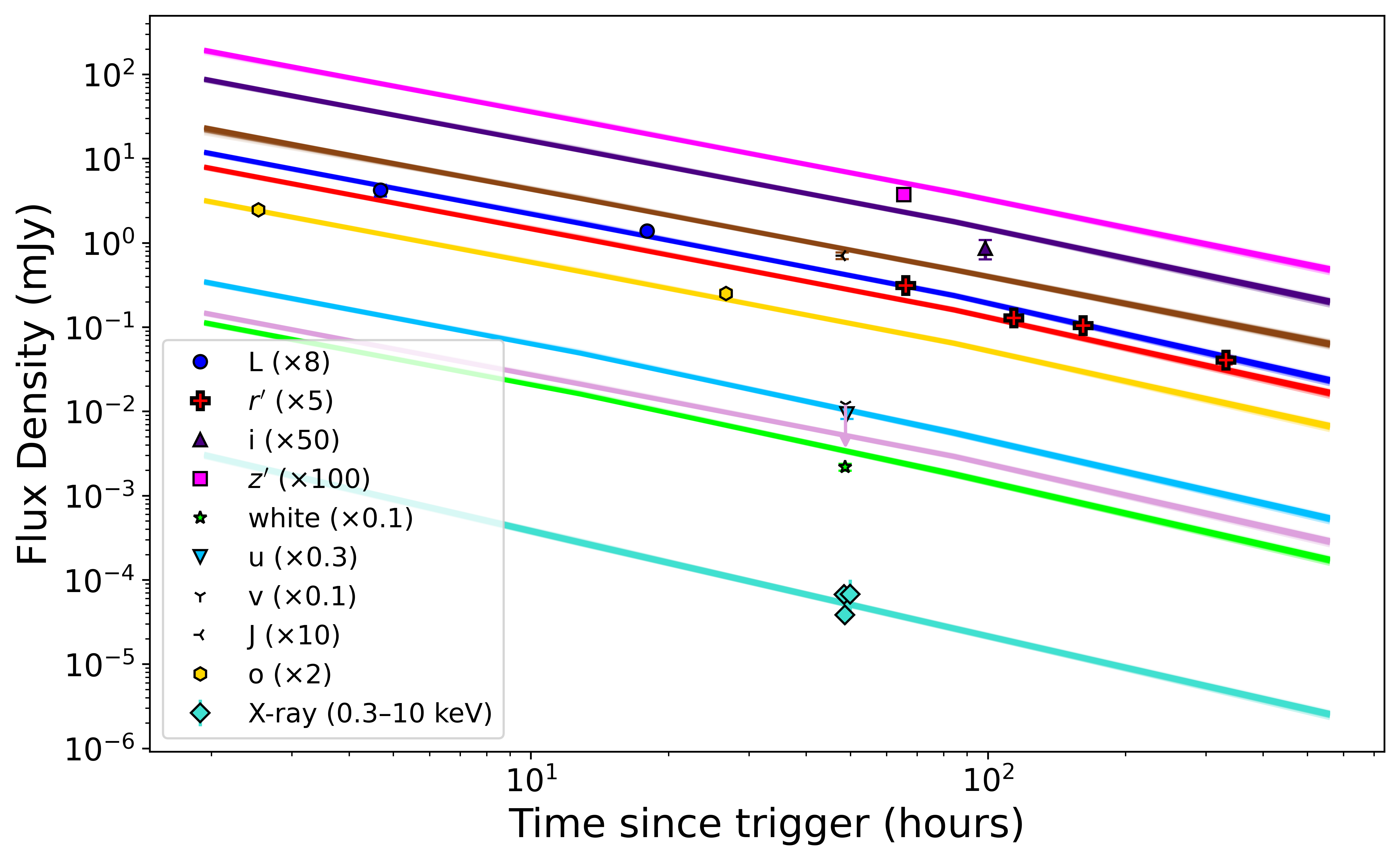}
\end{subfigure}
\hfill
\begin{subfigure}[t]{0.48\textwidth}
    \centering
    \captionsetup{justification=centering, font=small, skip=2pt}
    \caption*{GRB~240910A}
    \includegraphics[width=\textwidth]{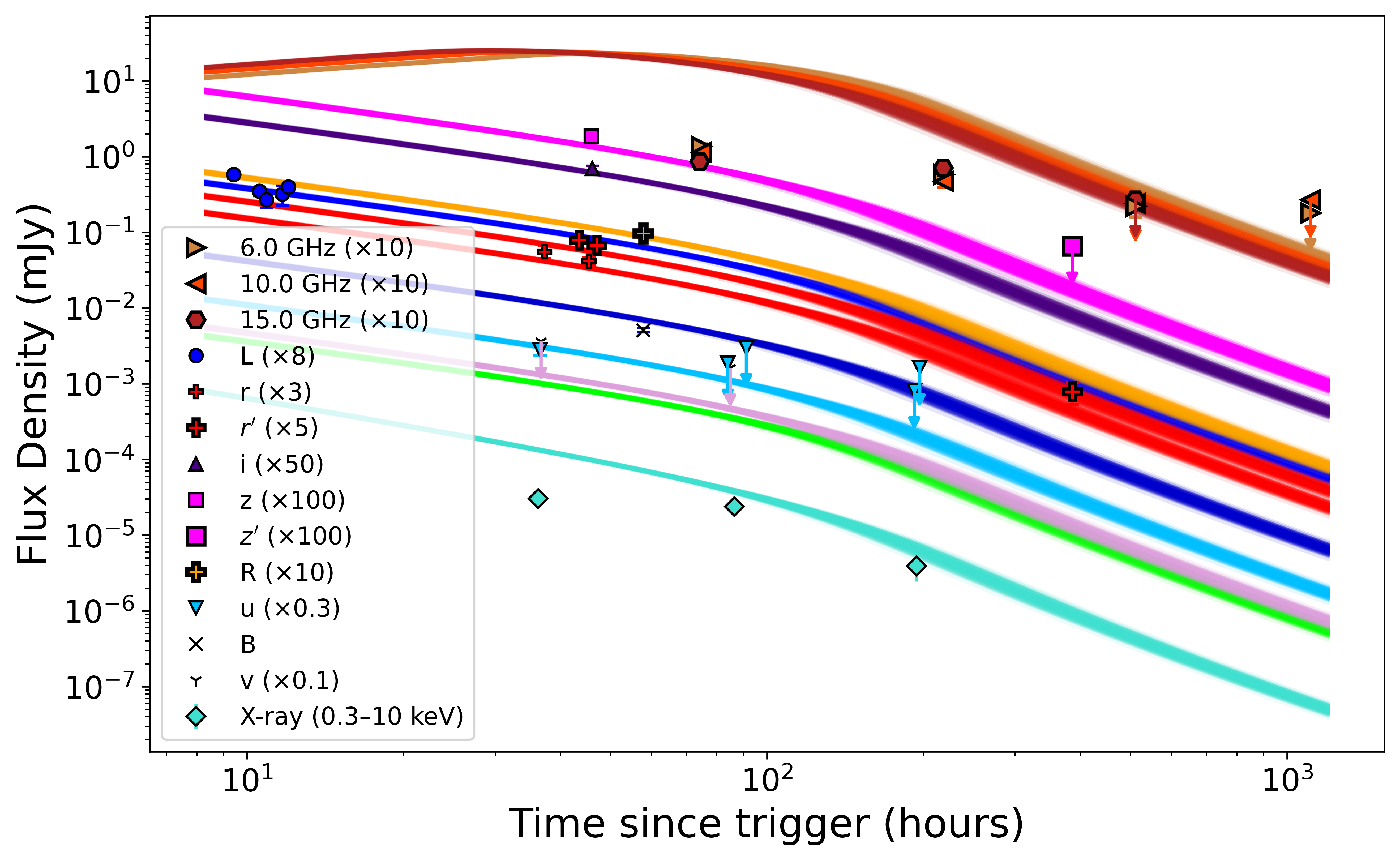}
\end{subfigure}
\vspace{1mm}
\begin{subfigure}[t]{0.48\textwidth}
    \centering
    \captionsetup{justification=centering, font=small, skip=2pt}
    \caption*{GRB~240916A}
    \includegraphics[width=\textwidth]{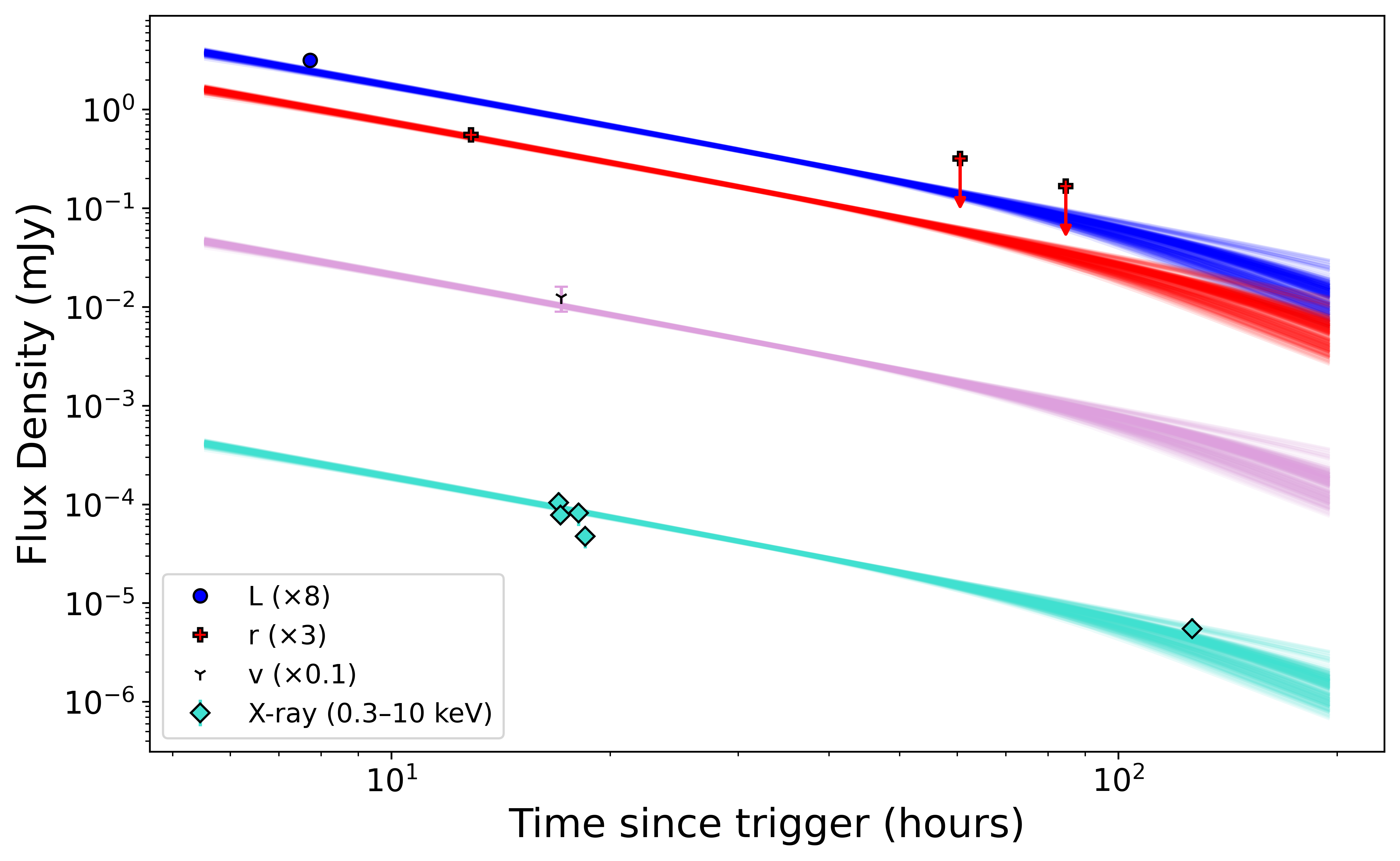}
\end{subfigure}
\hfill
\begin{subfigure}[t]{0.48\textwidth}
    \centering
    \captionsetup{justification=centering, font=small, skip=2pt}
    \caption*{GRB~241228B}
    \includegraphics[width=\textwidth]{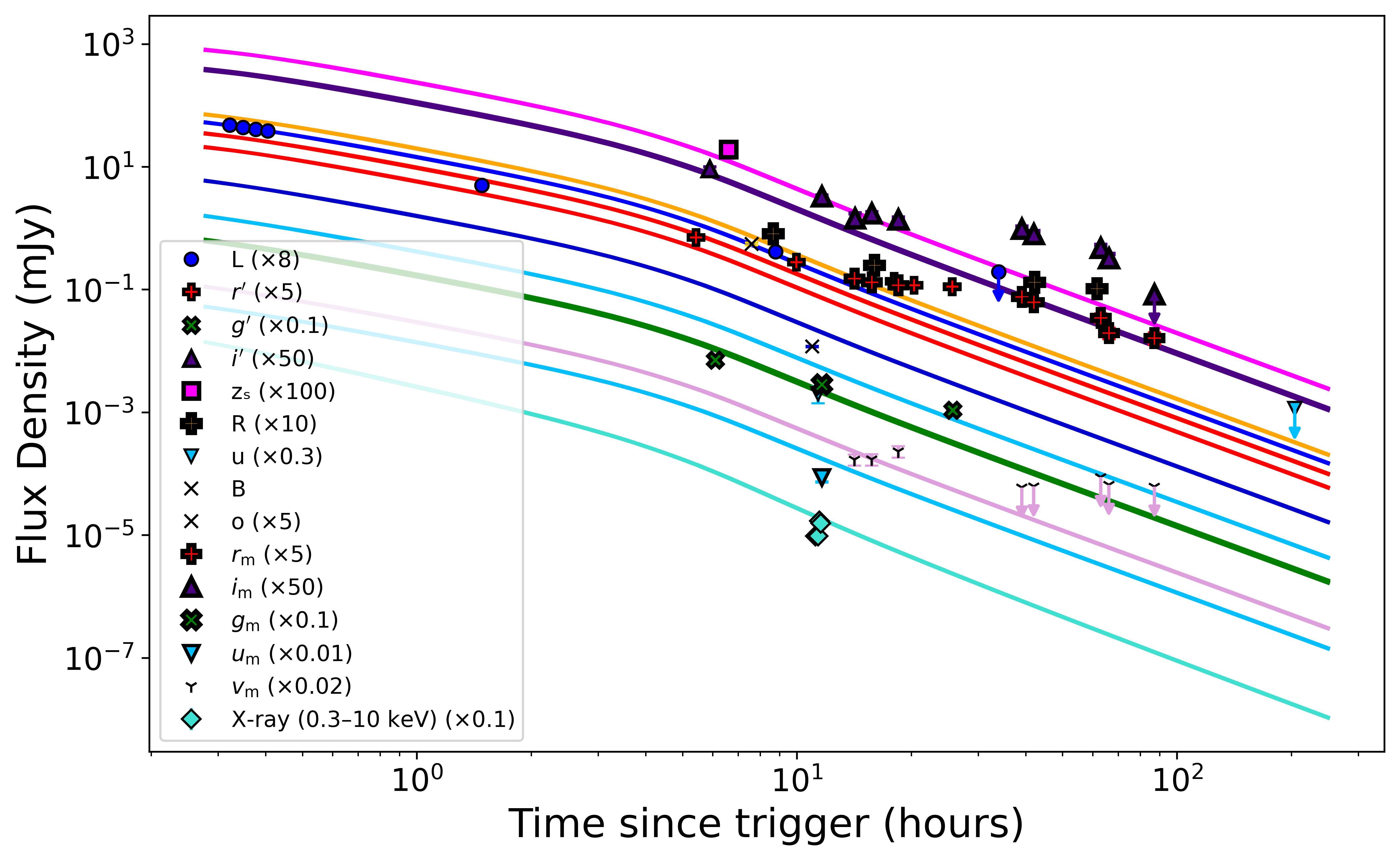}
\end{subfigure}
\caption{Multi-wavelength afterglow light curves for six GRBs modelled in this work: GRBs~240122A, 240225B, 240619A, 240910A, 240916A, and 241228B. Each panel shows the observed data points (UV/optical/NIR, X-ray, and radio, where available), overlaid with the best-fitting afterglow model using a TopHat (uniform) jet scenario (see the text for details). Photometric data are compiled from our observations and published GCN Circulars. Fluxes are rescaled for clarity as indicated in the legends.}
\label{fig:modelled_lcs}
\end{figure*}

\begin{figure*}
\centering
\begin{subfigure}[t]{0.48\textwidth}
    \centering
    \captionsetup{justification=centering, font=small, skip=2pt}
    \caption*{GRB~240122A}
    \includegraphics[width=0.75\textwidth]{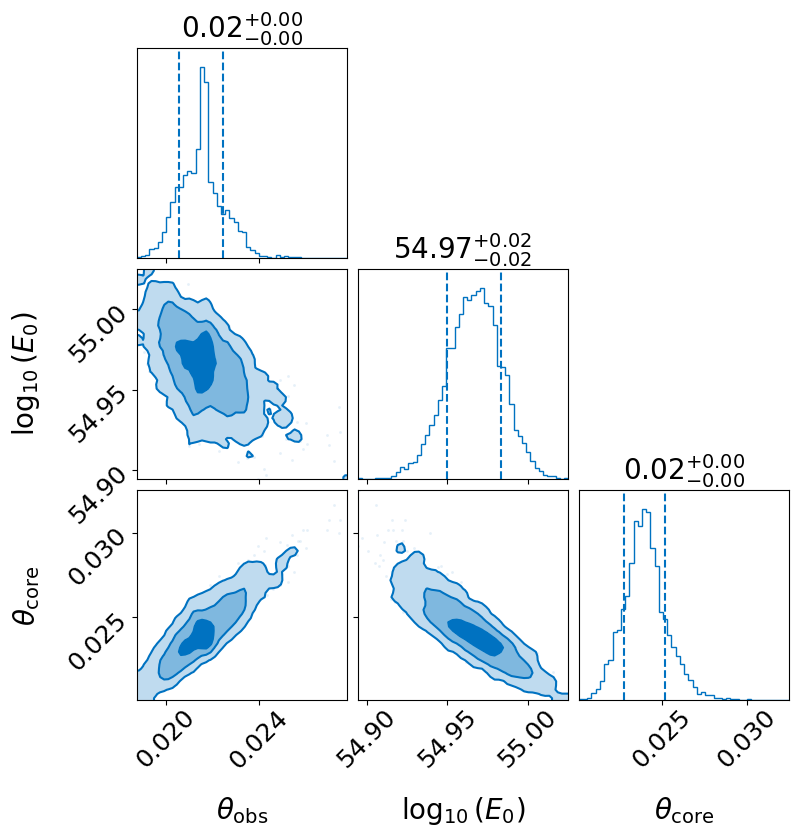}
\end{subfigure}
\hfill
\begin{subfigure}[t]{0.48\textwidth}
    \centering
    \captionsetup{justification=centering, font=small, skip=2pt}
    \caption*{GRB~240225B}
    \includegraphics[width=0.75\textwidth]{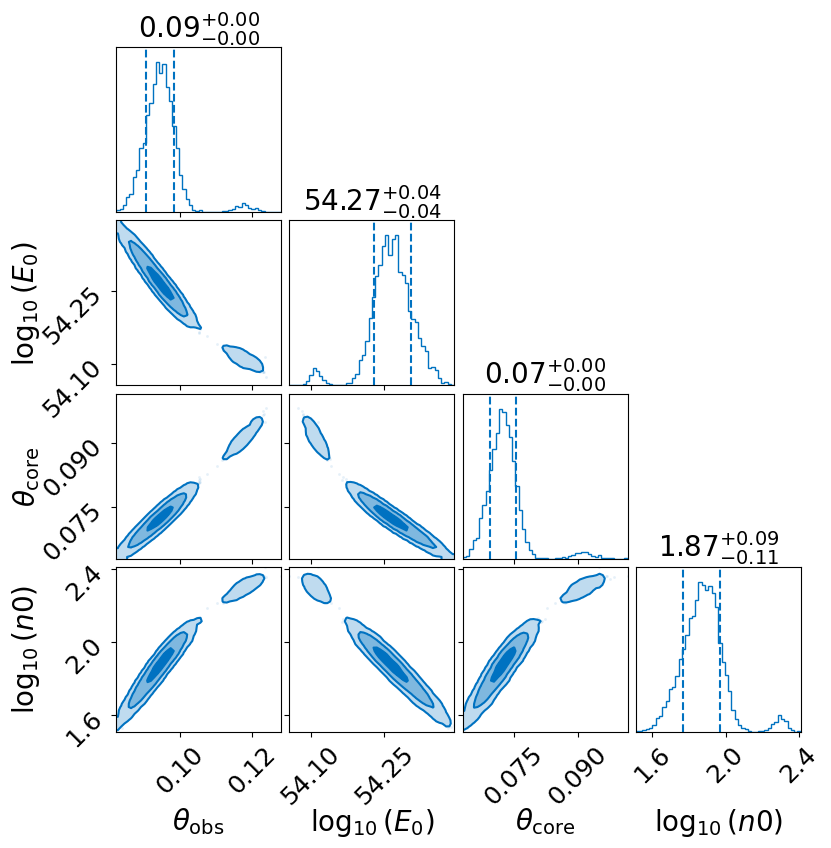}
\end{subfigure}
\vspace{1mm}
\begin{subfigure}[t]{0.48\textwidth}
    \centering
    \captionsetup{justification=centering, font=small, skip=2pt}
    \caption*{GRB~240619A}
    \includegraphics[width=0.75\textwidth]{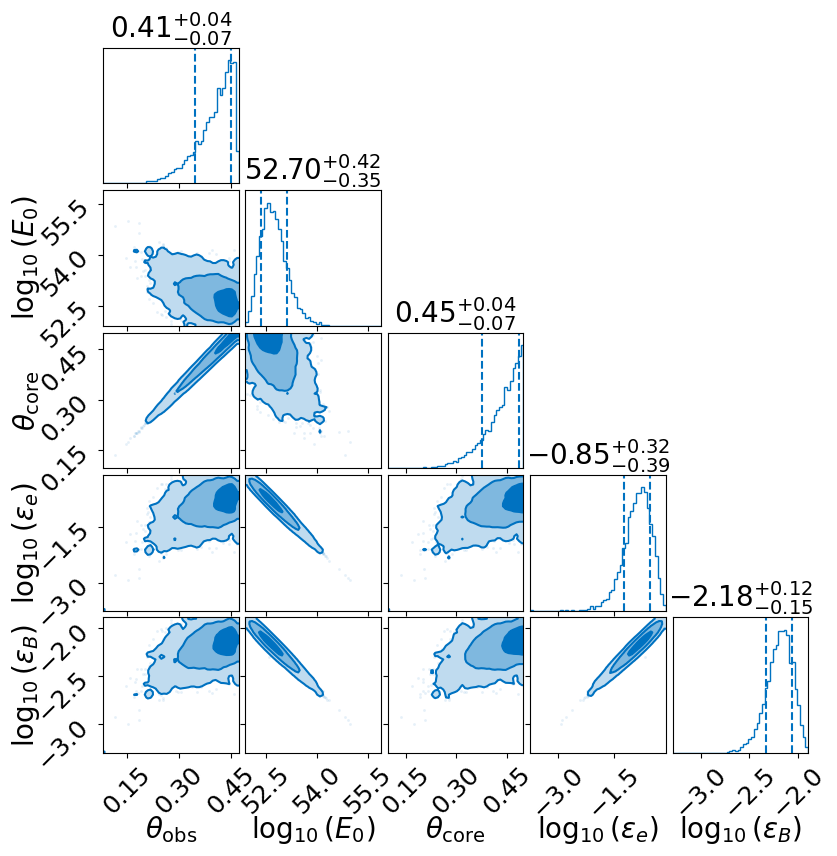}
\end{subfigure}
\hfill
\begin{subfigure}[t]{0.48\textwidth}
    \centering
    \captionsetup{justification=centering, font=small, skip=2pt}
    \caption*{GRB~240910A}
    \includegraphics[width=0.75\textwidth]{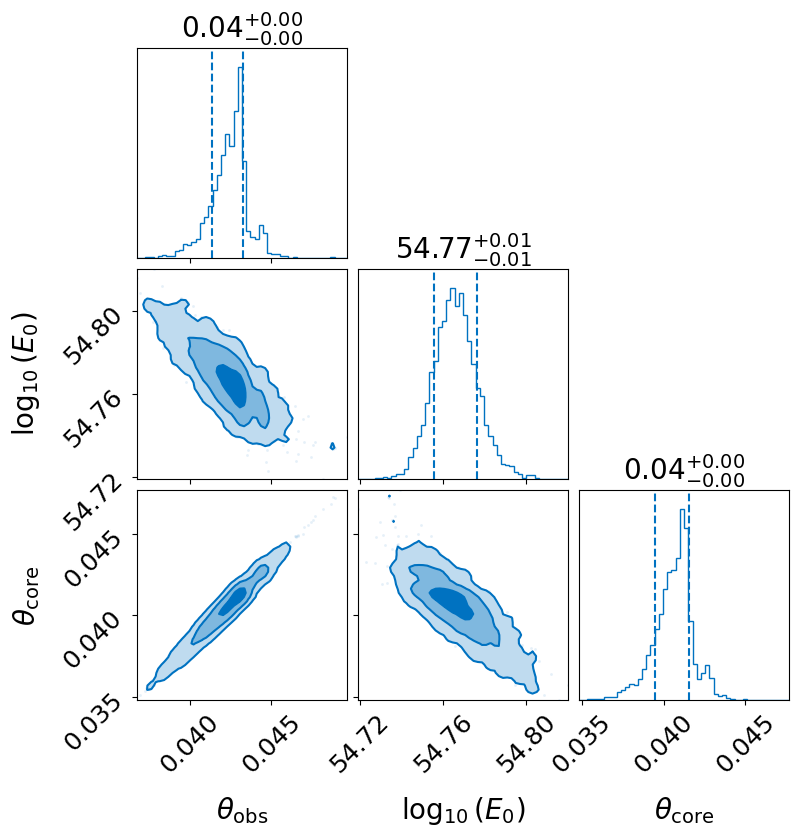}
\end{subfigure}
\vspace{1mm}
\begin{subfigure}[t]{0.48\textwidth}
    \centering
    \captionsetup{justification=centering, font=small, skip=2pt}
    \caption*{GRB~240916A}
    \includegraphics[width=0.75\textwidth]{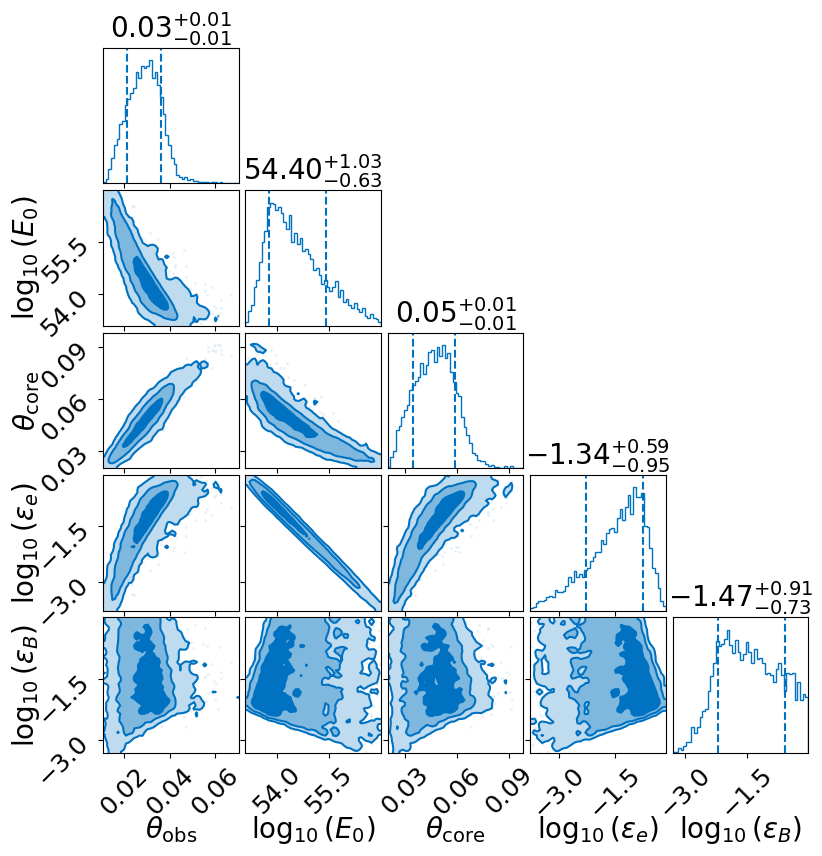}
\end{subfigure}
\hfill
\begin{subfigure}[t]{0.48\textwidth}
    \centering
    \captionsetup{justification=centering, font=small, skip=2pt}
    \caption*{GRB~241228B}
    \includegraphics[width=0.75\textwidth]{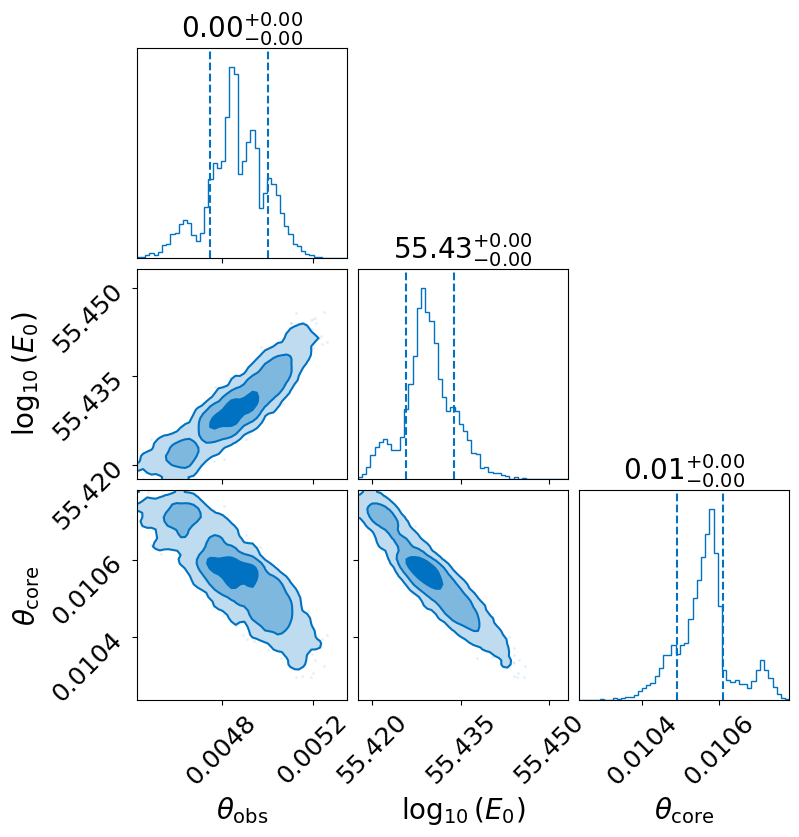}
\end{subfigure}
\caption{Posterior distributions for the TopHat-jet afterglow model parameters inferred for six GRBs analysed in this work: GRBs~240122A, 240225B, 240619A, 240910A, 240916A, and 241228B. Each panel shows the marginalised 1D and 2D posterior distributions from the \texttt{dynesty} nested-sampling run (via \texttt{Bilby}), with contours representing $68\%$ and $95\%$ credible regions. Inferred parameters include the observer angle, core angle, isotropic-equivalent energy, and microphysical quantities (see the text for details).}
\label{fig:modelled_cornerplots}
\end{figure*}

\subsection{Radio}
The radio light curves of GRBs~240122A, 240619A, 240910A, and 240916A are shown in Figure~\ref{fig:radio_comp_lcs}, plotted in flux density (left) and luminosity (right). For context, we compare these with the historical population of GRBs compiled at 8--10~GHz. The GOTO GRBs lie within the locus of LGRBs, showing flux densities and luminosities consistent with this population. None displays the systematically fainter or more rapidly fading behaviour typical of SGRBs. Within the sample, GRB~240122A is among the brightest radio afterglows, while GRB~240910A and GRB~240916A fall at the lower end of the distribution, illustrating the intrinsic spread in LGRB radio emission. Despite this variation, their temporal evolution remains broadly consistent with expectations for LGRB afterglows, reinforcing the conclusion that the radio properties of the GOTO events trace the same underlying population.

\section{Afterglow modelling}\label{sec:afterglow_modelling}

For the afterglow light curve modelling, we used the \texttt{afterglowpy} module (version 0.8.0;~\citealt{Ryan2020}). This Python-based tool utilises the single-shell approximation~\citep{vanEerten2010,vanEerten2018} to model GRB afterglow light curves by accounting for the effects of complex jet structures and an off-axis observer position. We modelled multi-band light curves for each of the GOTO-discovered GRBs presented in this paper. For our modelling, we assumed the simplest TopHat profile of the jet structure and fixed some parameters where required (see Table~\ref{tab:afterglow_modelling}).

All afterglow analyses were conducted using \texttt{dynesty} nested sampling within the \texttt{Bilby} framework (version 2.4.0;~\citealt{Ashton2019}). The dataset mostly consisted of relatively sparse data in optical bands, complemented by X-ray data from \swift/XRT. To ensure robustness and minimise bias, our priors were set to be broad, and the prior probabilities for most parameters were modelled using uniform distributions (see Table~\ref{tab:afterglow_modelling}). We used a Gaussian likelihood and the \texttt{dynesty} nested sampler with 1500 live points and a stopping tolerance of $\Delta\log\mathcal{Z}=0.1$.

When modelling GRBs with sparsely sampled afterglow data, we adopt a flexible strategy in which a subset of parameters is fixed to literature-informed medians that are representative of LGRBs drawn from previous population studies~\citep{Aksulu2022, Chrimes2022} to assess consistency with the typical long-GRB population. These works analysed large samples of LGRBs and reported values for key microphysical parameters: $\epsilon_e$~\citep{Aksulu2022}, and $\epsilon_B$, $n_0$, and $p$~\citep{Chrimes2022}. A value for $\log_{10}\epsilon_e$ is not reported in \citet{Chrimes2022}, from which we take most other fixed parameters owing to its larger sample. We therefore adopt $\epsilon_e$ from \citet{Aksulu2022}. We do not take all parameters from \citet{Aksulu2022} because its GRB sample is smaller; combining their $\epsilon_e$ with the broader \citet{Chrimes2022} set maximises coverage. Our adopted value is slightly below the peak $\epsilon_e\sim 0.13 - 0.15$ inferred from radio peaks by \citet{Beniamini2017}, but it lies within the $\epsilon_e\sim 0.01 - 0.16$ range for a homogeneous medium reported by \citet{Duncan2023}, who used radio peaks together with constraints from the prompt gamma-ray emission efficiency.

For GRBs where full sampling over all parameters led to unconstrained, multimodal, or non-convergent posteriors, we fixed one or more of these quantities to their literature-based mean values. This approach allows stable and interpretable modelling when the data cannot independently constrain all parameters. Fixing select values based on well-motivated priors reduces degeneracies, avoids overfitting, and maintains physical plausibility in the resulting fits.

\textbf{Results overview.} We modelled six GOTO-discovered GRBs (excluding GRB~241002B; no redshift) with the TopHat (uniform) jet model in \texttt{afterglowpy} using \texttt{dynesty} nested sampling via \texttt{Bilby}. Posterior summaries (medians with $16-84\%$ credible intervals) are listed in Table~\ref{tab:afterglow_modelling}; multi-band light curves and posterior corner plots are shown in Figures~\ref{fig:modelled_lcs} and \ref{fig:modelled_cornerplots}, respectively. Unless stated otherwise, we adopt $\xi_N=1$ (fraction of electrons accelerated) as our baseline; for sparsely constrained events, we fix a subset of microphysical parameters ($p$, $n_0$, and, where noted, $\epsilon_B$; see Table~\ref{tab:afterglow_modelling}) to population-informed values to suppress degeneracies. With this setup, the events are well described by narrow, near–on-axis geometries; expected covariances (e.g., $E_0 - \theta_c$) are present but posteriors are unimodal and not prior-bound.

For completeness, we provide short remarks on a subset of GRBs in our sample for which extra clarification is useful. These notes highlight only the key features or caveats, while the overall methodology and global results are presented above.

\noindent\textbf{GRB~240122A:} A three-parameter TopHat fit ($\theta_v$, $\theta_c$, $\log_{10}E_0$) with microphysics fixed as explained above reproduces the broadband evolution (Table~\ref{tab:afterglow_modelling}; Figures~\ref{fig:modelled_lcs},~\ref{fig:modelled_cornerplots}). Radio points lie slightly \emph{above} the model, while late-time X-ray points are slightly \emph{below}; since synchrotron self-absorption would further suppress early-time radio, it cannot explain the positive radio residuals—more plausible are a reverse shock, mild energy injection, a density bump, or calibration offsets.

\noindent\textbf{GRB~240910A:} A three-parameter TopHat fit ($\theta_v$, $\theta_c$, $\log_{10}E_0$) with the other microphysics held fixed reproduces the optical and X-ray light curves with a single parameter set (Table~\ref{tab:afterglow_modelling}; Figures~\ref{fig:modelled_lcs},~\ref{fig:modelled_cornerplots}). The model shows a modest, systematic overprediction in the radio bands. This behaviour is consistent with fixed microphysics --- at fixed $\epsilon_e$, adopting $\xi_N=1$ raises $F_{\nu,\max}$ and lowers $\nu_m$, which naturally boosts the radio while leaving higher-frequency bands close to the data. For completeness, we also explored fits in which additional microphysical parameters were allowed to vary, including $\xi_N$; these trials slightly reduced the radio residuals but degraded the X-ray agreement and produced broader, strongly correlated posteriors. For uniformity across the sample, we therefore retain the three-parameter fit and interpret the residual radio offsets as secondary systematics related to normalisation and propagation effects (e.g., synchrotron self-absorption;~\citealt{Sari1998,Granot2002}) and interstellar scintillation (\citealt{goodman_radio_1997,Frail1997}); possible host free–free absorption is also plausible (see, e.g., \citealt{weiler_radio_2002}).

\noindent\textbf{GRB~240916A.}
A five-parameter TopHat (uniform) jet fit ($\theta_v$, $\theta_c$, $\log_{10}E_0$, $\log_{10}\epsilon_e$, $\log_{10}\epsilon_B$) yields a clean broadband match with unimodal posteriors. The geometry is nearly on-axis with small angles. The microphysics favour a comparatively lower $\epsilon_e$ and a moderately higher $\epsilon_B$ within typical afterglow ranges. Magnetised internal–shock models for the \emph{prompt} phase indicate that magnetisation can alter radiative efficiency and shift the characteristic synchrotron/IC spectral peaks \cite[e.g.][]{Mimica2012}. However, the $\epsilon_B$ inferred here is the \emph{downstream} magnetic energy fraction of the \emph{afterglow} forward shock and is not directly comparable to the prompt–phase shell magnetisation; any putative link to $E_{p,i}$ is therefore model-dependent and not required by our data.

\noindent\textbf{GRB~241002B:} No secure redshift is available. We sampled $z$ with a broad prior, but the redshift posterior remained unconstrained; redshift-dependent quantities ($d_L$, $E_0$, and rest-frame times) track the priors, and the angles are only weakly informed. We therefore do not report parameter estimates and exclude this burst from population-level comparisons.

\noindent\textbf{GRB~241228B.}
We model the afterglow with a three–parameter TopHat jet, sampling $(\theta_v,\ \theta_c,\ \log_{10}E_0)$ while holding the microphysics fixed ($p$, $n_0$, $\epsilon_e$, $\epsilon_B$, $\xi_N$). This minimal configuration reproduces the optical and X–ray evolution at early--to--intermediate epochs (Table~\ref{tab:afterglow_modelling}; Figures~\ref{fig:modelled_lcs},~\ref{fig:modelled_cornerplots}). Small, band–dependent residuals appear around the X–ray band and the bluest optical filters, consistent with a cooling break lying close to the optical and/or modest host galaxy extinction; these offsets remain at a low level under reasonable microphysical choices. Because the microphysics are fixed, the fit can trade flux normalisation against geometry, and the posterior favours an effectively on–axis view with a very narrow core. We therefore regard the recovered $\theta_c$ as a model–dependent lower bound --- allowing, for example, $\epsilon_B$ to vary with a broad log–uniform prior would broaden the $\theta_c$ posterior and plausibly shift its median upward, at the cost of a higher $E_0$.

At late times, the model systematically underpredicts the flux across bands, indicating additional physics not captured by a single forward–shock component. Plausible explanations include mild, sustained energy injection (refreshed shocks), interaction with a local density enhancement, or an additional outflow component (e.g., a wider jet or cocoon). These effects can maintain the late–time emission above the single–jet prediction without disrupting the early–time agreement. Given the limited leverage to discriminate among scenarios, we retain the three–parameter TopHat fit for the main analysis and note the late–time excess as a likely secondary component.

\noindent\textbf{Beaming factor and jet energetics.}
For each GRB in our sample (see Table~\ref{tab:afterglow_modelling}) we compute the beaming factor and the beaming-corrected jet energy using the relations
\(
  f_b = 1 - \cos\theta_c,\qquad
  E_{\rm jet} = f_b\,E_0.
\)

\noindent\emph{Per-burst summary (medians with 16–84\% uncertainties):}

\emph{240122A}:\;
\(f_b = (2.86^{+0.31}_{-0.28})\times10^{-4}\),\;
\(E_{\rm jet} = (2.66^{+0.20}_{-0.19})\times10^{51}\,\mathrm{erg}\).

\emph{240225B}:\;
\(f_b = (2.61^{+0.22}_{-0.22})\times10^{-3}\),\;
\(E_{\rm jet} = (4.83^{+0.06}_{-0.05})\times10^{51}\,\mathrm{erg}\).

\emph{240619A}:\;
\(f_b = (9.79^{+1.76}_{-2.73})\times10^{-2}\),\;
\(E_{\rm jet} = (4.80^{+6.51}_{-2.51})\times10^{51}\,\mathrm{erg}\).

\emph{240910A}:\;
\(f_b = (8.29^{+0.33}_{-0.50})\times10^{-4}\),\;
\(E_{\rm jet} = (4.82^{+0.15}_{-0.21})\times10^{51}\,\mathrm{erg}\).

\emph{240916A}:\;
\(f_b = (1.12^{+0.62}_{-0.52})\times10^{-3}\),\;
\(E_{\rm jet} = (2.89^{+14.20}_{-1.93})\times10^{51}\,\mathrm{erg}\).

\emph{241228B}:\;
\(f_b = (5.58^{+0.05}_{-0.08})\times10^{-5}\),\;
\(E_{\rm jet} = (1.50^{+0.01}_{-0.01})\times10^{51}\,\mathrm{erg}\).

Overall, the parameters inferred from our TopHat (uniform) jet fits --- specifically the observer angle ($\theta_v$), the jet half-opening angle ($\theta_j \equiv \theta_c$ in this model), and the isotropic-equivalent kinetic energy ($E_0$) --- are consistent with the ranges reported in previous studies of long-duration GRB afterglows. In all cases, we recover $\theta_v$ and $\theta_j$ of order a few degrees, in line with the narrow jet geometries commonly found in broadband afterglow fits (e.g.~\citealt{Frail2001,Bloom2003,Friedman2005,Racusin2009}). We use $E_0$ to estimate the beaming-corrected jet kinetic energy, yielding values that cluster around $10^{51} - 10^{52}$\,erg, consistent with the canonical long-GRB energy scale (e.g.~\cite{Friedman2005}). This agreement in both angular geometry and energetics supports the robustness of our TopHat-jet modelling and places these events firmly within the established population of classical long-duration GRBs.

\section{Summary and Conclusion}\label{sec:summary_conclusion}
Over the past few years, the Gravitational-wave Optical Transient Observer (GOTO) has become instrumental in the search for, and rapid follow-up of, optical counterparts to poorly localised transients such as GRBs and GW events. Since achieving first light in June 2017, GOTO has steadily progressed from its prototype (GOTO-4) phase into a fully operational dual-site facility (GOTO-36). 

During the prototype era (2017--2020), the GOTO-4 system responded to 77 \fermi/GBM and 29 {\em Swift}/BAT triggers, securing its first optical afterglow detection with GRB~171205A. Following its expansion to GOTO-36, GOTO attempted follow-up observations of more than 257 \fermi, 43 {\em Swift}, 28 {\em EP}, and 7 {\em GECAM} triggers up to 31 December 2024. Whereas, to date, GOTO has issued nearly 80 GCN circulars and yielded $\approx28$ confirmed afterglow detections, ranging from rapid identifications such as GRB~230818A within $4.43$~min of the trigger, to wide-field discoveries of poorly localised events including GRB~230911A and the SGRB~241105A. Collectively, these results highlight GOTO’s ability to respond on timescales as short as 36 seconds and to cover hundreds of square degrees in order to identify optical afterglows under challenging localisation conditions.

Within this broader context, the present study focuses on the first systematic sample of LGRB afterglows detected by GOTO, discovered during 2024. Our sample comprises seven LGRBs (\grbssample), including two \maxi/GSC events (GRBs~240122A and 240225B), localised to arcminute precision and detected serendipitously during survey operations, and five \fermi/GBM events (GRBs~240619A, 240910A, 240916A, 241002B, and 241228B), recovered through rapid, targeted tiling of degree-scale localisation regions. For all seven LGRBs, GOTO provided the earliest optical detections, with response times ranging from $\sim19.3$~min to $9.4$~hr and with sky coverage exceeding $75\%$ of the $95\%$ probability regions for the GBM bursts. 

Notably, GRB~241228B provides an illustrative example: its afterglow was identified on the 94.5\% probability contour, outside the typical GBM $90\%$ localisation region. While the majority of GRB counterparts are recovered within the $90\%$ region, a small fraction are expected to lie beyond, making this a noteworthy case that highlights both the statistical nature of localisation regions and the importance of wide-field optical follow-up. These results highlight the adaptability and efficiency of the GOTO network in responding to both well-localised (\maxi/GSC) and more uncertain (\fermi/GBM) GRBs. Regardless of the size or shape of the localisation area, GOTO’s rapid tiling strategy enabled meaningful coverage and facilitated the identification of several optical afterglows within its fields.

These rapid identifications enabled immediate triggering of {\em Swift}/XRT and UVOT observations and coordinated multi-wavelength follow-up using facilities around the globe, underscoring the central role of optical discovery of poorly localised GRBs in constraining their properties. The follow-up campaign yielded detections in the X-ray, UV, optical, and radio bands for most of the events in our sample. {\em Swift}/XRT confirmed X-ray counterparts for all bursts. Optical photometry from multiple facilities provided light curves extending from minutes to days post-trigger, showing a broad range of brightnesses and decay rates. Spectroscopy for five events using the VLT/X-shooter (GRBs~240122A, 240225B, 240619A, 240916A, and 241228B) and for two events using GTC/OSIRIS (GRBs~240122A and 240910A) delivered precise redshifts spanning $z \approx 0.40-3.16$, along with absorption line diagnostics tracing both host galaxy interstellar media and, for the higher redshift bursts, intervening absorbers. Radio detections for four GRBs (240122A, 240619A, 240910A, and 240916A) utilising mainly ATCA and VLA confirmed long-lived synchrotron emission, most likely arising from forward shocks. Taken together, this multi-wavelength dataset has enabled robust classification and placed each burst in the broader context of the LGRB population.

Analysis of the prompt emission using \maxi/GSC and \fermi/GBM data revealed a spectrally hard sample, with four events yielding measurable $E_{\rm p}$ values (GRBs~240619A, 240910A, 240916A, and 241228B). Two bursts, GRBs~240916A and 241228B, stand out as $>3\sigma$ outliers to the Amati relation, while others displayed unusually hard low-energy photon indices, pointing to diversity in jet microphysics and, in some cases, potentially high magnetisation. The measured \tninty durations, ranging from $\sim$20\,s to over 270\,s, confirm all seven events as LGRBs, encompassing both short-engine and long-engine members of the class.

Comparisons with the broader GRB population reinforce this conclusion. The GOTO-detected afterglows occupy the established luminosity–time phase space of LGRBs in both X-rays and optical, while their radio detections likewise follow the known locus of synchrotron afterglows. Their redshifts ($z \sim 0.40 - 3.16$) span both nearby and distant events. Taken together, these results demonstrate that the GOTO sample is representative of the wider population, probing afterglows across X-ray, optical, and radio wavelengths and capturing their diversity in temporal evolution and redshift. This highlights GOTO’s capability to deliver well-localised optical counterparts that integrate seamlessly with multi-wavelength studies of GRBs. 

The contrast between typical afterglow behaviour and spectrally hard prompt emission in the GOTO sample likely reflects an observational bias: GRBs with higher $E_{\rm p}$ generally have larger $E_{\rm iso}$ and correspondingly brighter afterglows, making them easier to detect at optical wavelengths. While this tendency favours luminous events in poorly localised searches, it also provides a useful window into jet microphysics and central engine diversity. At the same time, it highlights the importance of wide-field optical facilities in complementing high-energy triggered samples and extending GRB studies across both nearby and high-redshift regimes.

We modelled the afterglows of six of seven GRBs in our sample (excluding GRB~241002B, which lacks a redshift). TopHat-jet parameters inferred here, observer angle ($\theta_{\mathrm{obs}}$), jet core angle ($\theta_c$), and isotropic-equivalent kinetic energy ($E_0$), are consistent with ranges typically found for LGRB afterglows. In all cases, we recover $\theta_{\mathrm{obs}}$ and $\theta_c$ of order a few degrees, in line with the narrow jet geometries commonly obtained from broadband afterglow fits. Using $E_0$ to estimate the true energy budget, the beaming-corrected jet kinetic energies cluster around $10^{51} - 10^{52}$~erg, consistent with the canonical LGRB energy scale after correcting for beaming. Microphysical posteriors are broadly consistent with expectations for external-shock synchrotron emission. One event, GRB~241228B, shows a late-time flux excess relative to the best-fitting TopHat model, suggestive of an additional emission component or prolonged central-engine activity. Taken together, the geometry, energetics, and microphysics inferred from our uniform fits place these GOTO-discovered GRBs squarely within the established population of classical long-duration bursts while clarifying the levers that drive diversity in light-curve morphology. GOTO’s early discovery and dense optical cadence provide key leverage for constraining pre-break behaviour and for enabling robust, comparable modelling across events.

In all, the results presented in this study clearly demonstrate that GOTO’s wide-field, dual-site, fully robotic design, combined with adaptive trigger-specific strategies, is highly effective for bridging the gap between poorly localised high-energy triggers and the precise positions needed for multi-wavelength follow-up. The detections presented here highlight GOTO’s ability to recover GRB afterglows in a wide range of redshifts, localisation scales, and intrinsic properties, spanning both representative events and rare, energetically extreme outliers. In the emerging era of multi-messenger astronomy, GOTO’s demonstrated capability for rapid, deep optical searches makes it a critical asset for identifying and characterising counterparts to both gravitational wave events and gamma-ray bursts, thus advancing our understanding of the most energetic explosions in the Universe.

\section*{Acknowledgements}
This work makes use of the Gravitational-wave Optical Transient Observer (GOTO) observations that led to the discovery of the optical afterglows presented in this study. GOTO project acknowledges support from the Science and Technology Facilities Council (STFC, grant numbers ST/T007184/1, ST/T003103/1, ST/T000406/1, ST/X001121/1 and ST/Z000165/1) and the GOTO consortium institutions; University of Warwick; Monash University; University of Sheffield; University of Leicester; Armagh Observatory \& Planetarium; the National Astronomical Research Institute of Thailand (NARIT); University of Manchester; Instituto de Astrofísica de Canarias (IAC); University of Portsmouth; University of Turku.

This work is partly based on observations collected at the European Organisation for Astronomical Research in the Southern Hemisphere under ESO programmes 110.24CF and 114.27PZ (Co-PIs Tanvir, Malesani, Vergani).

It is also partly based on data obtained with the instrument OSIRIS, built by a Consortium led by the Instituto de Astrof\'isica de Canarias in collaboration with the Instituto de Astronom\'ia of the Universidad Aut\'onoma de México. OSIRIS was funded by GRANTECAN and the National Plan of Astronomy and Astrophysics of the Spanish Government. 

It is also partly based on data obtained with the ATCA and we acknowledge the Gomeroi people as the traditional owners of the ATCA observatory site. The ATCA is part of the Australia Telescope National Facility, which is funded by the Australian Government for operation as a National Facility managed by CSIRO.

It is also partly based on observations made with the Nordic Optical Telescope, owned in collaboration by the University of Turku and Aarhus University, and operated jointly by Aarhus University, the University of Turku and the University of Oslo, representing Denmark, Finland and Norway, the University of Iceland and Stockholm University at the Observatorio del Roque de los Muchachos, La Palma, Spain, of the Instituto de Astrofisica de Canarias. The NOT data were obtained under program IDs 68-020, 69-023, and 70-507 (PIs Malesani, Fynbo, Xu).
The National Radio Astronomy Observatory and Green Bank Observatory are facilities of the U.S. National Science Foundation operated under cooperative agreement by Associated Universities, Inc.
Mephisto is developed at and operated by the South-Western Institute for Astronomy Research of Yunnan University (SWIFAR-YNU), funded by the ``Yunnan University Development Plan for World-Class University" and ``Yunnan University Development Plan for World-Class Astronomy Discipline".
Based on observations made at the Observatorio de Sierra Nevada (OSN), operated by the Instituto de Astrof\'isica de Andaluc\'ia (IAA-CSIC). Based on observations made with the Thai Robotic Telescopes under program ID TRTToO\_2024005, which is operated by the National Astronomical Research Institute of Thailand (Public Organisation).

AK and JRM are supported by the UK Science and Technology Facilities Council (STFC) Consolidated grant ST/V000853/1.
BPG acknowledges support from STFC grant No. ST/Y002253/1. BPG and DO acknowledge support from The Leverhulme Trust grant No. RPG-2024-117.
BS and SDV acknowledge the support of the French Agence Nationale de la Recherche ({\it ANR}), under grant ANR-23-CE31-0011 (project PEGaSUS).
DS is supported by the UK Science and Technology Facilities Council (STFC, grant numbers ST/T007184/1, ST/T003103/1, and ST/T000406/1).
SG acknowledges support from the Istituto Nazionale di Astrofisica (INAF), project number: 1.05.24.07.04.
BK acknowledges support from the ``Special Project for High-End Foreign Experts," Xingdian Funding from Yunnan Province; the Yunnan Key Laboratory of Survey Science (grant 202449CE340002), and National Key Research and Development Program of China (grant 2024YFA1611603).
BW is supported by the UKRI's STFC studentship grant funding, project reference ST/X508871/1.
RLCS and SM acknowledge support from The Leverhulme Trust grant RPG-2023-240.
NH, POB and NRT acknowledge support from UKRI/STFC grant ST/W000857/1.
JDL acknowledges support from a UK Research and Innovation Future Leaders Fellowship (MR/T020784/1).
TLK acknowledges a Warwick Astrophysics prize post-doctoral fellowship made possible thanks to a generous philanthropic donation.
MAA acknowledges support from grants PID2021-127495NB-I00 (funded by MCIN/AEI/10.13039/501100011033, EU), ASFAE/2022/026 (funded by MCIN, EU NextGenerationEU PRTR-C17.I1), and CIPROM/2022/13 (Generalitat Valenciana).  DLC acknowledges support from the UK Science and Technology Facilities Council (STFC) grant number ST/X001121/1. AS acknowledges support by a postdoctoral fellowship from the CNES.
AR acknowledges support from the INAF project Supporto Arizona \& Italia.
AMC and LC acknowledge support from the Irish Research Council Postgraduate Scholarship No. GOIPG/2022/1008.
MEW and IW are supported by the UKRI Science and Technology Facilities Council (STFC).
DBM is funded by the European Union (ERC, HEAVYMETAL, 101071865). Views and opinions expressed are, however, those of the authors only and do not necessarily reflect those of the European Union or the European Research Council. Neither the European Union nor the granting authority can be held responsible for them. The Cosmic Dawn Center (DAWN) is funded by the Danish National Research Foundation under grant DNRF140.
NSP and ASP are grateful to the Russian Science Foundation (project no. 23-12-00220) for their partial support of the data reduction, analysis of data.
EVK is grateful to the Ministry of Science and Higher Education of the Russian Federation for financial support of this work; the AZT-33IK telescope is part of the Center for Common Use "Angara".

\section*{Data Availability}
All datasets supporting this study are provided in the paper. Additional data are available from the corresponding author upon reasonable request.

\bibliographystyle{mnras}
\bibliography{mnras}

\begin{thebibliography}{}
\makeatletter
\relax
\def\mn@urlcharsother{\let\do\@makeother \do\$\do\&\do\#\do\^\do\_\do\%\do\~}
\def\mn@doi{\begingroup\mn@urlcharsother \@ifnextchar [ {\mn@doi@} {\mn@doi@[]}}
\def\mn@doi@[#1]#2{\def\@tempa{#1}\ifx\@tempa\@empty \href {http://dx.doi.org/#2} {doi:#2}\else \href {http://dx.doi.org/#2} {#1}\fi \endgroup}
\def\mn@eprint#1#2{\mn@eprint@#1:#2::\@nil}
\def\mn@eprint@arXiv#1{\href {http://arxiv.org/abs/#1} {{\tt arXiv:#1}}}
\def\mn@eprint@dblp#1{\href {http://dblp.uni-trier.de/rec/bibtex/#1.xml} {dblp:#1}}
\def\mn@eprint@#1:#2:#3:#4\@nil{\def\@tempa {#1}\def\@tempb {#2}\def\@tempc {#3}\ifx \@tempc \@empty \let \@tempc \@tempb \let \@tempb \@tempa \fi \ifx \@tempb \@empty \def\@tempb {arXiv}\fi \@ifundefined {mn@eprint@\@tempb}{\@tempb:\@tempc}{\expandafter \expandafter \csname mn@eprint@\@tempb\endcsname \expandafter{\@tempc}}}

\bibitem[\protect\citeauthoryear{{Abbott} et~al.,}{{Abbott} et~al.}{2017a}]{Abbott2017}
{Abbott} B.~P.,  et~al., 2017a, \mn@doi [\prl] {10.1103/PhysRevLett.119.161101}, \href {https://ui.adsabs.harvard.edu/abs/2017PhRvL.119p1101A} {119, 161101}

\bibitem[\protect\citeauthoryear{{Abbott} et~al.,}{{Abbott} et~al.}{2017b}]{Abbott2017b}
{Abbott} B.~P.,  et~al., 2017b, \mn@doi [\apjl] {10.3847/2041-8213/aa920c}, \href {https://ui.adsabs.harvard.edu/abs/2017ApJ...848L..13A} {848, L13}

\bibitem[\protect\citeauthoryear{{Ahumada} et~al.,}{{Ahumada} et~al.}{2021}]{Ahumada2021}
{Ahumada} T.,  et~al., 2021, \mn@doi [Nature Astronomy] {10.1038/s41550-021-01428-7}, \href {https://ui.adsabs.harvard.edu/abs/2021NatAs...5..917A} {5, 917}

\bibitem[\protect\citeauthoryear{{Akaike}}{{Akaike}}{1974}]{Akaike1974}
{Akaike} H.,  1974, \mn@doi [IEEE Transactions on Automatic Control] {10.1109/TAC.1974.1100705}, 19, 716

\bibitem[\protect\citeauthoryear{{Aksulu}, {Wijers}, {van Eerten}  \& {van der Horst}}{{Aksulu} et~al.}{2022}]{Aksulu2022}
{Aksulu} M.~D.,  {Wijers} R.~A.~M.~J.,  {van Eerten} H.~J.,   {van der Horst} A.~J.,  2022, \mn@doi [\mnras] {10.1093/mnras/stac246}, \href {https://ui.adsabs.harvard.edu/abs/2022MNRAS.511.2848A} {511, 2848}

\bibitem[\protect\citeauthoryear{{Aloy} \& {Obergaulinger}}{{Aloy} \& {Obergaulinger}}{2021}]{Aloy2021}
{Aloy} M.~{\'A}.,  {Obergaulinger} M.,  2021, \mn@doi [\mnras] {10.1093/mnras/staa3273}, \href {https://ui.adsabs.harvard.edu/abs/2021MNRAS.500.4365A} {500, 4365}

\bibitem[\protect\citeauthoryear{{Aloy}, {M{\"u}ller}, {Ib{\'a}{\~n}ez}, {Mart{\'\i}}  \& {MacFadyen}}{{Aloy} et~al.}{2000}]{Aloy2000}
{Aloy} M.~A.,  {M{\"u}ller} E.,  {Ib{\'a}{\~n}ez} J.~M.,  {Mart{\'\i}} J.~M.,   {MacFadyen} A.,  2000, \mn@doi [\apjl] {10.1086/312537}, \href {https://ui.adsabs.harvard.edu/abs/2000ApJ...531L.119A} {531, L119}

\bibitem[\protect\citeauthoryear{{Aloy}, {Cuesta-Mart{\'\i}nez}  \& {Obergaulinger}}{{Aloy} et~al.}{2018}]{Aloy2018}
{Aloy} M.~A.,  {Cuesta-Mart{\'\i}nez} C.,   {Obergaulinger} M.,  2018, \mn@doi [\mnras] {10.1093/mnras/sty1212}, \href {https://ui.adsabs.harvard.edu/abs/2018MNRAS.478.3576A} {478, 3576}

\bibitem[\protect\citeauthoryear{{Amati}}{{Amati}}{2006}]{Amati06}
{Amati} L.,  2006, \mn@doi [\mnras] {10.1111/j.1365-2966.2006.10840.x}, \href {https://ui.adsabs.harvard.edu/abs/2006MNRAS.372..233A} {372, 233}

\bibitem[\protect\citeauthoryear{{Amati} \& {Della Valle}}{{Amati} \& {Della Valle}}{2013}]{Amati2013}
{Amati} L.,  {Della Valle} M.,  2013, \mn@doi [International Journal of Modern Physics D] {10.1142/S0218271813300280}, \href {https://ui.adsabs.harvard.edu/abs/2013IJMPD..2230028A} {22, 1330028}

\bibitem[\protect\citeauthoryear{{Amati} et~al.,}{{Amati} et~al.}{2002}]{Amati02}
{Amati} L.,  et~al., 2002, \mn@doi [\aap] {10.1051/0004-6361:20020722}, \href {https://ui.adsabs.harvard.edu/abs/2002A&A...390...81A} {390, 81}

\bibitem[\protect\citeauthoryear{{An} et~al.,}{{An} et~al.}{2024}]{An2024GCN38704}
{An} J.,  et~al., 2024, GRB Coordinates Network, \href {https://ui.adsabs.harvard.edu/abs/2024GCN.38704....1A} {38704, 1}

\bibitem[\protect\citeauthoryear{{Anderson}, {Leung}, {Murphy}, {Lenc}, {Rhodes}, {van der Horst}  \& {Rowell}}{{Anderson} et~al.}{2023}]{2023GCN.33475....1A}
{Anderson} G.~E.,  {Leung} J.~K.,  {Murphy} T.,  {Lenc} E.,  {Rhodes} L.,  {van der Horst} A.~J.,   {Rowell} G.,  2023, GRB Coordinates Network, \href {https://ui.adsabs.harvard.edu/abs/2023GCN.33475....1A} {33475, 1}

\bibitem[\protect\citeauthoryear{{Anderson} et~al.,}{{Anderson} et~al.}{2024a}]{2024ApJ...975L..13A}
{Anderson} G.~E.,  et~al., 2024a, \mn@doi [\apjl] {10.3847/2041-8213/ad85e9}, \href {https://ui.adsabs.harvard.edu/abs/2024ApJ...975L..13A} {975, L13}

\bibitem[\protect\citeauthoryear{{Anderson}, {Gulati}, {Rhodes}, {Leung}, {van der Horst}, {Chastain}  \& {PanRadio GRB Collaboration}}{{Anderson} et~al.}{2024b}]{Anderson2024GCN35642}
{Anderson} G.~E.,  {Gulati} A.,  {Rhodes} L.,  {Leung} J.~K.,  {van der Horst} A.~J.,  {Chastain} S.,   {PanRadio GRB Collaboration} 2024b, GRB Coordinates Network, \href {https://ui.adsabs.harvard.edu/abs/2024GCN.35642....1A} {35642, 1}

\bibitem[\protect\citeauthoryear{{Anderson} et~al.,}{{Anderson} et~al.}{2025}]{anderson2025arXiv250814650A}
{Anderson} G.~E.,  et~al., 2025, \mn@doi [arXiv e-prints] {10.48550/arXiv.2508.14650}, \href {https://ui.adsabs.harvard.edu/abs/2025arXiv250814650A} {p. arXiv:2508.14650}

\bibitem[\protect\citeauthoryear{{Ashton} et~al.,}{{Ashton} et~al.}{2019}]{Ashton2019}
{Ashton} G.,  et~al., 2019, \mn@doi [\apjs] {10.3847/1538-4365/ab06fc}, \href {https://ui.adsabs.harvard.edu/abs/2019ApJS..241...27A} {241, 27}

\bibitem[\protect\citeauthoryear{{Band} et~al.,}{{Band} et~al.}{1993}]{Band1993}
{Band} D.,  et~al., 1993, \mn@doi [\apj] {10.1086/172995}, \href {https://ui.adsabs.harvard.edu/abs/1993ApJ...413..281B} {413, 281}

\bibitem[\protect\citeauthoryear{{Belkin} et~al.,}{{Belkin} et~al.}{2023}]{Belkin2023GCN34681}
{Belkin} S.,  et~al., 2023, GRB Coordinates Network, \href {https://ui.adsabs.harvard.edu/abs/2023GCN.34681....1P} {34681, 1}

\bibitem[\protect\citeauthoryear{{Belkin} et~al.,}{{Belkin} et~al.}{2024}]{Belkin2024RNAAS}
{Belkin} S.,  et~al., 2024, \mn@doi [Research Notes of the American Astronomical Society] {10.3847/2515-5172/ad1876}, \href {https://ui.adsabs.harvard.edu/abs/2024RNAAS...8....6B} {8, 6}

\bibitem[\protect\citeauthoryear{{Beniamini} \& {van der Horst}}{{Beniamini} \& {van der Horst}}{2017}]{Beniamini2017}
{Beniamini} P.,  {van der Horst} A.~J.,  2017, \mn@doi [\mnras] {10.1093/mnras/stx2203}, \href {https://ui.adsabs.harvard.edu/abs/2017MNRAS.472.3161B} {472, 3161}

\bibitem[\protect\citeauthoryear{{Beniamini}, {Granot}  \& {Gill}}{{Beniamini} et~al.}{2020}]{Beniamini2020}
{Beniamini} P.,  {Granot} J.,   {Gill} R.,  2020, \mn@doi [\mnras] {10.1093/mnras/staa538}, \href {https://ui.adsabs.harvard.edu/abs/2020MNRAS.493.3521B} {493, 3521}

\bibitem[\protect\citeauthoryear{Berger et~al.,}{Berger et~al.}{2000}]{berger_jet_2000}
Berger E.,  et~al., 2000, \mn@doi [The Astrophysical Journal] {10.1086/317814}, 545, 56

\bibitem[\protect\citeauthoryear{Berger et~al.,}{Berger et~al.}{2001a}]{berger_grb_2001}
Berger E.,  et~al., 2001a, \mn@doi [The Astrophysical Journal] {10.1086/321612}, 556, 556

\bibitem[\protect\citeauthoryear{Berger, Kulkarni  \& Frail}{Berger et~al.}{2001b}]{berger_host_2001}
Berger E.,  Kulkarni S.~R.,   Frail D.~A.,  2001b, \mn@doi [The Astrophysical Journal] {10.1086/322247}, 560, 652

\bibitem[\protect\citeauthoryear{{Berger} et~al.,}{{Berger} et~al.}{2005}]{2005Natur.438..988B}
{Berger} E.,  et~al., 2005, \mn@doi [\nat] {10.1038/nature04238}, \href {https://ui.adsabs.harvard.edu/abs/2005Natur.438..988B} {438, 988}

\bibitem[\protect\citeauthoryear{Bhat et~al.,}{Bhat et~al.}{2016}]{Bhat2016}
Bhat P.~N.,  et~al., 2016, \mn@doi [The Astrophysical Journal Supplement Series] {10.3847/0067-0049/223/2/28}, 223, 28

\bibitem[\protect\citeauthoryear{{Bj{\"o}rnsson}, {Gudmundsson}  \& {J{\'o}hannesson}}{{Bj{\"o}rnsson} et~al.}{2004}]{Bjornsson2004}
{Bj{\"o}rnsson} G.,  {Gudmundsson} E.~H.,   {J{\'o}hannesson} G.,  2004, \mn@doi [\apjl] {10.1086/426477}, \href {https://ui.adsabs.harvard.edu/abs/2004ApJ...615L..77B} {615, L77}

\bibitem[\protect\citeauthoryear{{Bloom}, {Frail}  \& {Kulkarni}}{{Bloom} et~al.}{2003}]{Bloom2003}
{Bloom} J.~S.,  {Frail} D.~A.,   {Kulkarni} S.~R.,  2003, \mn@doi [\apj] {10.1086/377125}, \href {https://ui.adsabs.harvard.edu/abs/2003ApJ...594..674B} {594, 674}

\bibitem[\protect\citeauthoryear{{Boch}, {Fernique}, {Bonnarel}, {Chaitra}, {Bot}, {Pineau}, {Baumann}  \& {Michel}}{{Boch} et~al.}{2020}]{Boch2020HIPS2FITS}
{Boch} T.,  {Fernique} P.,  {Bonnarel} F.,  {Chaitra} C.,  {Bot} C.,  {Pineau} F.~X.,  {Baumann} M.,   {Michel} L.,  2020, in {Pizzo} R.,  {Deul} E.~R.,  {Mol} J.~D.,  {de Plaa} J.,   {Verkouter} H.,  eds,  Astronomical Society of the Pacific Conference Series Vol. 527, Astronomical Data Analysis Software and Systems XXIX. p.~121

\bibitem[\protect\citeauthoryear{{Bradley} et~al.,}{{Bradley} et~al.}{2024}]{Larry2024}
{Bradley} L.,  et~al., 2024, {astropy/photutils: 2.0.2}, \mn@doi{10.5281/zenodo.13989456}

\bibitem[\protect\citeauthoryear{{Breeveld}, {Landsman}, {Holland}, {Roming}, {Kuin}  \& {Page}}{{Breeveld} et~al.}{2011}]{Breeveld11}
{Breeveld} A.~A.,  {Landsman} W.,  {Holland} S.~T.,  {Roming} P.,  {Kuin} N.~P.~M.,   {Page} M.~J.,  2011, in {McEnery} J.~E.,  {Racusin} J.~L.,   {Gehrels} N.,  eds,  American Institute of Physics Conference Series Vol. 1358, Gamma Ray Bursts 2010. AIP, pp 373--376 (\mn@eprint {arXiv} {1102.4717}), \mn@doi{10.1063/1.3621807}

\bibitem[\protect\citeauthoryear{{Bromberg}, {Nakar}, {Piran}  \& {Sari}}{{Bromberg} et~al.}{2013}]{Bromberg2013}
{Bromberg} O.,  {Nakar} E.,  {Piran} T.,   {Sari} R.,  2013, \mn@doi [\apj] {10.1088/0004-637X/764/2/179}, \href {https://ui.adsabs.harvard.edu/abs/2013ApJ...764..179B} {764, 179}

\bibitem[\protect\citeauthoryear{{Burrows} et~al.,}{{Burrows} et~al.}{2005}]{Burrows2005}
{Burrows} D.~N.,  et~al., 2005, \mn@doi [\ssr] {10.1007/s11214-005-5097-2}, \href {https://ui.adsabs.harvard.edu/abs/2005SSRv..120..165B} {120, 165}

\bibitem[\protect\citeauthoryear{{Cano}, {Wang}, {Dai}  \& {Wu}}{{Cano} et~al.}{2017}]{Cano2017}
{Cano} Z.,  {Wang} S.-Q.,  {Dai} Z.-G.,   {Wu} X.-F.,  2017, \mn@doi [Adv. Astron] {10.1155/2017/8929054}, \href {https://ui.adsabs.harvard.edu/abs/2017AdAst2017E...5C} {2017, 8929054}

\bibitem[\protect\citeauthoryear{{Cash}}{{Cash}}{1979}]{Cash1979}
{Cash} W.,  1979, \mn@doi [\apj] {10.1086/156922}, \href {https://ui.adsabs.harvard.edu/abs/1979ApJ...228..939C} {228, 939}

\bibitem[\protect\citeauthoryear{Cenko et~al.,}{Cenko et~al.}{2006}]{cenko_multiwavelength_2006}
Cenko S.~B.,  et~al., 2006, \mn@doi [The Astrophysical Journal] {10.1086/508149}, 652, 490

\bibitem[\protect\citeauthoryear{Cenko et~al.,}{Cenko et~al.}{2011}]{cenko_afterglow_2011}
Cenko S.~B.,  et~al., 2011, \mn@doi [The Astrophysical Journal] {10.1088/0004-637X/732/1/29}, 732, 29

\bibitem[\protect\citeauthoryear{Cenko et~al.,}{Cenko et~al.}{2012}]{cenko_swift_2012}
Cenko S.~B.,  et~al., 2012, \mn@doi [The Astrophysical Journal] {10.1088/0004-637X/753/1/77}, 753, 77

\bibitem[\protect\citeauthoryear{{Cepa}}{{Cepa}}{1998}]{Cepa1998}
{Cepa} J.,  1998, \mn@doi [\apss] {10.1023/A:1002144913887}, \href {https://ui.adsabs.harvard.edu/abs/1998Ap&SS.263..369C} {263, 369}

\bibitem[\protect\citeauthoryear{{Chambers} et~al.,}{{Chambers} et~al.}{2016}]{Chambers2016}
{Chambers} K.~C.,  et~al., 2016, \mn@doi [arXiv e-prints] {10.48550/arXiv.1612.05560}, \href {https://ui.adsabs.harvard.edu/abs/2016arXiv161205560C} {p. arXiv:1612.05560}

\bibitem[\protect\citeauthoryear{Chandra et~al.,}{Chandra et~al.}{2008}]{chandra_comprehensive_2008}
Chandra P.,  et~al., 2008, \mn@doi [The Astrophysical Journal] {10.1086/589807}, 683, 924

\bibitem[\protect\citeauthoryear{Chandra et~al.,}{Chandra et~al.}{2010}]{chandra_discovery_2010}
Chandra P.,  et~al., 2010, \mn@doi [The Astrophysical Journal] {10.1088/2041-8205/712/1/L31}, 712, L31

\bibitem[\protect\citeauthoryear{{Chen} et~al.,}{{Chen} et~al.}{2020}]{Chen2020}
{Chen} Y.,  et~al., 2020, \mn@doi [Scientia Sinica Physica, Mechanica \& Astronomica] {10.1360/SSPMA-2020-0120}, \href {https://ui.adsabs.harvard.edu/abs/2020SSPMA..50l9507C} {50, 129507}

\bibitem[\protect\citeauthoryear{{Chen} et~al.,}{{Chen} et~al.}{2024}]{Chen2024ApJ...971L...2C}
{Chen} X.,  et~al., 2024, \mn@doi [\apjl] {10.3847/2041-8213/ad62f7}, \href {https://ui.adsabs.harvard.edu/abs/2024ApJ...971L...2C} {971, L2}

\bibitem[\protect\citeauthoryear{{Cheng} et~al.,}{{Cheng} et~al.}{2025}]{Cheng2025ApJ}
{Cheng} Y.,  et~al., 2025, \mn@doi [\apj] {10.3847/1538-4357/ad9ea1}, \href {https://ui.adsabs.harvard.edu/abs/2025ApJ...979...38C} {979, 38}

\bibitem[\protect\citeauthoryear{{Cheung}, {Woolf}, {Kerr}, {Grove}, {Goldstein}, {Wilson-Hodge}, {Kocevski}  \& {Briggs}}{{Cheung} et~al.}{2024}]{Cheung2024GCN35848}
{Cheung} C.~C.,  {Woolf} R.,  {Kerr} M.,  {Grove} J.~E.,  {Goldstein} A.,  {Wilson-Hodge} C.~A.,  {Kocevski} D.,   {Briggs} M.~S.,  2024, GRB Coordinates Network, \href {https://ui.adsabs.harvard.edu/abs/2024GCN.35848....1C} {35848, 1}

\bibitem[\protect\citeauthoryear{{Chrimes} et~al.,}{{Chrimes} et~al.}{2022}]{Chrimes2022}
{Chrimes} A.~A.,  et~al., 2022, \mn@doi [\mnras] {10.1093/mnras/stac1796}, \href {https://ui.adsabs.harvard.edu/abs/2022MNRAS.515.2591C} {515, 2591}

\bibitem[\protect\citeauthoryear{{Christensen}, {Fynbo}, {Prochaska}, {Th{\"o}ne}, {de Ugarte Postigo}  \& {Jakobsson}}{{Christensen} et~al.}{2011}]{2011ApJ...727...73C}
{Christensen} L.,  {Fynbo} J.~P.~U.,  {Prochaska} J.~X.,  {Th{\"o}ne} C.~C.,  {de Ugarte Postigo} A.,   {Jakobsson} P.,  2011, \mn@doi [\apj] {10.1088/0004-637X/727/2/73}, \href {https://ui.adsabs.harvard.edu/abs/2011ApJ...727...73C} {727, 73}

\bibitem[\protect\citeauthoryear{{Clark}}{{Clark}}{1980}]{1980A&A....89..377C}
{Clark} B.~G.,  1980, \aap, \href {https://ui.adsabs.harvard.edu/abs/1980A&A....89..377C} {89, 377}

\bibitem[\protect\citeauthoryear{{Costa} et~al.,}{{Costa} et~al.}{1997}]{Costa1997}
{Costa} E.,  et~al., 1997, \mn@doi [\nat] {10.1038/42885}, \href {https://ui.adsabs.harvard.edu/abs/1997Natur.387..783C} {387, 783}

\bibitem[\protect\citeauthoryear{{Cotter} et~al.,}{{Cotter} et~al.}{2024}]{Cotter2024GCN36813}
{Cotter} L.,  et~al., 2024, GRB Coordinates Network, \href {https://ui.adsabs.harvard.edu/abs/2024GCN.36813....1C} {36813, 1}

\bibitem[\protect\citeauthoryear{Craig et~al.,}{Craig et~al.}{2017}]{matt_craig_2017_1069648}
Craig M.,  et~al., 2017, astropy/ccdproc: v1.3.0.post1, \mn@doi{10.5281/zenodo.1069648}, \url {https://doi.org/10.5281/zenodo.1069648}

\bibitem[\protect\citeauthoryear{{Dafcikova} et~al.,}{{Dafcikova} et~al.}{2024a}]{Dafcikova2024GCN36724}
{Dafcikova} M.,  et~al., 2024a, GRB Coordinates Network, \href {https://ui.adsabs.harvard.edu/abs/2024GCN.36724....1D} {36724, 1}

\bibitem[\protect\citeauthoryear{{Dafcikova} et~al.,}{{Dafcikova} et~al.}{2024b}]{Dafcikova2024GCN37543}
{Dafcikova} M.,  et~al., 2024b, GRB Coordinates Network, \href {https://ui.adsabs.harvard.edu/abs/2024GCN.37543....1D} {37543, 1}

\bibitem[\protect\citeauthoryear{{Dai}, {Guo}, {Zhang}, {Liu}  \& {Wang}}{{Dai} et~al.}{2024}]{Dai2024evidance}
{Dai} C.-Y.,  {Guo} C.-L.,  {Zhang} H.-M.,  {Liu} R.-Y.,   {Wang} X.-Y.,  2024, \mn@doi [\apjl] {10.3847/2041-8213/ad2680}, \href {https://ui.adsabs.harvard.edu/abs/2024ApJ...962L..37D} {962, L37}

\bibitem[\protect\citeauthoryear{{Dainotti} et~al.,}{{Dainotti} et~al.}{2024}]{Dainotti2024}
{Dainotti} M.~G.,  et~al., 2024, \mn@doi [\mnras] {10.1093/mnras/stae1484}, \href {https://ui.adsabs.harvard.edu/abs/2024MNRAS.533.4023D} {533, 4023}

\bibitem[\protect\citeauthoryear{{Dalessi}, {Meegan}  \& {Fermi Gamma-ray Burst Monitor Team}}{{Dalessi} et~al.}{2024}]{Dalessi2024GCN36717}
{Dalessi} S.,  {Meegan} C.,   {Fermi Gamma-ray Burst Monitor Team} 2024, GRB Coordinates Network, \href {https://ui.adsabs.harvard.edu/abs/2024GCN.36717....1D} {36717, 1}

\bibitem[\protect\citeauthoryear{{DeLaunay}, {Tohuvavohu}, {Ronchini}, {Raman}, {Kennea}  \& {Parsotan}}{{DeLaunay} et~al.}{2024a}]{DeLaunay2024GCN37704}
{DeLaunay} J.,  {Tohuvavohu} A.,  {Ronchini} S.,  {Raman} G.,  {Kennea} J.~A.,   {Parsotan} T.,  2024a, GRB Coordinates Network, \href {https://ui.adsabs.harvard.edu/abs/2024GCN.37704....1D} {37704, 1}

\bibitem[\protect\citeauthoryear{{DeLaunay}, {Ronchini}, {Tohuvavohu}, {Raman}, {Kennea}  \& {Parsotan}}{{DeLaunay} et~al.}{2024b}]{DeLaunay2024GCN38700}
{DeLaunay} J.,  {Ronchini} S.,  {Tohuvavohu} A.,  {Raman} G.,  {Kennea} J.~A.,   {Parsotan} T.,  2024b, GRB Coordinates Network, \href {https://ui.adsabs.harvard.edu/abs/2024GCN.38700....1D} {38700, 1}

\bibitem[\protect\citeauthoryear{{Demianski}, {Piedipalumbo}, {Sawant}  \& {Amati}}{{Demianski} et~al.}{2017}]{Demianski2017}
{Demianski} M.,  {Piedipalumbo} E.,  {Sawant} D.,   {Amati} L.,  2017, \mn@doi [\aap] {10.1051/0004-6361/201628909}, \href {https://ui.adsabs.harvard.edu/abs/2017A&A...598A.112D} {598, A112}

\bibitem[\protect\citeauthoryear{{Dereli}, {Bo{\"e}r}, {Gendre}, {Amati}, {Dichiara}  \& {Orange}}{{Dereli} et~al.}{2017}]{Dereli2017}
{Dereli} H.,  {Bo{\"e}r} M.,  {Gendre} B.,  {Amati} L.,  {Dichiara} S.,   {Orange} N.~B.,  2017, \mn@doi [\apj] {10.3847/1538-4357/aa947d}, \href {https://ui.adsabs.harvard.edu/abs/2017ApJ...850..117D} {850, 117}

\bibitem[\protect\citeauthoryear{{Devyatkin}, {Gorshanov}, {Kouprianov}  \& {Verestchagina}}{{Devyatkin} et~al.}{2010}]{Devyatkin2010}
{Devyatkin} A.~V.,  {Gorshanov} D.~L.,  {Kouprianov} V.~V.,   {Verestchagina} I.~A.,  2010, \mn@doi [Solar System Research] {10.1134/S0038094610010090}, \href {https://ui.adsabs.harvard.edu/abs/2010SoSyR..44...68D} {44, 68}

\bibitem[\protect\citeauthoryear{{Di Lalla}, {Holzmann Airasca}, {Khalil}, {Lopez}, {Depalo}, {Cheung}, {Bartolini}  \& {Fermi-LAT Collaboration}}{{Di Lalla} et~al.}{2025}]{DiLalla2025GCN38843}
{Di Lalla} N.,  {Holzmann Airasca} A.,  {Khalil} T.,  {Lopez} S.,  {Depalo} D.,  {Cheung} C.~C.,  {Bartolini} C.,   {Fermi-LAT Collaboration} 2025, GRB Coordinates Network, \href {https://ui.adsabs.harvard.edu/abs/2025GCN.38843....1D} {38843, 1}

\bibitem[\protect\citeauthoryear{{Dimple} et~al.,}{{Dimple} et~al.}{2025}]{Dimple2025}
{Dimple} et~al., 2025, \mn@doi [arXiv e-prints] {10.48550/arXiv.2507.15940}, \href {https://ui.adsabs.harvard.edu/abs/2025arXiv250715940D} {p. arXiv:2507.15940}

\bibitem[\protect\citeauthoryear{Djorgovski, Frail, Kulkarni, Bloom, Odewahn  \& Diercks}{Djorgovski et~al.}{2001}]{djorgovski_afterglow_2001}
Djorgovski S.~G.,  Frail D.~A.,  Kulkarni S.~R.,  Bloom J.~S.,  Odewahn S.~C.,   Diercks A.,  2001, \mn@doi [The Astrophysical Journal] {10.1086/323845}, 562, 654

\bibitem[\protect\citeauthoryear{{Djupvik} \& {Andersen}}{{Djupvik} \& {Andersen}}{2010}]{Djupvik2010}
{Djupvik} A.~A.,  {Andersen} J.,  2010, in {Diego} J.~M.,  {Goicoechea} L.~J.,  {Gonz{\'a}lez-Serrano} J.~I.,   {Gorgas} J.,  eds,  Astrophysics and Space Science Proceedings Vol. 14, Highlights of Spanish Astrophysics V. p.~211 (\mn@eprint {arXiv} {0901.4015}), \mn@doi{10.1007/978-3-642-11250-8_21}

\bibitem[\protect\citeauthoryear{{Duncan}, {van der Horst}  \& {Beniamini}}{{Duncan} et~al.}{2023}]{Duncan2023}
{Duncan} R.~A.,  {van der Horst} A.~J.,   {Beniamini} P.,  2023, \mn@doi [\mnras] {10.1093/mnras/stac3172}, \href {https://ui.adsabs.harvard.edu/abs/2023MNRAS.518.1522D} {518, 1522}

\bibitem[\protect\citeauthoryear{{Durbak}, {Guiffreda}, {Atri}, {Kutyrev}, {Troja}, {De}  \& {Cenko}}{{Durbak} et~al.}{2024a}]{Durbak2024GCN37700}
{Durbak} J.,  {Guiffreda} O.,  {Atri} S.,  {Kutyrev} A.~S.,  {Troja} E.,  {De} K.,   {Cenko} S.~B.,  2024a, GRB Coordinates Network, \href {https://ui.adsabs.harvard.edu/abs/2024GCN.37700....1D} {37700, 1}

\bibitem[\protect\citeauthoryear{{Durbak}, {Guiffreda}, {Atri}, {Kutyrev}, {Troja}, {De}  \& {Cenko}}{{Durbak} et~al.}{2024b}]{Durbak2024GCN37712}
{Durbak} J.,  {Guiffreda} O.,  {Atri} S.,  {Kutyrev} A.~S.,  {Troja} E.,  {De} K.,   {Cenko} S.~B.,  2024b, GRB Coordinates Network, \href {https://ui.adsabs.harvard.edu/abs/2024GCN.37712....1D} {37712, 1}

\bibitem[\protect\citeauthoryear{{Dyer} et~al.,}{{Dyer} et~al.}{2020}]{Dyer2020SPIE}
{Dyer} M.~J.,  et~al., 2020, in {Marshall} H.~K.,  {Spyromilio} J.,   {Usuda} T.,  eds,  Society of Photo-Optical Instrumentation Engineers (SPIE) Conference Series Vol. 11445, Ground-based and Airborne Telescopes VIII. p. 114457G (\mn@eprint {arXiv} {2012.02685}), \mn@doi{10.1117/12.2561008}

\bibitem[\protect\citeauthoryear{{Dyer} et~al.,}{{Dyer} et~al.}{2024}]{Dyer2024SPIE}
{Dyer} M.~J.,  et~al., 2024, in {Marshall} H.~K.,  {Spyromilio} J.,   {Usuda} T.,  eds,  Society of Photo-Optical Instrumentation Engineers (SPIE) Conference Series Vol. 13094, Ground-based and Airborne Telescopes X. p. 130941X (\mn@eprint {arXiv} {2407.17176}), \mn@doi{10.1117/12.3018305}

\bibitem[\protect\citeauthoryear{{Eichler}, {Livio}, {Piran}  \& {Schramm}}{{Eichler} et~al.}{1989}]{Eichler1989}
{Eichler} D.,  {Livio} M.,  {Piran} T.,   {Schramm} D.~N.,  1989, \mn@doi [\nat] {10.1038/340126a0}, \href {https://ui.adsabs.harvard.edu/abs/1989Natur.340..126E} {340, 126}

\bibitem[\protect\citeauthoryear{{Evans} et~al.,}{{Evans} et~al.}{2007}]{Evans2007}
{Evans} P.~A.,  et~al., 2007, \mn@doi [\aap] {10.1051/0004-6361:20077530}, \href {https://ui.adsabs.harvard.edu/abs/2007A&A...469..379E} {469, 379}

\bibitem[\protect\citeauthoryear{{Evans} et~al.,}{{Evans} et~al.}{2009}]{Evans2009}
{Evans} P.~A.,  et~al., 2009, \mn@doi [\mnras] {10.1111/j.1365-2966.2009.14913.x}, \href {https://ui.adsabs.harvard.edu/abs/2009MNRAS.397.1177E} {397, 1177}

\bibitem[\protect\citeauthoryear{{Fermi GBM Team}}{{Fermi GBM Team}}{2024a}]{FermiGCN36694}
{Fermi GBM Team} 2024a, GRB Coordinates Network, \href {https://ui.adsabs.harvard.edu/abs/2024GCN.36694....1F} {36694, 1}

\bibitem[\protect\citeauthoryear{{Fermi GBM Team}}{{Fermi GBM Team}}{2024b}]{Fermi2024GCN37441}
{Fermi GBM Team} 2024b, GRB Coordinates Network, \href {https://ui.adsabs.harvard.edu/abs/2024GCN.37441....1F} {37441, 1}

\bibitem[\protect\citeauthoryear{{Fermi GBM Team}}{{Fermi GBM Team}}{2024c}]{Fermi2024GCN37518}
{Fermi GBM Team} 2024c, GRB Coordinates Network, \href {https://ui.adsabs.harvard.edu/abs/2024GCN.37518....1F} {37518, 1}

\bibitem[\protect\citeauthoryear{{Fermi GBM Team}}{{Fermi GBM Team}}{2024d}]{Fermi2024GCN37668}
{Fermi GBM Team} 2024d, GRB Coordinates Network, \href {https://ui.adsabs.harvard.edu/abs/2024GCN.37668....1F} {37668, 1}

\bibitem[\protect\citeauthoryear{{Fermi GBM Team}}{{Fermi GBM Team}}{2024e}]{Fermi2024GCN38085}
{Fermi GBM Team} 2024e, GRB Coordinates Network, \href {https://gcn.nasa.gov/circulars/38085} {38085, 1}

\bibitem[\protect\citeauthoryear{{Fermi GBM Team}}{{Fermi GBM Team}}{2024f}]{Fermi2024GCN38682}
{Fermi GBM Team} 2024f, GRB Coordinates Network, \href {https://ui.adsabs.harvard.edu/abs/2024GCN.38682....1F} {38682, 1}

\bibitem[\protect\citeauthoryear{{Fiore}}{{Fiore}}{2001}]{Fiore2001}
{Fiore} F.,  2001, in {Inoue} H.,  {Kunieda} H.,  eds,  Astronomical Society of the Pacific Conference Series Vol. 251, New Century of X-ray Astronomy. p.~168 (\mn@eprint {arXiv} {astro-ph/0107276})

\bibitem[\protect\citeauthoryear{{Flewelling}}{{Flewelling}}{2018}]{Flewelling+2018}
{Flewelling} H.,  2018, in American Astronomical Society Meeting Abstracts \#231. p. 436.01

\bibitem[\protect\citeauthoryear{{Fong} et~al.,}{{Fong} et~al.}{2014}]{2014ApJ...780..118F}
{Fong} W.,  et~al., 2014, \mn@doi [\apj] {10.1088/0004-637X/780/2/118}, \href {https://ui.adsabs.harvard.edu/abs/2014ApJ...780..118F} {780, 118}

\bibitem[\protect\citeauthoryear{{Fong} et~al.,}{{Fong} et~al.}{2021}]{2021ApJ...906..127F}
{Fong} W.,  et~al., 2021, \mn@doi [\apj] {10.3847/1538-4357/abc74a}, \href {https://ui.adsabs.harvard.edu/abs/2021ApJ...906..127F} {906, 127}

\bibitem[\protect\citeauthoryear{{Fontana} et~al.,}{{Fontana} et~al.}{2014}]{Fontana2014a}
{Fontana} A.,  et~al., 2014, \mn@doi [\aap] {10.1051/0004-6361/201423543}, \href {https://ui.adsabs.harvard.edu/abs/2014A&A...570A..11F} {570, A11}

\bibitem[\protect\citeauthoryear{{Frail}, {Kulkarni}, {Nicastro}, {Feroci}  \& {Taylor}}{{Frail} et~al.}{1997}]{Frail1997}
{Frail} D.~A.,  {Kulkarni} S.~R.,  {Nicastro} L.,  {Feroci} M.,   {Taylor} G.~B.,  1997, \mn@doi [\nat] {10.1038/38451}, \href {https://ui.adsabs.harvard.edu/abs/1997Natur.389..261F} {389, 261}

\bibitem[\protect\citeauthoryear{Frail et~al.,}{Frail et~al.}{1999}]{frail_radio_1999}
Frail D.~A.,  et~al., 1999, \mn@doi [The Astrophysical Journal] {10.1086/312347}, 525, L81

\bibitem[\protect\citeauthoryear{Frail et~al.,}{Frail et~al.}{2000}]{frail_enigmatic_2000}
Frail D.~A.,  et~al., 2000, \mn@doi [The Astrophysical Journal] {10.1086/312807}, 538, L129

\bibitem[\protect\citeauthoryear{{Frail} et~al.,}{{Frail} et~al.}{2001}]{Frail2001}
{Frail} D.~A.,  et~al., 2001, \mn@doi [\apjl] {10.1086/338119}, \href {https://ui.adsabs.harvard.edu/abs/2001ApJ...562L..55F} {562, L55}

\bibitem[\protect\citeauthoryear{Frail, Soderberg, Kulkarni, Berger, Yost, Fox  \& Harrison}{Frail et~al.}{2005}]{frail_accurate_2005}
Frail D.~A.,  Soderberg A.~M.,  Kulkarni S.~R.,  Berger E.,  Yost S.,  Fox D.~W.,   Harrison F.~A.,  2005, \mn@doi [The Astrophysical Journal] {10.1086/426680}, 619, 994

\bibitem[\protect\citeauthoryear{Frail et~al.,}{Frail et~al.}{2006}]{frail_energetic_2006}
Frail D.~A.,  et~al., 2006, \mn@doi [The Astrophysical Journal] {10.1086/506934}, 646, L99

\bibitem[\protect\citeauthoryear{{Frederiks}, {Lysenko}, {Ridnaia}, {Svinkin}, {Tsvetkova}, {Ulanov}, {Cline}  \& {Konus-Wind Team}}{{Frederiks} et~al.}{2024}]{Frederiks2024GCN35835}
{Frederiks} D.,  {Lysenko} A.,  {Ridnaia} A.,  {Svinkin} D.,  {Tsvetkova} A.,  {Ulanov} M.,  {Cline} T.,   {Konus-Wind Team} 2024, GRB Coordinates Network, \href {https://ui.adsabs.harvard.edu/abs/2024GCN.35835....1F} {35835, 1}

\bibitem[\protect\citeauthoryear{{Friedman} \& {Bloom}}{{Friedman} \& {Bloom}}{2005}]{Friedman2005}
{Friedman} A.~S.,  {Bloom} J.~S.,  2005, \mn@doi [\apj] {10.1086/430292}, \href {https://ui.adsabs.harvard.edu/abs/2005ApJ...627....1F} {627, 1}

\bibitem[\protect\citeauthoryear{{Fynbo} et~al.,}{{Fynbo} et~al.}{2009}]{2009ApJS..185..526F}
{Fynbo} J.~P.~U.,  et~al., 2009, \mn@doi [\apjs] {10.1088/0067-0049/185/2/526}, \href {https://ui.adsabs.harvard.edu/abs/2009ApJS..185..526F} {185, 526}

\bibitem[\protect\citeauthoryear{{Galama} et~al.,}{{Galama} et~al.}{1998}]{Galama1998}
{Galama} T.~J.,  et~al., 1998, \iaucirc, \href {http://adsabs.harvard.edu/abs/1998IAUC.6895....1G} {6895}

\bibitem[\protect\citeauthoryear{Galama et~al.,}{Galama et~al.}{2000}]{galama_bright_2000}
Galama T.~J.,  et~al., 2000, \mn@doi [The Astrophysical Journal] {10.1086/312904}, 541, L45

\bibitem[\protect\citeauthoryear{Galama, Frail, Sari, Berger, Taylor  \& Kulkarni}{Galama et~al.}{2003}]{galama_continued_2003}
Galama T.~J.,  Frail D.~A.,  Sari R.,  Berger E.,  Taylor G.~B.,   Kulkarni S.~R.,  2003, \mn@doi [The Astrophysical Journal] {10.1086/346083}, 585, 899

\bibitem[\protect\citeauthoryear{{Gehrels} et~al.,}{{Gehrels} et~al.}{2004}]{Gehrels2004}
{Gehrels} N.,  et~al., 2004, \mn@doi [\apj] {10.1086/422091}, \href {https://ui.adsabs.harvard.edu/abs/2004ApJ...611.1005G} {611, 1005}

\bibitem[\protect\citeauthoryear{{Ghosh}, {Razzaque}, {Moskvitin}, {Sotnikova}, {Dukiya}  \& {Gupta}}{{Ghosh} et~al.}{2024}]{Ghosh2024GCN38702}
{Ghosh} A.,  {Razzaque} S.,  {Moskvitin} A.,  {Sotnikova} Y.,  {Dukiya} N.,   {Gupta} R.,  2024, GRB Coordinates Network, \href {https://ui.adsabs.harvard.edu/abs/2024GCN.38702....1G} {38702, 1}

\bibitem[\protect\citeauthoryear{{Giallongo} et~al.,}{{Giallongo} et~al.}{2008}]{Giallongo2008a}
{Giallongo} E.,  et~al., 2008, \mn@doi [\aap] {10.1051/0004-6361:20078402}, \href {https://ui.adsabs.harvard.edu/abs/2008A&A...482..349G} {482, 349}

\bibitem[\protect\citeauthoryear{{Giarratana}, {Giroletti}, {Ghirlanda}, {Di Lalla}, {Omodei}  \& {Salafia}}{{Giarratana} et~al.}{2024a}]{Giarratana2024GCN37569}
{Giarratana} S.,  {Giroletti} M.,  {Ghirlanda} G.,  {Di Lalla} N.,  {Omodei} N.,   {Salafia} O.~S.,  2024a, GRB Coordinates Network, \href {https://gcn.nasa.gov/circulars/37569} {37569, 1}

\bibitem[\protect\citeauthoryear{{Giarratana}, {Giroletti}, {Ghirlanda}, {Di Lalla}, {Omodei}  \& {Salafia}}{{Giarratana} et~al.}{2024b}]{Giarratana202437788}
{Giarratana} S.,  {Giroletti} M.,  {Ghirlanda} G.,  {Di Lalla} N.,  {Omodei} N.,   {Salafia} O.~S.,  2024b, GRB Coordinates Network, \href {https://ui.adsabs.harvard.edu/abs/2024GCN.37788....1G} {37788, 1}

\bibitem[\protect\citeauthoryear{Goldoni, Royer, Fran{\c c}ois, Horrobin, Blanc, Vernet, Modigliani  \& Larsen}{Goldoni et~al.}{2006}]{goldoni2006a}
Goldoni P.,  Royer F.,  Fran{\c c}ois P.,  Horrobin M.,  Blanc G.,  Vernet J.,  Modigliani A.,   Larsen J.,  2006, in Ground-Based and {{Airborne Instrumentation}} for {{Astronomy}}. SPIE, pp 822--832, \mn@doi{10.1117/12.669986}

\bibitem[\protect\citeauthoryear{{Goldstein} et~al.,}{{Goldstein} et~al.}{2017}]{Goldstein2017}
{Goldstein} A.,  et~al., 2017, \mn@doi [\apjl] {10.3847/2041-8213/aa8f41}, \href {https://ui.adsabs.harvard.edu/abs/2017ApJ...848L..14G} {848, L14}

\bibitem[\protect\citeauthoryear{Goldstein, Cleveland  \& Kocevski}{Goldstein et~al.}{2023}]{GDT-Fermi}
Goldstein A.,  Cleveland W.~H.,   Kocevski D.,  2023, Fermi Gamma-ray Data Tools: v2.0.0, \url {https://github.com/USRA-STI/gdt-fermi}

\bibitem[\protect\citeauthoryear{{Gompertz} et~al.,}{{Gompertz} et~al.}{2020}]{Gompertz2020}
{Gompertz} B.~P.,  et~al., 2020, \mn@doi [\mnras] {10.1093/mnras/staa1845}, \href {https://ui.adsabs.harvard.edu/abs/2020MNRAS.497..726G} {497, 726}

\bibitem[\protect\citeauthoryear{{Gompertz} et~al.,}{{Gompertz} et~al.}{2023a}]{Gompertz2023}
{Gompertz} B.~P.,  et~al., 2023a, \mn@doi [Nature Astronomy] {10.1038/s41550-022-01819-4}, \href {https://ui.adsabs.harvard.edu/abs/2023NatAs...7...67G} {7, 67}

\bibitem[\protect\citeauthoryear{{Gompertz} et~al.,}{{Gompertz} et~al.}{2023b}]{Gompertz2023GCN34023}
{Gompertz} B.,  et~al., 2023b, GRB Coordinates Network, \href {https://ui.adsabs.harvard.edu/abs/2023GCN.34023....1G} {34023, 1}

\bibitem[\protect\citeauthoryear{{Gompertz} et~al.,}{{Gompertz} et~al.}{2023c}]{Gompertz2023GCN34480}
{Gompertz} B.~P.,  et~al., 2023c, GRB Coordinates Network, \href {https://ui.adsabs.harvard.edu/abs/2023GCN.34480....1G} {34480, 1}

\bibitem[\protect\citeauthoryear{{Gompertz} et~al.,}{{Gompertz} et~al.}{2024a}]{Gompertz2024GCN35805}
{Gompertz} B.~P.,  et~al., 2024a, GRB Coordinates Network, \href {https://ui.adsabs.harvard.edu/abs/2024GCN.35805....1G} {35805, 1}

\bibitem[\protect\citeauthoryear{{Gompertz} et~al.,}{{Gompertz} et~al.}{2024b}]{Gompertz2024GCN36715}
{Gompertz} B.~P.,  et~al., 2024b, GRB Coordinates Network, \href {https://ui.adsabs.harvard.edu/abs/2024GCN.36715....1G} {36715, 1}

\bibitem[\protect\citeauthoryear{{Gompertz} et~al.,}{{Gompertz} et~al.}{2024c}]{Gompertz2024GCN37522}
{Gompertz} B.~P.,  et~al., 2024c, GRB Coordinates Network, \href {https://ui.adsabs.harvard.edu/abs/2024GCN.37522....1G} {37522, 1}

\bibitem[\protect\citeauthoryear{Goodman}{Goodman}{1997}]{goodman_radio_1997}
Goodman J.,  1997, \mn@doi [New Astronomy] {10.1016/S1384-1076(97)00031-6}, 2, 449

\bibitem[\protect\citeauthoryear{{Gordon} \& {Arnaud}}{{Gordon} \& {Arnaud}}{2021}]{PyXspec}
{Gordon} C.,  {Arnaud} K.,  2021, {PyXspec: Python interface to XSPEC spectral-fitting program}, Astrophysics Source Code Library, record ascl:2101.014

\bibitem[\protect\citeauthoryear{{G{\'o}rski}, {Hivon}, {Banday}, {Wandelt}, {Hansen}, {Reinecke}  \& {Bartelmann}}{{G{\'o}rski} et~al.}{2005}]{healpix}
{G{\'o}rski} K.~M.,  {Hivon} E.,  {Banday} A.~J.,  {Wandelt} B.~D.,  {Hansen} F.~K.,  {Reinecke} M.,   {Bartelmann} M.,  2005, \mn@doi [\apj] {10.1086/427976}, \href {https://ui.adsabs.harvard.edu/abs/2005ApJ...622..759G} {622, 759}

\bibitem[\protect\citeauthoryear{{Gottlieb}, {Bromberg}, {Singh}  \& {Nakar}}{{Gottlieb} et~al.}{2020}]{Gottlieb2020}
{Gottlieb} O.,  {Bromberg} O.,  {Singh} C.~B.,   {Nakar} E.,  2020, \mn@doi [\mnras] {10.1093/mnras/staa2567}, \href {https://ui.adsabs.harvard.edu/abs/2020MNRAS.498.3320G} {498, 3320}

\bibitem[\protect\citeauthoryear{{Granot} \& {Sari}}{{Granot} \& {Sari}}{2002}]{Granot2002}
{Granot} J.,  {Sari} R.,  2002, \mn@doi [\apj] {10.1086/338966}, \href {https://ui.adsabs.harvard.edu/abs/2002ApJ...568..820G} {568, 820}

\bibitem[\protect\citeauthoryear{Greiner et~al.,}{Greiner et~al.}{2013}]{greiner_unusual_2013}
Greiner J.,  et~al., 2013, \mn@doi [Astronomy and Astrophysics] {10.1051/0004-6361/201321284}, 560, A70

\bibitem[\protect\citeauthoryear{Gruber et~al.,}{Gruber et~al.}{2014}]{Gruber2014}
Gruber D.,  et~al., 2014, \mn@doi [The Astrophysical Journal Supplement Series] {10.1088/0067-0049/211/1/12}, 211, 12

\bibitem[\protect\citeauthoryear{{Guiffreda}, {Durbak}, {Atri}, {Kutyrev}, {Troja}  \& {Cenko}}{{Guiffreda} et~al.}{2024}]{Guiffreda2024GCN37736}
{Guiffreda} O.,  {Durbak} J.,  {Atri} S.,  {Kutyrev} A.~S.,  {Troja} E.,   {Cenko} S.~B.,  2024, GRB Coordinates Network, \href {https://ui.adsabs.harvard.edu/abs/2024GCN.37736....1G} {37736, 1}

\bibitem[\protect\citeauthoryear{{Hancock}, {Murphy}, {Gaensler}  \& {Zauderer}}{{Hancock} et~al.}{2012}]{2012GCN.12804....1H}
{Hancock} P.~J.,  {Murphy} T.,  {Gaensler} B.,   {Zauderer} A.,  2012, GRB Coordinates Network, \href {https://ui.adsabs.harvard.edu/abs/2012GCN.12804....1H} {12804, 1}

\bibitem[\protect\citeauthoryear{Harrison et~al.,}{Harrison et~al.}{1999}]{harrison_optical_1999}
Harrison F.~A.,  et~al., 1999, \mn@doi [The Astrophysical Journal] {10.1086/312282}, 523, L121

\bibitem[\protect\citeauthoryear{Harrison et~al.,}{Harrison et~al.}{2001}]{harrison_broadband_2001}
Harrison F.~A.,  et~al., 2001, \mn@doi [The Astrophysical Journal] {10.1086/322368}, 559, 123

\bibitem[\protect\citeauthoryear{{He} et~al.,}{{He} et~al.}{2025}]{He2025}
{He} J.,  et~al., 2025, \mn@doi [Experimental Astronomy] {10.1007/s10686-025-09983-x}, \href {https://ui.adsabs.harvard.edu/abs/2025ExA....59...15H} {59, 15}

\bibitem[\protect\citeauthoryear{Horesh, Cenko, Perley, Kulkarni, Hallinan  \& Bellm}{Horesh et~al.}{2015}]{horesh_unusual_2015}
Horesh A.,  Cenko S.~B.,  Perley D.~A.,  Kulkarni S.~R.,  Hallinan G.,   Bellm E.,  2015, \mn@doi [The Astrophysical Journal] {10.1088/0004-637X/812/1/86}, 812, 86

\bibitem[\protect\citeauthoryear{{Huang} et~al.,}{{Huang} et~al.}{2024}]{2024ApJS..271...13H}
{Huang} B.,  et~al., 2024, \mn@doi [\apjs] {10.3847/1538-4365/ad18b1}, \href {https://ui.adsabs.harvard.edu/abs/2024ApJS..271...13H} {271, 13}

\bibitem[\protect\citeauthoryear{{Izzo} et~al.,}{{Izzo} et~al.}{2019}]{Izzo2019}
{Izzo} L.,  et~al., 2019, \mn@doi [\nat] {10.1038/s41586-018-0826-3}, \href {https://ui.adsabs.harvard.edu/abs/2019Natur.565..324I} {565, 324}

\bibitem[\protect\citeauthoryear{{Izzo} et~al.,}{{Izzo} et~al.}{2024}]{Izzo2024GCN38167}
{Izzo} L.,  et~al., 2024, GRB Coordinates Network, \href {https://ui.adsabs.harvard.edu/abs/2024GCN.38167....1I} {38167, 1}

\bibitem[\protect\citeauthoryear{{Joshi}, {Waratkar}, {Vibhute}, {Bhalerao}, {Bhattacharya}, {Rao}, {Vadawale}  \& {AstroSat CZTI Collaboration}}{{Joshi} et~al.}{2024}]{Joshi2024GCN35798J}
{Joshi} J.,  {Waratkar} G.,  {Vibhute} A.,  {Bhalerao} V.,  {Bhattacharya} D.,  {Rao} A.~R.,  {Vadawale} S.,   {AstroSat CZTI Collaboration} 2024, GRB Coordinates Network, \href {https://ui.adsabs.harvard.edu/abs/2024GCN.35798....1J} {35798, 1}

\bibitem[\protect\citeauthoryear{{Julakanti} et~al.,}{{Julakanti} et~al.}{2024a}]{Julakanti2024GCN37459}
{Julakanti} Y.,  et~al., 2024a, GRB Coordinates Network, \href {https://ui.adsabs.harvard.edu/abs/2024GCN.37459....1J} {37459, 1}

\bibitem[\protect\citeauthoryear{{Julakanti} et~al.,}{{Julakanti} et~al.}{2024b}]{Julakanti2024GCN38088}
{Julakanti} Y.,  et~al., 2024b, GRB Coordinates Network, \href {https://gcn.nasa.gov/circulars/38088} {38088, 1}

\bibitem[\protect\citeauthoryear{{Kawakubo} et~al.,}{{Kawakubo} et~al.}{2024}]{Kawakubo2024GCN35811}
{Kawakubo} Y.,  et~al., 2024, GRB Coordinates Network, \href {https://ui.adsabs.harvard.edu/abs/2024GCN.35811....1K} {35811, 1}

\bibitem[\protect\citeauthoryear{{Killestein} et~al.,}{{Killestein} et~al.}{2021}]{Killestein2021}
{Killestein} T.~L.,  et~al., 2021, \mn@doi [\mnras] {10.1093/mnras/stab633}, \href {https://ui.adsabs.harvard.edu/abs/2021MNRAS.503.4838K} {503, 4838}

\bibitem[\protect\citeauthoryear{{Klebesadel}, {Strong}  \& {Olson}}{{Klebesadel} et~al.}{1973}]{Klebesadel1973}
{Klebesadel} R.~W.,  {Strong} I.~B.,   {Olson} R.~A.,  1973, \mn@doi [\apjl] {10.1086/181225}, \href {https://ui.adsabs.harvard.edu/abs/1973ApJ...182L..85K} {182, L85}

\bibitem[\protect\citeauthoryear{{Kouprianov}}{{Kouprianov}}{2012}]{Kouprianov2012}
{Kouprianov} V.,  2012, in 39th COSPAR Scientific Assembly. p.~974

\bibitem[\protect\citeauthoryear{{Kouveliotou}, {Meegan}, {Fishman}, {Bhat}, {Briggs}, {Koshut}, {Paciesas}  \& {Pendleton}}{{Kouveliotou} et~al.}{1993}]{Kouveliotou1993}
{Kouveliotou} C.,  {Meegan} C.~A.,  {Fishman} G.~J.,  {Bhat} N.~P.,  {Briggs} M.~S.,  {Koshut} T.~M.,  {Paciesas} W.~S.,   {Pendleton} G.~N.,  1993, \mn@doi [\apjl] {10.1086/186969}, \href {https://ui.adsabs.harvard.edu/abs/1993ApJ...413L.101K} {413, L101}

\bibitem[\protect\citeauthoryear{{Krogager}}{{Krogager}}{2018}]{2018arXiv180301187K}
{Krogager} J.-K.,  2018, \mn@doi [arXiv e-prints] {10.48550/arXiv.1803.01187}, \href {https://ui.adsabs.harvard.edu/abs/2018arXiv180301187K} {p. arXiv:1803.01187}

\bibitem[\protect\citeauthoryear{{Kulkarni} \& {Desai}}{{Kulkarni} \& {Desai}}{2017}]{Kulkarni2017}
{Kulkarni} S.,  {Desai} S.,  2017, \mn@doi [\apss] {10.1007/s10509-017-3047-6}, \href {https://ui.adsabs.harvard.edu/abs/2017Ap&SS.362...70K} {362, 70}

\bibitem[\protect\citeauthoryear{{Kumar} \& {Zhang}}{{Kumar} \& {Zhang}}{2015}]{Kumar2015}
{Kumar} P.,  {Zhang} B.,  2015, \mn@doi [\physrep] {10.1016/j.physrep.2014.09.008}, \href {https://ui.adsabs.harvard.edu/abs/2015PhR...561....1K} {561, 1}

\bibitem[\protect\citeauthoryear{{Kumar} et~al.,}{{Kumar} et~al.}{2024a}]{Kumar2024mnras}
{Kumar} A.,  et~al., 2024a, \mn@doi [\mnras] {10.1093/mnras/stae901}, \href {https://ui.adsabs.harvard.edu/abs/2024MNRAS.531.3297K} {531, 3297}

\bibitem[\protect\citeauthoryear{{Kumar} et~al.,}{{Kumar} et~al.}{2024b}]{Kumar2024GCN35596}
{Kumar} A.,  et~al., 2024b, GRB Coordinates Network, \href {https://ui.adsabs.harvard.edu/abs/2024GCN.35596....1P} {35596, 1}

\bibitem[\protect\citeauthoryear{{Kumar} et~al.,}{{Kumar} et~al.}{2024c}]{Kumar2024GCN37676}
{Kumar} A.,  et~al., 2024c, GRB Coordinates Network, \href {https://ui.adsabs.harvard.edu/abs/2024GCN.37676....1K} {37676, 1}

\bibitem[\protect\citeauthoryear{{Kumar} et~al.,}{{Kumar} et~al.}{2024d}]{Kumar2024GCN38684}
{Kumar} A.,  et~al., 2024d, GRB Coordinates Network, \href {https://ui.adsabs.harvard.edu/abs/2024GCN.38684....1K} {38684, 1}

\bibitem[\protect\citeauthoryear{{LHAASO Collaboration} et~al.,}{{LHAASO Collaboration} et~al.}{2023}]{LHAASO2023}
{LHAASO Collaboration} et~al., 2023, \mn@doi [Science] {10.1126/science.adg9328}, \href {https://ui.adsabs.harvard.edu/abs/2023Sci...380.1390L} {380, 1390}

\bibitem[\protect\citeauthoryear{{Lamb} \& {Kobayashi}}{{Lamb} \& {Kobayashi}}{2017}]{Lamb2017}
{Lamb} G.~P.,  {Kobayashi} S.,  2017, \mn@doi [\mnras] {10.1093/mnras/stx2345}, \href {https://ui.adsabs.harvard.edu/abs/2017MNRAS.472.4953L} {472, 4953}

\bibitem[\protect\citeauthoryear{{Lamb} et~al.,}{{Lamb} et~al.}{2019}]{2019ApJ...883...48L}
{Lamb} G.~P.,  et~al., 2019, \mn@doi [\apj] {10.3847/1538-4357/ab38bb}, \href {https://ui.adsabs.harvard.edu/abs/2019ApJ...883...48L} {883, 48}

\bibitem[\protect\citeauthoryear{{Lang}, {Hogg}, {Mierle}, {Blanton}  \& {Roweis}}{{Lang} et~al.}{2010}]{Lang+2010}
{Lang} D.,  {Hogg} D.~W.,  {Mierle} K.,  {Blanton} M.,   {Roweis} S.,  2010, \mn@doi [\aj] {10.1088/0004-6256/139/5/1782}, \href {https://ui.adsabs.harvard.edu/abs/2010AJ....139.1782L} {139, 1782}

\bibitem[\protect\citeauthoryear{{Laskar} et~al.,}{{Laskar} et~al.}{2013}]{Laskar2013}
{Laskar} T.,  et~al., 2013, \mn@doi [\apj] {10.1088/0004-637X/776/2/119}, \href {https://ui.adsabs.harvard.edu/abs/2013ApJ...776..119L} {776, 119}

\bibitem[\protect\citeauthoryear{{Laskar}, {Berger}, {Margutti}, {Perley}, {Zauderer}, {Sari}  \& {Fong}}{{Laskar} et~al.}{2015}]{Laskar2015}
{Laskar} T.,  {Berger} E.,  {Margutti} R.,  {Perley} D.,  {Zauderer} B.~A.,  {Sari} R.,   {Fong} W.-f.,  2015, \mn@doi [\apj] {10.1088/0004-637X/814/1/1}, \href {https://ui.adsabs.harvard.edu/abs/2015ApJ...814....1L} {814, 1}

\bibitem[\protect\citeauthoryear{Laskar et~al.,}{Laskar et~al.}{2016}]{laskar_reverse_2016}
Laskar T.,  et~al., 2016, \mn@doi [The Astrophysical Journal] {10.3847/1538-4357/833/1/88}, 833, 88

\bibitem[\protect\citeauthoryear{Laskar et~al.,}{Laskar et~al.}{2018}]{laskar_vla_2018}
Laskar T.,  et~al., 2018, \mn@doi [The Astrophysical Journal] {10.3847/1538-4357/aabfd8}, 859, 134

\bibitem[\protect\citeauthoryear{{Laskar} et~al.,}{{Laskar} et~al.}{2022}]{2022ApJ...935L..11L}
{Laskar} T.,  et~al., 2022, \mn@doi [\apjl] {10.3847/2041-8213/ac8421}, \href {https://ui.adsabs.harvard.edu/abs/2022ApJ...935L..11L} {935, L11}

\bibitem[\protect\citeauthoryear{Laskar et~al.,}{Laskar et~al.}{2023}]{laskar_radio_2023}
Laskar T.,  et~al., 2023, \mn@doi [The Astrophysical Journal] {10.3847/2041-8213/acbfad}, 946, L23

\bibitem[\protect\citeauthoryear{{Lazzati} \& {Begelman}}{{Lazzati} \& {Begelman}}{2005}]{Lazzati2005}
{Lazzati} D.,  {Begelman} M.~C.,  2005, \mn@doi [\apj] {10.1086/430877}, \href {https://ui.adsabs.harvard.edu/abs/2005ApJ...629..903L} {629, 903}

\bibitem[\protect\citeauthoryear{{Lazzati}, {Morsony}, {Margutti}  \& {Begelman}}{{Lazzati} et~al.}{2013}]{Lazzati2013}
{Lazzati} D.,  {Morsony} B.~J.,  {Margutti} R.,   {Begelman} M.~C.,  2013, \mn@doi [\apj] {10.1088/0004-637X/765/2/103}, \href {https://ui.adsabs.harvard.edu/abs/2013ApJ...765..103L} {765, 103}

\bibitem[\protect\citeauthoryear{{Levan}, {Crowther}, {de Grijs}, {Langer}, {Xu}  \& {Yoon}}{{Levan} et~al.}{2016}]{Levan2016}
{Levan} A.,  {Crowther} P.,  {de Grijs} R.,  {Langer} N.,  {Xu} D.,   {Yoon} S.-C.,  2016, \mn@doi [\ssr] {10.1007/s11214-016-0312-x}, \href {https://ui.adsabs.harvard.edu/abs/2016SSRv..202...33L} {202, 33}

\bibitem[\protect\citeauthoryear{{Levan} et~al.,}{{Levan} et~al.}{2024a}]{Levan2024}
{Levan} A.~J.,  et~al., 2024a, \mn@doi [\nat] {10.1038/s41586-023-06759-1}, \href {https://ui.adsabs.harvard.edu/abs/2024Natur.626..737L} {626, 737}

\bibitem[\protect\citeauthoryear{{Levan} et~al.,}{{Levan} et~al.}{2024b}]{2024Natur.626..737L}
{Levan} A.~J.,  et~al., 2024b, \mn@doi [\nat] {10.1038/s41586-023-06759-1}, \href {https://ui.adsabs.harvard.edu/abs/2024Natur.626..737L} {626, 737}

\bibitem[\protect\citeauthoryear{{Liang}, {Zhang}, {Virgili}  \& {Dai}}{{Liang} et~al.}{2007}]{Liang2007}
{Liang} E.,  {Zhang} B.,  {Virgili} F.,   {Dai} Z.~G.,  2007, \mn@doi [\apj] {10.1086/517959}, \href {https://ui.adsabs.harvard.edu/abs/2007ApJ...662.1111L} {662, 1111}

\bibitem[\protect\citeauthoryear{{Lundman}, {Pe'er}  \& {Ryde}}{{Lundman} et~al.}{2013}]{Lundman2013}
{Lundman} C.,  {Pe'er} A.,   {Ryde} F.,  2013, \mn@doi [\mnras] {10.1093/mnras/sts219}, \href {https://ui.adsabs.harvard.edu/abs/2013MNRAS.428.2430L} {428, 2430}

\bibitem[\protect\citeauthoryear{{Luongo} \& {Muccino}}{{Luongo} \& {Muccino}}{2021}]{Luongo2021}
{Luongo} O.,  {Muccino} M.,  2021, \mn@doi [Galaxies] {10.3390/galaxies9040077}, \href {https://ui.adsabs.harvard.edu/abs/2021Galax...9...77L} {9, 77}

\bibitem[\protect\citeauthoryear{{Lupton} et~al.,}{{Lupton} et~al.}{2005}]{Lupton2005}
{Lupton} R.~H.,  et~al., 2005, in American Astronomical Society Meeting Abstracts. p. 133.08

\bibitem[\protect\citeauthoryear{{Maeder} \& {Meynet}}{{Maeder} \& {Meynet}}{2012}]{Maeder2012}
{Maeder} A.,  {Meynet} G.,  2012, \mn@doi [Reviews of Modern Physics] {10.1103/RevModPhys.84.25}, \href {https://ui.adsabs.harvard.edu/abs/2012RvMP...84...25M} {84, 25}

\bibitem[\protect\citeauthoryear{{Margutti} et~al.,}{{Margutti} et~al.}{2013}]{Margutti2013}
{Margutti} R.,  et~al., 2013, \mn@doi [\apj] {10.1088/0004-637X/778/1/18}, \href {https://ui.adsabs.harvard.edu/abs/2013ApJ...778...18M} {778, 18}

\bibitem[\protect\citeauthoryear{{Matsuoka} et~al.,}{{Matsuoka} et~al.}{2009}]{Matsuoka2009}
{Matsuoka} M.,  et~al., 2009, \mn@doi [\pasj] {10.1093/pasj/61.5.999}, \href {https://ui.adsabs.harvard.edu/abs/2009PASJ...61..999M} {61, 999}

\bibitem[\protect\citeauthoryear{{Mazets}, {Golenetskii}, {Aptekar}, {Gurian}  \& {Ilinskii}}{{Mazets} et~al.}{1981}]{Mazets1981}
{Mazets} E.~P.,  {Golenetskii} S.~V.,  {Aptekar} R.~L.,  {Gurian} I.~A.,   {Ilinskii} V.~N.,  1981, \mn@doi [\nat] {10.1038/290378a0}, \href {https://ui.adsabs.harvard.edu/abs/1981Natur.290..378M} {290, 378}

\bibitem[\protect\citeauthoryear{{McMullin}, {Waters}, {Schiebel}, {Young}  \& {Golap}}{{McMullin} et~al.}{2007}]{McMullin2007}
{McMullin} J.~P.,  {Waters} B.,  {Schiebel} D.,  {Young} W.,   {Golap} K.,  2007, in {Shaw} R.~A.,  {Hill} F.,   {Bell} D.~J.,  eds,  Astronomical Society of the Pacific Conference Series Vol. 376, Astronomical Data Analysis Software and Systems XVI. p.~127

\bibitem[\protect\citeauthoryear{{Meegan}, {Fishman}, {Wilson}, {Paciesas}, {Pendleton}, {Horack}, {Brock}  \& {Kouveliotou}}{{Meegan} et~al.}{1992}]{Meegan1992}
{Meegan} C.~A.,  {Fishman} G.~J.,  {Wilson} R.~B.,  {Paciesas} W.~S.,  {Pendleton} G.~N.,  {Horack} J.~M.,  {Brock} M.~N.,   {Kouveliotou} C.,  1992, \mn@doi [\nat] {10.1038/355143a0}, \href {https://ui.adsabs.harvard.edu/abs/1992Natur.355..143M} {355, 143}

\bibitem[\protect\citeauthoryear{{Meegan} et~al.,}{{Meegan} et~al.}{2009}]{Meegan2009}
{Meegan} C.,  et~al., 2009, \mn@doi [\apj] {10.1088/0004-637X/702/1/791}, \href {https://ui.adsabs.harvard.edu/abs/2009ApJ...702..791M} {702, 791}

\bibitem[\protect\citeauthoryear{{M{\'e}sz{\'a}ros}}{{M{\'e}sz{\'a}ros}}{2013}]{Meszaros2013}
{M{\'e}sz{\'a}ros} P.,  2013, \mn@doi [Astroparticle Physics] {10.1016/j.astropartphys.2012.03.009}, \href {https://ui.adsabs.harvard.edu/abs/2013APh....43..134M} {43, 134}

\bibitem[\protect\citeauthoryear{{Metzger}}{{Metzger}}{2019}]{Metzger2019}
{Metzger} B.~D.,  2019, \mn@doi [Living Reviews in Relativity] {10.1007/s41114-019-0024-0}, \href {https://ui.adsabs.harvard.edu/abs/2019LRR....23....1M} {23, 1}

\bibitem[\protect\citeauthoryear{{Metzger}, {Djorgovski}, {Kulkarni}, {Steidel}, {Adelberger}, {Frail}, {Costa}  \& {Frontera}}{{Metzger} et~al.}{1997}]{Metzger1997}
{Metzger} M.~R.,  {Djorgovski} S.~G.,  {Kulkarni} S.~R.,  {Steidel} C.~C.,  {Adelberger} K.~L.,  {Frail} D.~A.,  {Costa} E.,   {Frontera} F.,  1997, \mn@doi [\nat] {10.1038/43132}, \href {https://ui.adsabs.harvard.edu/abs/1997Natur.387..878M} {387, 878}

\bibitem[\protect\citeauthoryear{{Miceli} \& {Nava}}{{Miceli} \& {Nava}}{2022}]{Miceli2022}
{Miceli} D.,  {Nava} L.,  2022, \mn@doi [Galaxies] {10.3390/galaxies10030066}, \href {https://ui.adsabs.harvard.edu/abs/2022Galax..10...66M} {10, 66}

\bibitem[\protect\citeauthoryear{{Mihara} et~al.,}{{Mihara} et~al.}{2011}]{Mihara2011PASJ}
{Mihara} T.,  et~al., 2011, \mn@doi [\pasj] {10.1093/pasj/63.sp3.S623}, \href {https://ui.adsabs.harvard.edu/abs/2011PASJ...63S.623M} {63, S623}

\bibitem[\protect\citeauthoryear{{Mimica} \& {Aloy}}{{Mimica} \& {Aloy}}{2012}]{Mimica2012}
{Mimica} P.,  {Aloy} M.~A.,  2012, \mn@doi [\mnras] {10.1111/j.1365-2966.2012.20495.x}, \href {https://ui.adsabs.harvard.edu/abs/2012MNRAS.421.2635M} {421, 2635}

\bibitem[\protect\citeauthoryear{{Mimica}, {Giannios}  \& {Aloy}}{{Mimica} et~al.}{2009}]{Mimica2009}
{Mimica} P.,  {Giannios} D.,   {Aloy} M.~A.,  2009, \mn@doi [\aap] {10.1051/0004-6361:200810756}, \href {https://ui.adsabs.harvard.edu/abs/2009A&A...494..879M} {494, 879}

\bibitem[\protect\citeauthoryear{{Mimica}, {Giannios}  \& {Aloy}}{{Mimica} et~al.}{2010}]{Mimica2010}
{Mimica} P.,  {Giannios} D.,   {Aloy} M.~A.,  2010, \mn@doi [\mnras] {10.1111/j.1365-2966.2010.17071.x}, \href {https://ui.adsabs.harvard.edu/abs/2010MNRAS.407.2501M} {407, 2501}

\bibitem[\protect\citeauthoryear{{Minaev} \& {Pozanenko}}{{Minaev} \& {Pozanenko}}{2020}]{Minaev20}
{Minaev} P.~Y.,  {Pozanenko} A.~S.,  2020, \mn@doi [\mnras] {10.1093/mnras/stz3611}, \href {https://ui.adsabs.harvard.edu/abs/2020MNRAS.492.1919M} {492, 1919}

\bibitem[\protect\citeauthoryear{{Mo} et~al.,}{{Mo} et~al.}{2023}]{Geoffrey2024GCN36739}
{Mo} G.,  et~al., 2023, GRB Coordinates Network, \href {https://ui.adsabs.harvard.edu/abs/2024GCN.36739....1M} {36739, 1}

\bibitem[\protect\citeauthoryear{{Mo}, {Karambelkar}, {Frostig}, {Stein}, {Lourie}, {Ahumada}, {Simcoe}  \& {Kasliwal}}{{Mo} et~al.}{2024}]{Mo2024GCN35829}
{Mo} G.,  {Karambelkar} V.,  {Frostig} D.,  {Stein} R.,  {Lourie} N.,  {Ahumada} T.,  {Simcoe} R.,   {Kasliwal} M.,  2024, GRB Coordinates Network, \href {https://ui.adsabs.harvard.edu/abs/2024GCN.35829....1M} {35829, 1}

\bibitem[\protect\citeauthoryear{Modigliani et~al.,}{Modigliani et~al.}{2010}]{modigliani2010a}
Modigliani A.,  et~al., 2010, in Observatory {{Operations}}: {{Strategies}}, {{Processes}}, and {{Systems III}}. SPIE, pp 572--583, \mn@doi{10.1117/12.857211}

\bibitem[\protect\citeauthoryear{Moin et~al.,}{Moin et~al.}{2013}]{moin_radio_2013}
Moin A.,  et~al., 2013, \mn@doi [The Astrophysical Journal] {10.1088/0004-637X/779/2/105}, 779, 105

\bibitem[\protect\citeauthoryear{{Mong} et~al.,}{{Mong} et~al.}{2021}]{Mong2021}
{Mong} Y.~L.,  et~al., 2021, \mn@doi [\mnras] {10.1093/mnras/stab2499}, \href {https://ui.adsabs.harvard.edu/abs/2021MNRAS.507.5463M} {507, 5463}

\bibitem[\protect\citeauthoryear{{Moresco} et~al.,}{{Moresco} et~al.}{2022}]{Moresco2022}
{Moresco} M.,  et~al., 2022, \mn@doi [Living Reviews in Relativity] {10.1007/s41114-022-00040-z}, \href {https://ui.adsabs.harvard.edu/abs/2022LRR....25....6M} {25, 6}

\bibitem[\protect\citeauthoryear{{Moskvitin}, {Spiridonova}  \& {GRB follow-up Team.}}{{Moskvitin} et~al.}{2024a}]{Moskvitin2024GCN35828}
{Moskvitin} A.~S.,  {Spiridonova} O.~I.,   {GRB follow-up Team.} 2024a, GRB Coordinates Network, \href {https://ui.adsabs.harvard.edu/abs/2024GCN.35828....1M} {35828, 1}

\bibitem[\protect\citeauthoryear{{Moskvitin}, {Spiridonova}  \& {GRB follow-up Team.}}{{Moskvitin} et~al.}{2024b}]{Moskvitin2024GCN35839}
{Moskvitin} A.~S.,  {Spiridonova} O.~I.,   {GRB follow-up Team.} 2024b, GRB Coordinates Network, \href {https://ui.adsabs.harvard.edu/abs/2024GCN.35839....1M} {35839, 1}

\bibitem[\protect\citeauthoryear{{Moskvitin}, {Spiridonova}, {Sotnikova}, {Volnova}, {Pozanenko}, {Ghosh}, {Razzaque}  \& {GRB follow-up Team}}{{Moskvitin} et~al.}{2024c}]{Moskvitin2024GCN38733}
{Moskvitin} A.~S.,  {Spiridonova} O.~I.,  {Sotnikova} Y.~V.,  {Volnova} A.,  {Pozanenko} A.,  {Ghosh} A.,  {Razzaque} S.,   {GRB follow-up Team} 2024c, GRB Coordinates Network, \href {https://ui.adsabs.harvard.edu/abs/2024GCN.38733....1M} {38733, 1}

\bibitem[\protect\citeauthoryear{{Mundell} et~al.,}{{Mundell} et~al.}{2007}]{Mundell2007}
{Mundell} C.~G.,  et~al., 2007, \mn@doi [\apj] {10.1086/512605}, \href {https://ui.adsabs.harvard.edu/abs/2007ApJ...660..489M} {660, 489}

\bibitem[\protect\citeauthoryear{{Nakajima} et~al.,}{{Nakajima} et~al.}{2024}]{Nakajima2024GCN35796}
{Nakajima} M.,  et~al., 2024, GRB Coordinates Network, \href {https://ui.adsabs.harvard.edu/abs/2024GCN.35796....1N} {35796, 1}

\bibitem[\protect\citeauthoryear{{Narayan}, {Paczynski}  \& {Piran}}{{Narayan} et~al.}{1992}]{Narayan1992}
{Narayan} R.,  {Paczynski} B.,   {Piran} T.,  1992, \mn@doi [\apjl] {10.1086/186493}, \href {https://ui.adsabs.harvard.edu/abs/1992ApJ...395L..83N} {395, L83}

\bibitem[\protect\citeauthoryear{{Nasa High Energy Astrophysics Science Archive Research Center (Heasarc)}}{{Nasa High Energy Astrophysics Science Archive Research Center (Heasarc)}}{2014}]{heasoft}
{Nasa High Energy Astrophysics Science Archive Research Center (Heasarc)} 2014, {HEAsoft: Unified Release of FTOOLS and XANADU}, Astrophysics Source Code Library, record ascl:1408.004

\bibitem[\protect\citeauthoryear{{Negoro} et~al.,}{{Negoro} et~al.}{2024}]{Negoro2024GCN35593}
{Negoro} H.,  et~al., 2024, GRB Coordinates Network, \href {https://ui.adsabs.harvard.edu/abs/2024GCN.35593....1N} {35593, 1}

\bibitem[\protect\citeauthoryear{{Nicholl} et~al.,}{{Nicholl} et~al.}{2023}]{Nicholl2023}
{Nicholl} M.,  et~al., 2023, \mn@doi [\apjl] {10.3847/2041-8213/acf0ba}, \href {https://ui.adsabs.harvard.edu/abs/2023ApJ...954L..28N} {954, L28}

\bibitem[\protect\citeauthoryear{O'Connor et~al.,}{O'Connor et~al.}{2023}]{oconnor_structured_2023}
O'Connor B.,  et~al., 2023, \mn@doi [Science Advances] {10.1126/sciadv.adi1405}, 9, eadi1405

\bibitem[\protect\citeauthoryear{{Obergaulinger} \& {Aloy}}{{Obergaulinger} \& {Aloy}}{2021}]{Obergaulinger2021}
{Obergaulinger} M.,  {Aloy} M.~{\'A}.,  2021, \mn@doi [\mnras] {10.1093/mnras/stab295}, \href {https://ui.adsabs.harvard.edu/abs/2021MNRAS.503.4942O} {503, 4942}

\bibitem[\protect\citeauthoryear{{Obergaulinger} \& {Aloy}}{{Obergaulinger} \& {Aloy}}{2022}]{Obergaulinger2022}
{Obergaulinger} M.,  {Aloy} M.~{\'A}.,  2022, \mn@doi [\mnras] {10.1093/mnras/stac613}, \href {https://ui.adsabs.harvard.edu/abs/2022MNRAS.512.2489O} {512, 2489}

\bibitem[\protect\citeauthoryear{{Oganesyan}, {Ascenzi}, {Branchesi}, {Salafia}, {Dall'Osso}  \& {Ghirlanda}}{{Oganesyan} et~al.}{2020}]{Oganesyan2020}
{Oganesyan} G.,  {Ascenzi} S.,  {Branchesi} M.,  {Salafia} O.~S.,  {Dall'Osso} S.,   {Ghirlanda} G.,  2020, \mn@doi [\apj] {10.3847/1538-4357/ab8221}, \href {https://ui.adsabs.harvard.edu/abs/2020ApJ...893...88O} {893, 88}

\bibitem[\protect\citeauthoryear{{Ortega-Casas} et~al.,}{{Ortega-Casas} et~al.}{2024}]{OrtegaCasas2024GCN38692}
{Ortega-Casas} I.,  et~al., 2024, GRB Coordinates Network, \href {https://ui.adsabs.harvard.edu/abs/2024GCN.38692....1O} {38692, 1}

\bibitem[\protect\citeauthoryear{{Panaitescu} \& {Kumar}}{{Panaitescu} \& {Kumar}}{2002}]{Panaitescu2002}
{Panaitescu} A.,  {Kumar} P.,  2002, \mn@doi [\apj] {10.1086/340094}, \href {https://ui.adsabs.harvard.edu/abs/2002ApJ...571..779P} {571, 779}

\bibitem[\protect\citeauthoryear{Pankov, Pozanenko, Kouprianov  \& Belkin}{Pankov et~al.}{2022}]{PankovCCIS2022}
Pankov N.,  Pozanenko A.,  Kouprianov V.,   Belkin S.,  2022, in Pozanenko A.,  Stupnikov S.,  Thalheim B.,  Mendez E.,   Kiselyova N.,  eds, Data Analytics and Management in Data Intensive Domains. Springer International Publishing, Cham, pp 104--134

\bibitem[\protect\citeauthoryear{{Pawar}}{{Pawar}}{2024}]{Pawar2024GCN37519}
{Pawar} D.,  2024, GRB Coordinates Network, \href {https://ui.adsabs.harvard.edu/abs/2024GCN.37519....1P} {37519, 1}

\bibitem[\protect\citeauthoryear{Perley et~al.,}{Perley et~al.}{2008}]{perley_grb_2008}
Perley D.~A.,  et~al., 2008, \mn@doi [The Astrophysical Journal] {10.1086/591961}, 688, 470

\bibitem[\protect\citeauthoryear{Perley et~al.,}{Perley et~al.}{2014}]{perley_afterglow_2014}
Perley D.~A.,  et~al., 2014, \mn@doi [The Astrophysical Journal] {10.1088/0004-637X/781/1/37}, 781, 37

\bibitem[\protect\citeauthoryear{{Petitjean} \& {Vergani}}{{Petitjean} \& {Vergani}}{2011}]{Petitjean2011}
{Petitjean} P.,  {Vergani} S.~D.,  2011, \mn@doi [Comptes Rendus Physique] {10.1016/j.crhy.2011.01.007}, \href {https://ui.adsabs.harvard.edu/abs/2011CRPhy..12..288P} {12, 288}

\bibitem[\protect\citeauthoryear{{Pian} et~al.,}{{Pian} et~al.}{2017}]{Pian2017}
{Pian} E.,  et~al., 2017, \mn@doi [\nat] {10.1038/nature24298}, \href {https://ui.adsabs.harvard.edu/abs/2017Natur.551...67P} {551, 67}

\bibitem[\protect\citeauthoryear{{Pieterse} et~al.,}{{Pieterse} et~al.}{2024}]{Pieterse2024GCN37532}
{Pieterse} D.,  et~al., 2024, GRB Coordinates Network, \href {https://ui.adsabs.harvard.edu/abs/2024GCN.37532....1P} {37532, 1}

\bibitem[\protect\citeauthoryear{{Price} et~al.,}{{Price} et~al.}{2002}]{2002ApJ...572L..51P}
{Price} P.~A.,  et~al., 2002, \mn@doi [\apjl] {10.1086/341552}, \href {https://ui.adsabs.harvard.edu/abs/2002ApJ...572L..51P} {572, L51}

\bibitem[\protect\citeauthoryear{{Racusin} et~al.,}{{Racusin} et~al.}{2009}]{Racusin2009}
{Racusin} J.~L.,  et~al., 2009, \mn@doi [\apj] {10.1088/0004-637X/698/1/43}, \href {https://ui.adsabs.harvard.edu/abs/2009ApJ...698...43R} {698, 43}

\bibitem[\protect\citeauthoryear{{Rastinejad} et~al.,}{{Rastinejad} et~al.}{2022}]{Rastinejad2022}
{Rastinejad} J.~C.,  et~al., 2022, \mn@doi [\nat] {10.1038/s41586-022-05390-w}, \href {https://ui.adsabs.harvard.edu/abs/2022Natur.612..223R} {612, 223}

\bibitem[\protect\citeauthoryear{{Ravasio}, {Oganesyan}, {Ghirlanda}, {Nava}, {Ghisellini}, {Pescalli}  \& {Celotti}}{{Ravasio} et~al.}{2018}]{Ravasio18}
{Ravasio} M.~E.,  {Oganesyan} G.,  {Ghirlanda} G.,  {Nava} L.,  {Ghisellini} G.,  {Pescalli} A.,   {Celotti} A.,  2018, \mn@doi [\aap] {10.1051/0004-6361/201732245}, \href {https://ui.adsabs.harvard.edu/abs/2018A&A...613A..16R} {613, A16}

\bibitem[\protect\citeauthoryear{{Ravasio}, {Ghirlanda}, {Nava}  \& {Ghisellini}}{{Ravasio} et~al.}{2019}]{Ravasio19}
{Ravasio} M.~E.,  {Ghirlanda} G.,  {Nava} L.,   {Ghisellini} G.,  2019, \mn@doi [\aap] {10.1051/0004-6361/201834987}, \href {https://ui.adsabs.harvard.edu/abs/2019A&A...625A..60R} {625, A60}

\bibitem[\protect\citeauthoryear{Reynolds}{Reynolds}{1994}]{Reynolds1994}
Reynolds J.,  1994, A Revised Flux Scale for the AT Compact Array, \url {https://www.atnf.csiro.au/observers/memos/d96783~1.pdf}

\bibitem[\protect\citeauthoryear{{Rhoads}}{{Rhoads}}{1999}]{Rhoads1999}
{Rhoads} J.~E.,  1999, \mn@doi [\apj] {10.1086/307907}, \href {https://ui.adsabs.harvard.edu/abs/1999ApJ...525..737R} {525, 737}

\bibitem[\protect\citeauthoryear{Rhodes et~al.,}{Rhodes et~al.}{2024a}]{rhodes_rocking_2024}
Rhodes L.,  et~al., 2024a, \mn@doi [Monthly Notices of the Royal Astronomical Society] {10.1093/mnras/stae2050}, 533, 4435

\bibitem[\protect\citeauthoryear{{Rhodes}, {Fender}, {Green}  \& {Titterington}}{{Rhodes} et~al.}{2024b}]{Rhodes2024GCN36744}
{Rhodes} L.,  {Fender} R.,  {Green} D.,   {Titterington} D.,  2024b, GRB Coordinates Network, \href {https://ui.adsabs.harvard.edu/abs/2024GCN.36744....1R} {36744, 1}

\bibitem[\protect\citeauthoryear{{Ripa} et~al.,}{{Ripa} et~al.}{2024}]{Ripa2024GCN37450}
{Ripa} J.,  et~al., 2024, GRB Coordinates Network, \href {https://ui.adsabs.harvard.edu/abs/2024GCN.37450....1R} {37450, 1}

\bibitem[\protect\citeauthoryear{{Roberts}, {Meegan}  \& {Fermi Gamma-ray Burst Monitor Team}}{{Roberts} et~al.}{2024a}]{Roberts2024GCN37535}
{Roberts} O.~J.,  {Meegan} C.,   {Fermi Gamma-ray Burst Monitor Team} 2024a, GRB Coordinates Network, \href {https://ui.adsabs.harvard.edu/abs/2024GCN.37535....1R} {37535, 1}

\bibitem[\protect\citeauthoryear{{Roberts}, {Hamburg}, {Meegan}  \& {Fermi Gamma-ray Burst Monitor Team}}{{Roberts} et~al.}{2024b}]{Roberts2024GCN37711}
{Roberts} O.~J.,  {Hamburg} R.,  {Meegan} C.,   {Fermi Gamma-ray Burst Monitor Team} 2024b, GRB Coordinates Network, \href {https://ui.adsabs.harvard.edu/abs/2024GCN.37711....1R} {37711, 1}

\bibitem[\protect\citeauthoryear{Rol et~al.,}{Rol et~al.}{2007}]{rol_grb_2007}
Rol E.,  et~al., 2007, \mn@doi [The Astrophysical Journal] {10.1086/521336}, 669, 1098

\bibitem[\protect\citeauthoryear{{Roming} et~al.,}{{Roming} et~al.}{2005}]{Roming2005}
{Roming} P. W.~A.,  et~al., 2005, \mn@doi [\ssr] {10.1007/s11214-005-5095-4}, \href {https://ui.adsabs.harvard.edu/abs/2005SSRv..120...95R} {120, 95}

\bibitem[\protect\citeauthoryear{{Ror}, {Pandey}, {Gupta}, {Aryan}  \& {Pandey}}{{Ror} et~al.}{2024}]{Ror2024GCN35830}
{Ror} A.~K.,  {Pandey} S.~B.,  {Gupta} R.,  {Aryan} A.,   {Pandey} S.,  2024, GRB Coordinates Network, \href {https://ui.adsabs.harvard.edu/abs/2024GCN.35830....1R} {35830, 1}

\bibitem[\protect\citeauthoryear{{Ror}, {Gupta}, {Pranshu}, {Pandey}  \& {Mishra}}{{Ror} et~al.}{2025}]{Ror2025GCN38816}
{Ror} A.~K.,  {Gupta} A.,  {Pranshu} {Pandey} S.~B.,   {Mishra} K.,  2025, GRB Coordinates Network, \href {https://ui.adsabs.harvard.edu/abs/2025GCN.38816....1R} {38816, 1}

\bibitem[\protect\citeauthoryear{{Rossi} et~al.,}{{Rossi} et~al.}{2022}]{Rossi2022}
{Rossi} A.,  et~al., 2022, \mn@doi [\apj] {10.3847/1538-4357/ac60a2}, \href {https://ui.adsabs.harvard.edu/abs/2022ApJ...932....1R} {932, 1}

\bibitem[\protect\citeauthoryear{{Ryan}, {van Eerten}, {Piro}  \& {Troja}}{{Ryan} et~al.}{2020}]{Ryan2020}
{Ryan} G.,  {van Eerten} H.,  {Piro} L.,   {Troja} E.,  2020, \mn@doi [\apj] {10.3847/1538-4357/ab93cf}, \href {https://ui.adsabs.harvard.edu/abs/2020ApJ...896..166R} {896, 166}

\bibitem[\protect\citeauthoryear{{SVOM/GRM Team} et~al.,}{{SVOM/GRM Team} et~al.}{2024}]{SVOM2024GCN37445}
{SVOM/GRM Team} et~al., 2024, GRB Coordinates Network, \href {https://ui.adsabs.harvard.edu/abs/2024GCN.37445....1S} {37445, 1}

\bibitem[\protect\citeauthoryear{{SVOM/VT Team} et~al.,}{{SVOM/VT Team} et~al.}{2024}]{SVOM2024GCN37503}
{SVOM/VT Team} et~al., 2024, GRB Coordinates Network, \href {https://ui.adsabs.harvard.edu/abs/2024GCN.37503....1S} {37503, 1}

\bibitem[\protect\citeauthoryear{{Saccardi} et~al.,}{{Saccardi} et~al.}{2023}]{Saccardi2023}
{Saccardi} A.,  et~al., 2023, \mn@doi [\aap] {10.1051/0004-6361/202244205}, \href {https://ui.adsabs.harvard.edu/abs/2023A&A...671A..84S} {671, A84}

\bibitem[\protect\citeauthoryear{{Saccardi} et~al.,}{{Saccardi} et~al.}{2024}]{Saccardi2024GCN35599}
{Saccardi} A.,  et~al., 2024, GRB Coordinates Network, \href {https://ui.adsabs.harvard.edu/abs/2024GCN.35599....1S} {35599, 1}

\bibitem[\protect\citeauthoryear{{Saccardi} et~al.,}{{Saccardi} et~al.}{2025}]{Saccardi2025}
{Saccardi} A.,  et~al., 2025, \mn@doi [arXiv e-prints] {10.48550/arXiv.2506.04340}, \href {https://ui.adsabs.harvard.edu/abs/2025arXiv250604340S} {p. arXiv:2506.04340}

\bibitem[\protect\citeauthoryear{{Sari} \& {Piran}}{{Sari} \& {Piran}}{1999}]{Sari1999}
{Sari} R.,  {Piran} T.,  1999, \mn@doi [\aaps] {10.1051/aas:1999342}, \href {https://ui.adsabs.harvard.edu/abs/1999A&AS..138..537S} {138, 537}

\bibitem[\protect\citeauthoryear{{Sari}, {Piran}  \& {Narayan}}{{Sari} et~al.}{1998}]{Sari1998}
{Sari} R.,  {Piran} T.,   {Narayan} R.,  1998, \mn@doi [\apjl] {10.1086/311269}, \href {https://ui.adsabs.harvard.edu/abs/1998ApJ...497L..17S} {497, L17}

\bibitem[\protect\citeauthoryear{{Sasada} et~al.,}{{Sasada} et~al.}{2024}]{Sasada2024GCN35831}
{Sasada} M.,  et~al., 2024, GRB Coordinates Network, \href {https://ui.adsabs.harvard.edu/abs/2024GCN.35831....1S} {35831, 1}

\bibitem[\protect\citeauthoryear{{Sault}, {Teuben}  \& {Wright}}{{Sault} et~al.}{1995}]{1995ASPC...77..433S}
{Sault} R.~J.,  {Teuben} P.~J.,   {Wright} M.~C.~H.,  1995, in {Shaw} R.~A.,  {Payne} H.~E.,   {Hayes} J.~J.~E.,  eds,  Astronomical Society of the Pacific Conference Series Vol. 77, Astronomical Data Analysis Software and Systems IV. p.~433 (\mn@eprint {arXiv} {astro-ph/0612759})

\bibitem[\protect\citeauthoryear{{Savaglio} \& {Fall}}{{Savaglio} \& {Fall}}{2004}]{Savaglio2004}
{Savaglio} S.,  {Fall} S.~M.,  2004, \mn@doi [\apj] {10.1086/423447}, \href {https://ui.adsabs.harvard.edu/abs/2004ApJ...614..293S} {614, 293}

\bibitem[\protect\citeauthoryear{{Schlafly} \& {Finkbeiner}}{{Schlafly} \& {Finkbeiner}}{2011}]{Schlafly2011}
{Schlafly} E.~F.,  {Finkbeiner} D.~P.,  2011, \mn@doi [\apj] {10.1088/0004-637X/737/2/103}, \href {https://ui.adsabs.harvard.edu/abs/2011ApJ...737..103S} {737, 103}

\bibitem[\protect\citeauthoryear{{Schneider} et~al.,}{{Schneider} et~al.}{2024}]{Schneider2024GCN35832}
{Schneider} B.,  et~al., 2024, GRB Coordinates Network, \href {https://ui.adsabs.harvard.edu/abs/2024GCN.35832....1S} {35832, 1}

\bibitem[\protect\citeauthoryear{Schroeder et~al.,}{Schroeder et~al.}{2024}]{schroeder_radio_2024}
Schroeder G.,  et~al., 2024, \mn@doi [The Astrophysical Journal] {10.3847/1538-4357/ad49ab}, 970, 139

\bibitem[\protect\citeauthoryear{{Schulze} et~al.,}{{Schulze} et~al.}{2011}]{Schulze2011}
{Schulze} S.,  et~al., 2011, \mn@doi [\aap] {10.1051/0004-6361/201015581}, \href {https://ui.adsabs.harvard.edu/abs/2011A&A...526A..23S} {526, A23}

\bibitem[\protect\citeauthoryear{{Scotton}, {Meegan}  \& {Fermi Gamma-ray Burst Monitor Team}}{{Scotton} et~al.}{2024}]{Scotton2024GCN38714}
{Scotton} L.,  {Meegan} C.,   {Fermi Gamma-ray Burst Monitor Team} 2024, GRB Coordinates Network, \href {https://ui.adsabs.harvard.edu/abs/2024GCN.38714....1S} {38714, 1}

\bibitem[\protect\citeauthoryear{Selsing et~al.,}{Selsing et~al.}{2019}]{selsing2019a}
Selsing J.,  et~al., 2019, \mn@doi [A\&A] {10.1051/0004-6361/201832835}, 623, A92

\bibitem[\protect\citeauthoryear{{Shingles} et~al.,}{{Shingles} et~al.}{2021}]{Shingles2021}
{Shingles} L.,  et~al., 2021, Transient Name Server AstroNote, \href {https://ui.adsabs.harvard.edu/abs/2021TNSAN...7....1S} {7, 1}

\bibitem[\protect\citeauthoryear{{Siegel}, {D'Elia}  \& {Swift/UVOT Team}}{{Siegel} et~al.}{2024}]{Siegel2024GCN35613}
{Siegel} M.~H.,  {D'Elia} V.,   {Swift/UVOT Team} 2024, GRB Coordinates Network, \href {https://ui.adsabs.harvard.edu/abs/2024GCN.35613....1S} {35613, 1}

\bibitem[\protect\citeauthoryear{Soderberg et~al.,}{Soderberg et~al.}{2004a}]{soderberg_redshift_2004}
Soderberg A.~M.,  et~al., 2004a, \mn@doi [The Astrophysical Journal] {10.1086/383082}, 606, 994

\bibitem[\protect\citeauthoryear{Soderberg, Frail  \& Wieringa}{Soderberg et~al.}{2004b}]{soderberg_constraints_2004}
Soderberg A.~M.,  Frail D.~A.,   Wieringa M.~H.,  2004b, \mn@doi [The Astrophysical Journal] {10.1086/421722}, 607, L13

\bibitem[\protect\citeauthoryear{{Soderberg} et~al.,}{{Soderberg} et~al.}{2006}]{2006ApJ...650..261S}
{Soderberg} A.~M.,  et~al., 2006, \mn@doi [\apj] {10.1086/506429}, \href {https://ui.adsabs.harvard.edu/abs/2006ApJ...650..261S} {650, 261}

\bibitem[\protect\citeauthoryear{{Steeghs} et~al.,}{{Steeghs} et~al.}{2017}]{Steeghs2017GCN22190}
{Steeghs} D.,  et~al., 2017, GRB Coordinates Network, \href {https://ui.adsabs.harvard.edu/abs/2017GCN.22190....1S} {22190, 1}

\bibitem[\protect\citeauthoryear{{Steeghs} et~al.,}{{Steeghs} et~al.}{2022}]{Steeghs2022}
{Steeghs} D.,  et~al., 2022, \mn@doi [\mnras] {10.1093/mnras/stac013}, \href {https://ui.adsabs.harvard.edu/abs/2022MNRAS.511.2405S} {511, 2405}

\bibitem[\protect\citeauthoryear{{Steele} et~al.,}{{Steele} et~al.}{2004}]{Steele2004}
{Steele} I.~A.,  et~al., 2004, in {Oschmann} Jacobus~M. J.,  ed.,  Society of Photo-Optical Instrumentation Engineers (SPIE) Conference Series Vol. 5489, Ground-based Telescopes. pp 679--692, \mn@doi{10.1117/12.551456}

\bibitem[\protect\citeauthoryear{{Strobl} \& {Jelinek}}{{Strobl} \& {Jelinek}}{2024}]{Strobl2024GCN38715}
{Strobl} J.,  {Jelinek} M.,  2024, GRB Coordinates Network, \href {https://ui.adsabs.harvard.edu/abs/2024GCN.38715....1S} {38715, 1}

\bibitem[\protect\citeauthoryear{{Strong}, {Klebesadel}  \& {Olson}}{{Strong} et~al.}{1974}]{Strong1974}
{Strong} I.~B.,  {Klebesadel} R.~W.,   {Olson} R.~A.,  1974, \mn@doi [\apjl] {10.1086/181415}, \href {https://ui.adsabs.harvard.edu/abs/1974ApJ...188L...1S} {188, L1}

\bibitem[\protect\citeauthoryear{{Sun} et~al.,}{{Sun} et~al.}{2025}]{Sun2025}
{Sun} H.,  et~al., 2025, \mn@doi [National Science Review] {10.1093/nsr/nwae401}, \href {https://ui.adsabs.harvard.edu/abs/2025NSRev..12..401S} {12, nwae401}

\bibitem[\protect\citeauthoryear{{Svinkin}, {Frederiks}, {Ulanov}, {Tsvetkova}, {Lysenko}, {Ridnaia}, {Cline}  \& {Konus-Wind Team}}{{Svinkin} et~al.}{2024}]{Svinkin2024GCN36768}
{Svinkin} D.,  {Frederiks} D.,  {Ulanov} M.,  {Tsvetkova} A.,  {Lysenko} A.,  {Ridnaia} A.,  {Cline} T.,   {Konus-Wind Team} 2024, GRB Coordinates Network, \href {https://ui.adsabs.harvard.edu/abs/2024GCN.36768....1S} {36768, 1}

\bibitem[\protect\citeauthoryear{{Tanaka}}{{Tanaka}}{2016}]{Tanaka2016}
{Tanaka} M.,  2016, \mn@doi [Advances in Astronomy] {10.1155/2016/6341974}, \href {https://ui.adsabs.harvard.edu/abs/2016AdAst2016E...8T} {2016, 634197}

\bibitem[\protect\citeauthoryear{{Tanvir} et~al.,}{{Tanvir} et~al.}{2009}]{Tanvir2009}
{Tanvir} N.~R.,  et~al., 2009, \mn@doi [\nat] {10.1038/nature08459}, \href {https://ui.adsabs.harvard.edu/abs/2009Natur.461.1254T} {461, 1254}

\bibitem[\protect\citeauthoryear{{Tanvir} et~al.,}{{Tanvir} et~al.}{2017}]{Tanvir2017}
{Tanvir} N.~R.,  et~al., 2017, \mn@doi [\apjl] {10.3847/2041-8213/aa90b6}, \href {https://ui.adsabs.harvard.edu/abs/2017ApJ...848L..27T} {848, L27}

\bibitem[\protect\citeauthoryear{Taylor, Frail, Kulkarni, Shepherd, Feroci  \& Frontera}{Taylor et~al.}{1998}]{taylor_discovery_1998}
Taylor G.~B.,  Frail D.~A.,  Kulkarni S.~R.,  Shepherd D.~S.,  Feroci M.,   Frontera F.,  1998, \mn@doi [The Astrophysical Journal] {10.1086/311513}, 502, L115

\bibitem[\protect\citeauthoryear{{Thoene} et~al.,}{{Thoene} et~al.}{2024}]{Thoene2024GCN35598}
{Thoene} C.~C.,  et~al., 2024, GRB Coordinates Network, \href {https://ui.adsabs.harvard.edu/abs/2024GCN.35598....1T} {35598, 1}

\bibitem[\protect\citeauthoryear{{Tingay}, {Jauncey}, {King}, {Tzioumis}, {Lovell}  \& {Edwards}}{{Tingay} et~al.}{2003}]{Tingay2003}
{Tingay} S.~J.,  {Jauncey} D.~L.,  {King} E.~A.,  {Tzioumis} A.~K.,  {Lovell} J. E.~J.,   {Edwards} P.~G.,  2003, \mn@doi [\pasj] {10.1093/pasj/55.2.351}, \href {https://ui.adsabs.harvard.edu/abs/2003PASJ...55..351T} {55, 351}

\bibitem[\protect\citeauthoryear{{Tody}}{{Tody}}{1986}]{Tody+1986}
{Tody} D.,  1986, in {Crawford} D.~L.,  ed.,  Society of Photo-Optical Instrumentation Engineers (SPIE) Conference Series Vol. 627, Instrumentation in astronomy VI. p.~733, \mn@doi{10.1117/12.968154}

\bibitem[\protect\citeauthoryear{{Tomaney} \& {Crotts}}{{Tomaney} \& {Crotts}}{1996}]{TomaneyAJ1996}
{Tomaney} A.~B.,  {Crotts} A. P.~S.,  1996, \mn@doi [\aj] {10.1086/118228}, \href {https://ui.adsabs.harvard.edu/abs/1996AJ....112.2872T} {112, 2872}

\bibitem[\protect\citeauthoryear{{Tonry} et~al.,}{{Tonry} et~al.}{2018}]{Tonry2018}
{Tonry} J.~L.,  et~al., 2018, \mn@doi [\pasp] {10.1088/1538-3873/aabadf}, \href {https://ui.adsabs.harvard.edu/abs/2018PASP..130f4505T} {130, 064505}

\bibitem[\protect\citeauthoryear{{Torii} et~al.,}{{Torii} et~al.}{2024}]{Torii2024GCN36719}
{Torii} S.,  et~al., 2024, GRB Coordinates Network, \href {https://ui.adsabs.harvard.edu/abs/2024GCN.36719....1T} {36719, 1}

\bibitem[\protect\citeauthoryear{{Torreiro Mart{\'\i}nez} et~al.,}{{Torreiro Mart{\'\i}nez} et~al.}{2024}]{Torreiro2024GCN37692}
{Torreiro Mart{\'\i}nez} M.,  et~al., 2024, GRB Coordinates Network, \href {https://ui.adsabs.harvard.edu/abs/2024GCN.37692....1T} {37692, 1}

\bibitem[\protect\citeauthoryear{{Troja} et~al.,}{{Troja} et~al.}{2017}]{Troja2017}
{Troja} E.,  et~al., 2017, \mn@doi [\nat] {10.1038/nature24290}, \href {https://ui.adsabs.harvard.edu/abs/2017Natur.551...71T} {551, 71}

\bibitem[\protect\citeauthoryear{{Troja} et~al.,}{{Troja} et~al.}{2022}]{Troja2022}
{Troja} E.,  et~al., 2022, \mn@doi [\nat] {10.1038/s41586-022-05327-3}, \href {https://ui.adsabs.harvard.edu/abs/2022Natur.612..228T} {612, 228}

\bibitem[\protect\citeauthoryear{{Valenti} et~al.,}{{Valenti} et~al.}{2017}]{Valenti2017}
{Valenti} S.,  et~al., 2017, \mn@doi [\apjl] {10.3847/2041-8213/aa8edf}, \href {https://ui.adsabs.harvard.edu/abs/2017ApJ...848L..24V} {848, L24}

\bibitem[\protect\citeauthoryear{{Vernet} et~al.,}{{Vernet} et~al.}{2011}]{vernet2011a}
{Vernet} J.,  et~al., 2011, \mn@doi [\aap] {10.1051/0004-6361/201117752}, \href {https://ui.adsabs.harvard.edu/abs/2011A&A...536A.105V} {536, A105}

\bibitem[\protect\citeauthoryear{{Wang} et~al.,}{{Wang} et~al.}{2017}]{Wang2017d}
{Wang} H.,  et~al., 2017, \mn@doi [\apjl] {10.3847/2041-8213/aa9e08}, \href {https://ui.adsabs.harvard.edu/abs/2017ApJ...851L..18W} {851, L18}

\bibitem[\protect\citeauthoryear{{Wei} et~al.,}{{Wei} et~al.}{2016}]{Wei2016}
{Wei} J.,  et~al., 2016, \mn@doi [arXiv e-prints] {10.48550/arXiv.1610.06892}, \href {https://ui.adsabs.harvard.edu/abs/2016arXiv161006892W} {p. arXiv:1610.06892}

\bibitem[\protect\citeauthoryear{Weiler, Panagia, Montes  \& Sramek}{Weiler et~al.}{2002}]{weiler_radio_2002}
Weiler K.~W.,  Panagia N.,  Montes M.~J.,   Sramek R.~A.,  2002, \mn@doi [Annual Review of Astronomy and Astrophysics] {10.1146/annurev.astro.40.060401.093744}, 40, 387

\bibitem[\protect\citeauthoryear{{Willingale}, {Starling}, {Beardmore}, {Tanvir}  \& {O'Brien}}{{Willingale} et~al.}{2013}]{Willingale2013}
{Willingale} R.,  {Starling} R.~L.~C.,  {Beardmore} A.~P.,  {Tanvir} N.~R.,   {O'Brien} P.~T.,  2013, \mn@doi [\mnras] {10.1093/mnras/stt175}, \href {https://ui.adsabs.harvard.edu/abs/2013MNRAS.431..394W} {431, 394}

\bibitem[\protect\citeauthoryear{Wilms, Allen  \& McCray}{Wilms et~al.}{2000}]{Wilms2000}
Wilms J.,  Allen A.,   McCray R.,  2000, \mn@doi [Astrophysical Journal] {10.1086/317016}, 542, 914

\bibitem[\protect\citeauthoryear{{Woosley}}{{Woosley}}{1993}]{Woosley1993}
{Woosley} S.~E.,  1993, \mn@doi [\apj] {10.1086/172359}, \href {https://ui.adsabs.harvard.edu/abs/1993ApJ...405..273W} {405, 273}

\bibitem[\protect\citeauthoryear{Woosley \& Bloom}{Woosley \& Bloom}{2006}]{woosley_supernova_2006}
Woosley S.~E.,  Bloom J.~S.,  2006, \mn@doi [Annual Review of Astronomy and Astrophysics] {10.1146/annurev.astro.43.072103.150558}, 44, 507

\bibitem[\protect\citeauthoryear{{Yang} et~al.,}{{Yang} et~al.}{2022}]{Yang2022}
{Yang} J.,  et~al., 2022, \mn@doi [\nat] {10.1038/s41586-022-05403-8}, \href {https://ui.adsabs.harvard.edu/abs/2022Natur.612..232Y} {612, 232}

\bibitem[\protect\citeauthoryear{{Yang} et~al.,}{{Yang} et~al.}{2024}]{Yang2024ApJ...969..126Y}
{Yang} Y.-P.,  et~al., 2024, \mn@doi [\apj] {10.3847/1538-4357/ad4be3}, \href {https://ui.adsabs.harvard.edu/abs/2024ApJ...969..126Y} {969, 126}

\bibitem[\protect\citeauthoryear{{Yi}, {Wu}, {Zou}  \& {Dai}}{{Yi} et~al.}{2020}]{Yi2020}
{Yi} S.-X.,  {Wu} X.-F.,  {Zou} Y.-C.,   {Dai} Z.-G.,  2020, \mn@doi [\apj] {10.3847/1538-4357/ab8a53}, \href {https://ui.adsabs.harvard.edu/abs/2020ApJ...895...94Y} {895, 94}

\bibitem[\protect\citeauthoryear{{York} et~al.,}{{York} et~al.}{2000}]{York2000}
{York} D.~G.,  et~al., 2000, \mn@doi [\aj] {10.1086/301513}, \href {https://ui.adsabs.harvard.edu/abs/2000AJ....120.1579Y} {120, 1579}

\bibitem[\protect\citeauthoryear{{Yuan} et~al.,}{{Yuan} et~al.}{2015}]{Yuan2015}
{Yuan} W.,  et~al., 2015, \mn@doi [arXiv e-prints] {10.48550/arXiv.1506.07735}, \href {https://ui.adsabs.harvard.edu/abs/2015arXiv150607735Y} {p. arXiv:1506.07735}

\bibitem[\protect\citeauthoryear{{Yuan} et~al.,}{{Yuan} et~al.}{2020}]{Yuan2020}
{Yuan} X.,  et~al., 2020, in {Marshall} H.~K.,  {Spyromilio} J.,   {Usuda} T.,  eds,  Society of Photo-Optical Instrumentation Engineers (SPIE) Conference Series Vol. 11445, Ground-based and Airborne Telescopes VIII. p. 114457M, \mn@doi{10.1117/12.2562334}

\bibitem[\protect\citeauthoryear{{Yuan}, {Zhang}, {Chen}  \& {Ling}}{{Yuan} et~al.}{2022}]{Yuan2022}
{Yuan} W.,  {Zhang} C.,  {Chen} Y.,   {Ling} Z.,  2022, in {Bambi} C.,  {Sangangelo} A.,  eds, , Handbook of X-ray and Gamma-ray Astrophysics.
Springer Living Reference Work, ISBN: 978-981-16-4544-0, 2022, id.86, p.~86, \mn@doi{10.1007/978-981-16-4544-0_151-1}

\bibitem[\protect\citeauthoryear{{Zhang}}{{Zhang}}{2019}]{Zhang2019book}
{Zhang} B.,  2019, {The physics of gamma-ray bursts}.
{Cambridge: Cambridge University Press. ISBN: 1-108-67683-9}

\bibitem[\protect\citeauthoryear{{Zhang} \& {Kobayashi}}{{Zhang} \& {Kobayashi}}{2005}]{Zhang2005}
{Zhang} B.,  {Kobayashi} S.,  2005, \mn@doi [\apj] {10.1086/429787}, \href {https://ui.adsabs.harvard.edu/abs/2005ApJ...628..315Z} {628, 315}

\bibitem[\protect\citeauthoryear{{Zhang}, {Woosley}  \& {Heger}}{{Zhang} et~al.}{2004}]{Zhang2004}
{Zhang} W.,  {Woosley} S.~E.,   {Heger} A.,  2004, \mn@doi [\apj] {10.1086/386300}, \href {https://ui.adsabs.harvard.edu/abs/2004ApJ...608..365Z} {608, 365}

\bibitem[\protect\citeauthoryear{{Zhang}, {Fan}, {Dyks}, {Kobayashi}, {M{\'e}sz{\'a}ros}, {Burrows}, {Nousek}  \& {Gehrels}}{{Zhang} et~al.}{2006}]{Zhang2006}
{Zhang} B.,  {Fan} Y.~Z.,  {Dyks} J.,  {Kobayashi} S.,  {M{\'e}sz{\'a}ros} P.,  {Burrows} D.~N.,  {Nousek} J.~A.,   {Gehrels} N.,  2006, \mn@doi [\apj] {10.1086/500723}, \href {https://ui.adsabs.harvard.edu/abs/2006ApJ...642..354Z} {642, 354}

\bibitem[\protect\citeauthoryear{{Zhang} et~al.,}{{Zhang} et~al.}{2009}]{Zhang2009}
{Zhang} B.,  et~al., 2009, \mn@doi [\apj] {10.1088/0004-637X/703/2/1696}, \href {https://ui.adsabs.harvard.edu/abs/2009ApJ...703.1696Z} {703, 1696}

\bibitem[\protect\citeauthoryear{{Zhang}, {Shao}, {Yan}  \& {Wei}}{{Zhang} et~al.}{2012}]{Zhang2012a}
{Zhang} F.-W.,  {Shao} L.,  {Yan} J.-Z.,   {Wei} D.-M.,  2012, \mn@doi [\apj] {10.1088/0004-637X/750/2/88}, \href {https://ui.adsabs.harvard.edu/abs/2012ApJ...750...88Z} {750, 88}

\bibitem[\protect\citeauthoryear{{Zhang} et~al.,}{{Zhang} et~al.}{2021}]{Zhang2021}
{Zhang} B.~B.,  et~al., 2021, \mn@doi [Nature Astronomy] {10.1038/s41550-021-01395-z}, \href {https://ui.adsabs.harvard.edu/abs/2021NatAs...5..911Z} {5, 911}

\bibitem[\protect\citeauthoryear{{Zhang}, {Zhong}, {Xin}  \& {Liang}}{{Zhang} et~al.}{2024}]{ZhangLu2024}
{Zhang} L.-L.,  {Zhong} S.-Q.,  {Xin} L.-P.,   {Liang} E.-W.,  2024, \mn@doi [\apj] {10.3847/1538-4357/ad5f92}, \href {https://ui.adsabs.harvard.edu/abs/2024ApJ...972..170Z} {972, 170}

\bibitem[\protect\citeauthoryear{{Zou} et~al.,}{{Zou} et~al.}{2025}]{ZouKumar2025}
{Zou} X.,  et~al., 2025, \mn@doi [arXiv e-prints] {10.48550/arXiv.2505.19831}, \href {https://ui.adsabs.harvard.edu/abs/2025arXiv250519831Z} {p. arXiv:2505.19831}

\bibitem[\protect\citeauthoryear{{de Ugarte Postigo} et~al.,}{{de Ugarte Postigo} et~al.}{2024}]{Postigo2024GCN37467}
{de Ugarte Postigo} A.,  et~al., 2024, GRB Coordinates Network, \href {https://ui.adsabs.harvard.edu/abs/2024GCN.37467....1D} {37467, 1}

\bibitem[\protect\citeauthoryear{{van Eerten}}{{van Eerten}}{2018}]{vanEerten2018}
{van Eerten} H.,  2018, \mn@doi [International Journal of Modern Physics D] {10.1142/S0218271818420026}, \href {https://ui.adsabs.harvard.edu/abs/2018IJMPD..2742002V} {27, 1842002}

\bibitem[\protect\citeauthoryear{{van Eerten}, {Zhang}  \& {MacFadyen}}{{van Eerten} et~al.}{2010}]{vanEerten2010}
{van Eerten} H.,  {Zhang} W.,   {MacFadyen} A.,  2010, \mn@doi [\apj] {10.1088/0004-637X/722/1/235}, \href {https://ui.adsabs.harvard.edu/abs/2010ApJ...722..235V} {722, 235}

\bibitem[\protect\citeauthoryear{{van Paradijs} et~al.,}{{van Paradijs} et~al.}{1997}]{vanParadijs1997}
{van Paradijs} J.,  et~al., 1997, \mn@doi [\nat] {10.1038/386686a0}, \href {https://ui.adsabs.harvard.edu/abs/1997Natur.386..686V} {386, 686}

\bibitem[\protect\citeauthoryear{van~der Horst et~al.,}{van~der Horst et~al.}{2008}]{van_der_horst_detailed_2008}
van~der Horst A.~J.,  et~al., 2008, \mn@doi [Astronomy and Astrophysics] {10.1051/0004-6361:20078051}, 480, 35

\bibitem[\protect\citeauthoryear{von Kienlin et~al.,}{von Kienlin et~al.}{2014}]{vonKienlin2014}
von Kienlin A.,  et~al., 2014, \mn@doi [The Astrophysical Journal Supplement Series] {10.1088/0067-0049/211/1/13}, 211, 13

\bibitem[\protect\citeauthoryear{von Kienlin et~al.,}{von Kienlin et~al.}{2020}]{vonKienlin2020}
von Kienlin A.,  et~al., 2020, \mn@doi [The Astrophysical Journal] {10.3847/1538-4357/ab7a18}, 893, 46

\makeatother
\end{thebibliography}

\appendix

\renewcommand{\thetable}{A\arabic{table}}
\setcounter{table}{0}

\renewcommand{\thefigure}{A\arabic{figure}}
\setcounter{figure}{0}

\section*{Appendix}

\onecolumn
\begin{longtable}{|c|c|c|c|c|c|c|} 
\caption{Optical afterglow observations of \grbssample{} compiled within this work, along with those collected from the reported GCNs. All the tabulated magnitudes are in AB system.}
\label{tab:phot_data}\\
\hline
$T-T_0$ (h) & Instrument/Telescope & Exp. Time & Filter & Mag & Mag err & Source \\
\hline
\endfirsthead

\multicolumn{7}{l}{\small \textit{Table \ref{tab:phot_data} continued from previous page.}}\\
\hline
$T-T_0$ (h) & Instrument/Telescope & Exp. Time & Filter & Mag & Mag err & Source \\
\hline
\endhead
\hline
\endfoot
\hline
\endlastfoot

\multicolumn{7}{|c|}{GRB 240122A (\maxiT = 2460331.93615); GOTO24eu} \\ 
\hline
0.728  & GOTO-S & $4\times45$\,s  &  $L$  & 17.581  &  0.037 &  This work   \\
9.800  & 0.6m BOOTES-2  & $14\times60$\,s & $clear$ & $>$17.81 &       & This work  \\
9.836  & UVOT/\swift           & $481$\,s   &  $u$  & $>$20.4 &        & \cite{Siegel2024GCN35613}   \\
12.318 & OSIRIS/GTC     & $30$\,s    &  $r^\prime$ & 20.431 & 0.115   & This work  \\
13.837 & 1.5m OSN & $10\times90$\,s & $V$ & 20.73    & 0.18   &   This work              \\
13.865 & 1.5m OSN & $11\times90$\,s & $B$ & $>$22.09    &        &   This work              \\
13.892 & 1.5m OSN & $10\times90$\,s & $I$ & 20.73    & 0.08   &   This work              \\
13.907 & 1.5m OSN & $11\times90$\,s & $R$ & 20.94    & 0.09   &   This work              \\
13.168 & IO:O/LT        & $3\times60$\,s    & $r$ & 20.396  &	0.207 & This work   \\
13.237 & IO:O/LT        & $3\times60$\,s    & $i$ & 20.489  &	0.200 & This work   \\
13.307 & IO:O/LT        & $3\times60$\,s    & $z$ & 20.461  &	0.255 & This work   \\
15.666 & X-shooter/VLT  & $11\times20$\,s  & $r^\prime$ & 20.652  &	0.033 & This work   \\
15.689 & X-shooter/VLT  & $3\times60$\,s  & $r^\prime$ & 20.610  &	0.046 & This work   \\
15.778 & X-shooter/VLT  & $3\times60$\,s  & $g^\prime$ & 21.277  &	0.026 & This work   \\
15.827 & X-shooter/VLT  & $3\times60$\,s  & $z^\prime$ & 20.494  &	0.029 & This work   \\
36.198 & 1.5m OSN & $10\times90$\,s & $B$ & $>$22.15   &       &   This work              \\
36.225 & 1.5m OSN & $10\times90$\,s & $V$ & 21.56    & 0.28   &   This work              \\
36.253 & 1.5m OSN & $10\times90$\,s & $R$ & 21.83    & 0.18   &   This work              \\
36.280 & 1.5m OSN & $10\times90$\,s & $I$ & 22.05    & 0.15   &   This work              \\
59.202 & 1.5m OSN & $12\times150$\,s & $B$ & $>$22.31    &      &   This work              \\
59.247 & 1.5m OSN & $12\times150$\,s & $V$ & $>$21.96    &      &   This work              \\
59.291 & 1.5m OSN & $12\times150$\,s & $R$ & 22.19  & 0.18  &   This work              \\
59.336 & 1.5m OSN & $12\times150$\,s & $I$ & 23.01  & 0.28  &   This work              \\
\hline
\multicolumn{7}{|c|}{GRB 240225B (\maxiT = 2460366.34428); GOTO24tz} \\ 
\hline
1.501  &  GOTO-N & $4\times45$\,s & $L$ & 17.118 &  0.043 & This work   \\
18.90  &  0.5m HMT & $30\times90$\,s & $R$ & 18.88   &  0.12   & This work   \\
25.29  &  0.5m HMT & $30\times90$\,s & $R$ & 19.81   &  0.3   & This work   \\
25.914 &  GOTO-N & $4\times45$\,s &$L$  &  19.694 &  0.178 & This work   \\
26.167 &  ALFOSC/NOT & $3\times300$\,s &$r^\prime$  & 19.159  &  0.016  &  This work  \\
28.096 &  IO:O/LT & $2\times75$\,s &$g$   & 19.971  &  0.123  &  This work   \\
28.150 &  IO:O/LT & $2\times75$\,s &$r$   &  19.513  &  0.151  &  This work   \\
28.204 &  IO:O/LT & $2\times75$\,s &$i$   &  19.442  &  0.105  &  This work   \\ 
28.258 &  IO:O/LT & $2\times75$\,s &$z$   &  19.281  &  0.143  &  This work   \\
39.312 &  50cm MITSuME-Akeno & $79\times60$\,s &$g$   &  19.78  &  0.13  &  \cite{Sasada2024GCN35831}    \\
39.312 &  50cm MITSuME-Akeno & $79\times60$\,s &$R$ &  19.53  &  0.10  &  \cite{Sasada2024GCN35831}    \\
39.312 &  50cm MITSuME-Akeno & $79\times60$\,s &$I$ &  $>$19.7  &    &  \cite{Sasada2024GCN35831}    \\
41.261 &  CMOS/AZT-33IK (Mondy) & $60\times60$\,s &$R$ & 19.93 & 0.07 & This work   \\
49.359 &  IO:O/LT & $15\times120$\,s &$r$   & 20.04  &  0.05  &  This work  \\
52.804 &  3.6m DOT & $60$\,s &$i$   & 19.96  &  0.04  & \cite{Ror2024GCN35830}   \\
57.656 &  WINTER/Palomar & $8\times120$\,s & $J$ & $>$18.8 &   & \cite{Mo2024GCN35829}   \\
57.656 &  WINTER/Palomar & $8\times120$\,s & $Y$ & $>$18.3 &   & \cite{Mo2024GCN35829}   \\
60.238 &  IO:O/LT        & $15\times120$\,s    & $r$ & 20.444  &	0.124 & This work   \\
68.194 &  CMOS/AZT-33IK (Mondy) & $60\times60$\,s &$R$ & 20.64 & 0.11 & This work   \\
70.337 &   Zeiss-1000 (SAO-RAS) & $8\times300$\,s &$R$ & 20.72  &  0.02  &  \cite{Moskvitin2024GCN35828}   \\
73.160 &  IO:O/LT                 & $15\times180$\,s    & $r$  & 21.303  &  0.126 & This work   \\
75.762 &  ALFOSC/NOT              & $3\times300$\,s     & $r^\prime$  & 20.660  &  0.052  &  This work  \\
77.280 &  X-shooter/VLT & $60$\,s   &$r^\prime$   & 20.879   &  0.031   &  This work   \\
77.324 &  X-shooter/VLT & $3\times60$\,s &$r^\prime$   & 20.956   &  0.035   &  This work   \\
77.410 &  X-shooter/VLT & $3\times60$\,s &$g^\prime$   & 21.055   &  0.082   &  This work   \\
77.477 &  X-shooter/VLT & $3\times60$\,s &$z^\prime$   & 20.79    &  0.129   &  This work   \\
93.367 &  Zeiss-1000 (SAO RAS) & $8\times300$\,s &$R$   & 21.32  &  0.04  &  \cite{Moskvitin2024GCN35839}   \\
113.026 &  CMOS/AZT-33IK (Mondy) & $60\times60$\,s &$R$ & 21.73 & 0.19 & This work   \\
137.128 &  CMOS/AZT-33IK (Mondy) & $60\times60$\,s &$R$ & 21.92 & 0.24 & This work   \\
235.750 &  CMOS/AZT-33IK (Mondy) & $90\times60$\,s &$R$ & $>$23.02 & & This work   \\
259.086 &  CMOS/AZT-33IK (Mondy) & $114\times60$\,s &$R$ & $>$23.92 & & This work   \\
283.934 &  CMOS/AZT-33IK (Mondy) & $116\times60$\,s &$R$ & $>$23.82 & & This work   \\
308.135 &  CMOS/AZT-33IK (Mondy) & $120\times60$\,s &$R$ & $>$23.82 & & This work   \\
379.454 &  CMOS/AZT-33IK (Mondy) & $150\times60$\,s &$R$ & $>$23.72 & & This work   \\
450.107 &  CMOS/AZT-33IK (Mondy) & $142\times60$\,s &$R$ & $>$23.31 & & This work   \\
476.372 &  CMOS/AZT-33IK (Mondy) & $150\times60$\,s &$R$ & $>$23.39 & & This work   \\
\hline
\multicolumn{7}{|c|}{GRB 240619A (\fermiT = 2460480.65522); GOTO24cvn} \\ 
\hline
2.532 & ATLAS & $30$\,s & $o$ & 16.242 & 0.014 & ATLAS FP   \\
4.689  & GOTO-S & $3\times90$\,s & $L$ & 17.171 & 0.170 & This work   \\
17.955 & GOTO-N & $4\times90$\,s & $L$ & 18.381 & 0.086 & This work   \\
26.705 & ATLAS & $30$\,s & $o$ & 18.724 & 0.125 & ATLAS FP   \\
48.010 & WINTER/1m Palomar & $30\times120$\,s & $J$ & 19.3   & ---   & \cite{Geoffrey2024GCN36739}   \\
48.810 & UVOT/\swift & $81.3$\,s & $v$ & $>$18.78 & & This work   \\
48.684 & UVOT/\swift & $628$\,s & $white$ & 20.64 & 0.10 & This work   \\
49.174 & UVOT/\swift & $1070.9$\,s & $u$ & 20.29 & 0.14 & This work   \\
66.054 & ALFOSC/NOT & $5\times300$\,s &$r^\prime$  & 19.489 & 0.030  &  This work  \\
65.482 & ALFOSC/NOT & $9\times200$\,s &$z^\prime$  & 20.01  & 0.04 &  This work  \\
98.758 & PS1-GPC1 & $300$\,s & $i$ & 20.86	& 0.28 & \href{https://www.wis-tns.org/object/2024lwv}{TNS, 215607}   \\
113.909 & ALFOSC/NOT & $5\times300$\,s & $r^\prime$  & 20.452  &  0.037  &  This work  \\
161.494 & ALFOSC/NOT & $5\times300$\,s & $r^\prime$  & 20.681  &  0.051  &  This work  \\
331.618 & X-shooter/VLT & $10\times10$\,s & $r^\prime$ & 21.705 & 0.101 & This work   \\
\hline
\multicolumn{7}{|c|}{GRB 240910A (\fermiT = 2460563.66718); GOTO24fvl} \\
\hline
9.430  & GOTO-S & $4\times90$\,s & $L$ & 19.329 & 0.130 & This work   \\
10.559 & GOTO-S & $4\times90$\,s & $L$ & 19.879 & 0.152 & This work   \\
10.885 & GOTO-S & $3\times90$\,s & $L$ & 20.161 & 0.243 & This work   \\
11.689 & GOTO-S & $4\times90$\,s & $L$ & 19.970 & 0.320 & This work   \\
12.014 & GOTO-S & $4\times90$\,s & $L$ & 19.740 & 0.123 & This work   \\
36.613 & UVOT/\swift & $1561.5$\,s & $u$ & 21.59 & 0.18 & This work   \\
36.758 & UVOT/\swift & $378.7$\,s & $v$ & $>$20.08 & & This work   \\ 
37.344 & ALT/100C & $10\times300$\,s & $r$ & 20.8 & 0.2 & This work   \\
43.504   & ALFOSC/NOT & $3\times300$\,s  & $r'$ & 20.985 & 0.047 & This work  \\
45.428 & IO:O/LT & $6\times180$\,s & $r$ & 21.12  & 0.13  & This work   \\
45.908 & IO:O/LT & $6\times180$\,s & $z$ & 20.76  & 0.09  & This work   \\
46.148 & IO:O/LT & $6\times180$\,s & $i$ & 21.09  & 0.10  & This work   \\
47.066 & OSIRIS/GTC     & $30$\,s    &  $r^\prime$ & 21.156 & 0.028   & This work  \\
57.841 & VT/SVOM &  & $R$ & 21.50 & 0.05 & \cite{SVOM2024GCN37503}   \\
57.841 & VT/SVOM &  & $B$ & 22.24 & 0.07 & \cite{SVOM2024GCN37503}   \\
84.022 & UVOT/\swift & $670.8$\,s & $u$ & $>$22.04 & & This work   \\
84.958 & UVOT/\swift & $1758.4$\,s & $v$ & $>$20.96 & & This work   \\
91.237 & UVOT/\swift & $315.7$\,s & $u$ & $>$21.53 & & This work   \\
191.928  & UVOT/\swift & $3313.0$\,s & $u$ & $>$22.95 & & This work   \\
196.643  & UVOT/\swift & $768.2$\,s & $u$ & $>$22.19 & & This work   \\
386.321 & LBC/LBT & $900$\,s & $z^\prime$ & $>$24.5 &   & This work   \\
386.321 & LBC/LBT & $900$\,s & $r^\prime$ & 25.99 & 0.35 & This work   \\
\hline
\multicolumn{7}{|c|}{GRB 240916A (\fermiT = 2460569.557581); GOTO24fzn} \\
\hline
7.731  & GOTO-S & $4\times90$\,s & $L$ & 17.801 & 0.055 & This work   \\
12.872 & ALT/100C & $4\times300$\,s & $r$ & 18.60 & 0.05 & This work   \\
17.064 & UVOT/\swift & $237$\,s & $white$ & 20.93 & 0.18 & This work   \\
17.126 & UVOT/\swift & $188.2$\,s & $v$   & 19.08 & 0.31 & This work   \\
18.339 & UVOT/\swift & $1185.6$\,s & $u$   & 21.14 & 0.24 & This work   \\
22.750 & X-shooter/VLT & $19\times30$\,s & $r^\prime$ & 19.437 & 0.023 & This work   \\
42.674 & ALFOSC/NOT & $3\times300$\,s & $r^\prime$ & 20.433 & 0.035  & This work   \\
46.373 & X-shooter/VLT & $7\times30$\,s & $r^\prime$ & 20.627 & 0.033 & This work   \\
60.595 & ALT/100C & $6\times300$\,s & $r$ & $>$19.2 & & This work   \\
84.684 & ALT/100B & $8\times180$\,s & $r$ & $>$19.9 & & This work   \\
101.853 & UVOT/\swift & $916.3$\,s & $white$ & $>$22.60 & & This work   \\
156.200 & UVOT/\swift & $1693.2$\,s & $white$ & $>$22.92 & & This work   \\
\hline
\multicolumn{7}{|c|}{GRB 241002B (\fermiT = 2460585.75993); GOTO24gpc} \\
\hline
3.051  & GOTO-S & $4\times90$\,s & $L$ & 19.53 & 0.09 & This work   \\
6.206 &  QHY600 CMOS/40cm LCOGT & $500$\,s & $g$ &  20.18 & 0.3 & \cite{Torreiro2024GCN37692}   \\
6.348 &  QHY600 CMOS/40cm LCOGT & $500$\,s & $r$  & 19.90 & 0.25 & \cite{Torreiro2024GCN37692}   \\
16    &  1.8m PRIME &  & $J$  &  20.0  &  0.2 & \cite{Durbak2024GCN37700}   \\
16    &  1.8m PRIME &  & $H$ &  19.6  &  0.1 & \cite{Durbak2024GCN37700}   \\
38.986 & UVOT/\swift & $2619.0$\,s & $u$ & 21.99 & 0.18 & This work   \\
40    &  1.8m PRIME &  & $J$ &  20.6  &  0.2 & \cite{Durbak2024GCN37712}   \\
40    &  1.8m PRIME &  & $H$ &  20.3  &  0.1 & \cite{Durbak2024GCN37712}   \\
64    &  1.8m PRIME &  & $H$ &  20.8  &  0.2 & \cite{Guiffreda2024GCN37736}   \\
172.080  & UVOT/\swift & $4302.1$\,s & $u$ & $>$23.17 &  & This work   \\
\hline
\multicolumn{7}{|c|}{GRB 241228B (\fermiT = 2460672.67575); GOTO24jmz} \\
\hline
0.322  & GOTO-N  & $4\times90$\,s & $L$  & 14.543 & 0.007 & This work   \\
0.349  & GOTO-N  & $4\times90$\,s & $L$  & 14.647 & 0.008 & This work   \\
0.377  & GOTO-N  & $4\times90$\,s & $L$  & 14.713 & 0.008 & This work   \\
0.405  & GOTO-N  & $4\times90$\,s & $L$  & 14.778 & 0.009 & This work   \\
1.483  & GOTO-N  & $4\times90$\,s & $L$  & 17.001 & 0.045 & This work   \\
5.420 & 1m LCOGT & $600$\,s  & $r^\prime$ & 18.62  & 0.06  & \cite{OrtegaCasas2024GCN38692}   \\
5.899 & 1m LCOGT & $600$\,s  & $i^\prime$ & 18.31  & 0.09  & \cite{OrtegaCasas2024GCN38692}   \\
6.095 & 1m LCOGT & $600$\,s  & $g^\prime$ & 19.40  & 0.06  & \cite{OrtegaCasas2024GCN38692}   \\
6.609 & 1m LCOGT & $600$\,s  & $z^\prime$ & 18.25  & 0.28  & \cite{OrtegaCasas2024GCN38692}   \\
7.602  & ATLAS   & $30$\,s   & $o$  & 18.861 & 0.057 & ATLAS FP   \\
8.664  & 0.7m TRT/SBO & $4\times300$\,s & $R$  & 19.41  & 0.04  & This work   \\
8.794  & GOTO-S & $4\times90$\,s  & $L$  & 19.702 & 0.099 & This work   \\
9.957  & 1m LCOGT & $600$\,s  & $r^\prime$ & 19.62 & 0.05 & \cite{OrtegaCasas2024GCN38692}   \\
10.976 & 1m LCOGT & $1200$\,s & $B$  & 21.35  & 0.03 & \cite{Ghosh2024GCN38702}  \\
11.362 & UVOT/\swift & $1778.5$\,s & $u$ & 21.98 & 0.33 & This work   \\
11.628 & 1.6m Mephisto & $300\times4$\,s & $u_m$ & 21.69 & 0.17 & This work   \\
11.628 & 1.6m Mephisto & $300\times4$\,s & $g_m$ & 20.36 & 0.08 & This work   \\
11.629 & 1.6m Mephisto & $300\times4$\,s & $i_m$ & 19.41 & 0.05 & This work   \\
14.175 & 1.6m Mephisto & $300\times3$\,s & $v_m$ & 21.71 & 0.22 & This work   \\
14.175 & 1.6m Mephisto & $300\times3$\,s & $r_m$ & 20.27 & 0.05 & This work   \\
14.212 & 1.6m Mephisto & $300\times2$\,s & $i_m$ & 20.30 & 0.24 & This work   \\
15.739 & 1.6m Mephisto & $300\times3$\,s & $v_m$ & 21.70 & 0.23 & This work   \\
15.739 & 1.6m Mephisto & $300\times3$\,s & $r_m$ & 20.42 & 0.07 & This work   \\
15.740 & 1.6m Mephisto & $300\times3$\,s & $i_m$ & 20.10 & 0.06 & This work   \\
15.991 & CMOS/AZT-33IK (Mondy) & $30\times120$\,s & $R$  & 20.49  & 0.04  & This work   \\
18     & D50   & $24\times120$\,s & $r^\prime$ & 20.4   & 0.1   & \cite{Strobl2024GCN38715}   \\
18.469 & 1.6m Mephisto & $300\times3$\,s & $v_m$ & 21.38 & 0.23 & This work   \\
18.469 & 1.6m Mephisto & $300\times3$\,s & $r_m$ & 20.54 & 0.08 & This work   \\
18.469 & 1.6m Mephisto & $300\times3$\,s & $i_m$ & 20.35 & 0.10 & This work   \\
20.327 & ALFOSC/NOT & $2\times150$\,s & $r^\prime$ & 20.560 & 0.061  & This work   \\
25.591 & X-shooter/VLT & $10$\,s   & $r^\prime$ & 20.626 & 0.029 & This work   \\
25.640 & X-shooter/VLT & $3\times60$\,s & $r^\prime$ & 20.617 & 0.032 & This work   \\
25.688 & X-shooter/VLT & $3\times60$\,s & $g^\prime$ & 21.448 & 0.011 & This work   \\
25.775 & X-shooter/VLT & $3\times60$\,s & $z^\prime$ & 20.267 & 0.023 & This work   \\
33.933 & GOTO-S & $4\times45$\,s  & $L$  &$>$20.52&       & This work   \\
39.066 & 1.6m Mephisto & $300\times4$\,s & $v_m$ & $>$22.85 &  & This work   \\
39.066 & 1.6m Mephisto & $300\times4$\,s & $r_m$ & 21.02 & 0.09 & This work   \\
39.068 & 1.6m Mephisto & $300\times4$\,s & $i_m$ & 20.72 & 0.10 & This work   \\
41.972 & 1.6m Mephisto & $300\times5$\,s & $v_m$ & $>$22.80 &  & This work   \\
41.972 & 1.6m Mephisto & $300\times5$\,s & $r_m$ & 21.24 & 0.11 & This work   \\
41.972 & 1.6m Mephisto & $300\times3$\,s & $i_m$ & 20.94 & 0.14 & This work   \\
42.175 & Zeiss-1000 (SAO RAS)  & $12\times300$\,s& $R$  & 21.18 & 0.06 & \cite{Moskvitin2024GCN38733}  \\
61.619 &1.3m DFOT&$24\times300$\,s& $R$  & 21.44  & 0.05  & \cite{Ror2025GCN38816}   \\
63.023 & 1.6m Mephisto & $300\times4$\,s & $v_m$ & $>$22.42 &  & This work   \\
63.023 & 1.6m Mephisto & $300\times4$\,s & $r_m$ & 21.87 & 0.15 & This work   \\
63.025 & 1.6m Mephisto & $300\times4$\,s & $i_m$ & 21.50 & 0.14 & This work   \\
66.200 & 1.6m Mephisto & $300\times4$\,s & $v_m$ & $>$22.74 &  & This work   \\
66.200 & 1.6m Mephisto & $300\times4$\,s & $r_m$ & 22.49 & 0.19 & This work   \\
66.203 & 1.6m Mephisto & $300\times4$\,s & $i_m$ & 21.92 & 0.19 & This work   \\
87.254 & 1.6m Mephisto & $300\times5$\,s & $v_m$ & $>$22.81 &  & This work   \\
87.254 & 1.6m Mephisto & $300\times5$\,s & $r_m$ & 22.70 & 0.20 & This work   \\
87.254 & 1.6m Mephisto & $300\times5$\,s & $i_m$ & $>$23.40 &  & This work   \\
204.161 & UVOT/\swift & $4273.9$\,s & $u$ & $>$22.59 &  & This work   \\
\end{longtable}

\twocolumn

\begin{table*}
\caption{Log of radio follow-up observations. The flux density errors are 1 sigma and the upper limits correspond to three times the image RMS.}
\addtolength{\tabcolsep}{11pt}
\label{tab:radio}
\begin{tabular}{|cccccc|}
\hline
  \multicolumn{1}{|c}{\begin{tabular}[c]{@{}c@{}}Observation Date\\ (UTC)\end{tabular}} &
  \multicolumn{1}{c}{\begin{tabular}[c]{@{}c@{}}Telescope\\ $ $\end{tabular}} &
  \multicolumn{1}{c}{\begin{tabular}[c]{@{}c@{}}Time Post-burst\\ (days)\end{tabular}} &
  \multicolumn{1}{c}{\begin{tabular}[c]{@{}c@{}}Frequency\\ (GHz)\end{tabular}} &
  \multicolumn{1}{c}{\begin{tabular}[c]{@{}c@{}}Flux Density\\ ($\mu$Jy)\end{tabular}} &
  Source \\ 
\hline
\multicolumn{6}{|c|}{GRB 240122A} \\
\hline
2024-01-24 12:01:00.0 UT & ATCA & 2.06  & 5.5  & \textless{}159.2  & \citet{Anderson2024GCN35642} \\
   & ATCA &    & 9.0  & 160.0 $\pm$ 20.0  & \citet{Anderson2024GCN35642} \\ 
\hline
2024-01-26 06:39:24.9 UT & ATCA & 3.84  & 5.5  & \textless{}1555.0 & This work \\
           & ATCA &    & 9.0  & \textless{}100.5  & This work \\
           & ATCA &    & 16.7 & \textless{}113.1  & This work \\
           & ATCA &    & 21.2 & \textless{}333.0  & This work \\ 
\hline
2024-01-28 10:37:34.9 UT & ATCA & 6.01  & 5.5  & \textless{}174.0  & This work \\
           & ATCA &    & 9.0  & 66.3 $\pm$ 19.7   & This work \\
           & ATCA &    & 16.7 & \textless{}55.5   & This work \\
           & ATCA &    & 21.2 & \textless{}111.3  & This work \\ 
\hline
2024-02-12 03:03:54.9 UT & ATCA & 20.69 & 5.5  & \textless{}45.0   & This work \\
           & ATCA &    & 9.0  & \textless{}33.6   & This work \\ 
\hline
\multicolumn{6}{|c|}{GRB 240619A} \\
\hline
2024-06-22 13:45:34 UT & AMI-LA & 3.42 & 15.5  & $1500 \pm 60$  & \cite{Rhodes2024GCN36744} \\
\hline
\multicolumn{6}{|c|}{GRB 240910A} \\
\hline
2024-09-13 06:34:25.6 UT & VLA &3.11  & 6  & $137 \pm 10$  & \cite{Giarratana2024GCN37569} \\
2024-09-13 06:56:22.1 UT & VLA &3.12       & 10 & $114 \pm 9$  & \cite{Giarratana2024GCN37569} \\
2024-09-13 06:16:03.7 UT & VLA &3.09       & 15 & $86 \pm 10$  & \cite{Giarratana2024GCN37569} \\
\hline
2024-09-19 06:33:15.8 UT & VLA &9.10  & 6  & $58 \pm 7$  &This work \\
2024-09-19 06:54:57.2 UT & VLA &9.12  & 10 & $47 \pm 8$  & This work \\
2024-09-19 06:14:53.8 UT & VLA &9.09  & 15 & $71 \pm 10$  & This work \\
\hline
2024-10-01 11:12:12.0 UT & VLA &21.30  & 6  & $22 \pm 6$  &This work \\
2024-10-01 11:34:09.0 UT & VLA &21.31  & 10 & \textless{}24  & This work \\
2024-10-01 10:53:51.0 UT & VLA &21.29  & 15 & \textless{}27  & This work \\
\hline
2024-10-26 09:15:46.0 UT & VLA &46.22  & 6  & \textless{}18  &This work \\
2024-10-26 09:38:12.0 UT & VLA &46.23  & 10 & \textless{}27  & This work \\
2024-10-26 08:55:00.0 UT & VLA &46.20  & 15 & \textless{}18  & This work \\
\hline
\multicolumn{6}{|c|}{GRB 240916A} \\
\hline
2024-09-18 00:43:57 UT & VLA & 1.97 & 6  & $35 \pm 8$  & \cite{Giarratana202437788} \\
                       & VLA &       & 10 & $44 \pm 8$  & \cite{Giarratana202437788} \\
                       & VLA &       & 15 & $135 \pm 8$  & \cite{Giarratana202437788} \\
\hline
\end{tabular}
\end{table*}

\begin{figure*}
\centering
\includegraphics[width=\hsize]{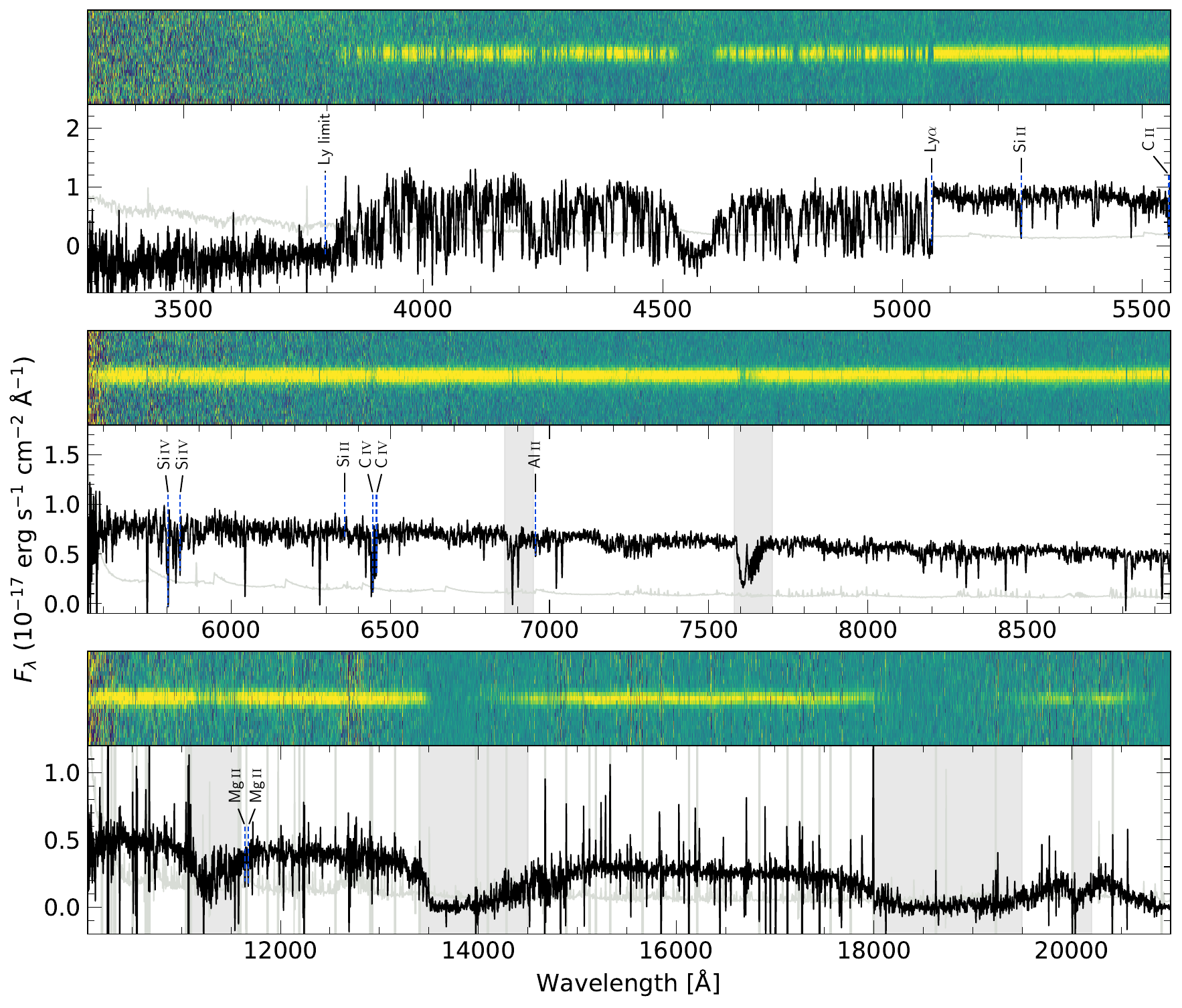}
\caption{X-shooter spectra of GRB~240122A at $z = 3.1634 \pm 0.0003$, observed at 16.130~hr post-trigger. The three panels show the spectra from the UVB, VIS, and NIR arms (top to bottom). Each panel displays the 2D spectrum (upper sub-panel) and the corresponding 1D extracted spectrum in black with the error spectrum in grey (lower sub-panel). Absorption lines identified at the redshift of the GRB are marked in blue and labelled, while grey-shaded regions indicate telluric absorption. The 1D spectra have been smoothed with a Savitzky–Golay filter to enhance the visibility of spectral features. The same colour scheme is adopted for all spectra presented throughout this work.}
\label{fig:VLT_spec_22A}
\end{figure*}

\begin{figure*}
\centering
\includegraphics[angle=0,scale=0.6]{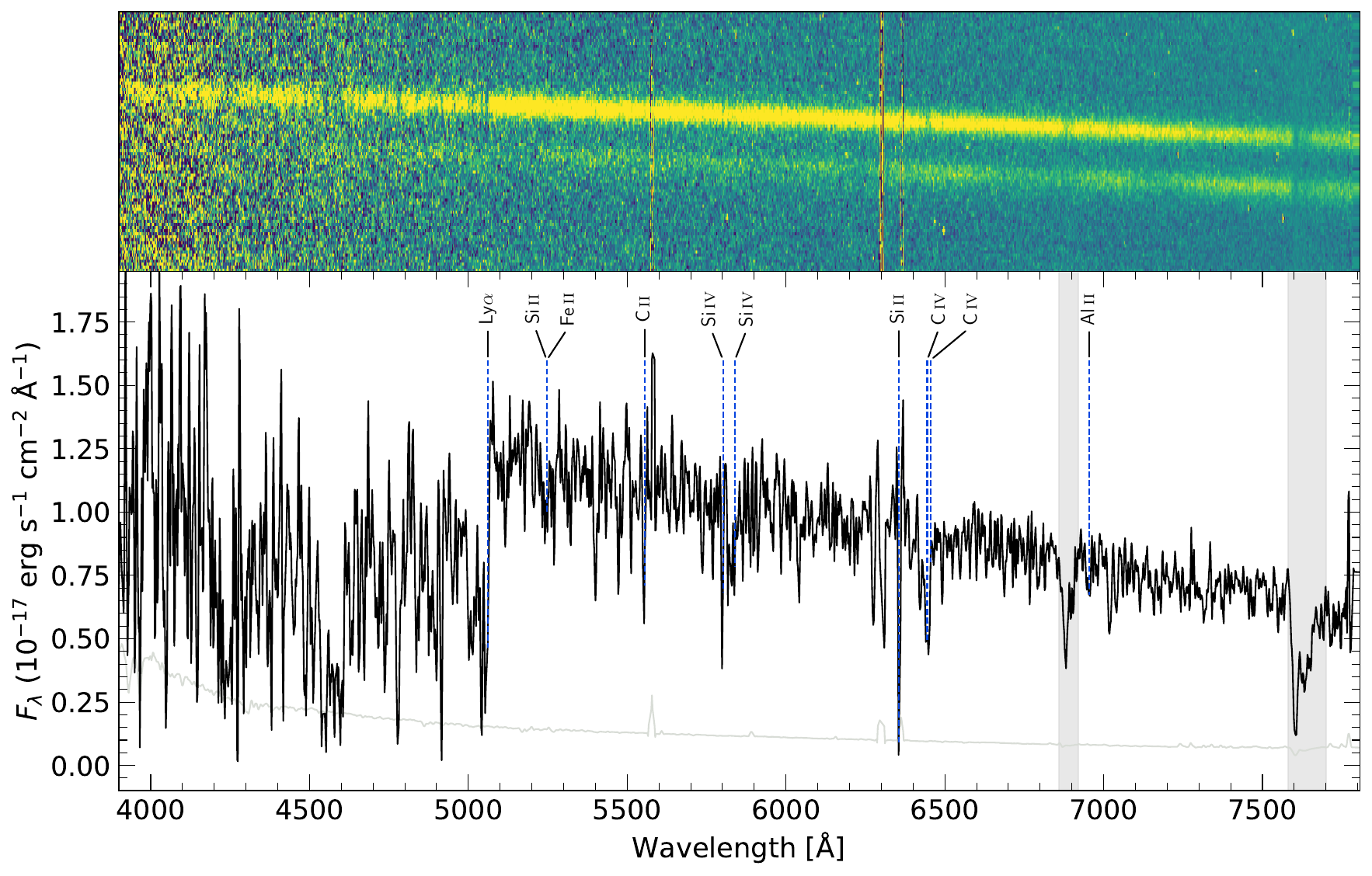}
\caption{GTC/OSIRIS spectrum of GRB~240122A at $z = 3.1634 \pm 0.0003$, obtained 12.857~hr after the trigger. The upper panel displays the 2D spectrum, and the lower panel shows the extracted 1D spectrum in black with its error spectrum in grey. Absorption features at the GRB redshift are highlighted in blue and labelled, while grey-shaded regions mark telluric absorption. The 1D spectra were smoothed using a Savitzky–Golay filter to improve the visibility of spectral features.}
\label{fig:GTC_spec_22A}
\end{figure*}

\begin{figure*}
\centering
\includegraphics[width=\hsize]{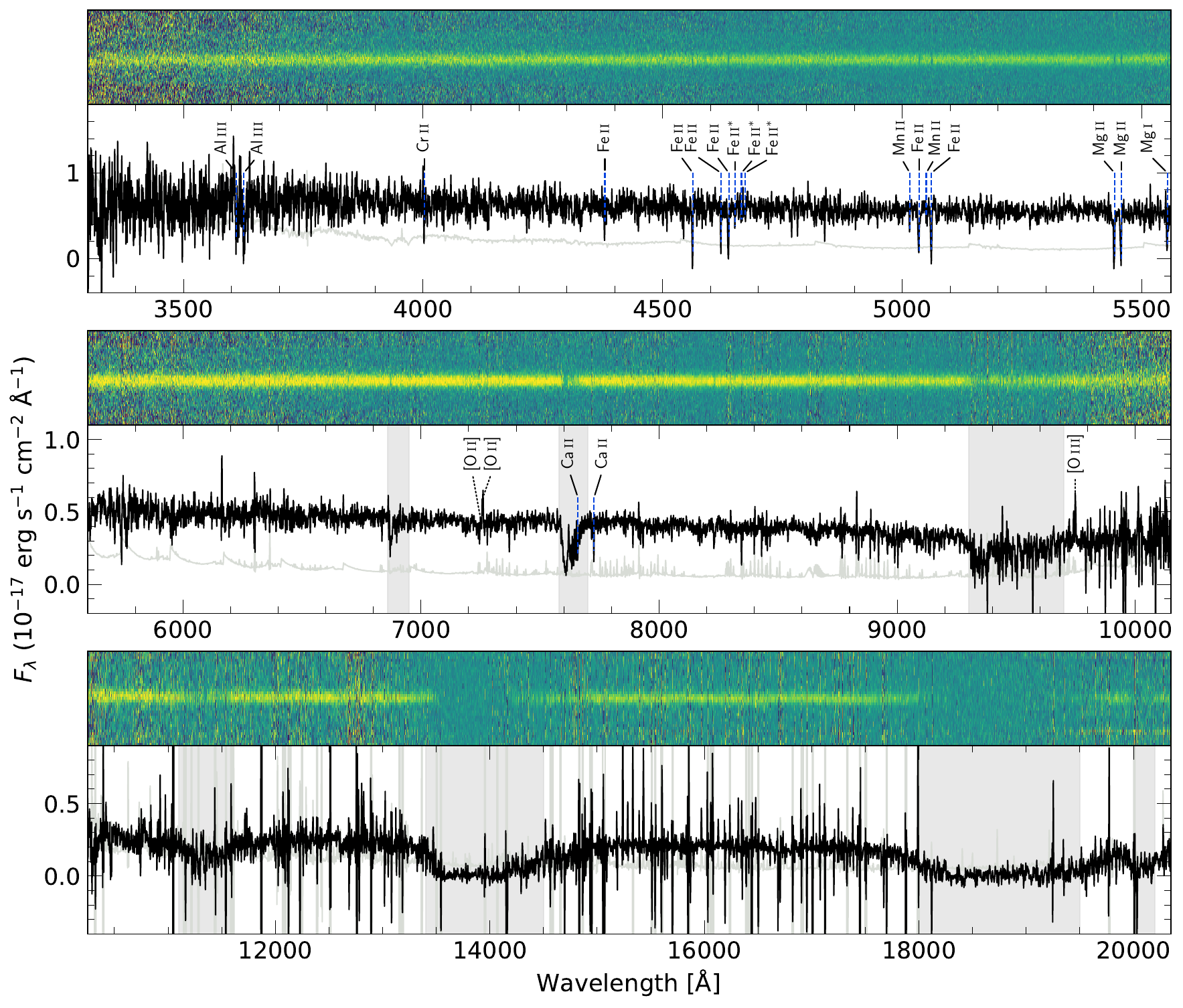}
\caption{VLT/X-shooter spectra of GRB 240225B at $z = 0.9462 \pm 0.0002$, taken at 3.237~d post-trigger. Absorption lines identified at the GRB redshift are marked in blue and labelled accordingly, while emission lines are indicated by black dotted vertical lines. A Savitzky-Golay filter was applied to the 1D spectra to enhance the clarity of spectral features.}
\label{fig:VLT_spec_25B}
\end{figure*}

\begin{figure*}
\centering
\includegraphics[width=\hsize]{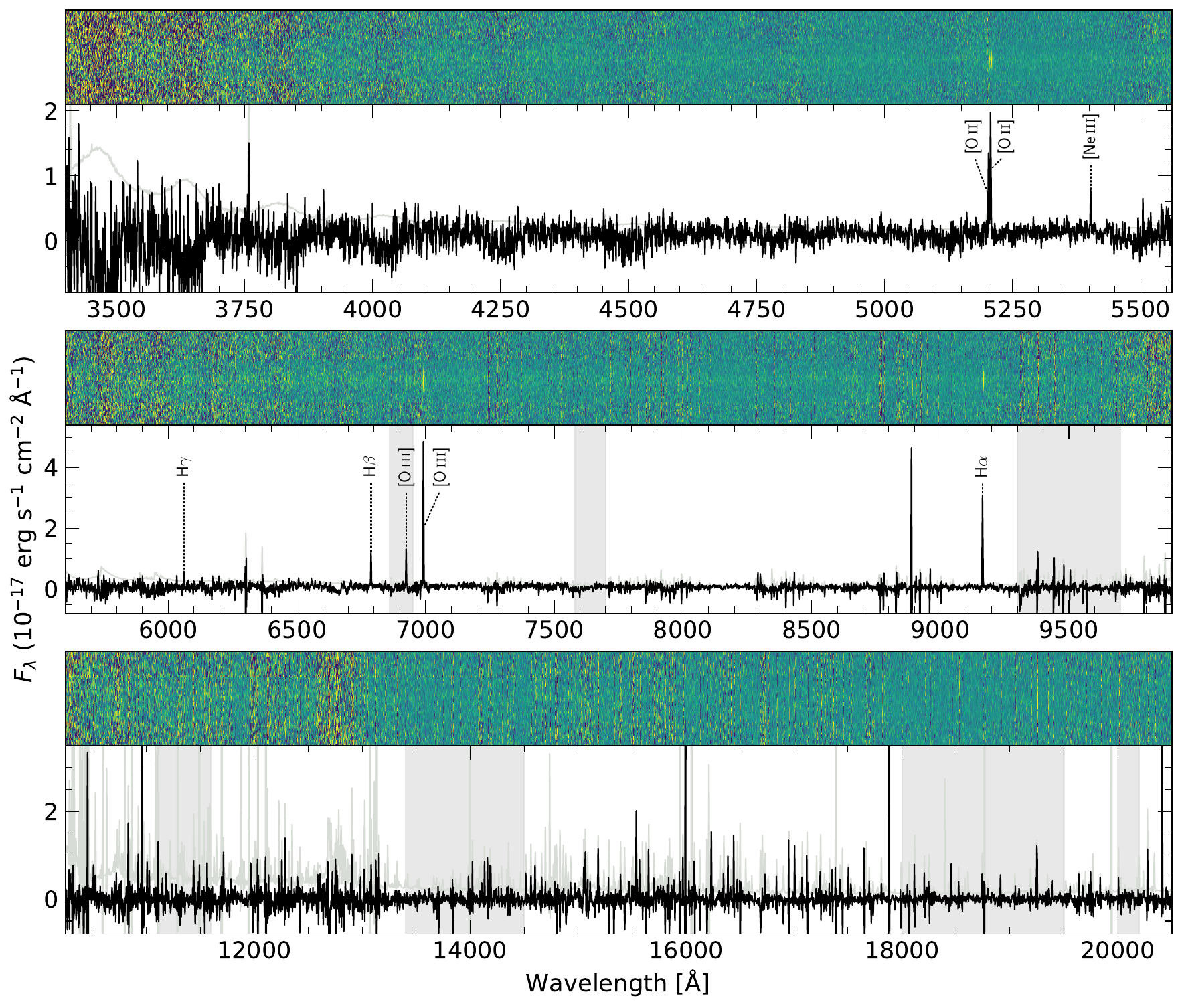}
\caption{VLT/X-shooter spectra of GRB 240619A at $z = 0.3960 \pm 0.0001$, observed at 13.826~d post-trigger. Emission lines are indicated by black dotted vertical lines and labelled accordingly. To highlight spectral features, the 1D spectra were smoothed with a Savitzky–Golay filter.}
\label{fig:VLT_spec_19A}
\end{figure*}

\begin{figure*}
\centering
\includegraphics[angle=0,scale=0.6]{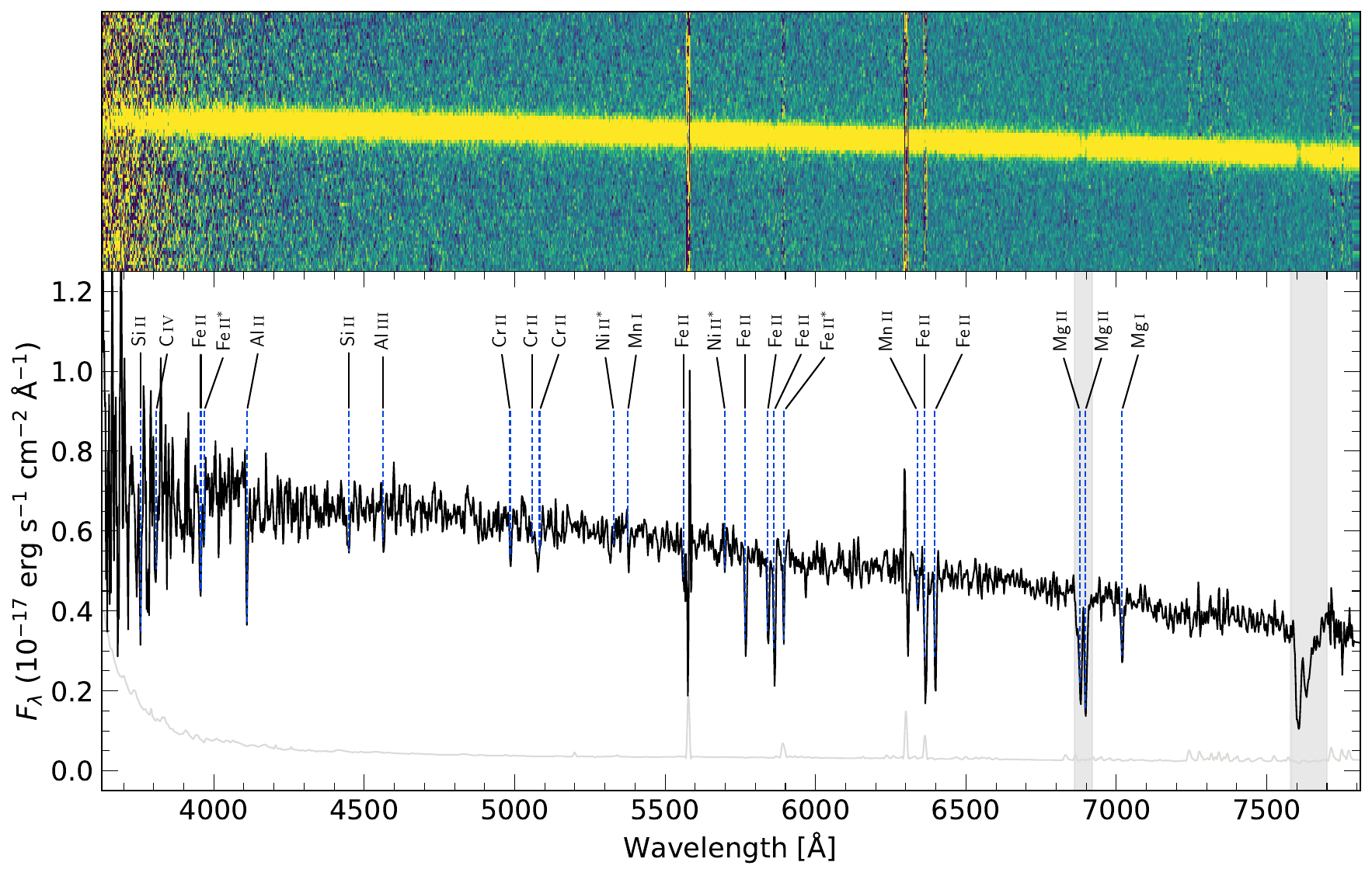}
\caption{GTC/OSIRIS spectrum of GRB 240910A at $z = 1.4605 \pm 0.0007$ at 1.964 days post-trigger. The upper panel presents the 2D spectrum, while the lower panel shows the extracted 1D spectrum in black with its associated error in grey. Absorption features at the GRB redshift are indicated in blue and labelled, and grey-shaded regions denote telluric absorption. For improved feature visibility, the 1D spectra were smoothed using a Savitzky-Golay filter.}
\label{fig:GTC_spec_10A}
\end{figure*}

\begin{figure*}
\centering
\includegraphics[width=\hsize]{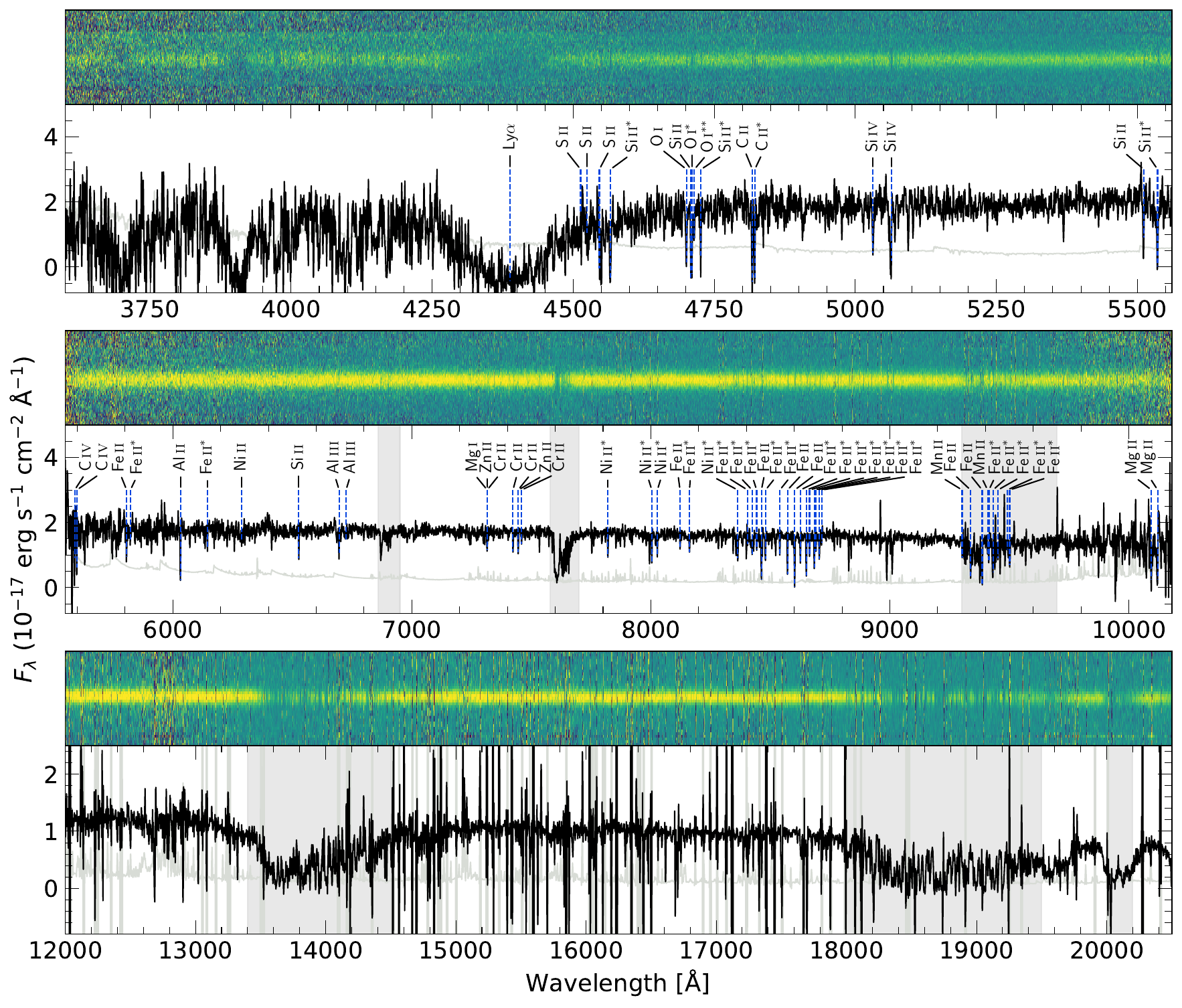}
\caption{VLT/X-shooter spectra of GRB 240916A at $z = 2.6100 \pm 0.0002$, taken at 22.962~hr post-trigger. Absorption lines identified at the redshift of the GRB are marked in blue and labelled accordingly. The 1D spectra were filtered with a Savitzky–Golay function to enhance the prominence of spectral features.}
\label{fig:VLT_spec_16A}
\end{figure*}

\begin{figure*}
\centering
\includegraphics[width=\hsize]{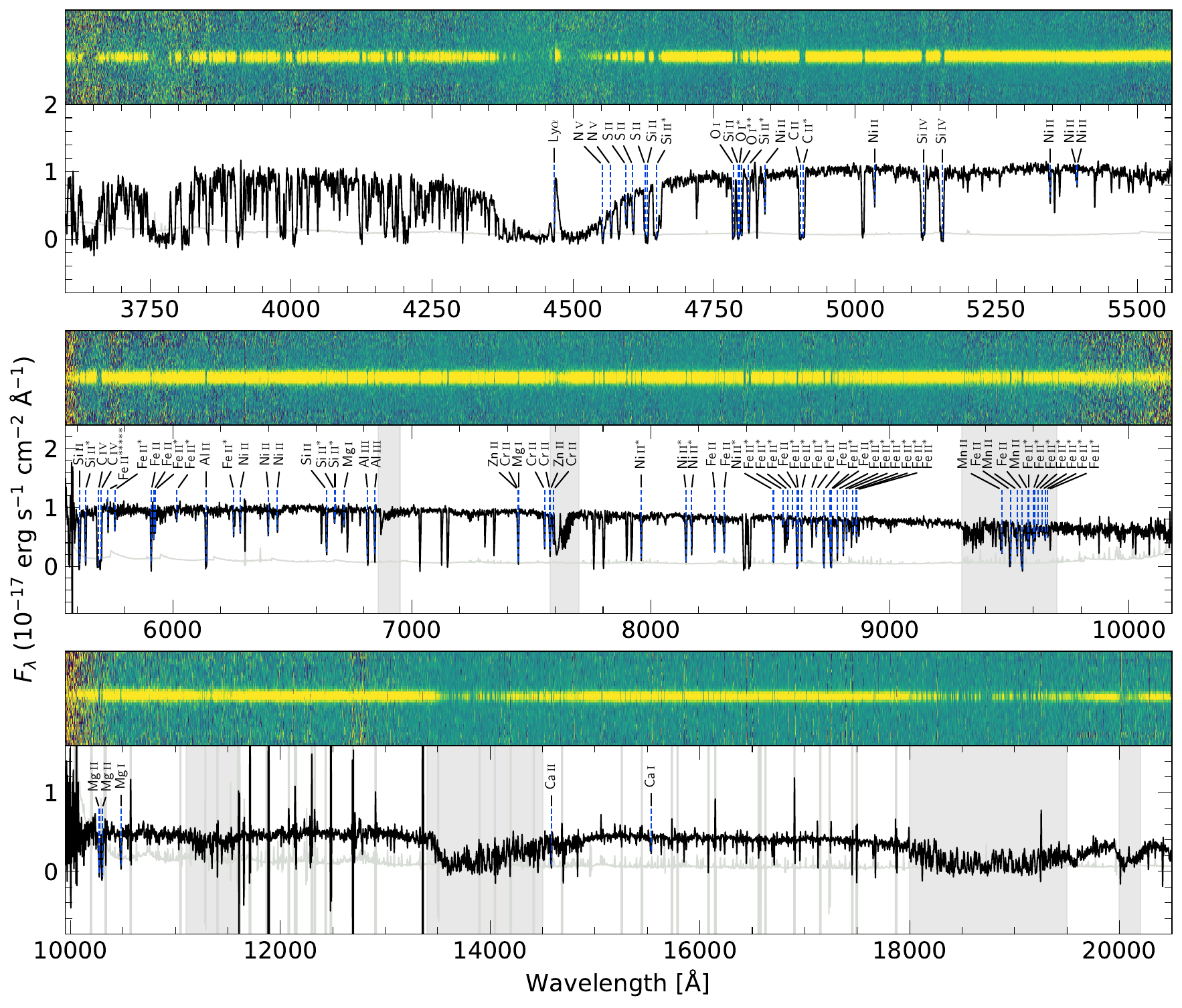}
\caption{VLT/X-shooter spectra of GRB 241228B at $z = 2.6745 \pm 0.0004$, obtained at 25.975~hr post-trigger. Absorption lines identified at the redshift of the GRB are marked in blue and labelled accordingly. The 1D spectra have been smoothed with a Savitzky–Golay filter to enhance the visibility of spectral features.}
\label{fig:VLT_spec_28B}
\end{figure*}

\begin{figure*}
\includegraphics[angle=0,scale=0.45]{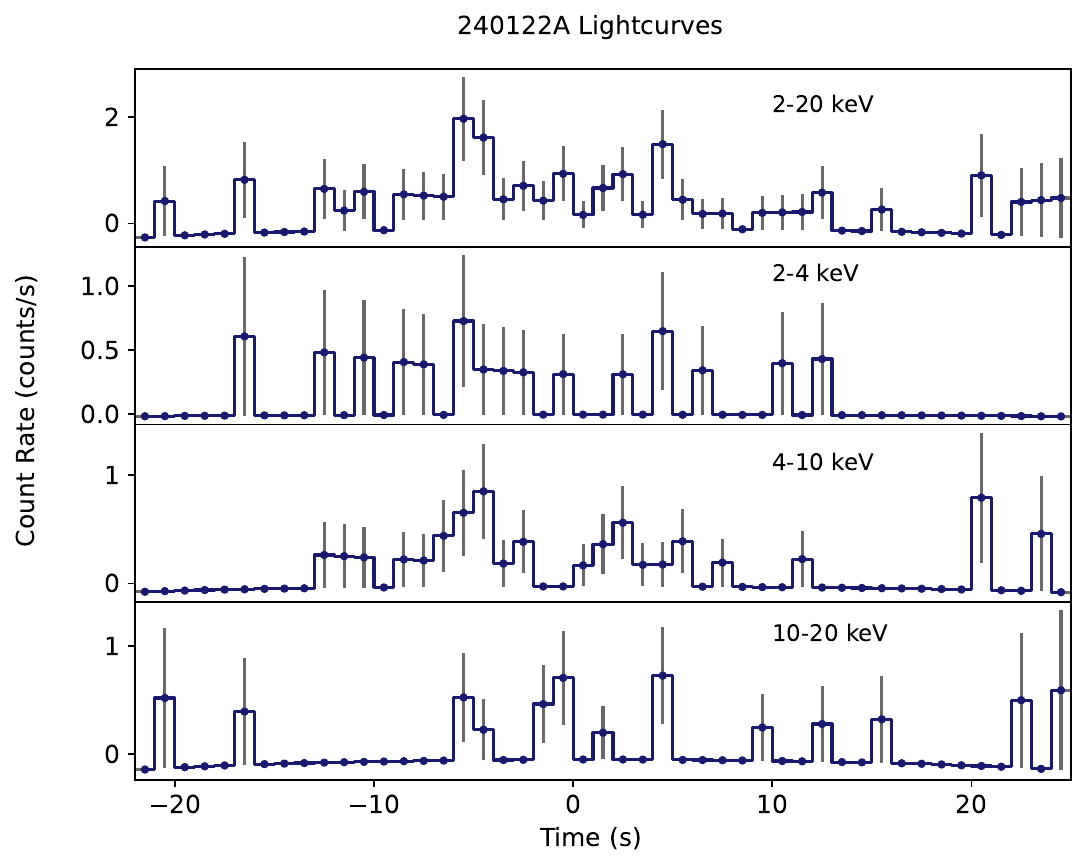}
\includegraphics[angle=0,scale=0.45]{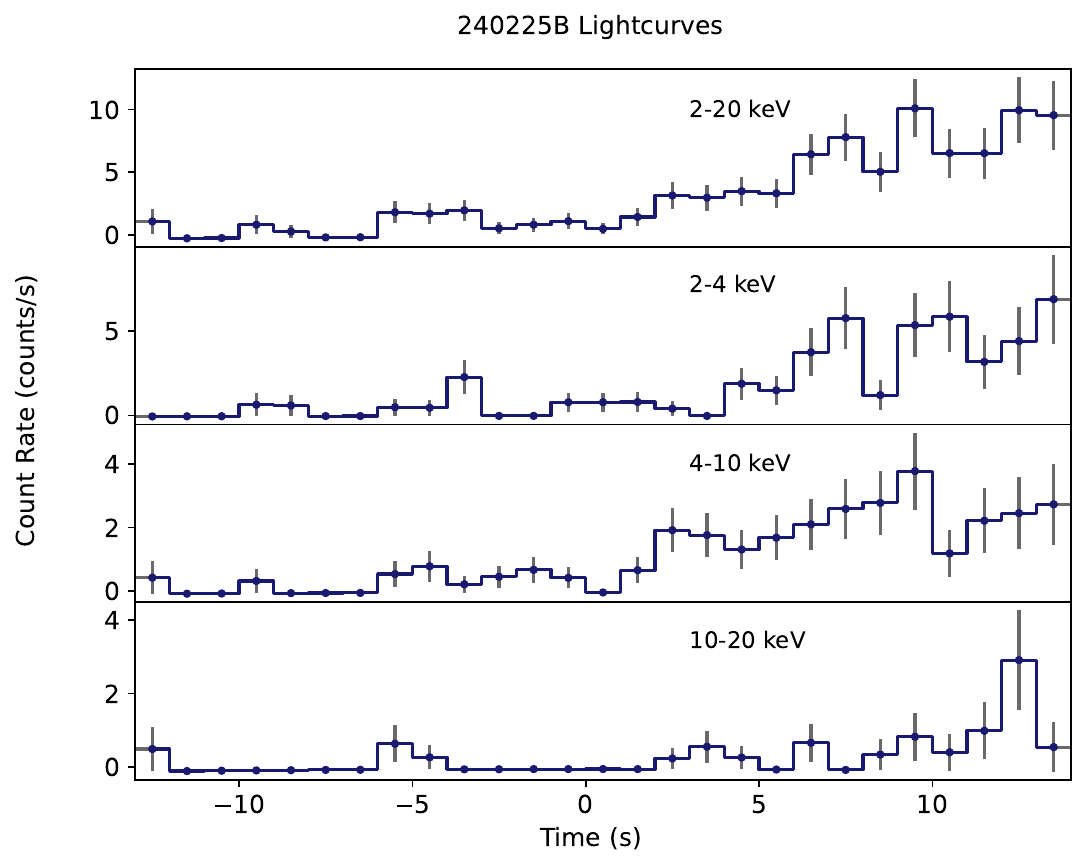}
\caption{\maxi/GSC light curves for GRBs in our sample (240122A, 240225B) are shown. Data are binned to 1-sec resolution.}
\label{fig:MAXI_LCs}
\end{figure*}

\begin{figure*}
\includegraphics[angle=0,scale=0.45]{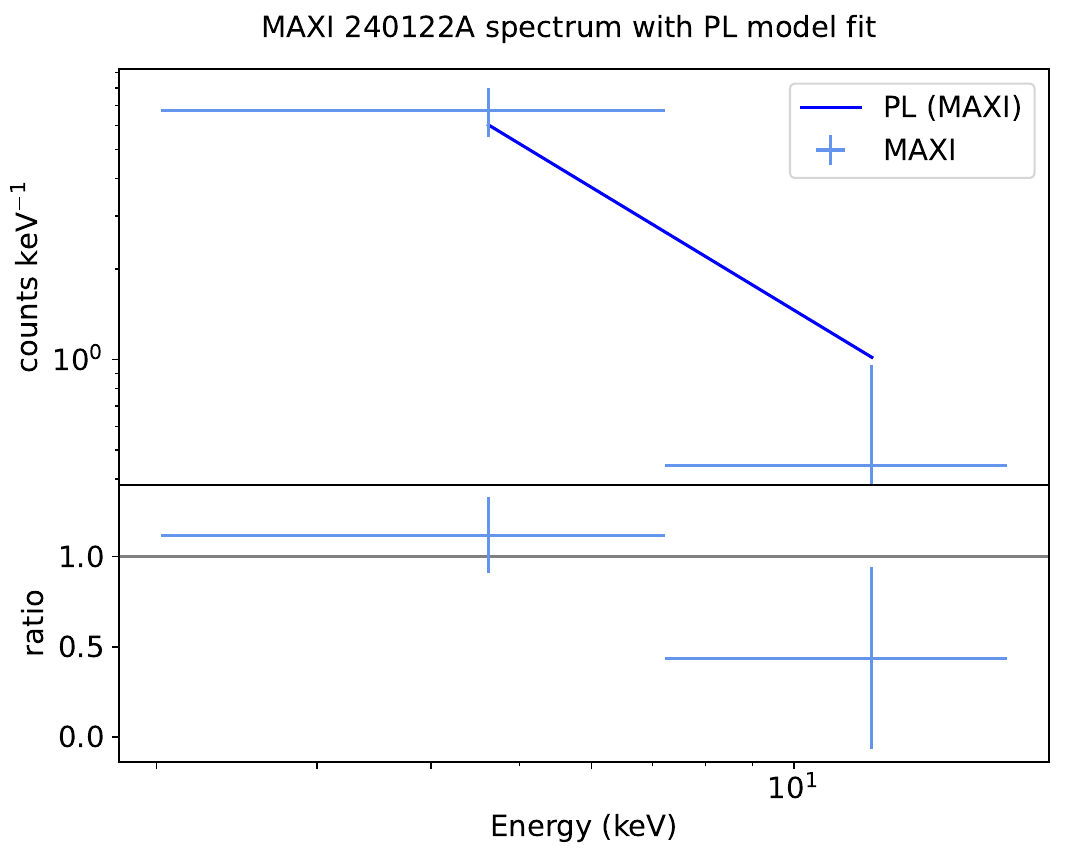}
\includegraphics[angle=0,scale=0.45]{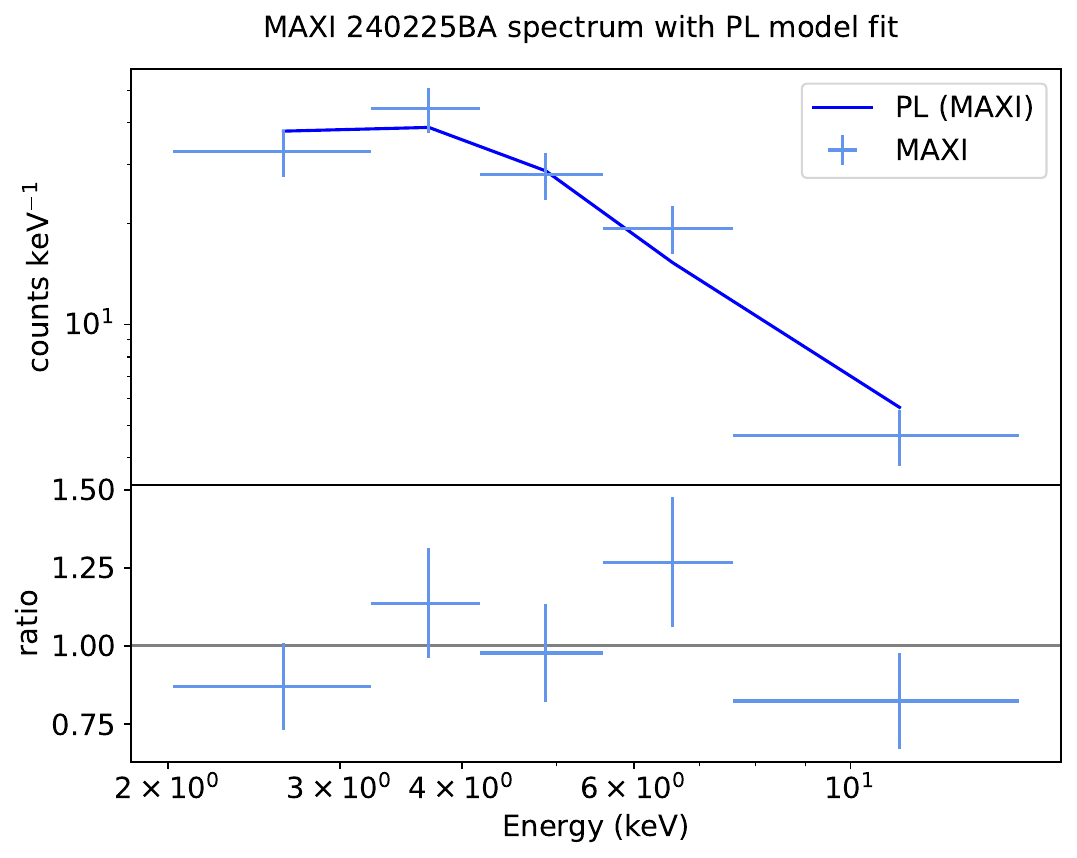}
\caption{\maxi~fitted spectra for GRBs 240122A and 240225B are shown. Models shown are the best-fitting models as described in Table~\ref{tab:fermi_prompt}.}
\label{fig:maxi-GBM_Specs}
\end{figure*}

\begin{figure*}
\includegraphics[angle=0,scale=0.45]{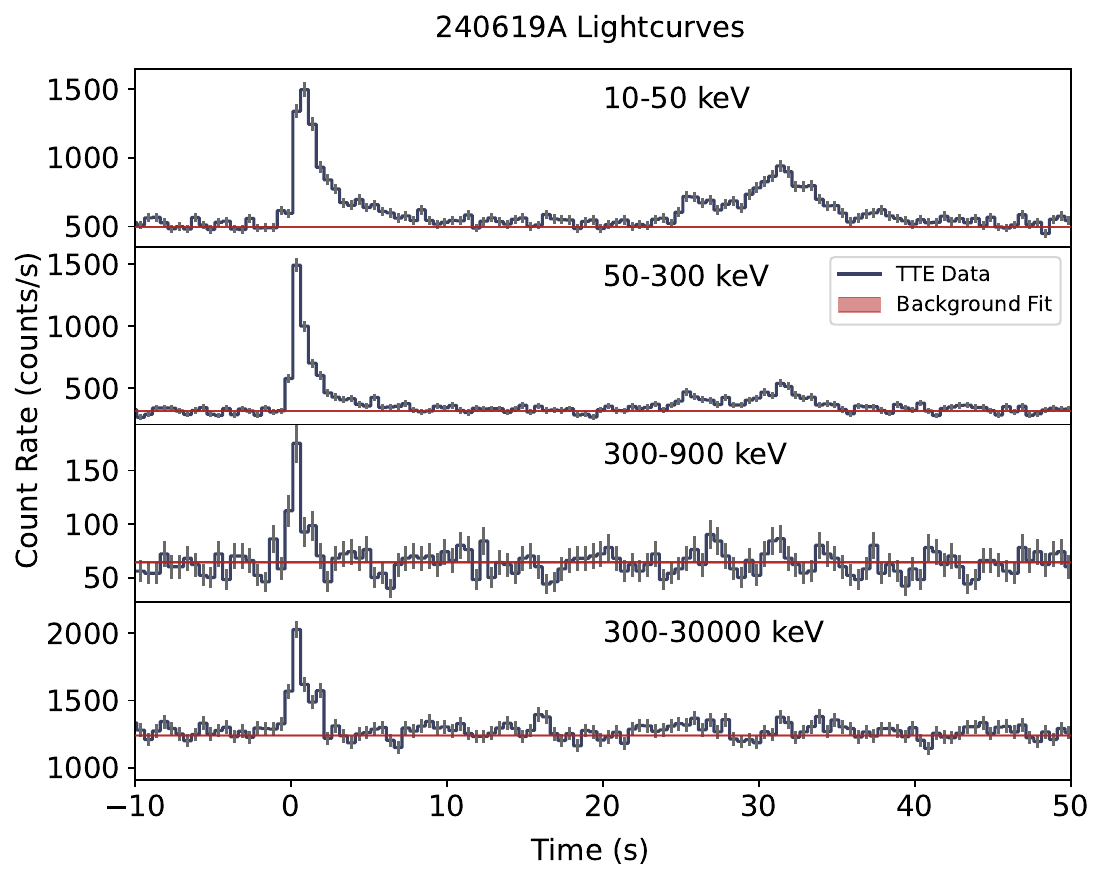}
\includegraphics[angle=0,scale=0.45]{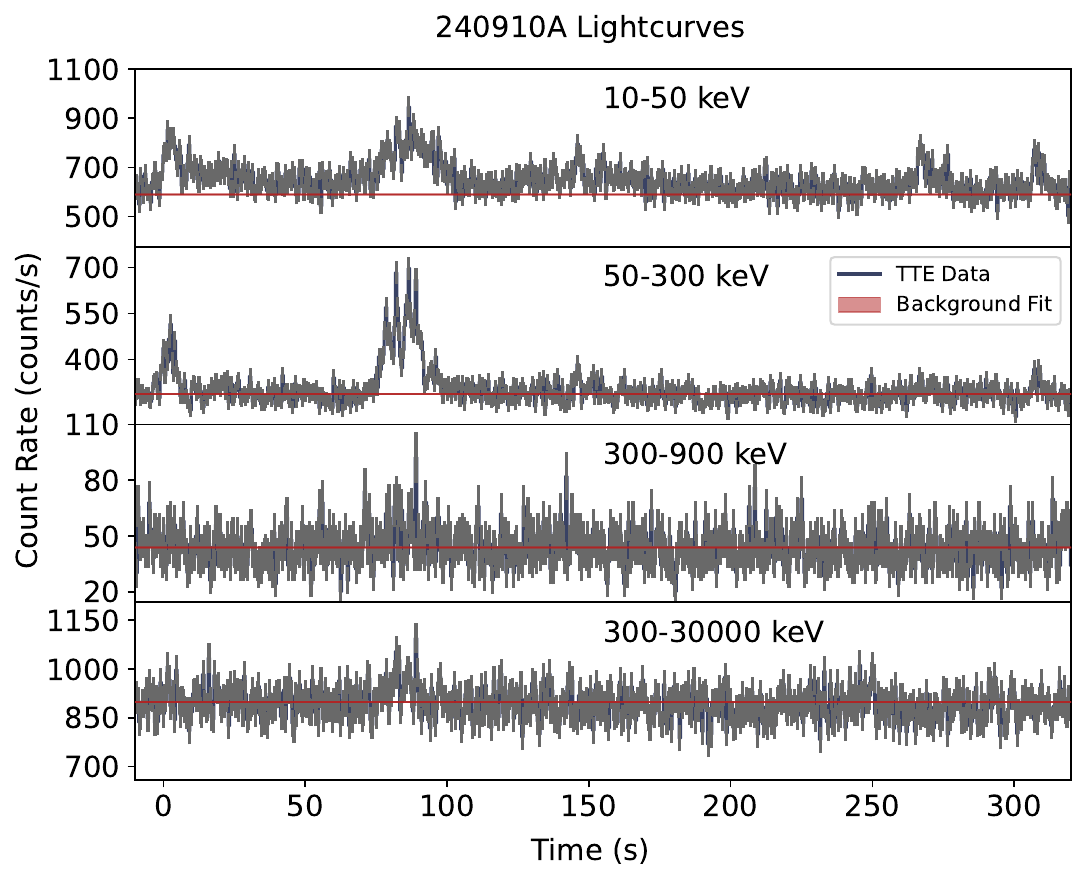}
\includegraphics[angle=0,scale=0.45]{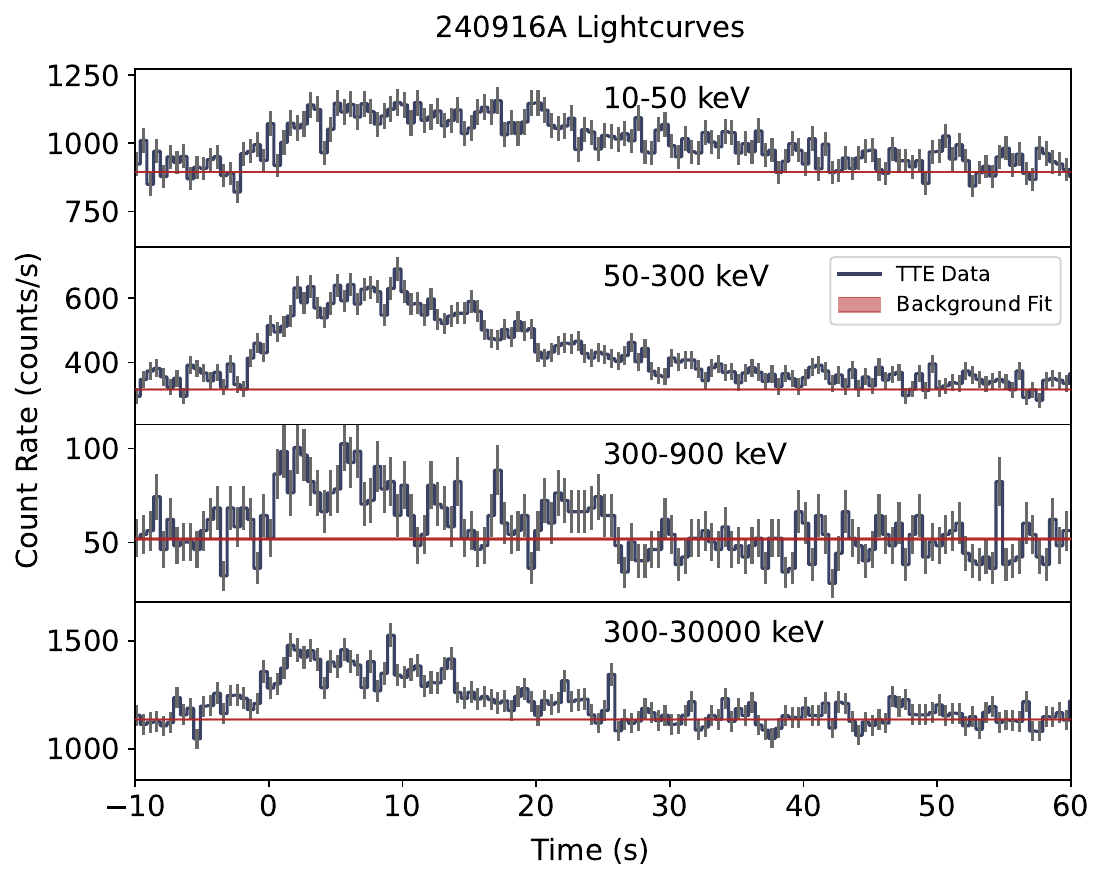}
\includegraphics[angle=0,scale=0.45]{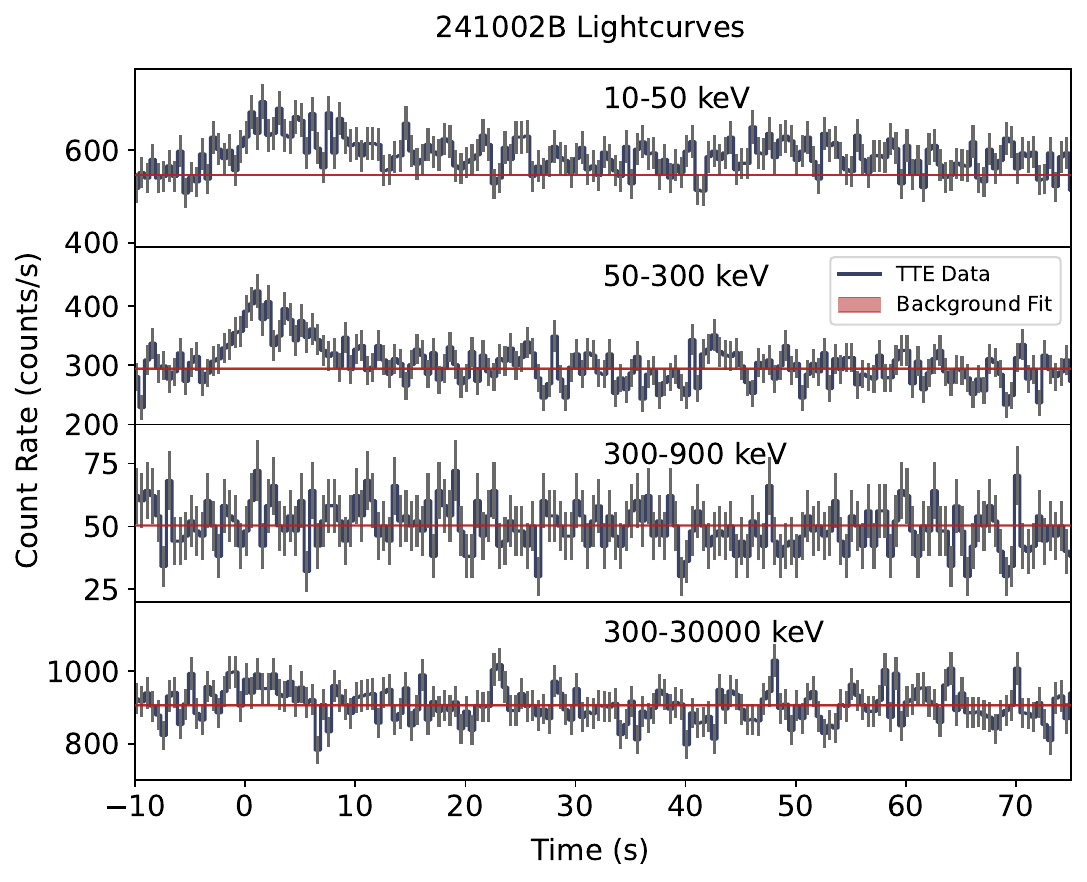}
\includegraphics[angle=0,scale=0.45]{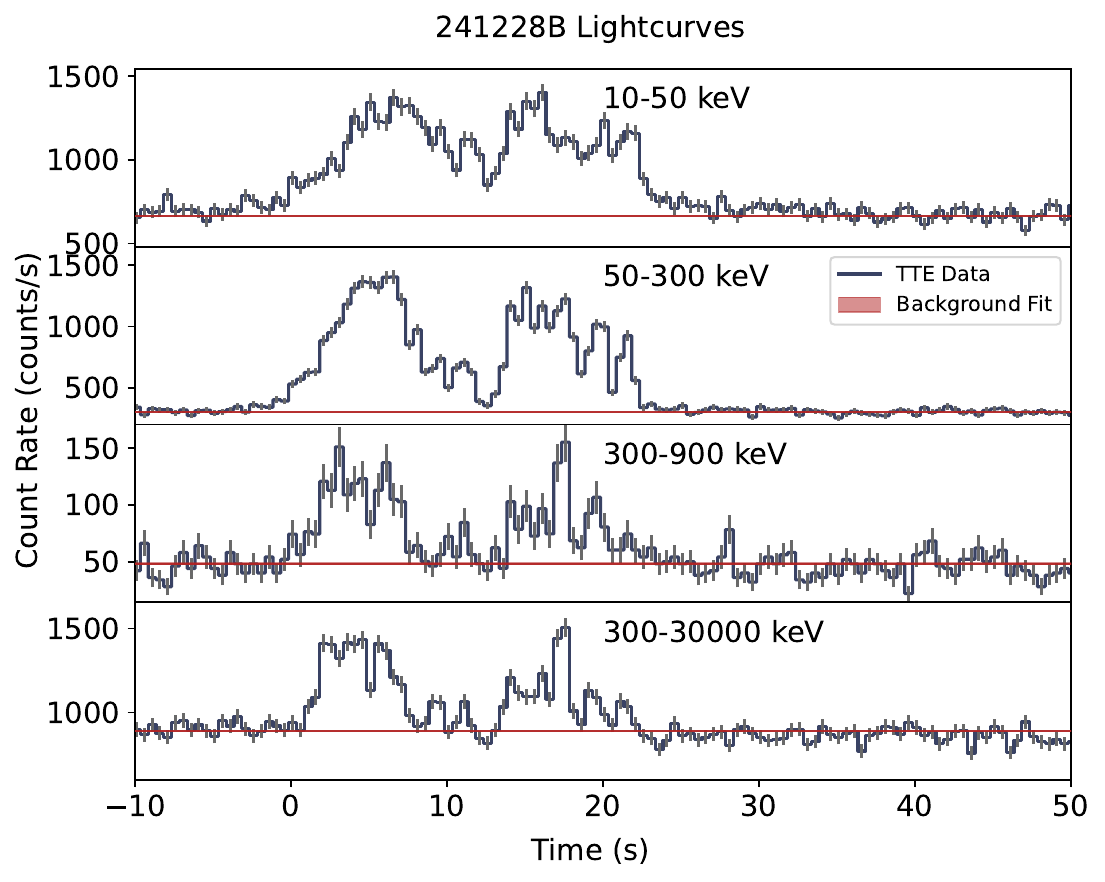}
\caption{\fermi/GBM light curves for \fermi GRBs in our sample are shown. TTE data are binned to 0.5s resolution.}
\label{fig:fermi-GBM_LCs}
\end{figure*}

\begin{figure*}
\includegraphics[angle=0,scale=0.45]{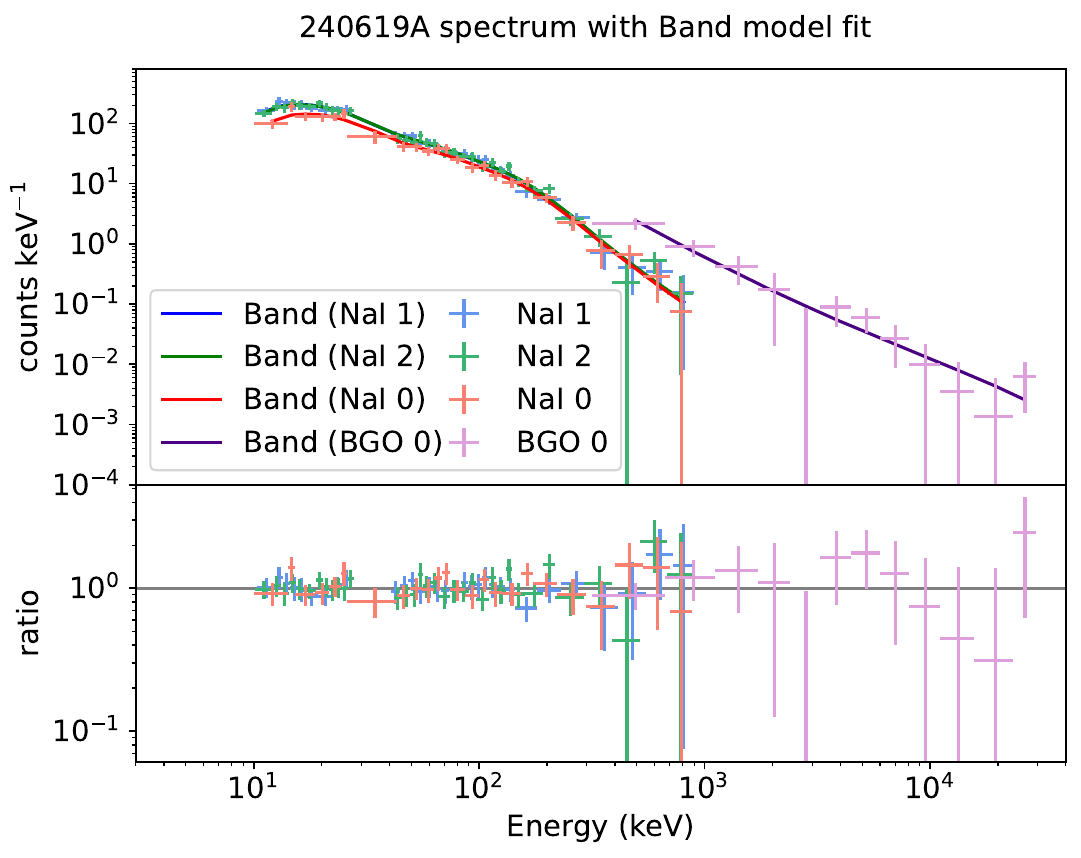}
\includegraphics[angle=0,scale=0.45]{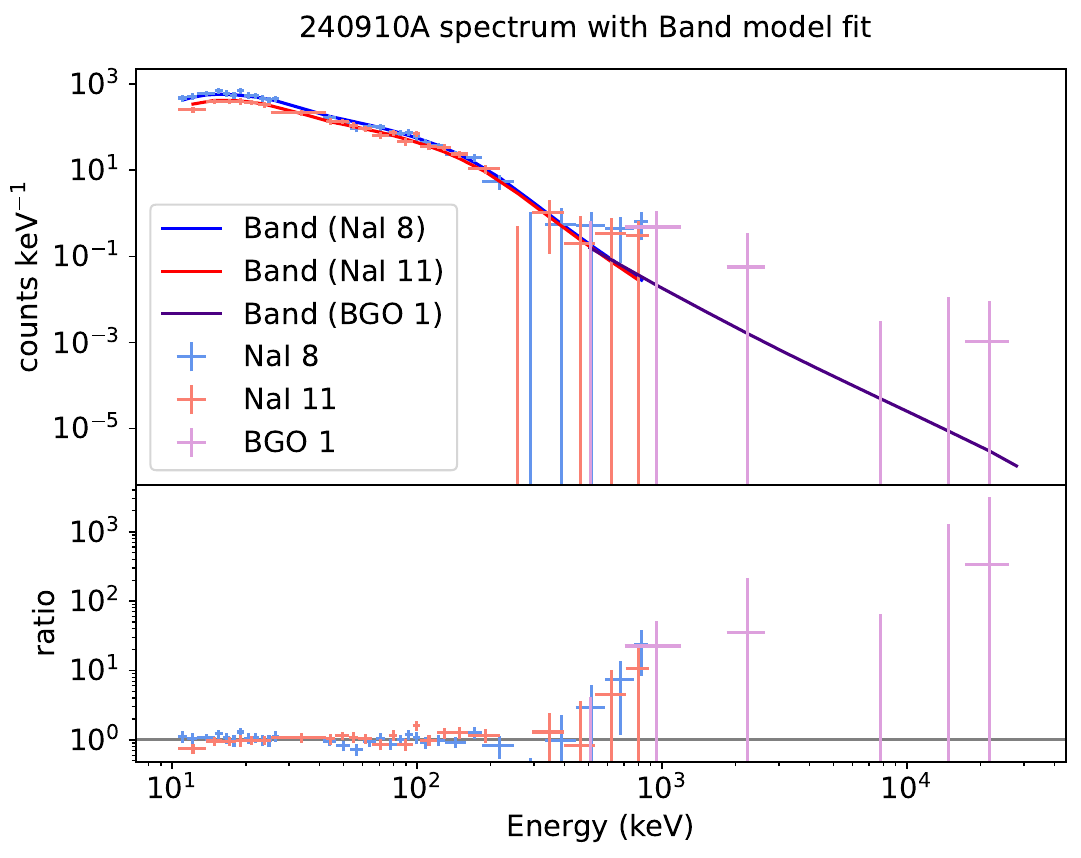}
\includegraphics[angle=0,scale=0.45]{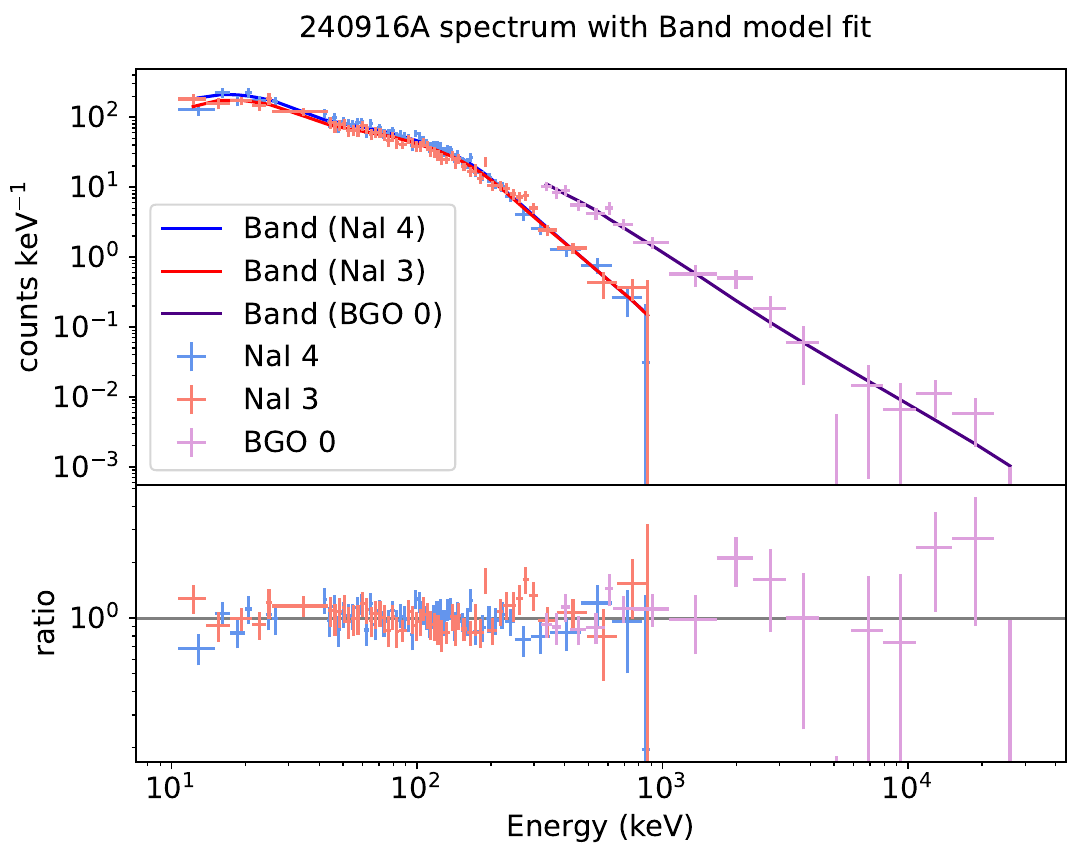}
\includegraphics[angle=0,scale=0.45]{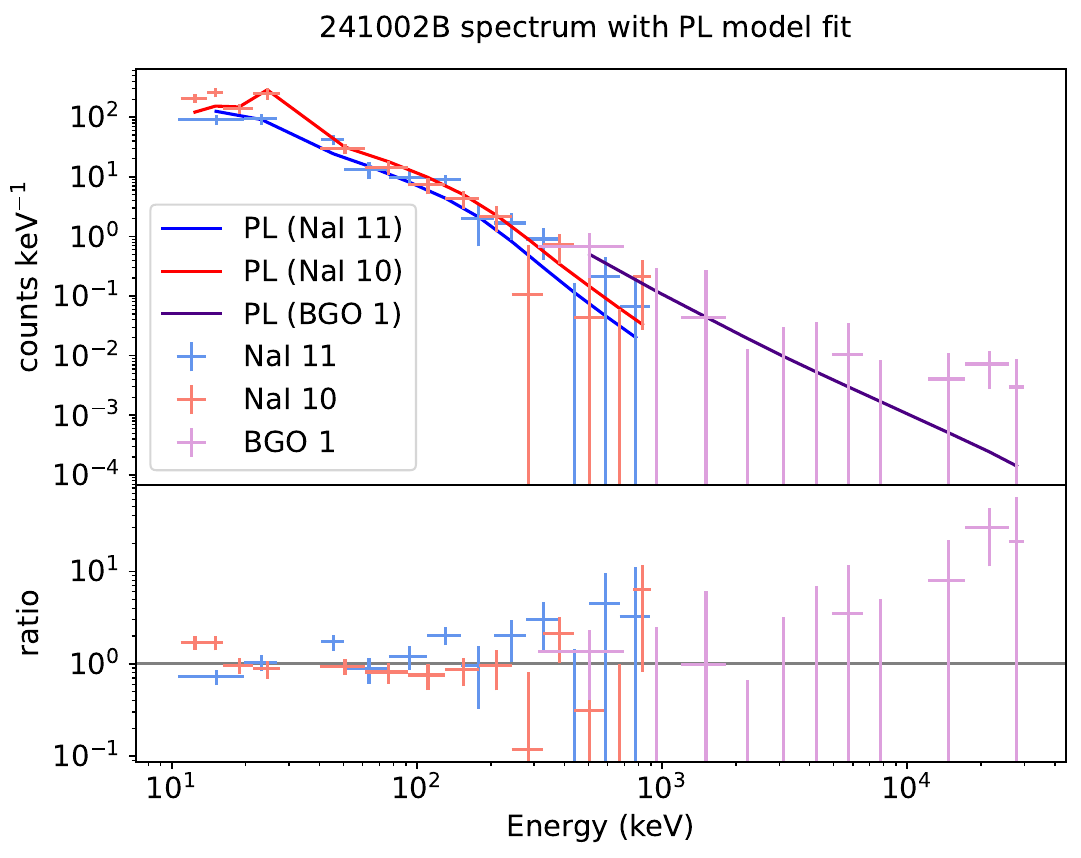}
\includegraphics[angle=0,scale=0.45]{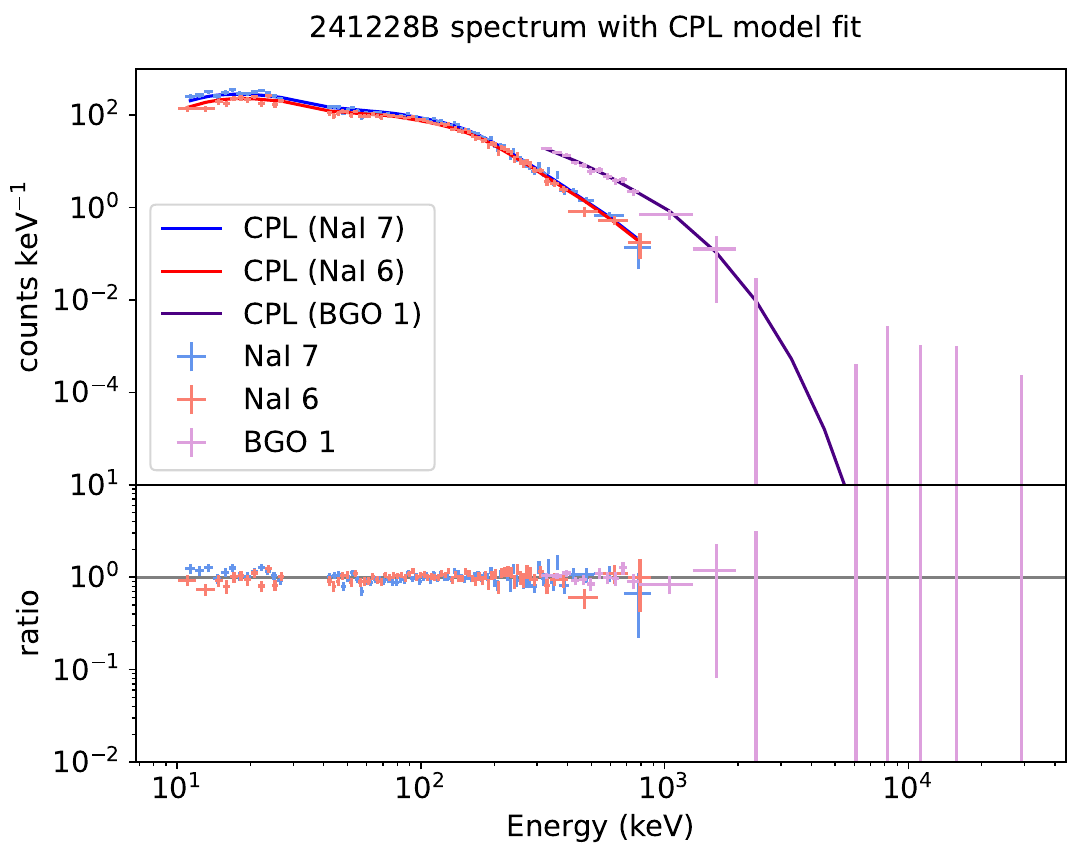}
\caption{\fermi/GBM fitted spectra for \fermi~GRBs in our sample are shown. Models shown are the best-fitting models as described in Table~\ref{tab:fermi_prompt}. GRB 241228B is poorly constrained in high energies, so the spectrum shown is zoomed in to show the behaviour at lower energies.}
\label{fig:fermi-GBM_Specs}
\end{figure*}

\begin{figure*}
\centering
\includegraphics[angle=0,scale=0.245]{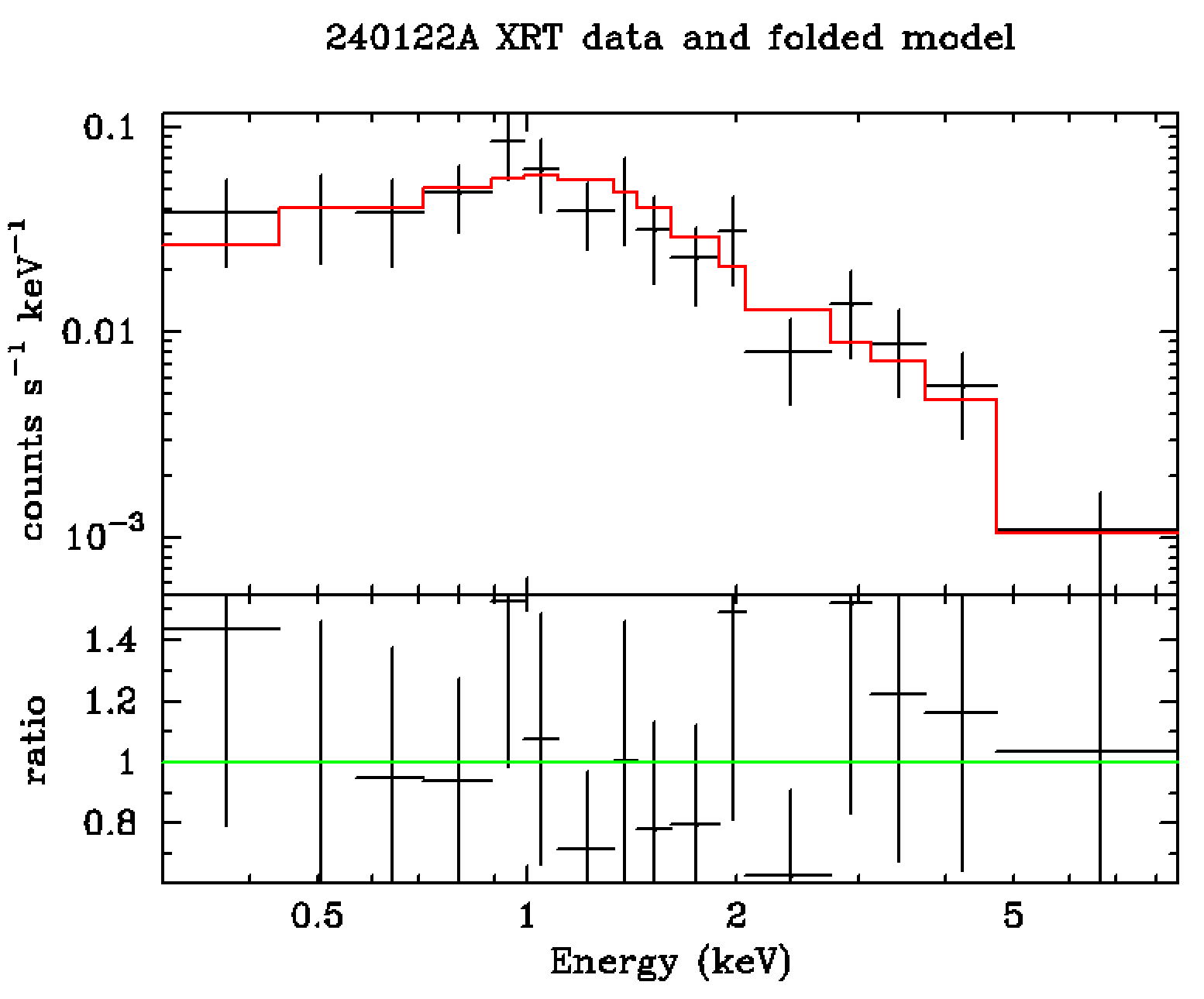}
\includegraphics[angle=0,scale=0.245]{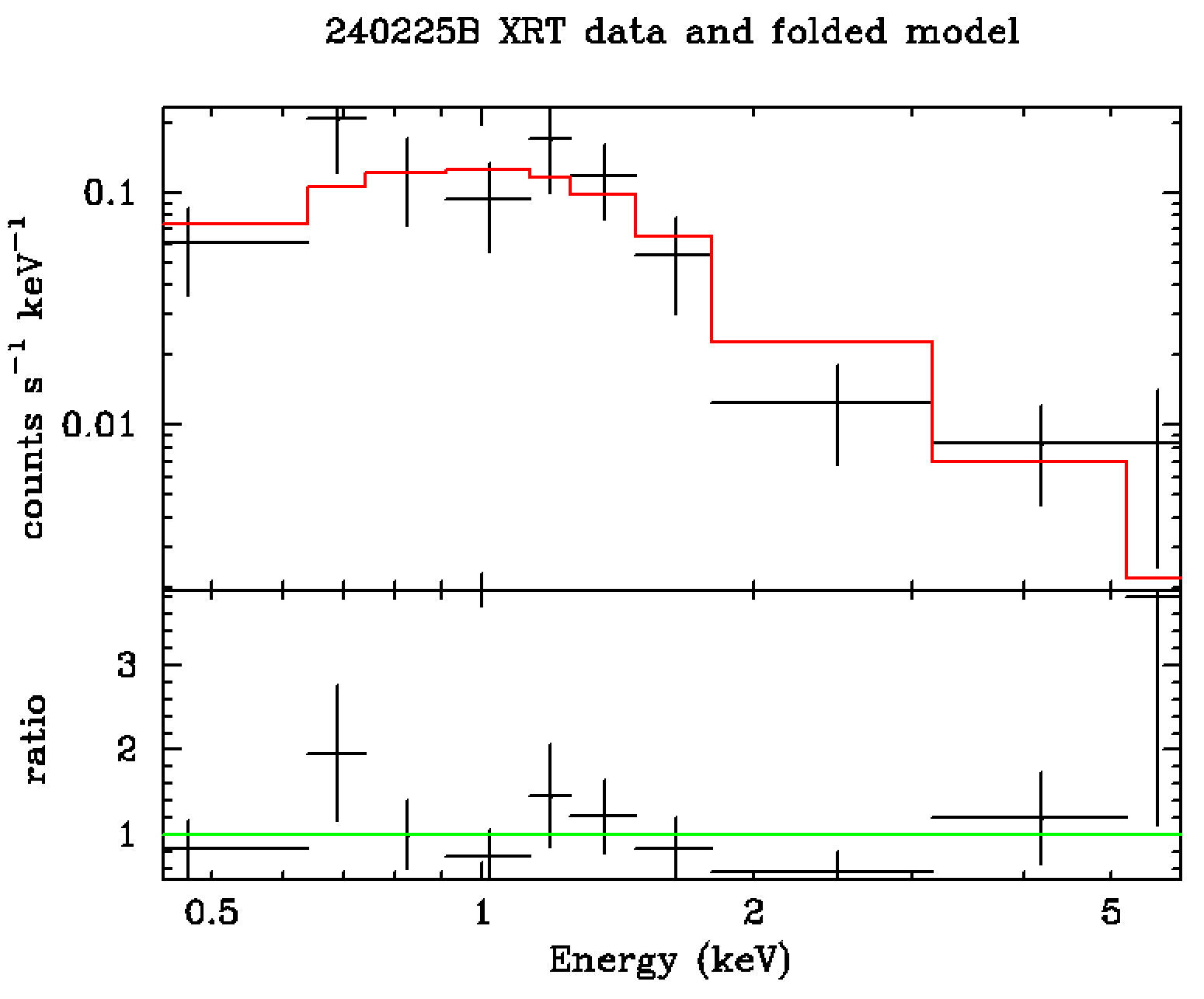}
\includegraphics[angle=0,scale=0.245]{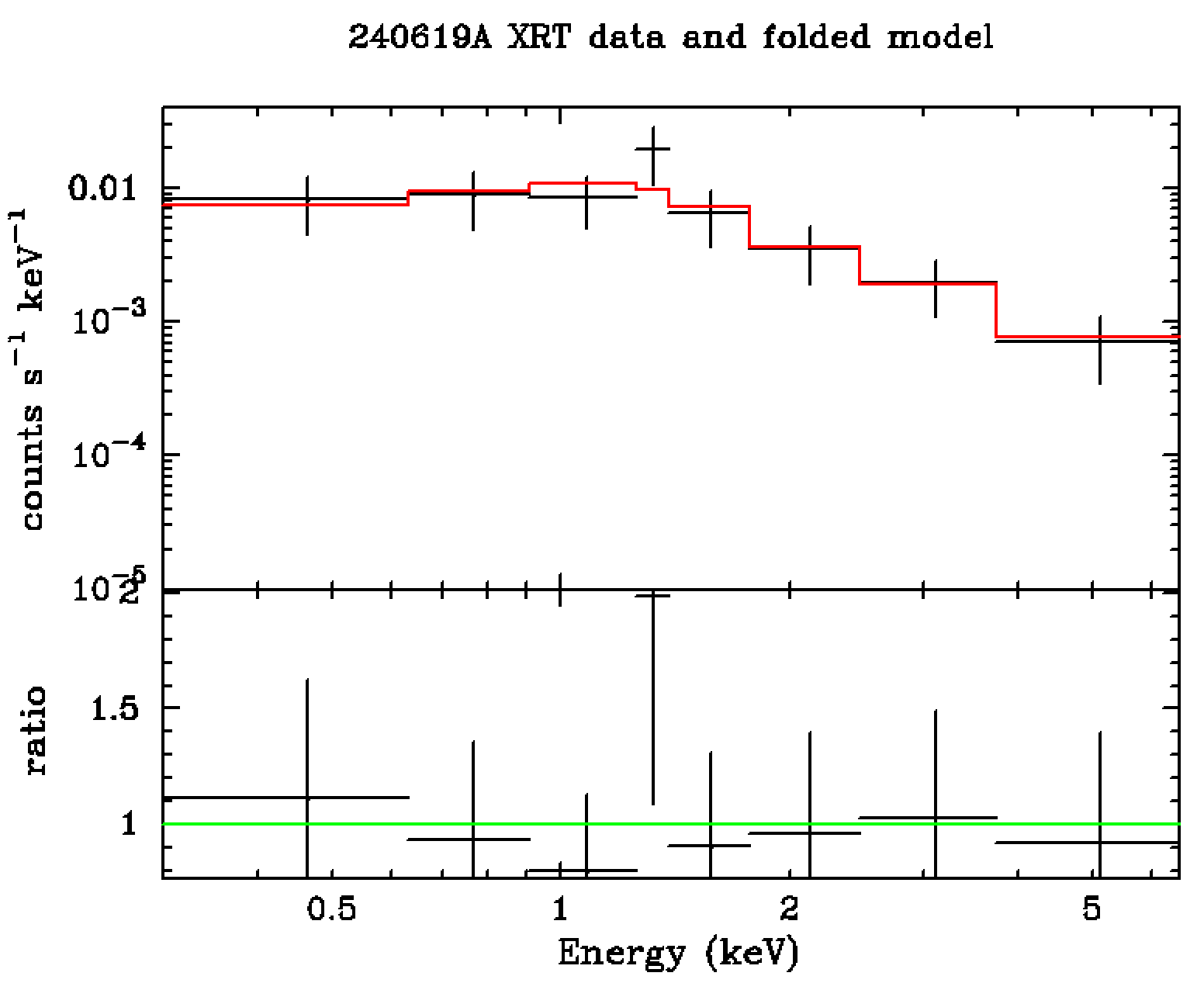}
\includegraphics[angle=0,scale=0.245]{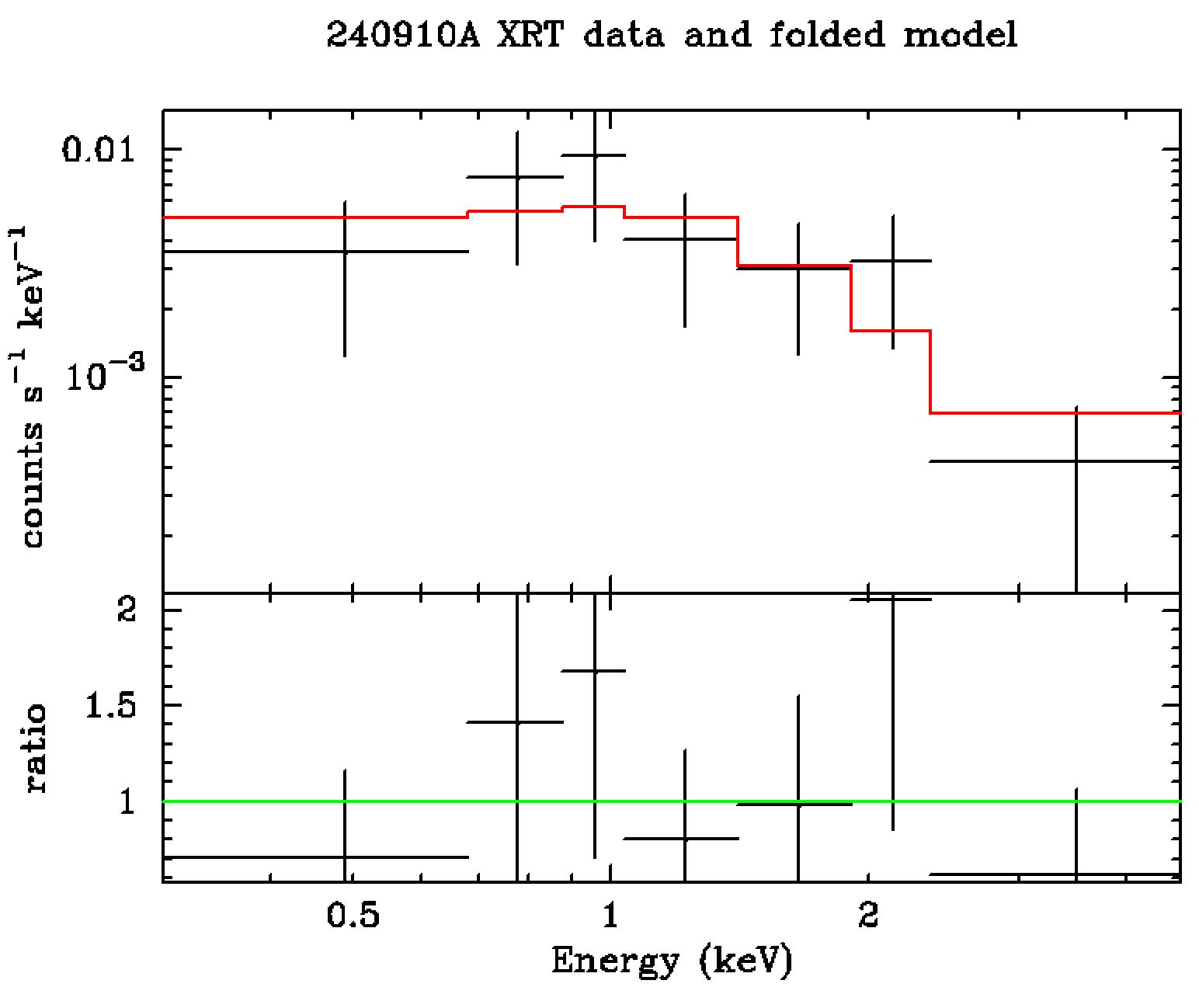}
\includegraphics[angle=0,scale=0.245]{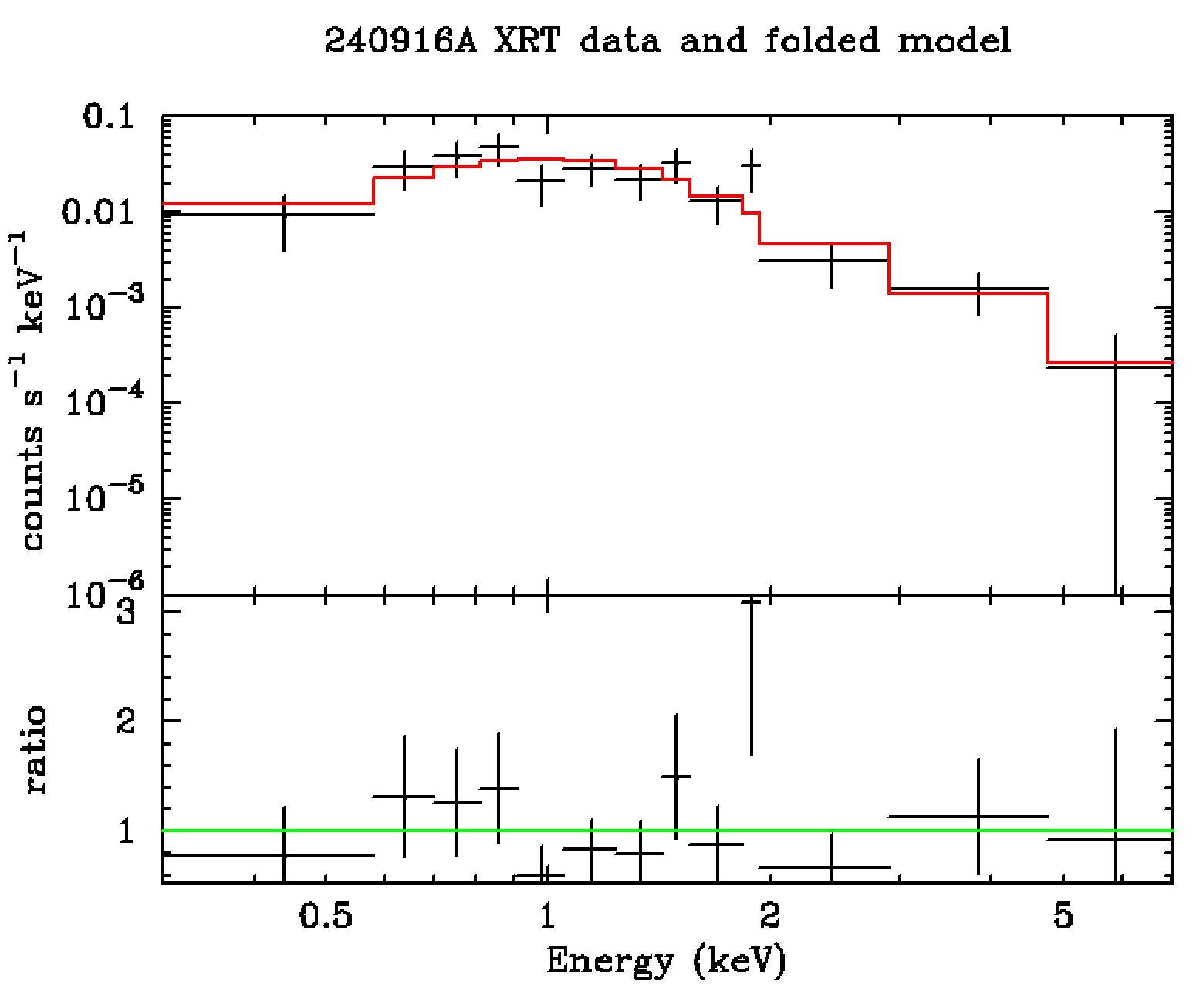}
\includegraphics[angle=0,scale=0.245]{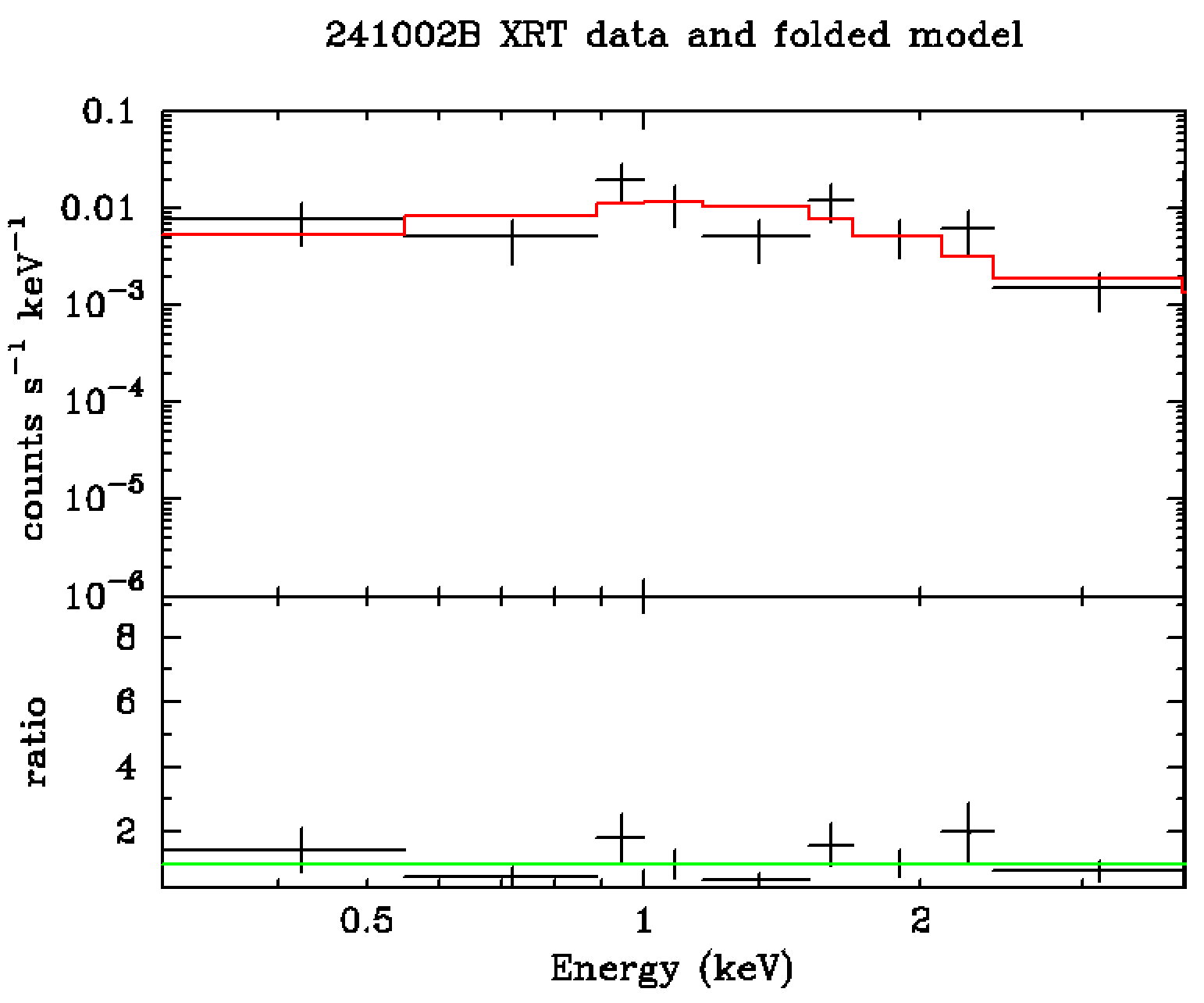}
\includegraphics[angle=0,scale=0.245]{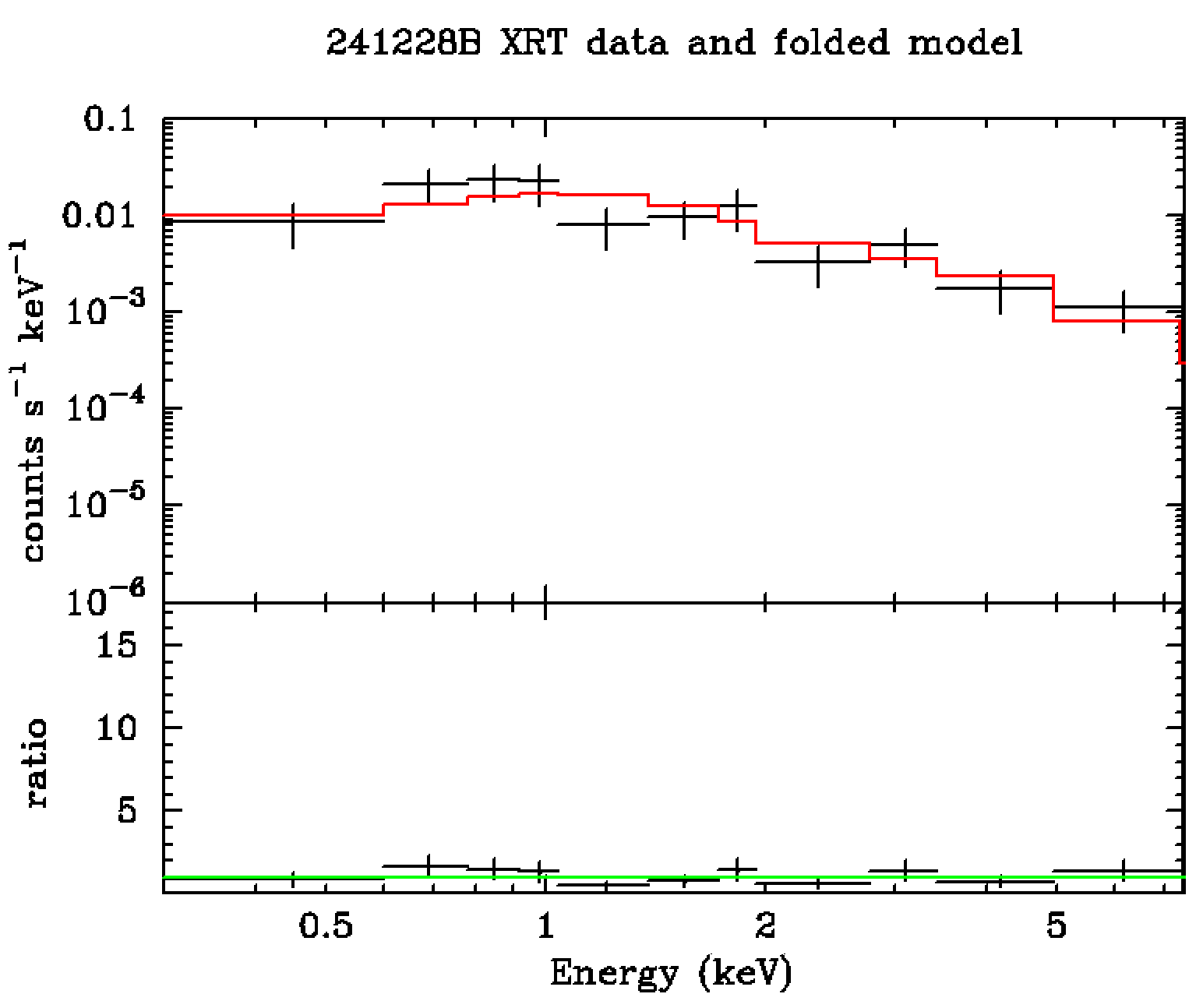}
\caption{\swift/XRT count-rate spectra and best-fitting X-ray continuum models for the GRBs in our sample. In each panel, the top plot displays observed spectral data with errors (black points) in the 0.3--10~keV range fitted with an absorbed power-law model (red line), using Cash statistics. The bottom plot shows the ratio of observed data to folded model predictions, used to assess the goodness of fit. These spectra were used to derive photon indices and absorption column densities ($N_\mathrm{H}$), contributing to the multi-wavelength characterisation of the afterglows.}
\label{fig:swift-XRT_Spec}
\end{figure*}

\begin{table}
\centering
\caption{Absorption features identified in the afterglow spectrum of GRB~240122A ($z = 3.1634 \pm 0.0003$), based on observations with VLT/X-shooter and GTC/OSIRIS. The UVB, VIS, and NIR designations refer to the respective arms of the VLT/X-shooter spectrograph.}
\label{tab:240122A}
\begin{tabular}{lcccc}
\hline
$\lambda_{\textrm{obs}}$ (\AA) & $\lambda_{\textrm{rest}}$ (\AA) & Feature & Type & Arm \\
\hline
3796.2 & 911.8 & Ly limit & abs & UVB \\
5061.6 & 1215.7 & Ly$\alpha$ & abs & UVB \\
5247.9 & 1260.4 & Si \textsc{ii} & abs & UVB \\
5556.5 & 1334.5 & C \textsc{ii} & abs & UVB \\
\hline
5803.1 & 1393.8 & Si \textsc{iv} & abs & VIS \\
5840.6 & 1402.8 & Si \textsc{iv} & abs & VIS \\
6356.6 & 1526.7 & Si \textsc{ii} & abs & VIS \\
6446.1 & 1548.2 & C \textsc{iv} & abs & VIS \\
6456.8 & 1550.8 & C \textsc{iv} & abs & VIS \\
6956.5 & 1670.8 & Al \textsc{ii} & abs & VIS \\
\hline
11642.9 & 2796.4 & Mg \textsc{ii} & abs & NIR \\
11672.8 & 2803.5 & Mg \textsc{ii} & abs & NIR \\
\hline \\
\end{tabular}
\end{table}

\begin{table}
\centering
\caption{Absorption and emission features identified in the afterglow spectrum of GRB~240225B ($z = 0.9462 \pm 0.0002$), obtained with VLT/X-shooter.}
\label{tab:240225B}
\begin{tabular}{lcccc}
\hline
$\lambda_{\textrm{obs}}$ (\AA) & $\lambda_{\textrm{rest}}$ (\AA) & Feature & Type & Arm \\
\hline
3610.2 & 1854.7 & Al \textsc{iii} & abs & UVB \\
3625.9 & 1862.8 & Al \textsc{iii} & abs & UVB \\
4002.5 & 2056.3 & Cr \textsc{ii} & abs & UVB \\
4379.4 & 2249.9 & Fe \textsc{ii} & abs & UVB \\
4563.0 & 2344.2 & Fe \textsc{ii} & abs & UVB \\
4621.9 & 2374.5 & Fe \textsc{ii} & abs & UVB \\
4638.1 & 2382.8 & Fe \textsc{ii} & abs & UVB \\
4650.9 & 2389.4 & Fe \textsc{ii}$^{*}$ & abs & UVB \\
4664.5 & 2396.4 & Fe \textsc{ii}$^{*}$ & abs & UVB \\
4671.6 & 2400.0 & Fe \textsc{ii}$^{*}$ & abs & UVB \\
5015.9 & 2576.9 & Mn \textsc{ii} & abs & UVB \\
5034.9 & 2586.6 & Fe \textsc{ii} & abs & UVB \\
5050.2 & 2594.5 & Mn \textsc{ii} & abs & UVB \\
5061.2 & 2600.2 & Fe \textsc{ii} & abs & UVB \\
5443.1 & 2796.4 & Mg \textsc{ii} & abs & UVB \\
5457.1 & 2803.5 & Mg \textsc{ii} & abs & UVB \\
5553.4 & 2853.0 & Mg \textsc{i} & abs & UVB \\
\hline
7254.8 & 3727.1 & [O \textsc{ii}] & em & VIS \\
7260.2 & 3729.9 & [O \textsc{ii}] & em & VIS \\
7659.0$^\textrm{a}$ & 3934.8 & Ca \textsc{ii} & abs & VIS \\
7726.8 & 3969.6 & Ca \textsc{ii} & abs & VIS \\
9748.5 & 5008.2 & [O \textsc{iii}] & em & VIS \\
\hline \\
\end{tabular}
\end{table}

\begin{table}
\centering
\caption{Absorption and emission features identified in the afterglow spectrum of GRB~240619A ($z = 0.3960 \pm 0.0001$), obtained with VLT/X-shooter. Here, $^{a}$ marks lines affected by telluric absorption.}
\label{tab:240619A}
\begin{tabular}{lcccc}
\hline
$\lambda_{\textrm{obs}}$ (\AA) & $\lambda_{\textrm{rest}}$ (\AA) & Feature & Type & Arm \\
\hline
5203.0 & 3727.1 & [O \textsc{ii}] & em & UVB \\
5206.9 & 3729.9 & [O \textsc{ii}] & em & UVB \\
5402.7 & 3870.2 & [Ne \textsc{iii}] & em & UVB \\
\hline
6061.0 & 4341.7  & H$\gamma$ & em & VIS \\
6788.3 & 4862.7  & H$\beta$ & em & VIS \\
6924.6$^{a}$ & 4960.3 & [O \textsc{iii}] & em & VIS \\
6991.5 & 5008.2 & [O \textsc{iii}] & em & VIS \\
9164.2 & 6564.6 & H$\alpha$ & em & VIS \\
\hline \\
\end{tabular}
\end{table}

\begin{table}
\centering
\caption{Absorption features identified in the afterglow spectrum of GRB~240910A ($z = 1.4605 \pm 0.0007$), obtained with GTC/OSIRIS. Here, $^{a}$ marks lines affected by telluric absorption.}
\label{tab:240910A}
\begin{tabular}{lccc}
\hline
$\lambda_{\textrm{obs}}$ (\AA) & $\lambda_{\textrm{rest}}$ (\AA) & Feature & Type \\
\hline
3755.5 & 1526.7 & Si \textsc{ii} & abs \\
3808.4 & 1548.2 & C \textsc{iv} & abs  \\
3956.8 & 1608.5 & Fe \textsc{ii} & abs  \\
3967.3 & 1612.8 & Fe \textsc{ii}$^{*}$ & abs  \\
4110.0 & 1670.8 & Al \textsc{ii} & abs  \\
4447.5 & 1808.0 & Si \textsc{ii} & abs  \\
4562.4 & 1854.7 & Al \textsc{iii} & abs  \\
4984.5 & 2026.3 & Cr \textsc{ii} & abs  \\
5058.3 & 2056.3 & Cr \textsc{ii} & abs  \\
5082.6 & 2066.2 & Cr \textsc{ii} & abs  \\
5328.6 & 2166.2 & Ni \textsc{ii}$^{*}$ & abs  \\
5376.4 & 2185.6 & Mn \textsc{i} & abs  \\
5561.3 & 2260.8 & Fe \textsc{ii} & abs  \\
5699.1 & 2316.8 & Ni \textsc{ii}$^{*}$ & abs  \\
5766.5 & 2344.2 & Fe \textsc{ii} & abs  \\
5841.0 & 2374.5 & Fe \textsc{ii} & abs  \\
5861.4 & 2382.8 & Fe \textsc{ii} & abs  \\
5894.7 & 2396.3 & Fe \textsc{ii}$^{*}$ & abs  \\
6338.9$^{a}$ & 2576.9 & Mn \textsc{ii} & abs  \\
6363.0$^{a}$ & 2586.7 & Fe \textsc{ii} & abs  \\
6396.2$^{a}$ & 2600.2 & Fe \textsc{ii} & abs  \\
6878.6 & 2796.3 & Mg \textsc{ii} & abs  \\
6896.3 & 2803.5 & Mg \textsc{ii} & abs  \\
7018.0 & 2852.7 & Mg \textsc{i} & abs \\
\hline \\
\end{tabular}
\end{table}

\begin{table}
\centering
\caption{Absorption features identified in the afterglow spectrum of GRB~240916A ($z = 2.6100 \pm 0.0002$), obtained with VLT/X-shooter. Superscript symbols indicate excited-state transitions ($^{*}$, $^{**}$), while $^{a}$ marks lines affected by telluric absorption.}
\label{tab:240916A}
\begin{tabular}{lcccc}
\hline
$\lambda_{\textrm{obs}}$ (\AA) & $\lambda_{\textrm{rest}}$ (\AA) & Feature & Type & Arm \\
\hline
4388.6 & 1215.7 & Ly$\alpha$ & abs & UVB \\
4513.2 & 1250.2 & S \textsc{ii} & abs & UVB \\
4524.8 & 1253.4 & S \textsc{ii} & abs & UVB \\
4546.4 & 1259.4 & S \textsc{ii} & abs & UVB \\
4566.1 & 1264.8 & Si \textsc{ii}$^{*}$ & abs & UVB \\
4700.8 & 1302.2 & O \textsc{i} & abs & UVB \\
4708.8 & 1304.4 & Si \textsc{ii} & abs & UVB \\
4710.5 & 1304.9 & O \textsc{i}$^{*}$ & abs & UVB \\
4714.8 & 1306.0 & O \textsc{i}$^{**}$ & abs & UVB \\
4726.5 & 1309.3 & Si \textsc{ii}$^{*}$ & abs & UVB \\
4817.7 & 1334.5 & C \textsc{ii} & abs & UVB \\
4821.9 & 1335.7 & C \textsc{ii}$^{*}$ & abs & UVB \\
5031.5 & 1393.8 & Si \textsc{iv} & abs & UVB \\
5064.0 & 1402.8 & Si \textsc{iv} & abs & UVB \\
5511.4 & 1526.7 & Si \textsc{ii} & abs & UVB \\
5535.7 & 1533.4 & Si \textsc{ii}$^{*}$ & abs & UVB \\
\hline
5589.0 & 1548.2 & C \textsc{iv} & abs & VIS \\
5598.3 & 1550.8 & C \textsc{iv} & abs & VIS \\
5806.5 & 1608.5 & Fe \textsc{ii} & abs & VIS \\
5822.2 & 1612.8 & Fe \textsc{ii}$^{*}$ & abs & VIS \\
6031.5 & 1670.8 & Al \textsc{ii} & abs & VIS \\
6144.2 & 1702.0 & Fe \textsc{ii}$^{*}$ & abs & VIS \\
6287.0 & 1741.6 & Ni \textsc{ii} & abs & VIS \\
6526.9 & 1808.0 & Si \textsc{ii} & abs & VIS \\
6695.5 & 1854.7 & Al \textsc{iii} & abs & VIS \\
6724.7 & 1862.8 & Al \textsc{iii} & abs & VIS \\
7315.6 & 2026.5 & Mg \textsc{i} & abs & VIS \\
7314.4 & 2026.1 & Zn \textsc{ii} & abs & VIS \\
7314.9 & 2026.3 & Cr \textsc{ii} & abs & VIS \\
7423.1 & 2056.3 & Cr \textsc{ii} & abs & VIS \\
7444.5 & 2062.2 & Cr \textsc{ii} & abs & VIS \\
7445.4 & 2062.4 & Zn \textsc{ii} & abs & VIS \\
7459.0 & 2066.2 & Cr \textsc{ii} & abs & VIS \\
7820.0 & 2166.2 & Ni \textsc{ii}$^{*}$ & abs & VIS \\
8004.1 & 2217.2 & Ni \textsc{ii}$^{*}$ & abs & VIS \\
8027.2 & 2223.6 & Ni \textsc{ii}$^{*}$ & abs & VIS \\
8122.1 & 2249.9 & Fe \textsc{ii} & abs & VIS \\
8161.5 & 2260.8 & Fe \textsc{ii}$^{*}$ & abs & VIS \\
8363.3 & 2316.7 & Ni \textsc{ii}$^{*}$ & abs & VIS \\
8404.4 & 2328.1 & Fe \textsc{ii}$^{*}$ & abs & VIS \\
8423.9 & 2333.5 & Fe \textsc{ii}$^{*}$ & abs & VIS \\
8442.7 & 2338.7 & Fe \textsc{ii}$^{*}$ & abs & VIS \\
8462.6 & 2344.2 & Fe \textsc{ii} & abs & VIS \\
8479.9 & 2349.0 & Fe \textsc{ii}$^{*}$ & abs & VIS \\
8539.8 & 2365.6 & Fe \textsc{ii}$^{*}$ & abs & VIS \\
8571.8 & 2374.5 & Fe \textsc{ii} & abs & VIS \\
8601.8 & 2382.8 & Fe \textsc{ii} & abs & VIS \\
8625.6 & 2389.4 & Fe \textsc{ii}$^{*}$ & abs & VIS \\
8650.8 & 2396.4 & Fe \textsc{ii}$^{*}$ & abs & VIS \\
8664.0 & 2400.0 & Fe \textsc{ii}$^{*}$ & abs & VIS \\
8684.2 & 2405.6 & Fe \textsc{ii}$^{*}$ & abs & VIS \\
8690.7 & 2407.4 & Fe \textsc{ii}$^{*}$ & abs & VIS \\
8704.8 & 2411.3 & Fe \textsc{ii}$^{*}$ & abs & VIS \\
8714.9 & 2414.1 & Fe \textsc{ii}$^{*}$ & abs & VIS \\
9302.5$^{a}$ & 2576.9 & Mn \textsc{ii} & abs & VIS \\
9337.8$^{a}$ & 2586.6 & Fe \textsc{ii} & abs & VIS \\
9386.6$^{a}$ & 2600.2 & Fe \textsc{ii} & abs & VIS \\
9409.3$^{a}$ & 2606.5 & Mn \textsc{ii} & abs & VIS \\
9414.5$^{a}$ & 2607.9 & Fe \textsc{ii}$^{*}$ & abs & VIS \\
9431.8$^{a}$ & 2612.7 & Fe \textsc{ii}$^{*}$ & abs & VIS \\
9452.4$^{a}$ & 2618.4 & Fe \textsc{ii}$^{*}$ & abs & VIS \\
9491.1$^{a}$ & 2629.1 & Fe \textsc{ii}$^{*}$ & abs & VIS \\
9501.3$^{a}$ & 2631.9 & Fe \textsc{ii}$^{*}$ & abs & VIS \\
10094.8 & 2796.4 & Mg \textsc{ii} & abs & VIS \\
10120.7 & 2803.5 & Mg \textsc{ii} & abs & VIS \\
\hline \\
\end{tabular}
\end{table}

\begin{table}
\centering
\caption{Absorption and emission features identified in the afterglow spectrum of GRB~241228B ($z = 2.6745 \pm 0.0004$), obtained with VLT/X-shooter. Superscript symbols denote excited-state transitions ($^{*}$, $^{**}$, etc.), while $^{a}$ marks features affected by telluric absorption.}
\label{tab:241228B}
\begin{tabular}{lcccc}
\hline
$\lambda_{\textrm{obs}}$ (\AA) & $\lambda_{\textrm{rest}}$ (\AA) & Feature & Type & Arm \\
\hline
4467.0 & 1215.7 & Ly$\alpha$ & abs & UVB \\
4551.8 & 1238.8 & N \textsc{v} & abs & UVB \\
4566.5 & 1242.8 & N \textsc{v} & abs & UVB \\
4593.7 & 1250.2 & S \textsc{ii} & abs & UVB \\
4605.5 & 1253.4 & S \textsc{ii} & abs & UVB \\
4627.5 & 1259.4 & S \textsc{ii} & abs & UVB \\
4631.2 & 1260.4 & Si \textsc{ii} & abs & UVB \\
4647.6 & 1264.8 & Si \textsc{ii}$^{*}$ & abs & UVB \\
4784.8 & 1302.2 & O \textsc{i} & abs & UVB \\
4792.9 & 1304.4 & Si \textsc{ii} & abs & UVB \\
4794.7 & 1304.9 & O \textsc{i}$^{*}$ & abs & UVB \\
4798.8 & 1306.0 & O \textsc{i}$^{**}$ & abs & UVB \\
4810.9 & 1309.3 & Si \textsc{ii}$^{*}$ & abs & UVB \\
4839.9 & 1317.2 & Ni \textsc{ii} & abs & UVB \\
4903.1 & 1334.4 & C \textsc{ii} & abs & UVB \\
4907.9 & 1335.7 & C \textsc{ii}$^{*}$ & abs & UVB \\
5034.3 & 1370.1 & Ni \textsc{ii} & abs & UVB \\
5121.2 & 1393.8 & Si \textsc{iv} & abs & UVB \\
5154.3 & 1402.8 & Si \textsc{iv} & abs & UVB \\
5345.5 & 1454.8 & Ni \textsc{ii} & abs & UVB \\
5391.4 & 1467.3 & Ni \textsc{ii} & abs & UVB \\
5393.3 & 1467.8 & Ni \textsc{ii} & abs & UVB \\
\hline
5609.7 & 1526.7 & Si \textsc{ii} & abs & VIS \\
5634.3 & 1533.4 & Si \textsc{ii}$^{*}$ & abs & VIS \\
5688.7 & 1548.2 & C \textsc{iv} & abs & VIS \\
5698.3 & 1550.8 & C \textsc{iv} & abs & VIS \\
5728.4 & 1559.0 & Fe \textsc{ii}$^{*****}$ & abs & VIS \\
5757.0 & 1566.8 & Fe \textsc{ii}$^{*}$ & abs & VIS \\
5910.1 & 1608.5 & Fe \textsc{ii} & abs & VIS \\
5920.2 & 1611.2 & Fe \textsc{ii} & abs & VIS \\
5926.1 & 1612.8 & Fe \textsc{ii}$^{*}$ & abs & VIS \\
6016.5 & 1637.4 & Fe \textsc{ii}$^{*}$ & abs & VIS \\
6139.0 & 1670.8 & Al \textsc{ii} & abs & VIS \\
6253.8 & 1702.0 & Fe \textsc{ii}$^{*}$ & abs & VIS \\
6281.8 & 1709.6 & Ni \textsc{ii} & abs & VIS \\
6399.3 & 1741.6 & Ni \textsc{ii} & abs & VIS \\
6437.2 & 1751.9 & Ni \textsc{ii} & abs & VIS \\
6643.3 & 1808.0 & Si \textsc{ii} & abs & VIS \\
6676.0 & 1816.9 & Si \textsc{ii}$^{*}$ & abs & VIS \\
6678.2 & 1817.5 & Si \textsc{ii}$^{*}$ & abs & VIS \\
6716.4 & 1827.9 & Mg \textsc{i} & abs & VIS \\
6814.9 & 1854.7 & Al \textsc{iii} & abs & VIS \\
6844.5 & 1862.8 & Al \textsc{iii} & abs & VIS \\
7444.7 & 2026.1 & Zn \textsc{ii} & abs & VIS \\
7445.4 & 2026.3 & Cr \textsc{ii} & abs & VIS \\
7446.2 & 2026.5 & Mg \textsc{i} & abs & VIS \\
7555.7 & 2056.3 & Cr \textsc{ii} & abs & VIS \\
7577.3 & 2062.2 & Cr \textsc{ii} & abs & VIS \\
7579.0$^{a}$ & 2062.7 & Zn \textsc{ii} & abs & VIS \\
7592.0$^{a}$ & 2066.2 & Cr \textsc{ii} & abs & VIS \\
7959.5 & 2166.2 & Ni \textsc{ii}$^{*}$ & abs & VIS \\
8146.9 & 2217.2 & Ni \textsc{ii}$^{*}$ & abs & VIS \\
8170.4 & 2223.6 & Ni \textsc{ii}$^{*}$ & abs & VIS \\
8267.0 & 2249.9 & Fe \textsc{ii} & abs & VIS \\
8307.1 & 2260.8 & Fe \textsc{ii} & abs & VIS \\
8512.8 & 2316.8 & Ni \textsc{ii}$^{*}$ & abs & VIS \\
8554.4 & 2328.1 & Fe \textsc{ii}$^{*}$ & abs & VIS \\
8574.2 & 2333.5 & Fe \textsc{ii}$^{*}$ & abs & VIS \\
8593.3 & 2338.7 & Fe \textsc{ii}$^{*}$ & abs & VIS \\
8613.5 & 2344.2 & Fe \textsc{ii} & abs & VIS \\
\hline \\
\end{tabular}
\end{table}

\begin{table}
\centering
\caption{GRB~241228B (continued)}
\label{tab:241228B-2}
\begin{tabular}{lcccc}
\hline
$\lambda_{\textrm{obs}}$ (\AA) & $\lambda_{\textrm{rest}}$ (\AA) & Feature & Type & Arm \\
\hline
8616.5 & 2345.0 & Fe \textsc{ii}$^{*}$ & abs & VIS \\
8631.2 & 2349.0 & Fe \textsc{ii}$^{*}$ & abs & VIS \\
8670.8 & 2359.8 & Fe \textsc{ii}$^{*}$ & abs & VIS \\
8692.2 & 2365.6 & Fe \textsc{ii}$^{*}$ & abs & VIS \\
8724.9 & 2374.5 & Fe \textsc{ii} & abs & VIS \\
8750.6 & 2381.5 & Fe \textsc{ii}$^{*}$ & abs & VIS \\
8755.4 & 2382.8 & Fe \textsc{ii} & abs & VIS \\
8779.6 & 2389.4 & Fe \textsc{ii}$^{*}$ & abs & VIS \\
8805.0 & 2396.3 & Fe \textsc{ii}$^{*}$ & abs & VIS \\
8818.6 & 2400.0 & Fe \textsc{ii}$^{*}$ & abs & VIS \\
8845.8 & 2407.4 & Fe \textsc{ii}$^{*}$ & abs & VIS \\
8859.9 & 2411.2 & Fe \textsc{ii}$^{*}$ & abs & VIS \\
8861.9 & 2411.8 & Fe \textsc{ii}$^{*}$ & abs & VIS \\
9468.6$^{a}$ & 2576.9 & Mn \textsc{ii} & abs & VIS \\
9504.4$^{a}$ & 2586.7 & Fe \textsc{ii} & abs & VIS \\
9533.2$^{a}$ & 2594.5 & Mn \textsc{ii} & abs & VIS \\
9554.2$^{a}$ & 2600.2 & Fe \textsc{ii} & abs & VIS \\
9577.3$^{a}$ & 2606.5 & Mn \textsc{ii} & abs & VIS \\
9582.5$^{a}$ & 2607.9 & Fe \textsc{ii}$^{*}$ & abs & VIS \\
9600.1$^{a}$ & 2612.7 & Fe \textsc{ii}$^{*}$ & abs & VIS \\
9607.1$^{a}$ & 2614.6 & Fe \textsc{ii}$^{*}$ & abs & VIS \\
9621.0$^{a}$ & 2618.4 & Fe \textsc{ii}$^{*}$ & abs & VIS \\
9636.1$^{a}$ & 2622.5 & Fe \textsc{ii}$^{*}$ & abs & VIS \\
9650.8$^{a}$ & 2626.5 & Fe \textsc{ii}$^{*}$ & abs & VIS \\
9660.4$^{a}$ & 2629.1 & Fe \textsc{ii}$^{*}$ & abs & VIS \\
\hline
10274.9 & 2796.3 & Mg \textsc{ii} & abs & NIR \\
10301.2 & 2803.5 & Mg \textsc{ii} & abs & NIR \\
10483.1 & 2853.0 & Mg \textsc{i} & abs & NIR \\
14585.9 & 3969.6 & Ca \textsc{ii} & abs & NIR \\
15535.0 & 4227.9 & Ca \textsc{i} & abs & NIR \\
\hline \\
\end{tabular}
\end{table}

\newpage
\twocolumn
\section*{Affiliations}
$^{1}$Department of Physics, Royal Holloway - University of London, Egham, TW20 0EX, UK\\
$^{2}$Department of Physics, University of Warwick, Gibbet Hill Road, Coventry CV4 7AL, UK\\
$^{3}$School of Physics and Astronomy, University of Birmingham, Edgbaston, Birmingham, B15 2TT, UK\\
$^{4}$Institute for Gravitational Wave Astronomy, University of Birmingham, Birmingham, B15 2TT, UK\\
$^{5}$Aix Marseille Univ., CNRS, CNES, LAM, Marseille, F-13388, France\\
$^{6}$School of Physics \& Astronomy, Monash University, Clayton VIC 3800, Australia\\
$^{7}$Université Paris-Saclay, Université Paris Cité, CEA, CNRS, AIM, 91191, Gif-sur-Yvette, France\\
$^{8}$School of Physics and Astronomy, University of Leicester, University Road,  Leicester, LE1 7RH, UK\\
$^{9}$Sydney Institute for Astronomy, School of Physics, The University of Sydney, NSW 2006, Australia\\
$^{10}$Centre of Excellence for Gravitational Wave Discovery (OzGrav), Hawthorn, VIC 3122, Australia\\
$^{11}$CSIRO Space and Astronomy, PO Box 76, Epping, NSW 1710, Australia\\
$^{12}$The Cosmic Dawn Centre (DAWN), R\r{a}dmandsgade 64, DK-2200, K{\o}benhavn N., Denmark\\
$^{13}$Niels Bohr Institute, University of Copenhagen, Jagtvej 155, DK-2200 Copenhagen N, Denmark \\      
$^{14}$Department of Astrophysics/IMAPP, Radboud University, NL-6525 AJ Nijmegen, the Netherlands\\
$^{15}$Astrophysics Research Cluster, School of Mathematical and Physical Sciences, University of Sheffield, Sheffield, S3 7RH, UK\\
$^{16}$Research Software Engineering, University of Sheffield, Sheffield, S1 4DP, UK\\
$^{17}$INAF Osservatorio Astronomico di Brera, Via E. Bianchi 46, I-23807 Merate, Italy\\
$^{18}$Department of Physical Science, Aoyama Gakuin University, 5-10-1 Fuchinobe, Chuo-ku, Sagamihara, Kanagawa 252-5258, Japan\\
$^{19}$South-Western Institute for Astronomy Research (SWIFAR), Yunnan University, Kunming, Yunnan 650500, People's Republic of China\\
$^{20}$Yunnan Key Laboratory of Survey Science, Yunnan University, Kunming, Yunnan 650500, People's Republic of China\\
$^{21}$National Astronomical Observatories, Chinese Academy of Sciences, Beijing 100101, China\\
$^{22}$Guangxi Key Laboratory for Relativistic Astrophysics, School of Physical Science and Technology, Guangxi University, Nanning 530004, China\\
$^{23}$Armagh Observatory and Planetarium, College Hill, Armagh, BT61 9DG, Northern Ireland, UK\\
$^{24}$Institute for Globally Distributed Open Research and Education (IGDORE)\\
$^{25}$Instituto de Astrofísica de Canarias, E-38205 La Laguna, Tenerife, Spain\\
$^{26}$National Astronomical Research Institute of Thailand, 260 Moo 4, T. Donkaew, A. Maerim, Chiangmai, 50180 Thailand\\
$^{27}$Department of Physics \& Astronomy, University of Turku, Vesilinnantie 5, Turku, FI-20014, Finland\\
$^{28}$Jodrell Bank Centre for Astrophysics, Department of Physics and Astronomy, The University of Manchester, Manchester, M13 9PL, UK\\
$^{29}$Institute of Cosmology and Gravitation, University of Portsmouth, Portsmouth, PO1 3FX, UK\\
$^{30}$Departamento de Astrof\'isica, Universidad de La Laguna, E-38206 La Laguna, Tenerife, Spain\\
$^{31}$School of Physics, University College Cork, Cork, T12 K8AF, Ireland\\
$^{32}$Centre for Astrophysics Research, University of Hertfordshire, College Lane, Hatfield AL10 9AB, UK\\
$^{33}$School of Physics and Centre for Space Research, University College Dublin, Belfield, Dublin 4, Ireland\\
$^{34}$National Research University `Higher School of Economics', Faculty of Physics, Myasnitskaya ul. 20, Moscow 101000, Russia\\
$^{35}$Space Research Institute of the Russian Academy of Sciences, Profsoyuznaya ul. 84/32, Moscow 117997, Russia\\
$^{36}$Centro Astron\'omico Hispano en Andaluc\'ia, Observatorio de Calar Alto, Sierra de los Filabres, G\'ergal, Almer\'ia, E-04550, Spain\\
$^{37}$Departament d'Astronom\'{\i}a i Astrof\'{\i}sica, Universitat de València, 46100 Burjassot, Spain\\
$^{38}$Observatori Astronòmic, Universitat de València, 46980 Paterna, Spain\\
$^{39}$International Centre for Radio Astronomy Research, Curtin University, GPO Box U1987, Perth, WA 6845, Australia\\ 
$^{40}$Astrophysics Research Institute, Liverpool John Moores University, 146 Brownlow Hill, Liverpool L3 5RF, UK\\
$^{41}$Instituto de Astrof\'isica de Andaluc\'ia (IAA-CSIC), Glorieta de la Astronom\'ia s/n, E-18008, Granada, Spain\\
$^{42}$Ingenier\'ia de Sistemas y Autom\'atica, Universidad de M\'alaga, Unidad Asociada al CSIC por el IAA, Escuela de Ingenier\'ias Industriales, Arquitecto Francisco Pe\~nalosa, 6, Campanillas, 29071 M\'alaga, Spain\\
$^{43}$Millennium Institute of Astrophysics (MAS), Nuncio Monsenor Sòtero Sanz 100, Providencia, Santiago RM, 8320000, Chile\\
$^{44}$MIFT Department, University of Messina, Via F.S. D’Alcontres 31, I-98166 Messina, Italy\\
$^{45}$Space Science Data Center (SSDC) – Agenzia Spaziale Italiana (ASI), Via del Politecnico snc, I-00133 Roma, Italy\\
$^{46}$Department of Astronomy, School of Physics, Huazhong University of Science and Technology, Wuhan 430074, China\\
$^{47}$Clemson University, Department of Physics and Astronomy, Clemson, SC 29634-0978, USA\\
$^{48}$Osservatorio Astronomico di Capodimonte, INAF, Salita Moiariello 16, I-80131 Napoli, Italy\\
$^{49}$Institute of Solar-Terrestrial Physics, Russian Academy of Sciences, Siberian Branch, Irkutsk 664033, Russia\\
$^{50}$GRANTECAN S.A., Cuesta de San Jos\'e s/n, E-38712 Bre\~na Baja, La Palma, Spain\\
$^{51}$INAF – Osservatorio di Astrofisica e Scienza dello Spazio, via Piero Gobetti 93/3, I-40129 Bologna, Italy\\
$^{52}$GEPI, Observatoire de Paris, Universite´ PSL, CNRS, 5 Place Jules Janssen, F-92190 Meudon, France\\
$^{53}$Institute of Physics and Technology, Institutskiy Pereulok, 9, Dolgoprudny 141701, Russia\\
$^{54}$Astronomical Institute Anton Pannekoek, University of Amsterdam, PO Box 94249, NL-1090 GE Amsterdam, the Netherlands\\
$^{55}$Materials Science and Applied Mathematics, Malmö University, SE-205 06 Malmö, Sweden\\
$^{56}$Nordic Optical Telescope, Rambla José Ana Fernández Pérez 7, ES-38711 Breña Baja, Spain\\
$^{57}$E. Kharadze Georgian National Astrophysical Observatory, Mt. Kanobili, Abastumani 0301, Adigeni, Georgia\\
$^{58}$Department of Physics, George Washington University, 725 21st St NW, Washington, DC 20052, USA\\
$^{59}$LUX, Observatoire de Paris, Universit\'e PSL, CNRS, Sorbonne Universit\'e, Meudon, 92190, France\\

\bsp
\label{lastpage}
\end{document}